\documentclass[a4paper,11pt]{article}
\pdfoutput=1

\usepackage{jheppub}

\usepackage{subfig}
\usepackage{xspace}
\usepackage[countmax]{subfloat}
\usepackage{slashed}

\usepackage{verbatim}

\allowdisplaybreaks[2]

\usepackage{empheq}

\usepackage{bbm}

\newcommand{\bn}{{\bar n}}

\newcommand{\sceti}{SCET$_{\rm I}$}

\newcommand{\nn}{\nonumber}

\newcommand{\fuse}{\text{Fuse}}

\newcommand{\cA}{{\cal I}}

\newcommand{\idop}{1\!\!1}

\DeclareRobustCommand{\Sec}[1]{Sec.~\ref{#1}}
\DeclareRobustCommand{\Secs}[2]{Secs.~\ref{#1} and \ref{#2}}
\DeclareRobustCommand{\App}[1]{App.~\ref{#1}}
\DeclareRobustCommand{\Tab}[1]{Table~\ref{#1}}

\DeclareRobustCommand{\Fig}[1]{Fig.~\ref{#1}}
\DeclareRobustCommand{\Figs}[2]{Figs.~\ref{#1} and \ref{#2}}
\DeclareRobustCommand{\Eq}[1]{Eq.~(\ref{#1})}
\DeclareRobustCommand{\Eqs}[2]{Eqs.~(\ref{#1}) and (\ref{#2})}
\DeclareRobustCommand{\Ref}[1]{Ref.~\cite{#1}}
\DeclareRobustCommand{\Refs}[1]{Refs.~\cite{#1}}

\newcommand{\Splitbar}{\overline{\text{Split}}}

\newcommand{\scetscale}{\lambda}

\newcommand{\sab}[2]{\langle #1  #2  \rangle}
\newcommand{\ssb}[2]{[ #1  #2  ]}

\preprint{MIT--CTP 4568}

\title{Soft Theorems from Effective Field Theory}

\author{Andrew J. Larkoski,}
\author{Duff Neill,}
\author{and Iain W. Stewart}

\affiliation{Center for Theoretical Physics, Massachusetts Institute of Technology, Cambridge, MA 02139, USA}

\emailAdd{larkoski@mit.edu}
\emailAdd{dneill@mit.edu}
\emailAdd{iains@mit.edu}

\abstract{
The singular limits of massless gauge theory amplitudes are described by an effective theory, called soft-collinear effective theory (SCET), which has been applied most successfully to make all-orders predictions for observables in collider physics and weak decays. At tree-level, the emission of a soft gauge boson at subleading order in its energy is given by the Low-Burnett-Kroll theorem, with the angular momentum operator acting on a lower-point amplitude.  For well separated particles at tree-level, we prove the Low-Burnett-Kroll theorem using matrix elements of subleading SCET Lagrangian and operator insertions which are individually gauge invariant.  These contributions are uniquely determined by gauge invariance and the reparametrization invariance (RPI) symmetry of SCET.  RPI in SCET is connected to the infinite-dimensional asymptotic symmetries of the S-matrix.  The Low-Burnett-Kroll theorem is generically spoiled by on-shell corrections, including collinear loops and collinear emissions.  We demonstrate this explicitly both at tree-level and at one-loop. The effective theory correctly describes these configurations, and we generalize the Low-Burnett-Kroll theorem into  a new {\em one-loop subleading soft theorem for amplitudes}.  Our analysis is presented in a manner that illustrates the wider utility of using effective theory techniques to understand the perturbative S-matrix.
}

\begin{document} 
\maketitle

\section{Introduction}
\label{sec:intro}

The modern study of the perturbative S-matrix is now a mature field which traces its roots to the Parke-Taylor formula for maximally-helicity-violating (MHV) amplitudes \cite{Parke:1986gb} in the 1980s, the unitarity methods of \Refs{Bern:1994zx,Bern:1994cg} in the early 1990s and the identification of perturbative gauge theory as a string theory in twistor space \cite{Witten:2003nn} in the early 2000s. This program has produced a wealth of results, including several on-shell methods for computing amplitudes both at tree-level and higher orders in perturbation theory (eg.~\cite{Cachazo:2004kj,Britto:2004ap,Britto:2005fq,ArkaniHamed:2010kv,Boels:2010nw}), an emergent infinite dimensional Yangian symmetry of ${\cal N}=4$ supersymmetric Yang-Mills (SYM) amplitudes (eg.~\cite{Drummond:2008vq,Beisert:2008iq,Brandhuber:2008pf,Drummond:2009fd}), the relationship between strong and weak coupling expansions of ${\cal N}=4$ SYM (eg.~\cite{Alday:2007hr,Drummond:2007aua,Brandhuber:2007yx,Drummond:2007cf,Berkovits:2008ic}), and significant surprises in the form and structure of amplitudes in gravity (eg.~\cite{Cachazo:2005ca,Bern:2006kd,Benincasa:2007qj,Bern:2008qj}).  The progress that has been made is encouraging for reaching the ultimate goal of a complete understanding of the perturbative S-matrix (a goal which is closer for highly supersymmetric theories).

An important part of this goal is an understanding of the kinematic limits of the S-matrix. As recognized long ago \cite{Low:1958sn, Weinberg:1965nx, Burnett:1967km,Gross:1968in,Jackiw:1968zza}, universal structures appear in perturbative gauge theory and gravity when the energy of external particles are taken soft. In gauge theories similar universal structures arise when external particles become collinear, leading to the definition of important quantities like the parton distribution function. Finally the universality of these soft and collinear limits has a profound interpretation in terms of factorization of the underlying space of states into physically realizable subprocesses \cite{Grammer:1973db,Ellis:1978sf,Sterman:1978bi,Sterman:1978bj,Collins:1981ta,Collins:1988ig,Amati:1978wx,Amati:1978by,Collins:1985ue}.

Weinberg~\cite{Weinberg:1965nx} in particular stressed that the universality of the soft limits comes from very general symmetry constraints, such as charge conservation and Lorentz invariance of the S-matrix, and pointed out the deep physical consequence that this forbids particles with spin greater than 2 from mediating long-range forces.  Schematically, Weinberg's version of the soft theorems for gravity and gauge theory amplitudes take the form
\begin{equation}
{\cal A}(1,\dotsc,N,s) \to S^{(0)}(s) {\cal A}(1,\dotsc,N) \ ,
\end{equation}
where ${\cal A}(1,\dotsc,N,s)$ is an $N+1$-point amplitude with the energy of particle $s$ taken to 0 and $S^{(0)}(s)$ is the leading term in the energy expansion.  Importantly, $S^{(0)}(s)$ is fully independent of the internal structure of the amplitude ${\cal A}(1,\dotsc,N)$.  More recently, studies in this direction led to defining an ``inverse soft'' construction of amplitudes \cite{ArkaniHamed:2009dn,Rajabi:2011eg,BoucherVeronneau:2011nm,Dunbar:2012aj,Nandan:2012rk} by which external particles are added to an amplitude systematically by undoing the universal soft limit.  

At tree level, it is known that this universality also extends to subleading terms in the expansion of amplitudes in gauge theory and gravity~\cite{Gross:1968in,Jackiw:1968zza,Laenen:2008gt,Laenen:2010uz,White:2011yy}. That is, in the soft limit, the amplitude takes the schematic form
\begin{equation}
{\cal A}(1,\dotsc,N,s) \to \left(S^{(0)}(s)+S^{(\text{sub})}(s)\right) {\cal A}(1,\dotsc,N) \ ,
\end{equation}
where now $S^{(\text{sub})}(s)$ is suppressed with respect to the leading soft factor, $S^{(0)}(s)$, by the energy of the soft particle over a hard scattering scale.  In contrast to the leading soft factor $S^{(0)}(s)$, the subleading soft factor $S^{(\text{sub})}(s)$ is a derivative operator that acts non-trivially on the lower-point amplitude ${\cal A}(1,\dotsc,N)$. Very recently, \Refs{Cachazo:2014fwa,Casali:2014xpa} identified an extension for these soft theorems for gravity and gauge tree amplitudes, corresponding to the next term in the soft limit expansion of the amplitude. Several papers \cite{Schwab:2014xua,Bern:2014oka,He:2014bga,Larkoski:2014hta,Cachazo:2014dia,Afkhami-Jeddi:2014fia,Adamo:2014yya,Geyer:2014lca,Schwab:2014fia,Bianchi:2014gla,Broedel:2014fsa,Bern:2014vva,White:2014qia,Zlotnikov:2014sva,Kalousios:2014uva,Du:2014eca,Liu:2014vva,Rao:2014zaa,Bonocore:2014wua,Luo:2014wea,Broedel:2014bza,Schwab:2014sla} have studied the consequences of the subleading soft theorems and have provided both insights and identifed puzzles of the soft limits of amplitudes.

In this paper, we will focus on the subleading soft amplitudes in gauge theory at both tree-level and one-loop order. At tree-level the leading soft factor $S^{(0)}(s)$ can be written as 
\begin{equation}
S^{(0)}(s)=\sum_{i=1}^N T_i\frac{\epsilon_s\cdot p_i}{p_i\cdot p_s} \ ,
\end{equation}
where the sum runs over all external particles in the amplitude, $p_{s\mu}$ is the momentum of the soft particle $s$, $\epsilon_{s\mu}$ is its polarization vector, and $T_i$ is the charge or appropriate color matrix of particle $i$.  The gauge invariance of $S^{(0)}(s)$ follows from the global conservation of charge: $\sum_i T_i = 0$.
The subleading soft factor $S^{\text{(sub)}}(s)$ can similarly be expressed as
\begin{equation}\label{eq:subsoft_intro}
S^{\text{(sub)}}(s)=\sum_{i=1}^N T_i\frac{\epsilon_{s\mu}p_{s\nu}J^{\mu\nu}_i}{p_i\cdot p_s} \ ,
\end{equation}
where $J^{\mu\nu}_i$ is the angular momentum operator associated with particle $i$ acting on the parent amplitude.  For gauge theory we will refer to this as the Low-Burnett-Kroll (LBK) theorem~\cite{Low:1958sn,Burnett:1967km}. A proof of the Low-Burnett-Kroll theorem at tree level for arbitrary numbers of external particles that only relies on gauge invariance and Lorentz invariance was given in \Ref{Bern:2014vva}.  Unlike the leading soft factor, $S^{\text{(sub)}}(s)$ is gauge invariant because for each $i$ the angular momentum factor $J^{\mu\nu}_i$ is anti-symmetric, and hence does not rely on a global symmetry.

In thinking about the soft limits of a quantum field theory, we should expect that the physics of the soft limit is described by a low-energy effective theory of the full theory.  In gauge theory, and in particular in QCD, this effective theory is Soft-Collinear Effective Theory (SCET) \cite{Bauer:2000ew,Bauer:2000yr,Bauer:2001ct,Bauer:2001yt} which was established in the early 2000's, around the same time as the modern amplitudes program.  SCET has been widely applied to flavor and collider physics including 
$B$-hadron decays (eg.~\cite{Bauer:2001cu,Chay:2002vy,Beneke:2002ph,Pirjol:2002km,Lunghi:2002ju,Mantry:2003uz,Beneke:2003pa,Lange:2003pk,Bauer:2004tj,Bosch:2004th,Neubert:2004dd,Ligeti:2008ac}),
Charmonia~(eg.~\cite{Fleming:2002rv,Fleming:2003gt,GarciaiTormo:2005ch}),
predictions of all-orders distributions of collider event observables (eg.~\cite{Bauer:2002nz,Manohar:2003vb,Idilbi:2005ni,Becher:2006mr,Schwartz:2007ib,Fleming:2007qr,Bauer:2008dt,Stewart:2009yx,Becher:2009th,Mantry:2009qz,Ellis:2010rwa,Becher:2010tm,Beneke:2010da,Berger:2010xi,GarciaEchevarria:2011rb,Chiu:2012ir,Larkoski:2014uqa}),
precise extractions of the strong coupling $\alpha_s$ (eg.~\cite{Becher:2008cf,Abbate:2010xh}), 
the structure of amplitudes 
(eg.~\cite{Becher:2009qa,Chiu:2009mg,Feige:2014wja}), 
Higgs physics 
(eg.~\cite{Berger:2010xi,Becher:2012qa,Stewart:2013faa,Moult:2014pja}), 
and for Heavy Ion Collisions 
(eg.~\cite{Idilbi:2008vm,DEramo:2010ak,Ovanesyan:2011xy}). 
	  As it is an effective theory of QCD (or of a gauge theory in general), SCET reproduces the physics in the infrared and collinear regions of phase space of the full theory.  Thus, we can utilize the significant power of SCET to study the issues of the soft theorems in gauge theory.  
SCET has also been formalized for gravity \cite{Beneke:2012xa}, and in that case is somewhat simpler than for gauge theory because there are no singular collinear limits in gravity.  In this paper, we will focus on the SCET of gauge theory and only briefly mention results in gravity.

The power of SCET is that it systematically organizes the factorization of the S-matrix into components that describe the different relevant physics that contribute to a process. The factorized amplitude ${\mathcal A}$ in SCET takes the form
\begin{equation}\label{eq:scetfact}
{\mathcal A} = \sum_{j} \Big[ C_H^{(j_c)} \otimes \prod_i\cA_i^{(j_I^i)}  \otimes S^{(j_s)} \Big]  
  = C_H^{(0)} \otimes \prod_i\cA_i^{(0)}  \otimes S^{(0)} + \ldots
  \,.
\end{equation}
Here the first term with $j_c=j_I^i=j_s=0$ is the leading order amplitude, and the sum over $j$ includes amplitudes that are power suppressed to order $j$ (where at each order $j_c+\sum_i j_I^i + j_s =j$). The $C_H$ are hard coefficients that describe the hard scattering event.  The $\cA_i$ are collinear amplitudes that describe states that include collinear emissions in the directions of each of the original particles in the hard scattering event.  $S$ are soft amplitude factors which describe the global soft radiation emitted from the particular configuration of hard, external particles in the process. Each $C_H^{(j_c)}$, $\cA_i^{(j_I^i)}$, and $S^{(j_s)}$ has a gauge coupling constant expansion. The symbol $\otimes$ denotes momentum space convolutions and global index contractions as the different functions will have some response to one another. Importantly, SCET tracks the correlated scales of all sectors by assigning a consistent power counting to the momenta of the hard, collinear and soft modes.  To study the soft factors in SCET requires a study of the amplitude at subleading power $j_s$.  The required formalism for studying these power corrections was worked out in the early SCET literature~\cite{Beneke:2002ni,Chay:2002vy,Manohar:2002fd,Pirjol:2002km,Beneke:2002ph,Bauer:2003mga}.

For an all-orders understanding of the soft limits, we need a consistent treatment of the soft and collinear singularities of a gauge theory.    Using SCET, we are able to characterize the structure of the subleading soft factor in gauge theory at tree-level, one-loop, and beyond.  The main results of this paper are as follows:
\begin{itemize}

\item {\bf Failure of the Low-Burnett-Kroll Theorem in QFT}\\
 In a quantum field theory with massless particles, the Low-Burnett-Kroll theorem is generically false, which has been shown by explicit counter-examples \cite{Bern:2014oka,He:2014bga}. The reason it is false is due to particles in the amplitude that become collinear resulting in a propagator going on-shell.\footnote{The importance of this region of phase space was noted as early as \Ref{DelDuca:1990gz}, which established an extension of the subleading soft theorem; see \Sec{sec:order_lambda_squ_one_loop} for a more detailed discussion.}  At tree-level, the collinear region of phase space can be avoided by judiciously choosing the external particles to be widely separated in angle.  At loop-level, however, we must integrate over all momentum regions of the particles in the loop, including those regions where the particle in the loop becomes collinear with an external particle.  There is no physical way to exclude this region of the loop integral.  In contrast, the structure of the subleading SCET Lagrangians and operators (valid to all orders in perturbation theory) imply that subleading soft effects can still be factored out of the hard interaction, and the collinear dynamics.

\item {\bf Enhanced Symmetries of the Effective Theory}\\  As an effective theory, SCET has enhanced symmetries with respect to the full theory.  These symmetries are manifested via the reorganization of the full-theory S-matrix in terms of factorized operators with soft and collinear fields. As an illustration of this, we discuss the reparametrization invariance (RPI) \cite{Manohar:2002fd} of SCET and show that it is a manifestation of the conjectured infinite-dimensional asymptotic symmetry of gauge theories and gravity~\cite{Strominger:2013jfa,He:2014laa,Strominger:2013lka,Kapec:2014opa}.\footnote{The RPI of gravity will be richer than gauge theory because Poincar\'e symmetry is gauged.  RPI in SCET for gravity should be related to those diffeomorphisms that are broken by the dominant directions of the external energetic particles.  Acting on the spacetime boundary, we expect these diffeomorphisms to be generated by a Virasoro algebra in four spacetime dimensions~\cite{Kapec:2014opa}. This connection deserves further study.}  At tree level we show that the subleading LBK soft factor $S^{\text{(sub)}}(s)$ in gauge theory is reproduced by matrix elements involving the subleading gauge invariant SCET Lagrangian and operators.  For massless particles the form of these Lagrangians and operators are fully constrained by the RPI and gauge symmetries of the effective theory, and these symmetries play an important role in deriving LBK. 

\item {\bf New Loop-Level Soft Theorems from the Effective Theory}\\ By its construction, SCET contains all of the physics of the infrared (IR) of the full gauge theory.   For a generic 1-loop amplitude we formulate a new subleading soft theorem for the emission of a soft gauge boson from $N$ well separated hard particles. This result involves an LBK contribution acting on a hard 1-loop amplitude, contributions from soft loops dressing the tree-level $N$-point amplitude ${\cal A}_N$, contributions where ${\cal A}_N$ is contracted with one-loop splitting amplitudes, and contributions involving the loop level fusion of collinear particles. We also formulate a subleading soft theorem for tree-level amplitudes which have two external collinear particles.

\end{itemize}

The outline of this paper is as follows.  Because we are bridging the fields of amplitudes and effective theory methods, we will provide a review of each in \Sec{sec:subsoft} attempting to be self-contained, focusing on the application of the subleading soft theorems.  In particular, we review the subleading soft theorems, modern amplitude techniques, and SCET.  We provide examples of explicit tree-level amplitudes, expand them in the soft limit, and check that the resulting subleading soft factors agree with LBK.  In our SCET review, we define the modes and operators of the theory and do some simple calculations, such as proving the leading power soft theorem in the presence of arbitrary loop corrections. We also review the subleading power SCET Lagrangian and the reparameterization invariance symmetry of SCET. Finally, we construct the subleading soft SCET hard-scattering operators that are relevant for $N$ point amplitudes with massless particles.
  
In \Sec{sec:subsoft_zero}, we study the subleading soft factor at tree-level and decompose the angular momentum operator into components with definite power counting in the effective theory.
We explicitly compute the subleading soft factor at the first few orders in the power counting and show that the SCET results yield the subleading soft factor given by LBK.  Finally we show that the reparametrization invariance of SCET is related to an effectively infinite-dimensional asymptotic symmetry of gauge theory.

In \Sec{sec:amps_nonzero}, we derive our main new results, including 
a subleading soft theorem that is valid at one-loop order. The corrections encoded in this soft theorem come from the region of the loop integral in which the loop momenta is collinear to external particles, and this situation violates the assumptions required in deriving the simpler Low-Burnett-Kroll result. The one-loop finite amplitude ${\mathcal A}(-,+,\cdots,+)$ provides a counterexample to the general validity of the Low-Burnett-Kroll theorem beyond tree-level~\cite{Bern:2014oka,He:2014bga}. We show that these amplitudes instead obey our generalized subleading soft theorem at the one-loop level.   
We also show by explicit calculation that the Low-Burnett-Kroll theorem is violated at tree-level if two of the external legs in the amplitude become collinear at a rate comparable to the rate that the soft momentum becomes soft.  For this situation, we derive a soft theorem for real emission graphs containing two collinear particles that are not well-separated in phase space.  The result includes both a direct emission contribution and an amplitude coupling to the soft limit of the $1\to3$ splitting amplitude.  This correlated soft-collinear scaling limit plays an important role in many physical cross sections, such as those for thrust in $e^+e^-\to $ jets, for jet mass predictions in $pp\to $ jets, or for small $p_T$ cross section in Higgs production. 

Finally, we conclude in \Sec{sec:conc} and comment on the potentially fruitful relationship between amplitudes and effective field theory methods. Further details and various calculations are included in appendices.

\section{Subleading Soft Theorems and Subleading SCET}
\label{sec:subsoft}

\subsection{Spinor Notation}  \label{sec:spinors}

In this section, we review the subleading soft theorems and the relevant modern amplitude techniques. Standard reviews of spinor helicity methods include~\Refs{Mangano:1990by,Dixon:1996wi}.  In four dimensions, the Lorentz group is locally isomorphic to SU(2)$\times$SU(2) and so a lightlike vector $p^\mu$
can be expressed as an outer product of two spinors $\lambda^a$ and $\tilde{\lambda}^{\dot{a}}$:
\begin{equation}
p^\mu = \sigma_{a\dot{a}}^\mu \lambda^a \tilde{\lambda}^{\dot{a}}  \ ,
\end{equation}
where $\sigma^0=\idop$ and $\sigma_{a\dot{a}}^\mu$ for $\mu=1,2,3$ are the Pauli spin matrices.  Depending on the signature of spacetime, the spinors $\lambda^a$ and $\tilde{\lambda}^{\dot{a}}$ are related to one another differently.  For example, in $3+1$ signature they are complex conjugates while in $2+2$ signature they are independent real spinors.  We will use the $3+1$ signature language and refer to $\lambda^a$ as the holomorphic spinor and $\tilde{\lambda}^{\dot{a}}$ as the antiholomorphic spinor.  Under the little group, $\lambda^a$ transforms as a $-$ helicity spinor and $\tilde{\lambda}^{\dot{a}}$ transforms as a $+$ helicity spinor.

The power of introducing the spinor notation is that helicity amplitudes in four dimensions are naturally expressed as functions of the spinors of the external particles.  The covariant inner product of the spinors for particles $i$ and $j$ is expressed as
\begin{equation}
\sab{i}{j}\equiv \epsilon_{ab}\lambda^a_i\lambda^b_j \ , \qquad
\ssb{i}{j}\equiv \epsilon_{\dot{a}\dot{b}}\tilde{\lambda}^{\dot{a}}_i\tilde{\lambda}^{\dot{b}}_j \ .
\end{equation}
Then, the dot product of two lightlike momenta $p_i$ and $p_j$ is
\begin{equation}
(p_i+p_j)^2 = 2p_i\cdot p_j = \sab{i}{j}\ssb{j}{i} \ .
\end{equation}
Thus, a helicity amplitude will be a function of the covariant spinor products $\sab{i}{j}$ and $\ssb{i}{j}$ for all particles $i$ and $j$ in the amplitude.  A little group transformation of particle $i$ can be expressed as a scaling of the helicity spinors.
\begin{equation}
\lambda_i\to t^{-1} \lambda_i \ , \qquad \tilde{\lambda}_i\to t\tilde{\lambda}_i \ .
\end{equation}
The amplitude\footnote{All of our amplitudes are matrix elements that are truncated by LSZ and stripped of the momentum conserving $\delta$-function.} must transform covariantly according to the helicity of particle $i$ under the little group as
\begin{equation}\label{eq:littlegroup}
{\cal A}(1,\dotsc,t\cdot i,\dotsc,N)=t^{2 h_i} {\cal A}(1,\dotsc,i,\dotsc,N) \ ,
\end{equation}
where $t\cdot i$ denotes the little group action on $i$ and $h_i$ is the helicity of particle $i$.  \Eq{eq:littlegroup} is a non-trivial constraint on the amplitude.

In a non-abelian gauge theory we can further simplify the amplitudes by exploiting color ordering.  For example, a pure gluon amplitude in a gauge theory can be decomposed at tree level into individual color orderings as
\begin{equation}
{\cal A}(1,\dotsc,N)=\sum_{\sigma\in S_N/\mathbb{Z}_N} \text{Tr}\left[
T_{\sigma(1)}\cdots T_{\sigma(N)} \right] {\cal A}(\sigma(1),\dotsc,\sigma(N )) \ ,
\end{equation}
where $\sigma$ is an element of the symmetric group $S_N$ modulo cyclic permutations $\mathbb{Z}_N$ and $T_i$ is the color matrix of gluon $i$.  For an amplitude that contains particles carrying color in a representation other than the adjoint, the trace will be replaced by the appropriate color index contractions.  Further, for simplicity we will often strip the overall numerical prefactor and factors of the coupling from the amplitude and consider the amplitude purely as a function of the kinematics of the scattering process.

\subsection{Soft factors of gauge theory and gravity amplitudes}
\label{subsec:subsoft}

As mentioned in the introduction, in the limit that the energy of particle $s$ becomes small, a tree-level amplitude in gauge theory or gravity should take the following form:
\begin{equation}
{\cal A}(1,\dotsc,N,s) \to \left(S^{(0)}(s)+S^{(\text{sub})}(s)\right) {\cal A}(1,\dotsc,N) \ ,
\end{equation}
where $S^{(0)}(s)$ is the leading soft factor and $S^{(\text{sub})}(s)$ is the subleading soft factor.  Higher order terms in the expansion have been dropped.  For a soft emission in gauge theory, $S^{(0)}(s)$ is
\begin{equation}\label{eq:weingauge}
S_{\text{gauge}}^{(0)}(s) = \sum_{i=1}^N T_i \frac{\epsilon_s\cdot p_{i}}{p_i\cdot p_s} \ ,
\end{equation}
where $\epsilon_s^\mu$ is the polarization vector of particle $s$ and $T_i$ is the charge, or appropriate color matrix, of particle $i$.  This soft factor is gauge invariant if charge/color is conserved:
\begin{equation}
\sum_{i=1}^N T_i = 0 \ .
\end{equation}
In gravity, the soft factor $S^{(0)}(s)$ is
\begin{equation}\label{eq:leadsoftgrav}
S_{\text{grav}}^{(0)}(s) = \sum_{i=1}^N Q_i \frac{\epsilon_s^{\mu\nu} p_{i\mu}p_{i\nu}}{p_i\cdot p_s} \ ,
\end{equation}
where $Q_i$ is the coupling of particle $i$ to the graviton and $\epsilon^{\mu\nu}_s$ is the soft graviton's polarization tensor.  This soft factor is gauge invariant if the graviton couples universally, $Q_i\equiv Q$, and if momentum is conserved:
\begin{equation}
\sum_{i=1}^N p_i^\mu = 0 \ .
\end{equation}
A celebrated consequence of Weinberg's soft theorems~\cite{Weinberg:1965nx} is that the soft factor of bosons with spin 3 or higher are only gauge invariant for non-generic kinematic configurations.  This implies that only spin 1 and spin 2 particles can mediate long range forces.

The subleading soft factor $S^{\text{(sub)}}(s)$ is known in gauge theory as the Low-Burnett-Kroll theorem \cite{Low:1958sn,Burnett:1967km}.  The Low-Burnett-Kroll theorem can be expressed in terms of the angular momentum operator as
\begin{equation}\label{eq:lbk}
S_\text{gauge}^{(\text{sub})}(s)\equiv  S^{\text{(sub)}}(s)= \sum_{i=1}^N T_i \frac{\epsilon_{s\mu} p_{s\nu}J_i^{\mu\nu}}{p_i\cdot p_s} \ ,
\end{equation}
where $J_i^{\mu\nu}$ is the angular momentum of particle $i$.  This is gauge invariant because $J_i^{\mu\nu}$ is an antisymmetric tensor.  Note therefore that unlike the leading soft factor in gauge theory, the gauge invariance of the subleading soft factor does not constrain particles' interactions.  The subleading soft factor for gravity takes a similar form \cite{Gross:1968in,Jackiw:1968zza,White:2011yy}:
\begin{equation}
S_\text{grav}^{(\text{sub})}(s)=   \sum_{i=1}^N Q \frac{\epsilon_{s\mu\nu} p_i^\mu p_{s\rho}J_i^{\nu\rho}}{p_i\cdot p_s} \ ,
\end{equation}
but here gauge invariance does follow from global angular momentum conservation:
\begin{equation}\label{eq:subsoftgravgaugeinv}
\sum_{i=1}^N J_i^{\mu\nu} = 0 \ .
\end{equation}

As shown in \Ref{Cachazo:2014fwa}, an efficient method for identifying the soft limit of tree-level helicity amplitudes is to scale the spinors of the soft particle appropriately and then expand in the scaling parameter. For the soft particle $s$ we can choose the scaling\footnote{\Ref{Cachazo:2014fwa} and papers since then employed a holomorphic scaling for positive helicity particles where only $\lambda_s$ was scaled while $\tilde{\lambda}_s$ remained unchanged.  These two scalings are related by a little group transformation and so result in identical physical content.  
}
\begin{equation}\label{eq:subscale}
\lambda_s\to \epsilon^{1/2} \lambda_s \ ,\qquad \tilde{\lambda}_s \to \epsilon^{1/2}\tilde{\lambda}_s \ .
\end{equation}
We choose to use this homogeneous scaling in order to more easily connect with the effective theory analysis that also has a homogeneous scaling. If the soft particle $s$ has $+$ helicity, then
gravity amplitudes have the expansion
\begin{equation}
{\cal A}(1,\dotsc,N,\{s^+,\epsilon\}) = \left(
\frac{1}{\epsilon}S_{\text{grav}}^{(0)}(s)+S_{\text{grav}}^{(\text{sub})}(s)
\right){\cal A}(1,\dotsc,N) + {\cal O}(\epsilon^{1}) \ .
\end{equation}
Expressed in terms of the helicity spinors and suppressing couplings, the soft factors are
\begin{equation}\label{eq:weingrav}
S_{\text{grav}}^{(0)}(s^+) 
= \sum_{i=1}^{N} \frac{\ssb{s}{i}\sab{x}{i}\sab{y}{i}}{\sab{s}{i}\sab{x}{s}\sab{y}{s}} 
 \ ,
\end{equation}
and 
\begin{equation}\label{eq:subsofthel}
S_{\text{grav}}^{(\text{sub})}(s^+) 
=\frac{1}{2} \sum_{i=1}^{N} 
\frac{\ssb{s}{i}}{\sab{s}{i}}\left(\!
\frac{\sab{x}{i}}{\sab{x}{s}}\!+\!\frac{\sab{y}{i}}{\sab{y}{s}} \! \right)
\tilde{\lambda}^{\dot{a}}_s\frac{\partial}{\partial \tilde{\lambda}^{\dot{a}}_i}
 \ .
\end{equation}
In the soft factors, $x$ and $y$ are arbitrary spinors representing the gauge redundancy.  This form of the subleading soft factor $S^{(\text{sub})}(s)$ makes it clear that it is an operator that acts non-trivially on the lower-point amplitude.  \Ref{Cachazo:2014fwa} showed that the subleading soft factor in \Eq{eq:subsofthel} holds to any number of external particles at tree-level using the BCFW recursion relations \cite{Britto:2004ap,Britto:2005fq}.  

Using similar techniques, \Ref{Casali:2014xpa} derived the subleading soft factor for color-ordered tree-level amplitudes in gauge theory.  Scaling the spinors as in \Eq{eq:subscale}, these gauge theory amplitudes with a soft gauge boson of positive helicity have the expansion
\begin{equation}\label{eq:gaugeexp}
{\cal A}(1,\dotsc,N,\{s^+,\epsilon\})) = \left(
\frac{1}{\epsilon}S_{\text{gauge}}^{(0)}(s)+
S^{\text{(sub)}}(s)
\right){\cal A}(1,\dotsc,N) + {\cal O}(\epsilon^{1}) \,,
\end{equation}
where the leading soft factor is
\begin{equation} \label{eq:Sgauge0s}
S_{\text{gauge}}^{(0)}(s^+) = \frac{\sab{N}{1}}{\sab{N}{s}\sab{s}{1}} \ ,
\end{equation}
and the subleading soft factor is 
\begin{equation}\label{eq:subsoft}
S^{\text{(sub)}}(s^+) = \frac{\tilde{\lambda}_s^{\dot{a}}}{\langle s1 \rangle}\frac{\partial}{\partial \tilde{\lambda}_1^{\dot{a}}}+\frac{\tilde{\lambda}_s^{\dot{a}}}{\langle Ns \rangle}\frac{\partial}{\partial \tilde{\lambda}_N^{\dot{a}}}\ .
\end{equation}
This subleading soft factor is composed of the total angular momentum operators of particles $N$ and $1$.
The soft factors for a minus helicity soft gauge boson $s^-$ are found by swapping all holomorphic and antiholomorphic spinors in \Eqs{eq:Sgauge0s}{eq:subsoft}.

It is important to note the assumptions implicit in these derivations of the soft factors. In the tree level derivations, \Refs{Low:1958sn,Burnett:1967km,Casali:2014xpa,Bern:2014vva} assume that a Laurent series of the amplitude in powers of the soft particle's momentum can be performed. This assumes that the soft momentum must not flow through a propagator that is itself becoming on-shell, such as for a collinear particle that  probes a pole in the amplitude not caused by the emission of the soft particle.  This is equivalent to assuming that all other external particles are both energetic and well-separated from each other in angle so that all Lorentz invariant products not involving the soft momentum are large. This constrains the expansion parameter $\epsilon$ as
\begin{align}\label{eq:lows_theorem_pc}
\epsilon\sim \frac{p_s\cdot p_k}{(p_i+p_j)^2} \ll 1 ,
\end{align}
for all particles $i,j,k$ in the amplitude, where $p_s$ is a soft momentum. 
 The requirement that $(p_i+p_j)^2$ does not vanish as fast as $p_s\cdot p_k$ can be enforced by a choice of kinematics for tree-level amplitudes.  However, \Eq{eq:lows_theorem_pc} alone should not be taken to define the region of soft emissions, in particular because collinear emissions play an important role in gauge theory, and soft emissions coupling through the leading factor $S^{(0)}$ are still well defined in this situation.
We will explain below in \Sec{sec:scet}  the more general power counting of SCET for soft interactions that works in the presence of collinear particles, and review a simple proof of the leading soft factor under these more general conditions in \Sec{subsec:simplecalcs}.

\subsection{Soft factors of explicit amplitude examples}
\label{subsec:subsoftexp}

To set the stage for our later discussion it is useful to study the soft expansion of amplitudes in several explicit examples, which we present in this subsection.  We will begin with an example in gravity that nicely illustrates the presence of the subleading soft factor. Examples of the same technique are then given for gauge theory amplitudes, to exhibit the Low-Burnett-Kroll theorem. In the final part of this subsection we turn to the one-loop finite single-minus amplitude ${\cal A}(-,+\cdots,+)$ in pure Yang-Mills theory. This amplitude provides a quantum mechanical counter-example to the Low-Burnett-Kroll theorem~\cite{Bern:2014oka,He:2014bga}.

\subsubsection{A Simple Gravity Amplitude}  \label{sec:graveg}

To see the subleading soft factor in gravity, we will consider the 5-graviton MHV tree-level amplitude which is
\begin{align}
{\cal A}(1^-,2^-,3^+,4^+,5^+) &= \frac{\ssb{1}{2}\ssb{2}{5}\sab{1}{2}^8}{\sab{1}{2}\sab{1}{3}\sab{1}{4}\sab{2}{5}\sab{3}{5}\sab{4}{5}\sab{3}{4}^2}+\frac{\ssb{1}{5}\ssb{2}{5}\sab{1}{2}^8}{\sab{1}{3}\sab{1}{4}\sab{1}{5}\sab{2}{3}\sab{2}{4}\sab{2}{5}\sab{3}{4}^2}\nonumber \\*
&\qquad+\frac{\ssb{1}{2}\ssb{1}{5}\sab{1}{2}^8}{\sab{1}{2}\sab{1}{5}\sab{2}{3}\sab{2}{4}\sab{3}{5}\sab{4}{5}\sab{3}{4}^2}
 \ ,
\end{align}
where the superscript denotes the helicities of the particles in the amplitude.  Scaling the spinors of particle $5$ as in \Eq{eq:subscale}, the amplitude has the following expansion:
\begin{align}  \label{eq:gravAmp}
{\cal A}(1^-,2^-,3^+,4^+,5^+_s) &= \frac{1}{\epsilon}\left(
\frac{\ssb{1}{5}\sab{1}{3}\sab{1}{4}}{\sab{1}{5}\sab{3}{5}\sab{4}{5}}+
\frac{\ssb{2}{5}\sab{2}{3}\sab{2}{4}}{\sab{2}{5}\sab{3}{5}\sab{4}{5}}
\right)\frac{\ssb{1}{2}\sab{1}{2}^8}{\sab{1}{2}\sab{1}{3}\sab{1}{4}\sab{2}{3}\sab{2}{4}\sab{3}{4}^2} \nonumber \\
&\qquad+\frac{\sab{1}{2}\ssb{1}{5}\ssb{2}{5}}{\ssb{1}{2}\sab{1}{5}\sab{2}{5}}\cdot \frac{\ssb{1}{2}\sab{1}{2}^8}{\sab{1}{2}\sab{1}{3}\sab{1}{4}\sab{2}{3}\sab{2}{4}\sab{3}{4}^2} \ .
\end{align}
The $\epsilon^{-1}$ term is immediately recognizable as the leading soft factor, \Eq{eq:weingrav}, times the 4-point amplitude.  The arbitrary spinors $x$ and $y$ in \Eq{eq:weingrav} have been set to the spinors of gravitons 3 and 4.  The $\epsilon^{0}$ term is precisely the action of the subleading soft factor on the 4-point amplitude, as shown in \Ref{Cachazo:2014fwa}.  In general, higher-point gravity amplitudes will have non-zero higher-order terms in the soft expansion, unlike the case here.

\subsubsection{Simple Gauge Theory Amplitudes}
\label{subsec:gauge_zero_sub}

We can do a similar exercise for gauge theory amplitudes, focusing on pure gluon scattering.  It is well-known in gauge theory that tree-level MHV-type amplitudes take the following form for any number of external gluons \cite{Parke:1986gb,Berends:1987me}:
\begin{equation}
{\cal A}^{[0]}(1,\dotsc,N) = \frac{\sab{i}{j}^4}{\sab{1}{2}\sab{2}{3}\cdots\sab{(N-1)}{N}\sab{N}{1}} \ ,
\end{equation}
where gluons $i$ and $j$ have $-$ helicity, and all other gluons have $+$ helicity. The superscript $[0]$ indicates that the amplitude is tree-level. If we consider any one of the $+$ helicity gluons to be soft, we immediately see that this full amplitude is exactly equal to the product of the corresponding $S^{(0)}$ times the $(N-1)$ point MHV-type amplitude ${\cal A}$ with that particle removed. So it is clear that there is no subleading soft factor.  This is exactly as predicted from the Low-Burnett-Kroll theorem, expressed with the operator $S^{\text{(sub)}}(s)$ in \Eq{eq:subsoft}, because MHV amplitudes are independent of the anti-holomorphic spinors $\tilde{\lambda}_i$. If we instead consider one of the $-$ helicity gluons to be soft, then this amplitude does not have a leading or subleading term, instead it is suppressed, ${\cal O}(\epsilon)$, and hence beyond the order that the soft theorems apply.

To study a non-trivial subleading soft factor we need to consider an amplitude beyond MHV.  The simplest such amplitude is the 6-gluon split-helicity next-to-MHV (NMHV) amplitude \cite{Mangano:1987xk,Mangano:1990by}:
\begin{align}\label{eq:6ptnmhv}
{\cal A}^{[0]}(1^-,2^-,3^-,4^+,5^+,6^+) = \frac{1}{\langle 5 |3+4|2]}&\left(
\frac{\langle 3|4+5|6]^3}{\ssb{6}{1}\ssb{1}{2}\sab{3}{4}\sab{4}{5}(3+4+5)^2}\right.\nonumber \\
&\quad\left.-\frac{\langle 1|5+6|4]^3}{\ssb{2}{3}\ssb{3}{4}\sab{5}{6}\sab{6}{1}(5+6+1)^2}
\right) \ .
\end{align}
Here $(i+j+k)^2 = (p_i+p_j+p_k)^2$ and
\begin{equation}
\langle i|j+k|l] \equiv \sab{i}{j}\ssb{j}{l}+ \sab{i}{k}\ssb{k}{l} \ .
\end{equation}
Because of the more complicated form of this amplitude, the form of the result obtained from the expansion will depend on how 6-point momentum conservation is applied to manipulate the original expression. Regardless of how this is done, the results obtained from the expansion are equal using 6-point momentum conservation. Hence, though the forms may look different, they all yield the same subleading soft factor. Thus we must always retain 6-point momentum conservation.  There is, however, some freedom in how we satisfy this momentum conservation, and in particular which hard particles carry momenta that balance the momentum of the soft particle. Here we let the split of this soft momentum be arbitrary amongst all the hard particles, and do not consider expanding the hard particle momenta into residual soft components.\footnote{We thank Zvi Bern for discussions about how momentum conservation was implemented in the literature.  It was shown in \Ref{Bern:2014oka} that the procedure we adopt here is identical to the original prescription presented in \Ref{Cachazo:2014fwa} where one solves the momentum conserving $\delta$-functions for the spinors of the particles that neighbor the soft particle. With the multipole expansion in SCET, the 6-point momentum conservation is often split into a 5-point momentum conservation for the large momenta of collinear particles, times an exact 6-point momentum conservation for the soft particle and residual momentum of the collinear particles. Hence the flow of soft momentum is also kept general, as we do here.}  For example, scaling the spinors of particle $5$ as in \Eq{eq:subscale}  the amplitude expands as
\begin{align}
& {\cal A}^{[0]}(1^-,2^-,3^-,4^+,5^+_s,6^+) = \!\Bigg\{\frac{1}{\epsilon}\frac{\sab{4}{6}}{\sab{4}{5}\sab{5}{6}} 
+\frac{1}{\langle 5|3\!+\!4|2]}\bigg(\langle 5|3\!+\!4|5]\frac{\ssb{3}{2}}{\ssb{3}{4}\sab{4}{5}}
+\langle 5| 6\!+\!1|5]\frac{\ssb{1}{2}}{\ssb{6}{1}\sab{5}{6}}
\nonumber \\
& \qquad
 + 3\langle 5|3|2]\frac{\ssb{6}{5}}{\ssb{4}{6}\sab{4}{5}}+3\langle 5|1|2]\frac{\ssb{4}{5}}{\ssb{4}{6}\sab{5}{6}}+\ssb{5}{2}\bigg)\Bigg\}\frac{\ssb{4}{6}^4}{\ssb{1}{2}\ssb{2}{3}\ssb{3}{4}\ssb{4}{6}\ssb{6}{1}}
 +{\cal O}(\epsilon^1) \ .
\end{align}
The presence of the leading soft factor is manifest, and, while not obvious, it can easily be shown that the  ${\cal O}(\epsilon^0)$ term is numerically identical to the action of the subleading soft factor, \Eq{eq:subsoft}, on the amplitude ${\cal A}(1^-,2^-,3^-,4^+,6^+)$.

\subsubsection{Single-minus amplitude in pure Yang-Mills}\label{sec:singminus}

The soft expansion of the single-minus helicity amplitude ${\mathcal A}^{[1]}(1^-,2^+,3^+,4^+,5^+)$ is particularly interesting. This amplitude is zero at tree level. At one loop it is nonzero and infrared finite, so na\"ively one might think one could expand it assuming the region of validity of the Low-Burnett-Kroll theorem for the external particles, \Eq{eq:lows_theorem_pc}. Taking the results from \Refs{Bern:2014oka,He:2014bga}, the large $N_c$ primitive amplitude\footnote{For gauge theories with only adjoint particles, all one loop amplitudes can be determined from the large $N_c$ color-ordered amplitudes alone, which are called the primitive amplitudes, see \Ref{Bern:1994zx}.} is
\begin{align}\label{eq:5_point_one_minus}
{\mathcal A}^{[1]}(1^-,2^+,3^+,4^+,5^+)&=\frac{i}{48\pi^2}\frac{1}{\sab{3}{4}^2}
  \left(
  -\frac{\sab{1}{3}^3\ssb{3}{2}\sab{4}{2}}{\sab{1}{5}\sab{5}{4}\sab{3}{2}^2}
  +\frac{\sab{1}{4}^3\ssb{4}{5}\sab{3}{5}}{\sab{1}{2}\sab{2}{3}\sab{4}{5}^2}
  -\frac{\ssb{2}{5}^3}{\ssb{1}{2}\ssb{5}{1}}
   \right) \,.
\end{align}
As particle 5 becomes soft, the first term contributes to the leading order eikonal soft factor, whereas the second gives a subleading contribution. As shown in \Refs{Bern:2014oka,He:2014bga} this does not take the form of the Low-Burnett-Kroll theorem acting on the four-point amplitude
\begin{align}
&{\mathcal A}^{[1]}(1^-,2^+,3^+,4^+)=\frac{i}{48\pi^2}\frac{\sab{1}{3}^3\sab{2}{4}\ssb{1}{2}}{\sab{2}{3}^2\sab{3}{4}^3}  \,,
\end{align}
since
\begin{align} \label{eq:S0S1A14}
\left(\frac{1}{\epsilon}S^{(0)}(5^+)+S^{\text{(sub)}}(5^+)\right)&{\mathcal A}^{[1]}(1^-,2^+,3^+,4^+)=
 \nonumber \\
&\qquad\frac{i}{48\pi^2}\frac{\sab{1}{3}^3\sab{2}{4}\ssb{1}{2}}{\sab{2}{3}^2\sab{3}{4}^3}\left(\frac{1}{\epsilon}\frac{\sab{4}{1}}{\sab{4}{5}\sab{5}{1}}+\frac{\ssb{5}{2}}{\sab{5}{1}\ssb{1}{2}}\right) \ .
\end{align}
Using momentum conservation in the form $\sab{3}{4}\ssb{2}{3} = \sab{1}{4}\ssb{1}{2}+\sab{4}{5}\ssb{2}{5}$, the two terms in \Eq{eq:S0S1A14} together exactly reproduces the first term in \Eq{eq:5_point_one_minus}. However the second term in \Eq{eq:5_point_one_minus} is not reproduced, so LBK is violated. 

The fact that the subleading soft behavior of this amplitude does not conform to the LBK theorem was analyzed in \Ref{Bern:2014vva}, which paid careful attention to the infrared structure at loop level, noting certain ``factorizing'' diagrams were responsible, similar to problems encountered in loop-level on-shell recursion. 

In \Sec{sec:amps_nonzero}, we will demonstrate that this behavior is quite generic at loop level, since it arises from a collinear region of the loop diagram that only depends on a single energetic external leg and hence can always be factorized from the remainder of the diagram. This factorization is manifest in SCET, and makes up one term in our loop-level soft theorem.

\subsection{Review of Soft-Collinear Effective Theory and Power Counting}
\label{sec:scet}

To analyze the soft limit of gauge theory amplitudes, we would like to write down an effective theory of soft emissions so that hard and soft physics are factorized from one another and can be studied independently. An eikonal effective field theory of this sort, known as LEET, was formulated in \Ref{Dugan:1990de}. As noted earlier, this is consistent at tree-level, where we can enforce all external particles to have large energy and be at large angles with respect to one another.  Thus, \Eq{eq:lows_theorem_pc} can be satisfied.  However, gauge theories in four dimensions also have collinear singularities and collinear particles, and it is well known that LEET is inconsistent beyond tree level for this reason.  At loop level, there is no way to avoid the region of the loop integral corresponding to a collinear virtual loop particle, and, generically, soft gluons can be sensitive to these collinear loops.  Therefore, for a consistent low-energy effective theory to all orders, we must include soft and collinear dynamics simultaneously.

SCET~\cite{Bauer:2000ew,Bauer:2000yr,Bauer:2001ct,Bauer:2001yt} is the effective theory of the soft and collinear dynamics of QCD, or more generally, any gauge theory with a weakly-coupled sector.  Because of collinear divergences, the dominant energy flow in any scattering event in a gauge theory will be along directions localized in space, so it is useful to define a coordinate system with respect to these directions.  We will refer to the directions of dominant energy flow as collinear or jet directions, and label each distinct direction $i$ by a dimensionless light-like vector $n_i^\mu$.\footnote{When referring to an arbitrary collinear direction, we will often drop the $i$ subscript and just call it the $n$-collinear direction.}  To fully decompose a four vector using $n_i^\mu$ as a basis vector, we need to specify the component along $n_i$ using another dimensionless light-like basis vector $\bn_i^\mu$, and we adopt the normalization convention $n_i\cdot \bn_i = 2$.  For a four-vector $p^\mu$ we then have contributions along $n_i$, $\bn_i$, and in the transverse directions as
\begin{equation}
p^\mu = \frac{\bar{n}_i\cdot p}{2}\, n_i^\mu + \frac{n_i\cdot p}{2}\, \bar{n_i}^\mu + p_\perp^\mu \,.
\end{equation}
It is sometimes convenient to denote $n_i\cdot p \equiv p^+$ and $\bar{n}_i\cdot p \equiv p^-$.  Here $p^2 = p^+p^- + p_\perp^2$.

\begin{figure}
\begin{center}
\subfloat[]{\label{fig:njet}
\includegraphics[scale = 0.9]{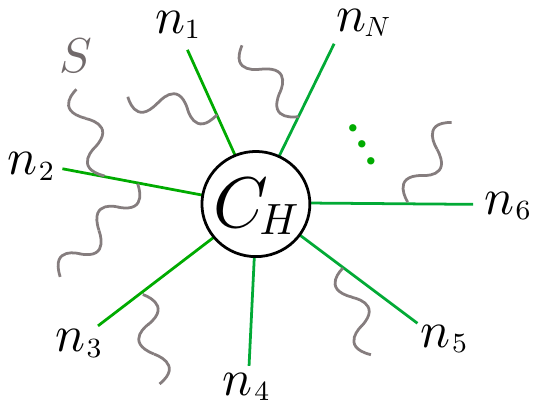}
}
$\qquad$
\subfloat[]{\label{fig:2jet}
\includegraphics[scale = 0.9]{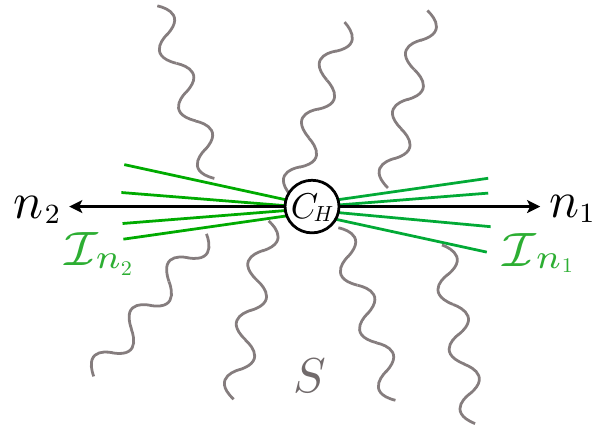}
}
\end{center}
\vspace{-0.2cm}
\caption{a) Illustration of an amplitude with $N$ energetic external lines with soft gluon attachments encoded by a soft amplitude $S$.  b)  Illustration of the dominant modes for a two-jet event, where we have a hard amplitude $C_H$, two directions with splittings generating collinear amplitudes ${\cal I}_{n_1}$ and ${\cal I}_{n_2}$, and a soft amplitude $S$.  
}
\label{fig:scet_2jet}
\end{figure}

As a simple example of an amplitude that will be relevant to our discussion later on,  consider $N$ massless energetic particles that are well-separated in phase space as shown in \Fig{fig:njet}. The external particles in this situation will satisfy \Eq{eq:lows_theorem_pc}.  If their momenta are labelled $p_i$ with $i=1,\ldots,N$, then by well-separated we mean that if $i\ne j$ then $p_i\cdot p_j \gg \lambda^2$ where $\lambda$ is related to the scaling parameter for a soft momentum, which is either $p_s^\mu\sim \lambda^2$ or $p_s^\mu \sim \lambda$. For this amplitude we can use the momenta themselves as the $n_i$ basis vectors, so that $p_i^\mu = n_i^\mu \bn_i\cdot p_i /2 $.   The well-separated condition is then $n_i \cdot n_j \gg \lambda^2$ which is precisely the condition for having independent collinear sectors in SCET~\cite{Bauer:2002nz}.

The above situation is violated when we have more than one particle in a collinear sector.  For example, we may have an amplitude with two large momenta $p_1$ and $p_1'$ that are parametrically close in phase space, and a third momentum $p_2$ that is well-separated. In this case we can have $p_1\cdot p_1' \sim \lambda^2$ even though $p_1\cdot p_2 \sim p_1'\cdot p_2 \gg \lambda^2$, so that neither $p_{1}$, nor $p_1'$, nor $p_2$ are soft.   In this situation we say that both $p_1$ and $p_1'$ are collinear, and that they belong to the same $n_1$ collinear sector, while $p_2$ belongs to a different collinear sector.  Here the momenta $p_1$ and $p_1'$ violate the LBK condition in \Eq{eq:lows_theorem_pc}.  In this situation there will be propagators for energetic particles where it is not possible to Taylor expand all soft momenta out of the denominator of the propagator and into the numerator. 

As a simple example of a process with nontrivial collinear dynamics, consider an initial state in a scattering process that is not charged under the gauge theory, which then decays with a large energy release into gauge theory particles. Here there must be at least two energetic (collinear) particles in the final state. The amplitude that describes the dominant infrared contributions for the simplest final state with two jets in directions $n_1$ and $n_2$ is shown in \Fig{fig:2jet}, and there are two sets of collinear modes corresponding to these two directions as well as nontrivial collinear amplitudes ${\cal I}_{n_1}$ and ${\cal I}_{n_2}$ caused by splittings.  In addition there is a soft amplitude $S$, which is generated by the dipole formed from the collinear particles.

\begin{table}[t]
\begin{center}
\begin{tabular}{c|c|c|c}
Mode &$n_i\cdot p$ & $\bn_i\cdot p$ & $p_\perp$ \\ 
\hline
hard & $Q$ & $Q$ & $Q$ \\
$n_i$-collinear & $\scetscale^2 Q$ & $Q$ &  $\scetscale Q$ \\
soft$_1$  & $\scetscale^2 Q$ & $\scetscale^2 Q$ &  $\scetscale^2 Q$ \\
soft$_2$  & $\scetscale Q$ & $\scetscale Q$ &  $\scetscale Q$ \\
\end{tabular}
\end{center}
\caption{Scaling of momentum components of the hard, $n_i$-collinear, and soft modes with respect to the total scattering energy $Q$.  $\scetscale$ is the small power-counting parameter in SCET. For the soft modes two common scalings are shown.  If there is a need to distinguish then they are often referred to as soft with $p_s\sim \lambda$, and ultrasoft with $p_s\sim\lambda^2$.}
\label{tab:scaling}
\end{table}
In a general gauge theory, the most important on-shell modes correspond to soft and collinear physics that dominate the dynamics of a scattering event, as illustrated in \Fig{fig:scet_2jet}.  These modes have momenta that scale  with a small parameter $\lambda \ll 1$ relative to a large momentum scale $Q$ that sets the dimensions, and are summarized in \Tab{tab:scaling}.\footnote{When the initial state of the scattering process is not given by well separated particles that participate in the primary hard scattering, an off-shell mode referred as a Glauber mode may also play an important role in infrared gauge theory dynamics~\cite{Bodwin:1981fv,Collins:1988ig}. Throughout this paper, we assume that the initial state satisfies this and that Glauber modes can be ignored.}  
Collinear modes have a large component of momentum in one direction, and parametrically smaller momenta in the others. For example, any energetic external particle will be referred to as collinear.  Two $n$-collinear momenta will satisfy $p_i \cdot p_i' \sim \lambda^2$.   Soft modes are isotropic, with all components of their momenta small with respect to $Q$.  By contrast, the collinear modes' momenta is not isotropic; for example an 
$n$-collinear momentum is predominantly in the $n$ direction, spread about $n$ by a small angle $\theta_n \sim p_\perp/p^- \sim \lambda$.  We will generically refer to any momentum that does not scale with $\lambda$ as hard.   In particular the sum of momenta from two distinct collinear sectors, $p_i+p_j$, is hard.

The definition of on-shell we adopt is broader than the amplitude literature, in that we consider both particles that are exactly on-shell $p^2 = p^+p^- + p_\perp^2=0$, and those that are parametrically close to the on-shell region  of momentum space with a homogeneous scaling, so that $p^+ p^- \sim p_\perp^2$. Here $\sim$ means that the two expressions have the same scaling in $\scetscale$, and this condition is satisfied for all soft and collinear modes  in \Tab{tab:scaling}.  For example, since we adopt $p_\perp \sim \lambda$ for a $n$-collinear particle this implies that its momentum $n\cdot p\sim \lambda^2$.  The definition of on-shell in the effective theory implies that a propagator with momenta from a single sector is not expanded in powers of $\scetscale$.

To see the differences between the two soft momentum scalings listed in  \Tab{tab:scaling} requires comparisons between different momentum types.
Both types of soft momenta act in the same manner when compared to hard momenta, or in a situation with only one collinear particle in each sector. In these cases \Eq{eq:lows_theorem_pc} is satisfied.  When compared to two  momenta, $p_i$ and $p_i'$, both with $n$-collinear scaling, the two soft scalings act differently. Soft$_1$ scales such that $p_s\cdot p_i \sim p_i\cdot p_i'$, whereas the scaling for soft$_2$ implies $p_s\cdot p_i \gg p_i\cdot p_i'$.  Thus together these two soft modes cover both ways in which the LBK condition in \Eq{eq:lows_theorem_pc} can be parametrically violated. In the literature, physical examples that require soft$_1$ modes are referred to as SCET$_{\rm I}$, whereas those that require soft$_2$ modes are referred to as SCET$_{\rm II}$~\cite{Bauer:2002aj}. 

Our focus here will be on the soft$_1$ modes, and from here on we will  use the name ``soft'' to simply refer to the scaling associated to these modes.\footnote{In the literature the soft$_1$ modes are often called ultrasoft to distinguish them from the soft=soft$_2$ modes.  Different relative scalings of the soft modes with respect to the collinear modes are important for the study of specific observables \cite{Becher:2012qa,Stewart:2013faa,Chiu:2012ir,Larkoski:2014uqa}, but a full discussion of the differences is beyond the scope of our paper. Some of our results can be immediately applied to the case of soft$_2$ modes by performing what is called a SCET$_{\rm I}$ to SCET$_{\rm II}$ matching, and we will briefly mention when this is the case, mostly with footnotes.}
Any collinear mode that absorbs a soft particle remains collinear: 
\begin{align}\label{eq:merging_soft_collinear_momenta}
p_c+p_s\sim Q(\lambda^2,\lambda^2,\lambda^2)+Q(\lambda^2,1,\lambda)\sim Q(\lambda^2,1,\lambda) \ ,
\end{align}
where the components are the $(+,-,\perp)$ momenta.  Since $n\cdot p_c\sim n\cdot p_s$ this component of $p_s$ is not expanded in powers of $\scetscale$ in a propagator carrying momentum $p_s+p_c$. 

When relating $\lambda$ to the scaling parameter $\epsilon$ used to identify the subleading soft factor in our review of amplitudes, we note that $p_s^\mu \sim \lambda^2$ and $p_s^\mu\sim \epsilon$, and hence take $\lambda^2 \sim \epsilon$ throughout the rest of this paper.
As mentioned earlier, the power counting in \Eq{eq:lows_theorem_pc} used for the Low-Burnett-Kroll theorem must be extended for an analysis at loop level, or in the presence of non-trivial collinear states.  The SCET power counting presented here is consistent to all orders in both the coupling constant and $\lambda$ expansion, and can mix soft and collinear momentum components.  Even for cases in which \Eq{eq:lows_theorem_pc} is violated, SCET correctly describes the soft limits.

\subsection{SCET Lagrangian and Operators}

The SCET Lagrangian ${\cal L}_{\rm SCET} $ governs the dynamics of soft and collinear particles, while the physics of a hard collision is encoded in external operators ${\cal O}$ that connect together different collinear sectors.  Both of these have a power expansion in $\lambda$:
\begin{align}
  {\cal L}_{\rm SCET}  & = {\cal L}^{(0)} + {\cal L}^{(1)} + \ldots \,,
  & {\cal O}   & = {\cal O}^{(0)} + {\cal O}^{(1)} + \ldots \,,
\end{align}
where the superscript denotes the order in the power expansion in $\lambda$. The power counting theorem of SCET~\cite{Bauer:2002uv} implies that we can simply add these exponents to determine the relative size of various time ordered product contributions.
In this section we will briefly review the leading order terms, leaving the review of the subleading soft operators relevant to our analysis to \Secs{subsec:scetsublag}{sec:sub_ope} below.  For the leading order SCET Lagrangian we have
\begin{align}
   {\cal L}^{(0)}  &= {\cal L}_{\rm soft}^{(0)} +  \sum_{n} \left( {\cal L}_{\xi_{n}}^{(0)} + {\cal L}_{A_{n}}^{(0)} \right) \,.
\end{align}
The sum is over distinct collinear equivalence classes $\{n\}$, determined by $n_i\cdot n_j \gg \lambda^2$.
This enforces that collinear emissions within a given sector are at parametrically smaller angles than emissions described by two disinct collinear sectors.
Here ${\cal L}_{\rm soft}^{(0)}$ is simply the Yang-Mills Lagrangian with soft fermion fields $\psi_s$ and soft gluon fields $A_s^\mu$.  For the $n$-collinear fields we have $\phi_n$ for scalars, $\xi_n$ for fermions and $A_n^\mu$ for gluons and the leading-power collinear scalar, fermion, and gauge boson Lagrangians are~\cite{Bauer:2000yr,Bauer:2001yt}
\begin{align}\label{eq:scetlag}
{\cal L}_{\phi_n}^{(0)} 
 &= 2\,{\rm Tr}\big[
 \phi_n^*\big( \bar{n}\cdot D_n\: n\cdot D_{ns} + D_{n\perp}^2 \big)\phi_n 
 \big]
  \ ,\nonumber \\
{\cal L}_{\xi_n}^{(0)} 
 &= \bar{\xi}_n \Big( i n\cdot D_{ns} + 
  i\slashed{D}_{n\perp} \frac{1}{i\bar{n}\cdot D_n}i\slashed{D}_{n\perp} 
  \Big ) \frac{\slashed{\bar{n}}}{2} \xi_n 
   \,,\\
 {\cal L}_{A_n}^{(0)}
 & = \frac{1}{2g^2}  {\rm Tr} \Big( \big[ i D_{ns}^\mu , i D_{ns}^\nu \big]  \big[ i D_{ns}^\mu , i D_{ns}^\nu \big]\Big) +  {\cal L}_{A_n,\text{gf}}^{(0)} 
  \,. 
  \nonumber
\end{align}
In a covariant gauge, the gauge fixing terms in the collinear gluon Lagrangian are
\begin{equation}
 {\cal L}_{A_n,\text{gf}}^{(0)} = \frac{1}{\alpha}\text{Tr}\Big(
 \big[
 i \partial_{ns}^\mu, A_{n\mu}
 \big]
 \big[
 i \partial_{ns}^\nu, A_{n\nu}
 \big]
 \Big) + \text{ghosts} \,,
\end{equation}
where we omit the ghost Lagrangian~\cite{Bauer:2001yt} as it is not needed for the analysis in this paper.
Here $iD_n^\mu = i\partial_n^\mu + g A_n^\mu$ is the $n$-collinear covariant derivative, whose components are 
\begin{align}\label{eq:covdercomp1}
in\cdot D_n &= in\cdot\partial_n+g n\cdot A_n 
\,,
% \ ,\\
& i \bn\cdot D_n &= i\bn\cdot\partial_n + g\bn\cdot A_n 
\,,
%  \ , \nonumber\\
& i D_{n\perp}^\mu &= i\partial_{n\perp}^\mu+g A_{n\perp}^\mu 
%\ .\nonumber 
\,.
\end{align}
In addition when the soft field that is the same order in $\lambda$ is included we use
\begin{align}\label{eq:covdercomp2}
iD_{ns}^\mu &= i D_n^\mu + \frac{\bn^\mu}{2} g n\cdot A_s
\ ,
 & in\cdot D_{ns} &= in\cdot D_n + g n\cdot A_s 
 \ ,
 \nn\\
 i \partial_{ns}^\mu & = i\partial_n^\mu + \frac{\bn^\mu}{2} g n\cdot A_s
 \ .
\end{align}
We provide a schematic derivation of the collinear fermion Lagrangian in \App{app:colq}. We also define the soft covariant derivative as
\begin{align}
 iD_{s\perp}^\mu & = i\partial^\mu_{s\perp}+g A_{s\perp}^\mu \,,
\end{align} 
noting that $i\partial_s\sim \lambda^2$ and $i n\cdot \partial_s = in\cdot \partial_n$. Due to the SCET multipole expansion $i \partial_{n\perp}^\mu$ and $i \bn\cdot \partial_n$ do not act on soft fields. When we need to refer to the terms in ${\cal L}_n^{(0)}$ that depend on the soft gluon field, we will refer to it as ${\cal L}_{n,{\rm soft}}^{(0)}$.

\begin{figure}
 	\begin{center}
 		\includegraphics[width=15cm]{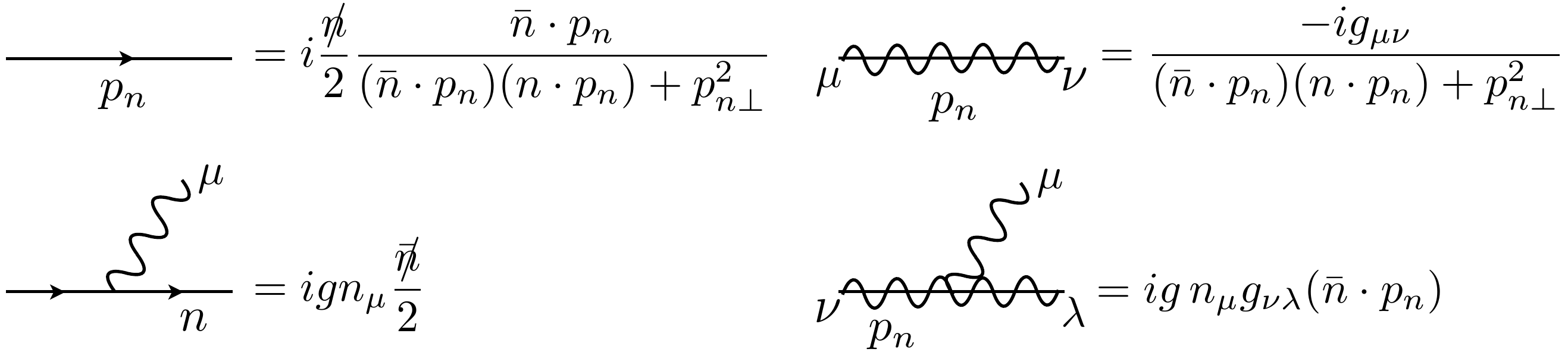}
 	\end{center}
 	\caption{
 		The leading-power color-ordered SCET Feynman rules for emission of a soft gluon off of an $n$-collinear fermion or gluon in Feynman-`t Hooft gauge.  The collinear fermion is denoted by the solid line, the collinear gluon by the wave with a line through it, and the soft gluon is the wavy line.
 	}
 	\label{fig:scet0_feyn}
 \end{figure}

At this stage a few illustrative observations can be made.  First, from \Eq{eq:scetlag} note that the propagator for a collinear fermion with $\bn\cdot p>0$ is
\begin{equation}
\frac{i}{\left(n\cdot p +\frac{p_\perp^2}{\bar{n}\cdot p} +i0\right)\frac{\slashed{\bar{n}}}{2}} = \frac{i\slashed{n}}{2}\frac{\bar{n}\cdot p}{p^2+i0} \ ,
\end{equation}
which is just the full fermion propagator expanded to leading power.  We can also use \Eq{eq:scetlag} to immediately write down a few relevant Feynman rules, such as the coupling of a soft gluon to a collinear fermion or gluon. The color-ordered SCET Feynman rules are presented in \Fig{fig:scet0_feyn}. The propagators for a collinear fermion and gluon are also included.  
The interactions in the leading power SCET Lagrangian for $n$-collinear fermions preserve helicity defined about the $n$-direction (just like the full QCD Lagrangian with massless fermions does with a common fixed direction).  This is easy to see, since the helicity projection operator ${\cal P}_{L,R}$ commutes with the derivative operator in the collinear fermion Lagrangian:
\begin{equation}
\left[{\cal P}_{L,R},i n\cdot D_{ns} + i\slashed{D}_{n\perp} \frac{1}{i\bar{n}\cdot D_n}i\slashed{D}_{n\perp}\right]
 =\left[\frac{1\pm \gamma_5}{2},i n\cdot D_{ns} + i\slashed{D}_{n\perp} \frac{1}{i\bar{n}\cdot D_n}i\slashed{D}_{n\perp}\right]=0 \ .
\end{equation}
For a pure collinear gluon splitting the SCET Lagrangian only preserves the total angular momentum in the $n$ direction (just like QCD). 

Also, we can determine the scaling of the collinear fields in $\scetscale$ by demanding that the leading order action scales as $\scetscale^0$.  With $p\cdot x \sim \scetscale^0$, we have $d^4 x \sim \scetscale^{-4}$.  We also know that $n\cdot D_{ns} \sim p^+ \sim \scetscale^2$, and so it follows that $\xi_n\sim \scetscale$.  That is, the scaling of the collinear fermion field is not its engineering dimension $(=3/2)$, but rather its twist, $3/2-1/2 = 1$. Indeed the power counting of SCET corresponds to the dynamic twist expansion in cases where a twist expansion exists such as deep-inelastic scattering~\cite{Bauer:2002uv}.  A summary of the power counting for the SCET fields  is given in \Tab{tab:powercount}. Since $\bn\cdot A_n\sim \lambda^0$ it is convenient to trade it for the collinear Wilson line
\begin{align}  \label{eq:W}
   W_n &  =\text{P}\exp\left[ig\int_0^\infty ds\, \bar{n}\cdot A_n(x+s\bar{n})\right] \ ,
\end{align}
where P denotes path-ordering. This is done with $i\bn\cdot D_n = W_n i\bn\cdot \partial_n W_n^\dagger$ and associated relations. If $W_n$ is in the fundamental representation, then the adjoint Wilson line ${\cal W}_n^{AB}$ obeys $W_n^\dagger T^A W_n = {\cal W}_n^{AB}T^B$.

\begin{table}[t]
\begin{center}
\begin{tabular}{c|c|c|c|c}
 \hline
$\phi_n\sim \lambda$&$\xi_n \sim \lambda$ &  $(n\cdot A_n, \bn\cdot A_n , A_n^\perp) \sim (\lambda^2,\lambda^0,\lambda)$ 
   &  $\psi_s\sim \lambda^3$ &  $A_s^\mu \sim \lambda^2$\\
\hline
\end{tabular}
\end{center}
\vspace{-0.4cm}
\caption{Power counting for the $n$-collinear fields and soft fields.}
\label{tab:powercount}
\end{table}

The description of hard scattering in SCET makes use of external operators constructed from the fields. For example, the leading order quark operator for deep inelastic scattering is ${\cal O}^{(0)} = \bar\xi_n W_n \delta(\bar n\cdot Q - \bn\cdot i\partial_n) W_n^\dagger \xi_n$ which is $\sim \lambda^2$.   For hard scattering processes a popular set of operators are those that couple together multiple collinear directions, which can be referred to as an $N$-jet operator. At leading power they are
\begin{align}  \label{eq:ON}
  {\mathcal O}_N^{(0)} &=  C_N\big( \{ Q_i  \}\big) \otimes \prod_{i=1}^N  \Big[\delta(\bn_i\cdot Q_i - \bn\cdot i\partial_n)  X_{n_i}^{\kappa_i}(0)  \Big] \,,
\end{align}
where the hard scattering takes place at the origin, and $\kappa_i$ denotes whether the operator corresponds to a fundamental fermion, adjoint gluon, etc., and includes helicity.  All operators in SCET are constructed out of the fundamental building blocks $X_{n_i}$. For a fermion, gluon, or adjoint scalar $\phi_{n_i}$ the building blocks are $X_{n_i}=\{\chi_{n_i},B_{n_i\perp}^\mu,\Phi_{n_i}\}$ respectively, where:
\begin{align}\label{eq:Collinear_Field_Operators}
   \chi_{n_i} &= W_{n_i}^\dagger \xi_{n_i} \,,
  & B_{{n_i}\perp}^\mu & = \big[ W_{n_i}^\dagger i D_{{n_i}\perp}^\mu W_{n_i} \big] = B_{n_i\perp}^{A\mu} T^A 
  \,, \\
  \Phi_{n_i} &= W_{n_i}^\dagger  \phi_{n_i}^A T^A W_{n_i} 
   = \phi_{n_i}^A {\cal W}_n^{AB} T^B   
   \,. \nn
\end{align}
Here the square bracket here indicates that the covariant derivative acts only on the objects inside.
The perpendicular derivatives $i\partial_{n\perp}^\mu$ are an additional building block for collinear operators. The $W_{n_i}$ Wilson lines in the building blocks ensure collinear gauge invariance in each sector, and encode emissions of $\bn_i\cdot A_{n_i}$ collinear gluons off of all other collinear and hard modes. The $\delta$-function in \Eq{eq:ON} constrains the large light-cone component of the collinear sectors' momenta to be the jet momentum $\bar{n}_i\cdot Q_i$. The $C_N$ is the Wilson coefficient obtained by matching order by order in the gauge coupling, and depends only on the large momenta $Q_i$.   Finally, $\otimes$ denotes color and helicity contractions between $C_N$ and the final state particles, while
for simplicity additional indices like those for fermion flavor are suppressed.  The distinction between a generic hard scattering operator in SCET and the example of the $N$-jet operator ${\cal O}_N^{(0)}$, is that in \Eq{eq:ON} only one building block appears for each collinear sector, and no soft fields appear. At the amplitude level the operators ${\cal O}_N^{(0)}$ plus soft attachments describe precisely the situation pictured in \Fig{fig:njet} above.  Collinear splitting amplitudes can be defined as matrix elements of the $1$-jet operator ${\cal O}_1^{(0)}$ with ${\cal L}_n^{(0)}$ insertions and will also be considered in the following sections.

The relationship between SCET and the full theory takes the form of an operator expansion where the S-matrix is encoded by SCET operators with a power series in $\lambda$. That is, integrating out the hard modes of the full theory induces operators that couple multiple jet directions together. In general, one writes down all possible operators, organizing their relative importance via power counting. For later purposes we also introduce the following short-hand notation
\begin{align} \label{eq:iDc}
  i {\cal D}_{n\perp}^\mu &= W_n^\dagger\, i D_{n\perp}^\mu W_n \,,
  & i {\cal D}_{ns}^\mu &= W_n^\dagger\, i D_{ns}^\mu\, W_n \,.
\end{align}
Note that $i {\cal D}_{n\perp}^\mu = i\partial_{n\perp}^\mu + g B_{n\perp}^\mu$.   

It is important to note that a given $N$-point scattering amplitude does not correspond to a unique SCET operator.  For example, a given $N$-point amplitude and a related $N+1$-point amplitude with the addition of a soft particle match to the same $N$-jet operator with the same hard matching coefficient, just taken with different matrix elements. This occurs because the soft emission is described by the Lagrangian of SCET rather than the operator alone.  Also if several emissions in the full theory amplitude were to be determined to be in the same collinear sector, the leading contribution would come from a lower-point $N$-jet operator plus additional collinear emissions from the leading SCET Lagrangian.

%%%%%%%%%%%%%%%%%%%%%%%%%%%%%%%%%%%%%%%%%%%%%%%%%%%%%%%%%%%%
\subsection{The Leading Soft Factor from SCET with Arbitrary Loops}
\label{subsec:simplecalcs}
%%%%%%%%%%%%%%%%%%%%%%%%%%%%%%%%%%%%%%%%%%%%%%%%%%%%%%%%%%%%

At this point, it is illustrative to review a simple calculation involving soft gluons in the SCET framework.  We will therefore reproduce the leading power eikonal soft factor $S^{(0)}(s)$ and demonstrate that it holds both for tree amplitudes, as well as for amplitudes involving arbitrary hard and collinear loop corrections. The final result also gives the generalization to an arbitrary number of simultaneous soft emissions, and allows for leading power soft loops.

Let's start at tree level and consider the amplitude for a single soft gluon emission at leading power emitted in the presence of $N$ energetic particles.  At leading power, the matrix element of the operator \Eq{eq:ON} with $N$ collinear states describes the scattering amplitude:
\begin{equation}\label{eq:leadwilsoncoeff}
\big\langle 0 \big| {\cal O}_N^{(0)}\big| p_1,\dotsc,p_N\big\rangle 
   = C_N^{[0]} \: e_1 \cdots e_N
   = {\cal A}_N^{[0]} \big[1 + {\cal O}(g^2) \big] \ ,
\end{equation}
where $C_N^{[0]}$ is the Wilson coefficient, $e_i=\{1,u^\pm,\epsilon^\pm\}$ are scalar, fermion, or gluon polarizations, ${\cal A}_N^{[0]}$ is the tree-level amplitude for $N$ final state particles with momenta $p_1,\dotsc,p_N$. 
With an additional soft particle in the state, $| p_1,\dotsc,p_N,p_s\rangle$, we wish to show that the matrix element is 
\begin{align}\label{eq:operatorfact}
\big\langle 0 \big| {\cal O}_N^{(0)} \big| p_1,\dotsc,p_N,p_s\big\rangle 
  & =\big\langle 0 \big| T \big\{ {\cal O}_N^{(0)}, \sum_i {\cal L}^{\rm (0)}_{n_i,{\rm soft}}\big\} \big| p_1,\dotsc,p_N,p_s\big\rangle_{\rm int} 
   + \ldots 
   \nonumber \\ 
  & = S^{(0)}(s) \, {\cal A}_N^{[0]} +\dotsc \ ,
\end{align}
where the dots represent higher order perturbative corrections.
In the first matrix element the fields are in the Heisenberg picture, so factors from the leading power SCET Lagrangian are implicit. In the second matrix element (subscript ``int'') we are in the interaction picture with the soft interaction Lagrangian explicit. 
Here ${\cal L}_{n_i,{\rm soft}}^{(0)}$ are the terms in the leading $n_i$-collinear Lagrangian that involve the soft field, $n_i\cdot A_s$. 
Since the collinear fields in ${\cal L}_{n_i,{\rm soft}}^{(0)}$ can only be contracted with the  corresponding $n_i$-collinear field in ${\cal O}_N^{(0)}$ we immediately get a sum of contributions, one for each of the $N$ particles in ${\cal O}_N^{(0)}$ just like we have in the leading power universal soft factor $S^{(0)}(s)$ in \Eq{eq:weingauge}.  

The final step needed to prove the last equality in \Eq{eq:operatorfact} is to demonstrate that the result is eikonal with the correct form.  Using the Feynman rules from Fig.~\ref{fig:scet0_feyn} we can calculate the amplitude for the soft emission from a collinear fermion or gluon. For a collinear fermion in the $n_i$ direction we have
\begin{equation} \label{eq:scetleadsfotq}
\raisebox{-0.23\height}{\includegraphics[width=2.5cm]{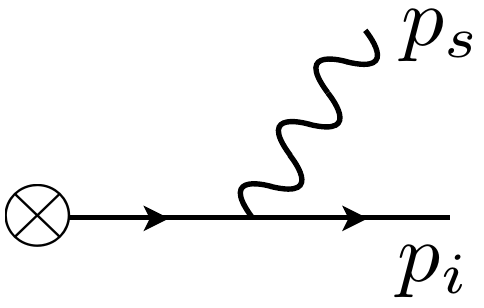}}
= \bar u(p_i) \Big( i g T_i\,
n_i\cdot \epsilon_s \frac{\slashed{\bar{n}}_i}{2} \Big)
\frac{i\frac{\slashed{n}_i}{2} p_i^-}{p_i^-(n_i\cdot p_s)+i0} 
=\bar u(p_i) \cdot \bigg[-g T_i\,
\frac{(p_i^- n_i)\cdot \epsilon_s}{(p_i^- n_i)\cdot p_s} \bigg]
 \ .
\end{equation}
where $T_i$ is the color matrix for the $i$'th fermion.  Here, the $\otimes$ symbol denotes the hard matching coefficient, which, from \Eq{eq:leadwilsoncoeff}, is just the tree-level amplitude at lowest order.  So we indeed get the expected eikonal soft factor that appears in $S^{(0)}(s)$. 
Similarly, the amplitude for emission from a collinear gluon in the $n_i$ direction is
\begin{align}\label{eq:scetleadsfotg}
\raisebox{-0.23\height}{\includegraphics[width=2.5cm]{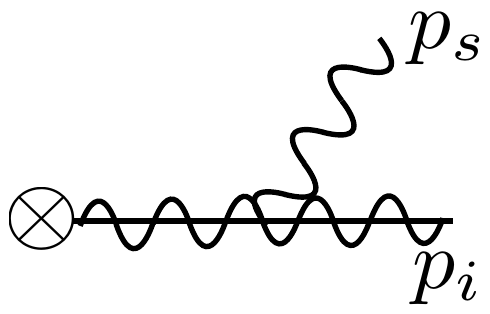}}
&=  ( ig T_i n_i\cdot \epsilon_s p_i^- \epsilon_i^\nu ) 
  \frac{-i g_{\mu \nu}}{p_i^-(n_i\cdot p_s)+i0}
= \epsilon_{i\mu} \bigg[-g T_i\,
\frac{(p_i^-n_i)\cdot \epsilon_s}{( p_i^-n_i)\cdot p_s} \bigg] \ .
\end{align} 
Again the universal form for the leading soft factor is obtained. Note that with the power counting we have defined, this factor scales like $\lambda^{-2}$, exactly as expected from the amplitude analysis in \Eq{eq:gaugeexp} with the replacement $\epsilon\to \lambda^2$.

It is worth emphasizing that once the SCET Lagrangian was derived with its explicit power counting, that proving \Eqs{eq:scetleadsfotq}{eq:scetleadsfotg} required no further expansions.  Also one may ask what happened to soft gluons emitted from internal propagators in the hard scattering amplitude ${\cal A}_N$. These propagators enter in the factor $C_N$ in \Eq{eq:ON}, and the fact that soft gluon attachments to these propagators are power suppressed is already made explicit in the leading power SCET Lagrangian and operators. In SCET such soft gluon attachments are represented by power suppressed operators ${\cal O}^{(i)}$ in which an explicit soft operator appears, such as a $D_{s}^\mu$. For the \sceti~situation considered here the power counting immediately implies that soft fields first enter as ${\cal O}(\lambda^2)$ corrections in hard-scattering operators.

The machinery of the effective theory makes it simple to extend the above leading power analysis to include an arbitrary number of soft gluon emissions as well as arbitrary loops in the amplitudes. In particular, we can make a field redefinition in SCET, known as the BPS field redefinition~\cite{Bauer:2001yt}, which decouples the soft gluons completely from the leading order collinear Lagrangian, sending ${\cal L}_{n_i,{\rm soft}}^{(0)}\to 0$. This field redefinition does change the form of the hard scattering operators, such as ${\cal O}_N^{(0)}$, and the modified form of these operators will encode all leading power contributions from soft gluons. 
Each distinct collinear sector $n_i$ has a different field redefinition, which is~\cite{Bauer:2001yt}
\begin{align}\label{eq:bps}
A_n^{A\mu} \to {\cal Y}^{AB}_n A_n^{B\mu}\ ,
& & \xi_{n}^\alpha \to Y_n^{\alpha\beta} \xi_{n}^\beta \ ,
& & \phi_{n}^A \to {\cal Y}^{AB}_n \phi_{n}^B \ ,
\end{align}
where the $Y$s are Wilson lines that appear in the  color representation appropriate to each collinear field. Thus here $Y_n^{\alpha\beta}$ is the soft Wilson line in the fundamental representation, and ${\cal Y}_n^{AB}$ is in the adjoint representation. In a generic representation 
\begin{equation}
Y_n(x) = \text{P}\exp\left[
ig \int_0^\infty ds \, n\cdot A_s(x+sn)
\right] \,,
\end{equation}
where P denotes path-ordering.\footnote{In the original work of Collins, Soper, and Sterman \cite{Collins:1988ig,Collins:1985ue,Collins:1989gx}, the decoupling of soft radiation from collinear interactions proceeds by use of a Ward identity. This Ward identity is implemented order-by-order in perturbation theory, and builds up these same soft Wilson lines.} \Eq{eq:bps} and the SCET multipole expansion also imply that under the field redefinition 
\begin{align}\label{eq:BPS}
  B_{n\perp}^{A\mu} \to {\cal Y}_n^{AB}  B_{n\perp}^{B\mu} \,, 
 & & \chi_n^\alpha \to Y_n^{\alpha\beta} \chi_{n}^\beta \,,
 & & \Phi_n^{A} \to {\cal Y}_n^{AB}  \Phi_n^{B} \,.
\end{align}

When we consider loop level amplitudes they can still be decomposed following the factorization formula in \Eq{eq:scetfact}. The leading power part of the loop level amplitudes can therefore be decomposed into factorized loop level amplitudes in $C_N^{(0)}$, $\cA_i^{(0)}$, and $S^{(0)}$ respectively. The hard amplitude $C_N^{(0)}$ is infrared finite, while the collinear and soft amplitudes generically have IR singularities.  
The field redefinition in \Eq{eq:bps} commutes with $\delta(\bn_i\cdot Q_i - \bn\cdot i\partial_n)$ since the $Y$s do not carry $\bn\cdot i\partial_n$ momenta due to the SCET multipole expansion. So the field redefinition does not change the form of the hard Wilson coefficient $C_N$ even when this Wilson coefficient contains loop level hard amplitude corrections.  Furthermore, the field redefinition works in the presence of collinear loops at any order in the coupling expansion, where it implies that at leading power the form of the soft interactions are entirely determined by {\em external} collinear particles. Thus the field redefinition does not change the collinear loop amplitudes $\cA_i^{(0)}$. Effectively what the field redefinition does is give a simpler operator from which to calculate the soft amplitude $S^{(0)}$ at loop level. The soft amplitude carries color indices and can share momenta with collinear particles.   All of these properties~\cite{Bauer:2001yt} are commonly exploited in the SCET literature.  

After performing the field redefinition the $N$-jet operators become:
\begin{align}  \label{eq:ON_soft_factorized}
{\mathcal O}_N^{(0)} &=  C_N^{(0)}\big( \{ Q_i  \}\big) \otimes \prod_{i=1}^N  \Big[\delta(\bn_i\cdot Q_i - \bn\cdot i\partial_n)  X_{n_i}^{\kappa_i}(0)  \Big] \otimes \text{T}\Big\{ \prod_{i=1}^{N}Y_{n_i}^{\kappa_i}(0)\Big\}\,,
\end{align}
where T time orders all soft gluon contributions and the superscript $\kappa_i$ on the $Y_{n_i}$ Wilson lines indicates the appropriate color representation. Here all leading power soft interactions (including mixed soft-collinear loop corrections, and soft gluon self-interactions) have been removed from the collinear sectors. For one gluon emission the Feynman rules from the $Y_{n_i}^{\kappa_i}$ Wilson lines immediately give the results in \Eqs{eq:scetleadsfotq}{eq:scetleadsfotg} above. Furthermore, \Eq{eq:ON_soft_factorized} encodes the fact that the emission of an arbitrary number of soft gluons from an energetic gauge charged particle is described in terms of a Wilson line that extends from the origin (where the hard particle was created) to infinity (since for simplicity our energetic particles are all outgoing).  
Once we have performed the field redefinitions, the new Lagrangians confine all modes and loop corrections involving \Eq{eq:ON_soft_factorized} to their own collinear or soft sector. This result leads directly to the fact that the leading power amplitude with arbitrary collinear loops and soft loops  factorizes in the form shown as the first term in \Eq{eq:scetfact}.\footnote{Although we have only discussed soft couplings for SCET$_{\rm I}$ in this section, the same 
result in \Eq{eq:ON_soft_factorized} is also valid in SCET$_{\rm II}$, where the soft momenta are $\sim \lambda$ rather than $\sim \lambda^2$. The simplest proof for this case first matches QCD to SCET$_{\rm I}$ giving the result in \Eq{eq:ON_soft_factorized}, then lowers the value of $\lambda$ for the collinear fields to match SCET$_{\rm I}$ to SCET$_{\rm II}$. Due to the simplicity of the operator, this second stage of matching simply replaces ultrasoft fields by the soft fields of SCET$_{\rm II}$~\cite{Bauer:2003mga}, thus yielding the desired result.}  The SCET amplitude result in \Eq{eq:ON_soft_factorized} has also been explicitly derived from full QCD without relying on the SCET machinery at both tree level and loop level~\cite{Feige:2013zla,Feige:2014wja}.  

For future use, we will define the hatted $\hat {\cal O}_N^{(0)}$ to contain just the collinear fields, where we pull out the Wilson coefficient and the soft fields after the field redefinition:
\begin{align} \label{eq:Onhat}
{\mathcal O}_N^{(0)} 
 = C_N^{(0)}\big( \{ Q_i  \}\big)
 \otimes \hat{{\cal O}}_N^{(0)} 
  \otimes \text{T}\Big\{ \prod_{i=1}^{N}Y_{n_i}^{\kappa_i}(0)\Big\} \,.
\end{align}

\subsection{SCET Lagrangian at Subleading Power}
\label{subsec:scetsublag}

As mentioned in the SCET calculation of the leading soft factor, the relationship between the SCET power counting parameter $\lambda$ and the amplitude scaling factor $\epsilon$ is $\lambda^2\sim\epsilon$.  
Therefore the leading soft factor in SCET scales like $\lambda^{-2}$ and the subleading soft factor from the Low-Burnett-Kroll theorem scales like $\lambda^0$.  In the effective theory, this is therefore an ${\cal O}(\lambda^2)$ correction to the interaction of soft emissions (that is, second order in the power expansion).  Thus in SCET one must first consider terms that 
contribute ${\cal O}(\lambda)$ soft corrections, which could come from an ${\cal L}^{(1)}$ Lagrangian or an ${\cal O}_N^{(1)}$ hard scattering operator.  Although there exist no ${\cal O}_N^{(1)}$ hard scattering operators which have explicit soft fields, we must still consider these operators together with an insertion of ${\cal L}^{(0)}_{\rm soft}$. (We will later show that both of the ${\cal O}(\lambda)$ contributions vanish at tree level under the conditions of the Low-Burnett-Kroll theorem.)
In this section, we will present the SCET Lagrangians at both ${\cal O}(\lambda)$ and ${\cal O}(\lambda^2)$ and in \Sec{sec:sub_ope}, we will discuss the power suppressed operators that contribute to $N$-point scattering up through ${\cal O}(\lambda^2)$.

Starting at ${\cal O}(\lambda)$ we have several contributions to the subleading SCET Lagrangian
\begin{align}
  {\cal L}^{(1)} &=  {\cal L}_{\xi_n}^{(1)}  + {\cal L}_{A_n}^{(1)} 
  + {\cal L}_{\phi_n}^{(1)} + {\cal L}_{\xi_n \psi_s}^{(1)} 
  + {\cal L}_{\phi_n\phi_s}^{(1)} \,.
\end{align}
The operators that contribute to these Lagrangians scale as $\lambda^5$. 
The collinear fermion and gluon Lagrangians at ${\cal O}(\lambda)$  in our notation are~\cite{Chay:2002vy,Manohar:2002fd,Bauer:2003mga} 
\begin{align}\label{eq:sublagcollq}
{\cal L}_{\xi_n}^{(1)} &= \bar \chi_n \Big(
  i \slashed{D}_{s\perp}\frac{1}{ i\bar{n}\cdot \partial_n} 
    i \slashed{\cal D}_{n\perp}
  +i \slashed{\cal D}_{n\perp}\, \frac{1}{i\bar{n}\cdot \partial_n} 
    i \slashed{D}_{s\perp} 
 \Big)\frac{\slashed{\bar{n}}}{2} \chi_n \ ,
\nonumber \\ 
{\cal L}_{A_n}^{(1)}&= \frac{2}{g^2}\text{Tr}\Big(
 \big[ i {\cal D}_{ns}^\mu,i {\cal D}_{n\perp}^\nu \big]\big[
  i {\cal D}_{ns\mu},iD_{s\perp \,\nu} 
\big]
\Big)+ {\cal L}_{A_n,\text{gf}}^{(1)}  \ .  
\end{align}  
In a covariant gauge, the gauge fixing terms at subleading power are \cite{Pirjol:2002km}
\begin{equation}
{\cal L}_{A_n,\text{gf}}^{(1)} = \frac{2}{\alpha} \text{Tr}\Big(
\big[
iD_{s\perp}^{\mu},A_{n\perp\,\mu}
\big]
\big[
i \partial_{ns}^\nu, A_{n\nu}
\big]
\Big)+\text{ghosts} \,.
\end{equation}
Here, the covariant derivative $iD_{s\perp}^\mu = i\partial^\mu_{s\perp}+g A_{s\perp}^\mu$ and the other covariant derivatives contain Wilson lines that ensure collinear gauge invariance, and were defined in \Eq{eq:iDc}. It is straightforward to work out the analogous power suppressed SCET Lagrangian involving collinear scalars, which is
\begin{align}
{\cal L}_{\phi_n}^{(1)} &= 
 2 {\rm Tr}\:\big[ \Phi_n^* \big( D_{s\perp}^\mu {\cal D}^\perp_{n\mu} 
 +{\cal D}^\perp_{n\mu}  D_{s\perp}^\mu
 \big) \Phi_n \big] \,.
\end{align}
The complete subleading SCET Lagrangian at ${\cal O}(\lambda)$ for fermions,  gauge bosons and scalars also contains an additional Lagrangian that permits soft fermion or soft scalar emission,
\begin{align} \label{eq:Lxiq}
{\cal L}_{\xi_n \psi_s}^{(1)} 
  &= (\bar{\xi}_n W_n) \frac{1}{i\bar{n}\cdot \partial_n}  g \slashed{B}_{n\perp} \psi_s+\text{h.c.,}
  \\
{\cal L}_{\phi_n\phi_s}^{(1)} 
  & = -2ig\, {\rm Tr}\big[ \big(\Phi_n^* {\cal D}_{n\perp}^\mu  B_{n\mu}^\perp \big) \phi_s \big]+\text{h.c.,}
\end{align}
where $\psi_s\sim \lambda^3$ is the soft fermion field and $\phi_s= \phi_s^A T^A\sim \lambda^2 $ is the adjoint soft scalar field. Since these operators involve a soft fermion or soft scalar rather than a soft gauge boson the study of the general structure of the amplitudes they generate is beyond the goals of this paper. However, we do note that ${\cal L}_{\xi_n\psi_s}^{(1)}$ can play a crucial role in the study of physical QCD processes, for  example \Ref{Mantry:2003uz}. Eq.~(\ref{eq:Lxiq}) is also interesting for a supersymmetric theory like ${\cal N}=4$ SYM, where these Lagrangians should play a role in understanding the soft limits of superfields.

At ${\cal O}(\lambda^2)$ the sub-subleading SCET Lagrangian contains the terms, 
\begin{align}
  {\cal L}^{(2)} &= {\cal L}_{\xi_n}^{(2)}  + {\cal L}_{A_n}^{(2)} 
    + {\cal L}_{\phi_n}^{(2)} + {\cal L}_{\xi_n \psi_s}^{(2)} 
  + {\cal L}_{\phi_n\phi_s}^{(2)} \,.
\end{align}
The operators that contribute to these Lagrangians scale as $\lambda^6$. 
At this order the collinear fermion and gluon Lagrangians in our notation are~\cite{Manohar:2002fd,Bauer:2003mga}
\begin{align}\label{eq:subsublag}
{\cal L}_{\xi_n}^{(2)} &= \bar\chi_n \bigg(
 i \slashed{D}_{s\perp}\frac{1}{ i\bar{n}\cdot \partial_n}i \slashed{D}_{s\perp}
  - i \slashed{\cal D}_{n\perp} \frac{i\bar{n}\cdot D_s}{(i\bar{n}\cdot \partial_n)^2}  i \slashed{\cal D}_{n\perp}
 \bigg)\frac{\slashed{\bar{n}}}{2} \chi_n \,,  \\[5pt]
{\cal L}_{A_n}^{(2)}&=  \frac{1}{g^2}\text{Tr}\Big(
\big[
i {\cal D}_{ns}^\mu, i D_s^{\perp \nu} 
\big]\big[
i {\cal D}_{ns\mu}, i D^\perp_{s\nu}
\big]
\Big)
+\frac{1}{g^2}\text{Tr}
\Big( \big[
i D_{s\perp}^{\mu},  iD_{s\perp}^{\nu} 
\big]\big[
i {\cal D}_{n\mu}^{\perp},  i {\cal D}^\perp_{n\nu}
\big] \Big)
\nonumber \\
& 
+\frac{1}{g^2}\text{Tr}
\Big( \big[
i {\cal D}_{ns}^{\mu},  in\cdot {\cal D}_{ns}
\big]\big[
i {\cal D}_{ns\mu},  i \bar{n}\cdot D_{s}
\big] \Big)
+\frac{1}{g^2}\text{Tr}
\Big( \big[
 i D_{s\perp}^{\mu},  i{\cal D}_{n\perp}^{\nu} 
\big]\big[
i {\cal D}_{n\mu}^{\perp},  i D^\perp_{s\nu}
\big] \Big)\nonumber \\
&+ {\cal L}_{A_n,\text{gf}}^{(2)}
\,. \nonumber
\end{align}
In a covariant gauge, the gauge fixing terms at sub-subleading power are \cite{Pirjol:2002km}
\begin{equation}
{\cal L}_{A_n,\text{gf}}^{(2)}=\frac{1}{\alpha}\text{Tr}
\Big(
\big[
 i D_{s\perp}^{\mu}, A_{n\perp \mu}
\big]
\big[
 i D_{s\perp}^{\nu}, A_{n\perp \nu}
\big]
\Big)+
\frac{1}{\alpha}\text{Tr}
\Big(
\big[
 i \bar{n}\cdot D_{s},n\cdot A_n
\big]
\big[
 i \partial_{ns}^{\mu},A_{n\mu}
\big]
\Big)+\,\text{ghosts} \, .
\end{equation}
It is straightforward to write down the analogous power suppressed SCET Lagrangian involving collinear scalars, which is
\begin{align}
{\cal L}_{\phi_n}^{(2)} &= 2 \, {\rm Tr} \Big[
\Phi_n^* \Big( 
\frac12  n\cdot {\cal D}_{ns}\, \bn\cdot D_s  
+\frac12  \bn\cdot D_s\, n\cdot {\cal D}_{ns}
+  D_{s\perp}^2  
\Big) \Phi_n \Big]  \,.
\end{align}
At this order there are also ${\cal O}(\lambda^2)$ Lagrangians involving soft fermions and soft scalars, ${\cal L}_{\xi_n \psi_s}^{(2)} 
  + {\cal L}_{\phi_n\phi_s}^{(2)}$, which are higher order versions of \Eq{eq:Lxiq}. Though ${\cal L}_{\xi_n \psi_s}^{(2)}$ is known in the literature, and it is straightforward to determine ${\cal L}_{\phi_n\phi_s}^{(2)}$, neither of these soft fermion/scalar Lagrangians will be needed for our analysis here. Note that all of the Lagrangians discussed in this section are individually gauge invariant.

For later purposes it will also be useful to consider the form that the subleading Lagrangians take after the BPS field redefinition in \Eq{eq:BPS}. The field redefinition introduces Wilson lines $Y_n$ which factor from the collinear fields in a manner so that they always sandwich the soft covariant derivatives, $Y_n^\dagger D_s^\mu Y_n$. In the process use of $in\cdot D_s Y_n =0$ causes soft gauge fields to drop out of the mixed covariant derivatives, $iD_{ns}^\mu \to i D_n^\mu$ and $i {\cal D}_{ns}^\mu \to i {\cal D}_n^\mu$.  In order to fully factor the soft and collinear fields we also want to separate out terms where the derivative in $D_s^\mu$ acts on collinear fields, which we can do with the identity
\begin{align} \label{eq:YDY}
 Y_n^\dagger i D_s^\mu Y_n & =   i \partial_s^\mu  + \big[ Y_n^\dagger i D_s^\mu Y_n \big] 
     =  i \partial_s^\mu +  T^A g B_{s(n)}^{A\mu} \,,
\end{align}
where the covariant derivative acts only on terms within the square brackets, and 
\begin{align}
 g B_{s(n)}^{A\mu} &= \Big[ \frac{1}{in\cdot \partial_s} n_\nu i F_s^{B\nu\mu} {\cal Y}_n^{BA}\Big] \,.
\end{align}
Here $F_s^{B\nu\mu}$ is the soft field strength, and ${\cal Y}_n$ is the soft Wilson line in the adjoint representation. This allows us to write the sum of all subleading Lagrangians in a factorized form as
\begin{align} \label{eq:LKB}
{\cal L}^{(1)} &=  \sum_n \Big[  \hat K_n^{(1)} 
  + \hat K_{n\mu}^{(1) \kappa} T^{\kappa A} g B_{s(n)}^{A\mu} \Big] \,,
 \\
 {\cal L}^{(2)} &=  \sum_n \Big[ \hat K_n^{(2)} + \hat K_{n\mu}^{(2) \kappa} T^{\kappa A}  g B_{s(n)}^{A\mu}  
   + \hat K_{n\mu\nu}^{(2)\kappa\kappa'} T^{\kappa A} T^{\kappa' B} g B_{s(n)}^{A\mu} g B_{s(n)}^{B\nu}\Big] 
   \,. \nn
\end{align} 
Here $T^{\kappa A}$ is the $A$'th component of the color generator in the $\kappa$ representation upon which the $i D_s^\mu$ acted. The various $\hat K_n^{(1)}$ and $\hat K_n^{(2)}$ terms contain only $n$-collinear quark, gauge boson, and scalar fields, plus $i\partial_s^\mu$ derivatives, and can be written down explicitly with the results given above.  All terms involving soft fields have been made explicit in the $gB_{s(n)}^{A\mu}$ factors.\footnote{Note that here the superscripts $(1)$ or $(2)$ on the $\hat K_n$s denote the Lagrangian that these terms came from rather than their power of $\lambda$.} 

\subsection{Reparametrization Invariance of the Effective Theory}
\label{subsec:rpi}

Because a jet or collinear particle in SCET is defined with a light-like direction $n$, there is a preferred coordinate system (also involving $\bn$) which partially breaks the Lorentz symmetry.  Rotations in the $\perp$ plane are unbroken, while the transformations that correspond to the 5 Lorentz generators  $n_\mu M^{\mu\nu}$ and $\bar{n}_\mu M^{\mu\nu}$ are broken by the presence of the auxiliary vecors $n$ and $\bn$.  
Even for these broken generators there is a residual symmetry, namely a reparameterization invaraiance (RPI)~\cite{Chay:2002vy,Manohar:2002fd}. One part of the RPI encodes our ability to make different choices for $n$ and $\bn$ without changing the physics, while the other part encodes our ability to shift small contributions between $i\bn\cdot\partial_n$ and $i\bn\cdot\partial_s$, and between $i\partial_{n\perp}^\mu$ and $i\partial_{s\perp}^\mu$ which are parametrically different, and are treated with a multipole expansion in SCET.

When considering choices for the vectors $n$ and $\bar{n}$, they only must satisfy
\begin{equation}\label{eq:rpiconsts}
n^2=0\ ,\qquad \bar{n}^2 = 0 \ , \qquad n\cdot \bar{n} = 2 \ ,
\end{equation}
plus the constraint that $n$ must be parametrically close, by ${\cal O}(\lambda)$, to the physical collinear particle or jet direction.
We are free to choose $n$ and $\bar{n}$ arbitrarily as long as these constraints are satisfied.  There are three possible sets of RPI transformations that maintain \Eq{eq:rpiconsts}:
\begin{equation}\label{eq:RPI_Trans}
\begin{aligned}
&{\bf RPI\mbox{-}I}\\
&n_\mu\to n_\mu+\Delta^\perp_\mu  \\
&\bar{n}_\mu\to \bar{n}_\mu 
\end{aligned}
\qquad
\qquad
\begin{aligned}
&{\bf RPI\mbox{-}II}\\
&n_\mu \to n_\mu  \\
&\bar{n}_\mu\to \bar{n}_\mu + \epsilon^\perp_\mu 
\end{aligned}
\qquad
\qquad
\begin{aligned}
&{\bf RPI\mbox{-}III}\\
&n_\mu \to e^\alpha n_\mu  \\
&\bar{n}_\mu\to e^{-\alpha}\bar{n}_\mu
\end{aligned}
\end{equation}
For RPI-I the size of change is constrained by the collinear power counting, $\Delta^\perp\sim \lambda $, whereas the transformations for RPI-II and RPI-III are unconstrained, $\epsilon^\perp\sim \lambda^0$ and $\alpha\sim \lambda^0$. In \Eq{eq:RPI_Trans} the parameters $\Delta_\mu^\perp$ and $\epsilon_\mu^\perp$ are infinitesimal, and satisfy $n\cdot \Delta^\perp = \bar{n}\cdot \Delta^\perp = n\cdot \epsilon^\perp =\bar{n}\cdot \epsilon^\perp = 0$.\footnote{\Ref{Baumgart:2010qf} discusses finite RPI transformations.  }
These RPI-I and RPI-II transformations correspond to transverse translations of the vectors $n$ and $\bar{n}$ respectively.  If $n$ and $\bn$ are back-to-back vectors then the RPI-III transformation corresponds to a boost in the $n$ direction by a finite parameter $\alpha$. 

Another part of the RPI is a connection between collinear and soft derivatives. By RPI every large collinear derivative must also come together with the analogous smaller soft derivative, $i\partial_{n\perp}+i\partial_{s\perp}$ and $i\bn\cdot\partial_n+ i\bn\cdot\partial_s$. Combining this with gauge invariance implies that we always have the combinations
\begin{align}\label{eq:RPI_to_soft_covariant_der_I}
  & i D_{n\perp}^\mu + W_n iD_{s\perp}^\mu W_n^\dagger \,,
  & i \bn\cdot D_n + W_n i\bn\cdot D_s W_n^\dagger \,.
\end{align}
Using the identities obeyed by the Wilson line, these combinations can be written as
\begin{align}\label{eq:RPI_to_soft_covariant_der_II}
  i {\cal D}_{n\perp}^\mu + i D_{s\perp}^\mu \,,
 &&   i \bn\cdot \partial_n  + i \bn\cdot D_s  \,.
\end{align}
For later purposes we list the available SCET field objects which are themselves RPI invariant that are relevant for our analysis
\begin{align} \label{eq:rpiops}
 & \Psi_{n_i}^{\rm RPI} = {\cal W}_n^\dagger \psi_n \,,  
  & & {\cal G}_n^{\mu\nu\,{\rm RPI}} = {\cal W}_n^\dagger G_n^{\mu\nu} {\cal W}_n  \,, 
  & \Phi_n^{\rm RPI} &= {\cal W}_n^\dagger \phi_n {\cal W}_n \,,
  \nn\\
  & i\partial_n^\mu \,,
  & & \delta^4(Q^\mu - i\partial_n^\mu) \,, 
  & F_s^{\mu\nu} & \,.
\end{align}
Here $\psi_n$ and $G_n^{\mu\nu}$ are the RPI collinear fermion field and gluon field strength, ${\cal W}_n$ is the RPI Wilson line operator, and explicit formulas can be found in Ref.~\cite{Marcantonini:2008qn}. We take these objects to be invariant under both types of RPI transformations by always including soft derivatives in the combinations appearing in \Eq{eq:RPI_to_soft_covariant_der_II}.  (For processes with additional external vectors, $q^\mu$, we may also form additional invariants like $\delta(\omega - q\cdot i \partial_n)$.)

Accounting for both of these RPI relationships, we have connections between the leading and subleading SCET Lagrangians and operators. For example, RPI implies that there are no nontrivial Wilson coefficients for the subleading Lagrangians in \Eqs{eq:sublagcollq}{eq:Lxiq} to all orders in the coupling expansion~\cite{Beneke:2002ni,Manohar:2002fd}. 

\subsection{SCET $N$-Jet Operators to Subleading Power}
\label{sec:sub_ope}

The SCET Lagrangian describes soft and collinear emissions from external particles and in particular, do not correspond to modifications of the hard interaction.  Beyond leading power, there will in general be modifications to the hard interaction, and these will include the emission of soft particles from internal lines in a diagram.  The physics of the hard interaction in the effective theory is described by operators localized at the origin, where the hard scattering takes place.  By contrast, the SCET Lagrangian describes the physics far from the hard interaction, at a scale set by the virtuality of the external particles.  Thus, for a complete description of the subleading soft factor in SCET, we need to identify the operators at subleading power that could contribute. 

Recall the $N$-jet operator that creates $N$ hard particles at leading power from \Eq{eq:ON_soft_factorized}:
\begin{align}
{\mathcal O}_N^{(0)} &=  C_N^{(0)}\big( \{ Q_i  \}\big) \otimes \prod_{i=1}^N  \Big[\delta(\bn_i\cdot Q_i - \bn_i\cdot i\partial_{n_i})  X_{n_i}^{\kappa_i}(0)  \Big] \otimes \text{T}\Big\{ \prod_{i=1}^{N}Y_{n_i}^{\kappa_i}(0)\Big\} \nn\\
&= C_N^{(0)}\big( \{ Q_i  \}\big) \otimes 
  \hat{{\mathcal O}}_N^{(0)} \otimes \text{T}\Big\{ \prod_{i=1}^{N}Y_{n_i}^{\kappa_i}(0)\Big\} \
\,,
\end{align}
where the $X_{n_i}$ are fields that create scalar, fermion, or vector excitations, the $Y_{n_i}$s are soft gluon Wilson lines, and $\otimes$ indicates color and Lorentz index contractions. For an operator with $N$-jets ($N$ collinear sectors) the only possibilities for obtaining an ${\cal O}(\lambda)$ suppression to get a ${\mathcal O}_N^{(1)}$ are from having two $X_{n_i}$ factors in one of the collinear sectors which we denote as ${\mathcal O}_N^{(1,X)}$, or from having one $i\partial_{n_i\perp}^\mu X_{n_i}$ which we denote as ${\mathcal O}_N^{(1,\partial)}$,
\begin{align} \label{eq:sub_ops}
&{\mathcal O}_{N}^{(1,\partial)} 
= C_{N\alpha}^{(1\partial)}\big( \{ Q_i  \}\big)\otimes\sum_{k=1}^N \prod_{i=1,i\neq k}^{N}  \Big[\delta(\bn_i\cdot Q_i - \bn_i\cdot i\partial_{n_i})  X_{n_i}^{\kappa_i}(0)  \Big]
 \\*
&\qquad\qquad \times 
\Big[\delta(\bn_k\cdot Q_k - \bn_k\cdot i\partial_{n_k}) 
  \: i \partial_{n_k\perp}^\alpha
 X_{n_k}^{\kappa_k}(0) \Big]
\otimes \text{T}\Big\{ \prod_{i=1}^{N}Y_{n_i}^{\kappa_i}(0)\Big\}
 \,, \nn
 \\
&{\mathcal O}_{N}^{(1,X)} 
= C_N^{(1X)}\big( \{ Q_i  \}\big)\otimes\sum_{k=1}^N \prod_{i=1,i\neq k}^{N}  \Big[\delta(\bn_i\cdot Q_i - \bn_i\cdot i\partial_{n_i})  X_{n_i}^{\kappa_i}(0)  \Big]
 \nn \\
&\qquad\qquad \times 
\Big[\delta(\bn_k\cdot Q_k - \bn_k\cdot i\partial_{n_k}) 
  X_{n_k}^{\kappa_k}(0) 
    \delta(\bn_k\cdot Q_k' - \bn_k\cdot i\partial_{n_k})
   X_{n_k}^{\kappa_k'}(0) \Big]
\otimes \text{T}\Big\{ Y_{n_k}^{\kappa_k'} \prod_{i=1}^{N}Y_{n_i}^{\kappa_i}(0)\Big\}
 \,. \nn
\end{align} 
In the following, we will use the notation
\begin{equation}
   \delta(\bn_k\cdot Q_k' - \bn_k\cdot i\partial_{n_k})
   X_{n_k}^{\kappa_k'}(0) \equiv X_{n_k,\omega_k'}^{\kappa_k'}(0)
\end{equation}
to express the large momentum fraction of the fields in a single collinear sector.  These operators describe the subleading collinear limits of two particles. At tree level ${\mathcal O}_N^{(1,X)}$ must produce two energetic collinear particles. Neither of these operators directly produces a soft gluon.    The operator ${\mathcal O}_N^{(1,\partial)}$ will be connected by RPI to the $N$-jet operator ${\cal O}_N^{(0)}$, but for some of the allowed Lorentz and color combinations ${\mathcal O}_N^{(1,X)}$ will be not be connected.   For the application of fixed-order amplitudes, we can exploit RPI to set the total $\perp$ component of momentum of each collinear sector to zero. Since $ i \partial_{n_k\perp}^\alpha X_{n_k}^{\kappa_k}(0)$ is proportional to the total $\perp$ momenta of the $n_k$ collinear direction we then have $ {\mathcal O}_{N}^{(1,\partial)} =0$. Since the operators ${\mathcal O}_{N}^{(1,X)} $ are not all connected by RPI, in general the matching coefficient $C_N^{(1X)}$ must be determined by considering collinear limits of full theory amplitudes.  

At ${\cal O}(\lambda^2)$ there are several distinct sources for operators ${\cal O}_N^{(2)}$. One possibility are purely collinear operators involving 3 $X_{n_i}$s, two $X_{n_i}$s and two $X_{n_j}$s, or cases where one or more of these $X_{n_i}$s are replaced by a $\partial_{n_i\perp}^\mu$. This gives seven types of field content for these operators:
\begin{align} \label{eq:sub_ops2}
&{\mathcal O}_{N}^{(2,\partial^2)} 
= C_{N\alpha\beta}^{(2\partial^2)}\big( \{ Q_i  \}\big)\otimes
  \sum_{k=1}^N \prod_{i=1,i\neq k}^{N}  \Big[
  \delta(\bn_i\!\cdot\! Q_i - \bn_i\!\cdot\! i\partial_{n_i})  
   X_{n_i}^{\kappa_i}(0)  \Big]
 \\
&\qquad\qquad \times 
\Big[\delta(\bn_k\!\cdot\! Q_k - \bn_k\!\cdot\! i\partial_{n_k}) 
  i \partial_{n_k\perp}^\alpha\, i \partial_{n_k\perp}^\beta 
 X_{n_k}^{\kappa_k}(0) \Big]
\otimes \text{T}\Big\{ \prod_{i=1}^{N}Y_{n_i}^{\kappa_i}(0)\Big\}
 \,, \nn
 \\
&{\mathcal O}_{N}^{(2,\partial,\partial)} 
= C_{N\alpha\beta}^{(2\partial\partial)}\big( \{ Q_i  \}\big)\otimes
\sum_{k=1}^N\sum_{j=1}^N \prod_{i=1,i\neq k,j}^{N}  
 \Big[\delta(\bn_i\!\cdot\! Q_i - \bn_i\!\cdot\! i\partial_{n_i})  X_{n_i}^{\kappa_i}(0)  \Big]
 \otimes \text{T}\Big\{ \prod_{i=1}^{N}Y_{n_i}^{\kappa_i}(0)\Big\}
 \nn \\*
&\qquad\qquad \otimes 
\Big[\delta(\bn_k\!\cdot\! Q_k - \bn_k\!\cdot\! i\partial_{n_k}) 
  i \partial_{n_k\perp}^\alpha 
 X_{n_k}^{\kappa_k}(0) \Big]
 \Big[\delta(\bn_{j}\!\cdot\! Q_{j} - \bn_{j}\!\cdot\! i\partial_{n_{j}}) 
  i \partial_{n_{j}\perp}^\beta
 X_{n_{j}}^{\kappa_{j}}(0) \Big]
 , \nn \\
&{\mathcal O}_{N}^{(2,\partial X)} 
= C_{N\alpha}^{(2\partial X)}\big( \{ Q_i  \}\big)\otimes
  \sum_{k=1}^N \prod_{i=1,i\neq k}^{N}  \Big[
  \delta(\bn_i\!\cdot\! Q_i - \bn_i\!\cdot\! i\partial_{n_i})  
   X_{n_i}^{\kappa_i}(0)  \Big]
 \otimes \text{T}\Big\{ Y_{n_k}^{\kappa_k'}
  \prod_{i=1}^{N}Y_{n_i}^{\kappa_i}(0)\Big\}
 \nn \\
&\qquad\qquad \otimes 
\Big[\delta(\bn_k\!\cdot\! Q_k - \bn_k\!\cdot\! i\partial_{n_k}) 
  i \partial_{n_k\perp}^\alpha
 \Big(X_{n_k}^{\kappa_k}(0) X_{n_k,\omega_k'}^{\kappa_k'}(0)\Big) \Big]
 \,, \nn \\
&{\mathcal O}_{N}^{(2,X\partial )} 
=   C_{N\alpha}^{(2X\partial )}\big( \{ Q_i  \}\big)\otimes
  \sum_{k=1}^N \prod_{i=1,i\neq k}^{N}  \Big[
  \delta(\bn_i\!\cdot\! Q_i - \bn_i\!\cdot\! i\partial_{n_i})  
   X_{n_i}^{\kappa_i}(0)  \Big]
 \otimes \text{T}\Big\{ Y_{n_k}^{\kappa_k'}
  \prod_{i=1}^{N}Y_{n_i}^{\kappa_i}(0)\Big\}
  \nn \\*
&\qquad\qquad \otimes 
\Big[\delta(\bn_k\!\cdot\! Q_k - \bn_k\!\cdot\! i\partial_{n_k}) 
  X_{n_k,\omega_k'}^{\kappa_k'}(0)\, i \partial_{n_k\perp}^\alpha
 X_{n_k}^{\kappa_k}(0)  \Big]
 \,, \nn \\
&{\mathcal O}_{N}^{(2,X,\partial )} 
=   C_{N\alpha}^{(2X,\partial )}\big( \{ Q_i  \}\big)\otimes
  \sum_{k=1}^N \sum_{j=1,j\ne k}^N  \prod_{i=1,i\neq k,j}^{N}  \Big[
  \delta(\bn_i\!\cdot\! Q_i - \bn_i\!\cdot\! i\partial_{n_i})  
   X_{n_i}^{\kappa_i}(0)  \Big]
\otimes \text{T}\Big\{ Y_{n_k}^{\kappa_k'}
  \prod_{i=1}^{N}Y_{n_i}^{\kappa_i}(0)\Big\}
 \nn \\*
&\qquad\qquad \otimes 
\Big[\delta(\bn_k\!\cdot\! Q_k - \bn_k\!\cdot\! i\partial_{n_k})  
 X_{n_k}^{\kappa_k}(0) X_{n_k,\omega_k'}^{\kappa_k'}(0) \Big]
 \Big[\delta(\bn_{j}\!\cdot\! Q_{j} - \bn_{j}\!\cdot\! i\partial_{n_{j}}) 
  i \partial_{n_{j}\perp}^\alpha
 X_{n_{j}}^{\kappa_{j}}(0) \Big]
 , \nn \\
&{\mathcal O}_{N}^{(2,X^2)} 
= C_N^{(2X^2)}\big( \{ Q_i  \}\big)\otimes\sum_{k=1}^N \prod_{i=1,i\neq k}^{N}  \Big[\delta(\bn_i\!\cdot\! Q_i - \bn_i\!\cdot\! i\partial_{n_i})  X_{n_i}^{\kappa_i}(0)  \Big]
 \nn \\*
&\qquad\qquad \times 
\Big[\delta(\bn_k\!\cdot\! Q_k - \bn_k\!\cdot\! i\partial_{n_k})
  X_{n_k}^{\kappa_k}(0) X_{n_k,\omega_k'}^{\kappa_k'}(0) X_{n_k,\omega_k''}^{\kappa_k''}(0) \Big]
\otimes \text{T}\Big\{ Y_{n_k}^{\kappa_k'}  Y_{n_k}^{\kappa_k''}  \prod_{i=1}^{N}Y_{n_i}^{\kappa_i}(0)\Big\}
 , \nn \\
&{\mathcal O}_{N}^{(2,X,X)} 
=   C_{N}^{(2XX )}\big( \{ Q_i  \}\big)\otimes
  \sum_{k=1}^N \sum_{j=1,j\ne k}^N  \prod_{i=1,i\neq k,j}^{N}  \Big[
  \delta(\bn_i\!\cdot\! Q_i - \bn_i\!\cdot\! i\partial_{n_i})  
   X_{n_i}^{\kappa_i}(0)  \Big]
 \nn \\*
&\qquad\qquad \times 
\Big[\delta(\bn_k\!\cdot\! Q_k - \bn_k\!\cdot\! i\partial_{n_k})  
 X_{n_k}^{\kappa_k}(0) X_{n_k,\omega_k'}^{\kappa_k'}(0) \Big]
 \Big[\delta(\bn_{j}\!\cdot\! Q_{j} - \bn_{j}\!\cdot\! i\partial_{n_{j}}) 
 X_{n_{j}}^{\kappa_{j}}(0)  X_{n_{j},\omega_j'}^{\kappa_j'}(0)  \Big]
  \nn\\*
& \qquad\qquad  \otimes 
 \text{T}\Big\{ Y_{n_k}^{\kappa_k'} Y_{n_j}^{\kappa_j'}
  \prod_{i=1}^{N}Y_{n_i}^{\kappa_i}(0)\Big\}
\,. \nn
\end{align} 
When referring to these terms we will denote all of these operators as ${\cal O}_N^{(2X/\partial)}$.\footnote{Since $\partial (A B)=(\partial A)B+A\partial B$, it suffices to only consider the two operators ${\mathcal O}_{N}^{(2,\partial X)},\,{\mathcal O}_{N}^{(2,X\partial)}$ given above for the cases that have both multiple collinear fields and derivatives in a single sector.} Again these operators do not directly produce soft gluons. Similar to the operator ${\mathcal O}_N^{(1,\partial)}$, we can use RPI to set the total $\perp$ momenta in each collinear sector to zero.  This means that the operators $ {\mathcal O}_{N}^{(2,\partial^2)}$, ${\mathcal O}_{N}^{(2,\partial,\partial)}$, ${\mathcal O}_{N}^{(2,\partial X)}$, and ${\mathcal O}_{N}^{(2,X,\partial )}$ can all be set to zero by RPI, and from the operators given in \Eqs{eq:sub_ops}{eq:sub_ops2} only
\begin{align} \label{eq:O2nonrpi}
  {\mathcal O}_{N}^{(1,X)} \,, \qquad
  {\mathcal O}_{N}^{(2,X\partial )} \,, \qquad
  {\mathcal O}_{N}^{(2,X^2)} \,, \qquad
  {\mathcal O}_{N}^{(2,X,X)} \,,
\end{align}
must be considered in our analysis. These operators describe subleading collinear limits and can have matching coefficients that are not fixed by RPI, and hence unrelated to that of ${\cal O}_N^{(0)}$. There can also be RPI relations between the operators in  \Eq{eq:O2nonrpi} with or without ${\cal O}_N^{(0)}$.  In general the terms in \Eq{eq:O2nonrpi} also involve convolution integrals between the Wilson coefficients and the collinear operators. This occurs because only the total momentum in a collinear sector is fixed externally by momentum conservation, so when there are two or more collinear building block operators present we have a convolution integral over the fraction of this total hard momentum that each of them carries.

In addition we can have operators which directly produce a soft gluon along with  $N$ jets, involving a $D_s^\mu\sim\lambda^2$.  We present the calculation that shows how these operators arise in \App{app:rpi}, including the demonstration that they all are uniquely determined by the RPI symmetry at this order. They arise from two sources.    One source is the expansion of the momentum constraints on the collinear sectors that feed into the hard interaction, which is constrained by RPI and can be obtained following the logic in~\cite{Marcantonini:2008qn}. This produces derivatives that act on the $\delta$-functions constraining the collinear sectors, which can be transformed into derivatives on the hard Wilson coefficient.  After the BPS field redefinition this results in the ${\cal O}(\lambda^2)$ operator
\begin{align}\label{eq:subsub_op_delta}
{\mathcal O}_{N}^{(2,\delta)}
 &=  - \sum_{k=1}^N
  \frac{\partial}{\partial \bar{n}_k\cdot Q_k} C_N^{(0)}\big( \{ Q_i  \}\big) 
  \otimes \prod_{i=1}^{N}  \Big[\delta(\bn_i\cdot Q_i - \bn\cdot i\partial_n)  X_{n_i}^{\kappa_i}(0)  \Big]
 \nonumber  \\ 
&\qquad 
  \otimes \text{T}\Big\{ \bn_k \cdot  g B_{s}^{(n_k)A} T^{\kappa_k A} \prod_{i=1}^{N}Y_{n_i}^{\kappa_i}(0) \Big\}
\nn\\
&= - \sum_k  
  \frac{\partial C_N^{(0)}\big( \{ Q_i  \}\big) }{\partial \bar{n}_k\cdot Q_k} 
  \otimes 
\hat{\cal O}_N^{(0)}\, \otimes
\text{T}\Big\{ \bn_k \cdot  g B_{s}^{(n_k)A}(0) T^{\kappa_k A}  \prod_{i=1}^{N}Y_{n_i}^{\kappa_i}(0) \Big\}
% g B_s^{(n_k)}\cdot \bar{n}_k
 \,.
\end{align}
The explicit derivative acts only on the Wilson coefficient, $C_N^{(0)}$. By expanding out the soft momenta from the $k$-th collinear sector's constraint $\delta(\bn_k\cdot Q_k - \bn\cdot i\partial_n)$ we obtained a $\bn_k \cdot i D_s$ acting on the $k$-th soft Wilson line. This was converted to a $g B_s^{(n_k)A\mu}$ with \Eq{eq:YDY}, and the term with $\bn_k\cdot i\partial_s X_{n_k}^{\kappa_k}$ was dropped using RPI.  We also wrote the result in terms of the purely collinear hatted operator $\hat{\cal O}_N^{(0)}$.

The next source of subleading $N$-jet operators involving a $D_s^\mu$ comes from the RPI completion of the collinear field operators in the leading order $N$-jet operator. This completion will contain both soft and collinear power suppressed operators, but here we only need the soft components. To the second order in the RPI completion of the leading order operator~\cite{Marcantonini:2008qn}, we have for the $X_{n_i}^{\kappa_i}(0)$ field operators (before the BPS field redefinition \Eq{eq:bps}) the relevant terms:
\begin{align}\label{eq:subsubfields}
\Psi_{n_i}^{\rm RPI}
& = \chi_{n_i} + \ldots 
 + \frac{1}{i \bar{n}_i\cdot \partial_{n_i}} i\slashed{D}_s^\perp \frac{\slashed{\bar{n}}_i}{2} \chi_{n_i} +... 
  \,, \\ 
 {\cal G}^{\mu \nu \text{RPI}}_{n_i}
&=  i\bn_i\cdot \partial_n\, n_i^{[\mu} B_{n_i\perp}^{\nu]} 
   + \ldots
  \,, \nn \\
 \Phi_{n_i}^{\rm RPI} &= \Phi_{n_i} + \ldots 
   \,. \nonumber
\end{align} 
In the ellipsis, we have dropped both the explicitly collinear subleading operators, which are necessary for complete RPI invariance but which do not induce direct couplings to soft gluons (the ellipses also include higher order terms beyond ${\cal O}(\lambda^3)$).  These non-displayed collinear terms that are ${\cal O}(\lambda)$ or ${\cal O}(\lambda^2)$ will induce RPI connections to some of the operators in \Eqs{eq:sub_ops}{eq:sub_ops2}, but we will not exploit these relations for our analysis, and hence they are not discussed here.
At this order it turns out that both $\Phi_n$ and ${B}_{n\perp}^\mu$ are not connected to operators with a soft gluon by RPI. Performing the BPS field redefinition, and using \Eq{eq:YDY}, we can then write down the other ${\cal O}(\lambda^2)$ $N$-jet operator:
\begin{align}\label{eq:subsub_op_rpi}
{\mathcal O}_{N}^{(2,r)} 
& = C_N^{(0)}\big( \{ Q_i  \}\big)\otimes\sum_{k=1}^N \prod_{i=1,i\neq k}^{N}  \Big[\delta(\bn_i\cdot Q_i - \bn\cdot i\partial_n)  X_{n_i}^{\kappa_i}(0)  \Big]
 \nn \\
&\qquad \times
\Big[\delta(\bn_k\cdot Q_k - \bn\cdot i\partial_n) \frac{t_k^\mu}{\bn_k\cdot Q_k} X_{n_k}^{\kappa_k}(0) \Big]
  \otimes \text{T}\Big\{   g B_{s\mu}^{(n_k)A} T^{\kappa_k A}  \prod_{i=1}^{N}Y_{n_i}^{\kappa_i}(0) \Big\} 
 \, . 
\end{align}
Again the BPS field redefinition gave the $Y_{n}^\kappa$s and allowed the soft fields to be factorized from the collinear fields.
Depending on the identity of the field in the collinear sector, the vector $t_k^\mu$ is
\begin{align} \label{eq:tk}
t_k^\mu=\begin{cases}
  \gamma_{\perp}^\mu\frac{\slashed{\bar{n}}_k}{2}\, &\text{collinear fermion ($X_n=\chi_n$)}\\
   0\, &\text{collinear gluon or scalar ($X_n={B}_{n\perp}, \Phi_n$)}
\end{cases}
\end{align}
Note that these operators ${\mathcal O}_{N}^{(2,\delta)}$, ${\mathcal O}_{N}^{(2,r)}$ are both gauge invariant by themselves, and are suppressed by $\lambda^2$ with respect to ${\cal O}_N^{(0)}$ because of the explicit soft derivative, $D_s^\mu$, that acts on a soft Wilson line. These operators can be used to analyze subleading soft effects both at tree-level and including loop corrections.

As we did with the leading power operator $\hat{{\cal O}}_N^{(0)}$ in \Eq{eq:Onhat},  we implicitly define a hatted notation to denote the collinear components of the operators at subleading power, which are thus independent of the Wilson coefficient and soft fields. These operators are determined after the BPS field redefinition. For the operators that can not be set to zero by RPI, these hatted  operators include
\begin{align}
 & \hat O_N^{(1,X)}\,,
 & \hat O_N^{(2,X\partial)} &\,,
 & \hat O_N^{(2,X^2)} &\,,
 & \hat O_N^{(2,X,X)} &\,,
 & \hat O_N^{(0,\delta)}= \hat O_N^{(0)} &\,,
 & \hat O_N^{(0,r)} &\,.
\end{align}
Here the number in the exponent indicates the power suppression in $\lambda$ for these collinear operators. For example we define the operator $\hat{\cal O}_{N\mu}^{(2,r_k)}$ via
\begin{align}\label{eq:hardcoll_subsoft}
{\cal O}_N^{(2,r)} 
&= C_N^{(0)}\big( \{ Q_i  \}\big)\otimes
 \sum_k \hat{\cal O}_{N\mu}^{(0,r_k)}\, \otimes 
\text{T}\Big\{  g B_{s}^{(n_k)A\mu}(0)T^{\kappa_k A}   \prod_{i=1}^{N}Y_{n_i}^{\kappa_i}(0) \Big\}
\,,\nn
\end{align}
where $\otimes$ denotes color-index contractions.

\section{Effective Theory Analysis of Subleading Soft Factor}
\label{sec:subsoft_zero}

In this section, we will present a detailed study of the subleading soft factor in gauge theory at tree-level in the framework of SCET.  In particular, we will show how the SCET Lagrangian ${\cal L}^{(2)}$ and the $N$-jet operators ${\cal O}_N^{(2\delta,2r)}$ defined in \Eqs{eq:subsub_op_delta}{eq:subsub_op_rpi} reproduce the Low-Burnett-Kroll theorem.  Unlike the Low-Burnett-Kroll theorem, the effective theory structures are well-defined to all-orders in perturbation theory, a point which we will utilize to derive a loop-level soft theorem in \Sec{sec:amps_nonzero}.  

\subsection{Power Counting Angular Momentum in the Subleading Soft Factor}
\label{sec:orbital}

From the power counting of SCET, we can make precise statements about various contributions to the subleading soft theorem in gauge theory.  Since it is gauge invariant by itself, let us just consider a single term in the subleading soft factor for the $i$th particle in the amplitude:
\begin{equation}\label{eq:subsoft_term}
  S_i^{\text{(sub)}}(s) = T^i\frac{\epsilon_{s\mu} p_{s\nu}J_i^{\mu\nu} }{p_i\cdot p_s} \ .
\end{equation}
When $p_i$ has a collinear scaling, $S_i^{\text{(sub)}}(s)$ has a Taylor series in $\lambda$, and the effective theory is setup to address the terms at each order in this $\lambda$ expansion. So, we must expand \Eq{eq:subsoft_term} in $\lambda$ to determine the precise subleading soft factor and the terms that contribute to it at tree-level in gauge theory.

Recall that all components of $p_{s}^{\nu}\sim \lambda^2$ and the power counting for the fields is $A_{us}^\mu\sim \lambda^2$. For the soft gluon state we have $|g_{us}\rangle\sim \lambda^{-2}$, since with relativistic normalization 
\begin{equation}
\langle g_{us}(p)|g_{us}(p')\rangle= 2 E_p \delta^3(\vec p-\vec p{\,}')\sim \lambda^{-4} \ .
\end{equation}
This implies that the soft polarization vector $\epsilon_{s}^\mu\sim \lambda^0$ since $\epsilon_s^\mu \sim A_{us}^\mu |g_{us}\rangle$.  

The power-counting of the dot product $p_i\cdot p_s$ is a bit more subtle as we need to know the power-counting of the momentum of particle $i$.  The correct power counting that does not require further assumptions is to consider particle $i$ to be a collinear particle with the momentum scaling as listed in \Tab{tab:scaling}. It may seem somewhat odd that we consider an external particle at tree-level to be ``collinear'' and at the same time require it to be well-separated in angle to all other particles in the amplitude.  We do this because the external particles must be described by on-shell fields in the effective theory. By assigning the collinear power counting to the momentum of particle $i$, and stating that it is the only collinear particle in this collinear sector, we are defining a region of phase space about the particle that scales with $\lambda$ where other particles are forbidden,  thus ensuring that external particles are at large angles to one another.  

Then, the dot product can be expanded as
\begin{equation}\label{eq:prop_exp_sub}
p_i\cdot p_s = 
 \underbrace{\frac{(\bar{n}\cdot p_i)(n\cdot p_s)}{2}}_{\sim \lambda^2}
 +\underbrace{p_{i\perp }\cdot p_{s\perp }}_{\sim \lambda^3} 
 +\underbrace{\frac{(n\cdot p_i)(\bar{n}\cdot p_s)}{2}}_{\sim \lambda^4} \ ,
\end{equation}
assuming that particle $i$ is in the $n$ direction.  It is then straightforward to determine the first few terms in the expansion of the propagator factor of the subleading soft term, \Eq{eq:subsoft_term}:
\begin{equation}
\frac{1}{p_i\cdot p_s} = 
 \frac{2}{(\bar{n}\cdot p_i)(n\cdot p_s)}
-\frac{4p_{i\perp }\cdot p_{s\perp }}{(\bar{n}\cdot p_i)^2(n\cdot p_s)^2}
+{\cal O}(\lambda^0) \ ,
\end{equation}
where the ${\cal O}(\lambda^0)$ terms not enhanced by inverse powers of $\lambda$ can be ignored for the analysis at the order at which LBK applies.  To complete the power counting, we need to expand the angular momentum $J_i^{\mu\nu}$ assuming it acts on a particle with collinear scaling.

To determine the expansion of the angular momentum in the subleading soft factor we first note that $J_i^{\mu\nu}$ is an operator that acts on an amplitude; it is not the field-valued operator.  One can formulate the angular momentum as an operator that acts on Green's functions within the LSZ reduction formula.  Doing this then gives a precise field theoretic interpretation of the angular momentum appearing in \Eq{eq:subsoft_term} and we can power expand $J_i^{\mu\nu}$ assuming it acts on matrix elements that create scalar, fermion or gluon excitations.  In \App{app:LSZ}, we provide the definition of the Low-Burnett-Kroll theorem from the LSZ reduction formula.  Importantly, in the Low-Burnett-Kroll theorem the angular momentum acts on free fields, whose power counting can be taken to be collinear corresponding to the momentum 
of the external states. (This can be only guaranteed for certain tree-level amplitudes.  We will return to this point when discussing other configuration which violate the Low-Burnett-Kroll theorem in \Sec{sec:amps_nonzero}.)

With this setup, we can write the angular momentum operator for a free particle as
\begin{equation}\label{eq:angmom_exp}
J_{i\mu\nu}=p_{i[\mu}\frac{\partial}{\partial p_i^{\nu]}} + \Sigma_{i\mu\nu} \ ,
\end{equation}
where $p_i$ is the momentum carried by the field and $ \Sigma_i^{\mu\nu}$ is the spin component of angular momentum.  Of course, for a collinear scalar field $ \Sigma_i^{\mu\nu}=0$, but for collinear fermions or gluons it is non-zero. In a gauge theory the physical spin components are independent of the momentum and scale like $\lambda^0$, because $\Sigma_i^{\mu\nu}$ is constructed from gamma matrices (for fermions) or the flat spacetime metric (for gluons). 
On the other hand, the first term of \Eq{eq:angmom_exp} is the orbital angular momentum, and when it acts on a collinear particle it must be expanded appropriately in powers of $\lambda$ according to the scaling of the collinear momentum.

In light-cone coordinates, the orbital component of angular momentum can be written as:
\begin{align}  \label{eq:orbexpn}
p_{i[\mu}\frac{\partial}{\partial p_i^{\nu]}}&=
% O(lam^-1)
\bigg\{ p_{i\perp [\mu} n_{\nu]} \frac{\partial}{\partial (n\cdot p_i)} 
+ n_{[\mu} \frac{\bn\cdot p_i}{2} \frac{\partial}{\partial p_{i\perp}^{\nu]}}
\bigg\}
\nonumber \\
&\ 
% O(lam^0)
+ \bigg\{ p_{i\perp [\mu}  \frac{\partial}{\partial p_{i\perp}^{\nu]}}
+\bar{n}_{[\mu} n_{\nu]} \frac{n\cdot p_i}{2} \frac{\partial}{\partial (n\cdot p_i)}+ n_{[\mu}\bar{n}_{\nu]} \frac{\bar{n}\cdot p_i}{2} \frac{\partial}{\partial (\bar{n}\cdot p_i)} 
\bigg\}
\nonumber \\
&\ 
% O(lam^1)
+ \bigg\{ 
p_{i\perp [\mu} \bn_{\nu]} \frac{\partial}{\partial (\bar{n}\cdot p_i)}
+ \bar{n}_{[\mu} \frac{n\cdot p_i}{2}\frac{\partial}{\partial p_{i\perp}^{\nu]}}
\bigg\} \ ,
\end{align}
where two other terms in this decomposition vanished due to the antisymmetry.
Using the power counting for a collinear particle, $(n\cdot p_i,\bn\cdot p_i,p_{i\perp})\sim (\lambda^2,\lambda^0,\lambda)$, the three groups of terms in \Eq{eq:orbexpn} are ${\cal O}(\lambda^{-1})$, ${\cal O}(\lambda^{0})$, and ${\cal O}(\lambda^{1})$ respectively.  Including spin, the total angular momentum operator expanded in powers of $\lambda$ is then
\begin{align}  \label{eq:orbexpn_j}
J_{i\mu\nu}&=
% O(lam^-1)
\bigg\{ p_{i\perp [\mu} n_{\nu]} \frac{\partial}{\partial (n\cdot p_i)} 
+ n_{[\mu} \frac{\bn\cdot p_i}{2} \frac{\partial}{\partial p_{i\perp}^{\nu]}}
\bigg\}
\nonumber \\
&\ 
% O(lam^0)
+ \bigg\{ p_{i\perp [\mu}  \frac{\partial}{\partial p_{i\perp}^{\nu]}}
+\bar{n}_{[\mu} n_{\nu]} \frac{n\cdot p_i}{2} \frac{\partial}{\partial (n\cdot p_i)}+ n_{[\mu}\bar{n}_{\nu]} \frac{\bar{n}\cdot p_i}{2} \frac{\partial}{\partial (\bar{n}\cdot p_i)} + \Sigma_{i\mu\nu}
\bigg\}
\nonumber \\*
&\ 
% O(lam^1)
+ {\cal O}(\lambda^1) \ ,
\end{align}
where again, the terms are associated by their relative power counting and we have ignored the contribution suppressed by a positive power of $\lambda$.

Putting the pieces together, the subleading soft factor in \Eq{eq:subsoft_term} has the following power expansion:
\begin{align}\label{eq:sub_soft_scale_exp}
S_i^{\text{(sub)}} 
&=T^i\frac{2\epsilon_{s\mu} p_{s\nu} }{(\bar{n}\cdot p_i)(n\cdot p_s)}\bigg\{ p_{i\perp}^{[\mu} n^{\nu]} \frac{\partial}{\partial (n\cdot p_i)} 
+ n^{[\mu} \frac{\bn\cdot p_i}{2} \frac{\partial}{\partial p_{i\perp \nu]}}
\bigg\}\nonumber \\
&\!+T^i\frac{2\epsilon_{s\mu} p_{s\nu} }{(\bar{n}\cdot p_i)(n\cdot p_s)}\left[\bigg\{ p_{i\perp}^{[\mu}  \frac{\partial}{\partial p_{i\perp \nu]}}
+\bar{n}^{[\mu} n^{\nu]} \frac{n\cdot p_i}{2} \frac{\partial}{\partial (n\cdot p_i)}+ n^{[\mu}\bar{n}^{\nu]} \frac{\bar{n}\cdot p_i}{2} \frac{\partial}{\partial (\bar{n}\cdot p_i)} + \Sigma_{i}^{\mu\nu}
\bigg\}\right.\nonumber \\
&\left.\qquad\qquad\qquad\qquad\qquad-
\frac{2p_{\perp i}\cdot p_{\perp s}}{(\bar{n}\cdot p_i)(n\cdot p_s)}\bigg\{ p_{i\perp}^{[\mu} n^{\nu]} \frac{\partial}{\partial (n\cdot p_i)} 
+ n^{[\mu} \frac{\bn\cdot p_i}{2} \frac{\partial}{\partial p_{i\perp \nu]}}
\bigg\}
\right]\nonumber\\*
&\!+{\cal O}(\lambda^1) \ ,
\end{align}
where the first line is ${\cal O}(\lambda^{-1})$, the terms on the second two lines are ${\cal O}(\lambda^0)$,  and higher order terms are dropped.
Recall that in terms of the $\lambda$ scaling, we expect from the amplitude analysis that the subleading soft factor scales like $\lambda^0$.  However, in the expansion of \Eq{eq:sub_soft_scale_exp}, there are terms that scale like $\lambda^{-1}$, which do not correspond to the terms in the Low-Burnett-Kroll theorem at tree-level.  These terms are actually generated by the general choice of basis used to represent the leading order soft theorem, and different choices are connected by the symmetry of RPI. 
If we choose from the start coordinates for each collinear field such that $p_{i\perp}^\mu = 0$ for each $i$, then all these terms at ${\cal O}(\lambda^{-1})$ manifestly vanish, and the first nonzero subleading soft factor scales as $\lambda^0$.   Then, with enforcing the collinear particle to be on-shell, the subleading soft factor 
acting on the amplitude is
\begin{align}\label{eq:lambda0_exp}
\left. S_i^{\text{(sub)}}\right|_{p_{i\perp} =\, 0} {\cal A}_N\Big|_{p_{i\perp} =\, 0} 
 &= \left.T^i\frac{\epsilon_{s\mu} p_{s\nu}J_i^{\mu\nu} }{p_i\cdot p_s}\right|_{p_{i\perp} =\, 0} 
   {\cal A}_N\Big|_{p_{i\perp} =\, 0}
\nonumber \\
&\simeq T^i\frac{2\epsilon_{s\mu} p_{s\nu} }{(\bar{n}\cdot p_i)(n\cdot p_s)}\bigg\{n^{[\mu}\bar{n}^{\nu]} \frac{\bar{n}\cdot p_i}{2} \frac{\partial}{\partial (\bar{n}\cdot p_i)} + \Sigma_{i}^{\mu\nu}
\bigg\} {\cal A}_N\Big|_{p_{i\perp} = 0} \,.
\end{align}
Here $S_i^{\text{(sub)}}$ scales like $\lambda^0$, as expected, so we will also use a notation that tracks the power suppression by $\lambda^2$ relative to $S^{(0)}$, defining
\begin{align}
  S^{(2)}_i &\equiv S_i^{\text{(sub)}} \,\Big|_{p_{i\perp} = \,0}  \,.
\end{align}
We will discuss this point further in the following section and show that the terms at ${\cal O}(\lambda^0)$ that survive in the $p_{i\perp}^\mu = 0$ limit are described in SCET.

\subsection{Correspondence of LBK with SCET at Tree-Level}
\label{subsec:proofsub}

Having identified the components at different orders in the power counting parameter $\lambda$ in the subleading soft factor, in this section we will show explicitly that the effective theory reproduces the Low-Burnett-Kroll theorem at tree level.  The leading order soft terms are ${\cal O}(\lambda^{-2})$ and were addressed in \Sec{subsec:simplecalcs}.   To identify the subleading soft factor in SCET requires three pieces.  First, we must show that the possible terms at ${\cal O}(\lambda^{-1})$ in the expansion of $S^{\text{(sub)}}$ vanish at tree-level.  These contributions are described by the subleading hard-scattering operators ${\cal O}_N^{(1X/\partial)}$ and the ${\cal L}^{(1)}$ SCET Lagrangians in \Eq{eq:sublagcollq}.  Next we consider ${\cal O}(\lambda^0)$. Here there are several possible contributions to the subleading soft factor, from: two insertions of ${\cal L}^{(1)}$, the ${\cal L}^{(2)}$  SCET Lagrangian in \Eq{eq:subsublag}, an ${\cal O}_N^{(1X/\partial)}$ together with an ${\cal L}^{(1)}$, ${\cal O}_N^{(2X/\partial)}$, and the ${\cal O}_N^{(2\delta,2r)}$ subleading-soft $N$-jet operators in \Eqs{eq:subsub_op_delta}{eq:subsub_op_rpi}.  We will show that the sum of the matrix elements of ${\cal O}_N^{(2\delta,2r)}$ and the time ordered products $T\{ {\cal O}_N^{(0)} {\cal L}^{(2)} \}$ reproduce the LBK result given by  \Eq{eq:lambda0_exp}, while the remaining terms vanish.  In this section, we will only present detailed calculations for soft gluon emission from fermions, but throughout, we will comment on emission from scalars and gluons.  We present the details of the SCET calculation of the LBK soft factor for gluons in \App{app:gluonLBK}.

\subsubsection{Vanishing of ${\cal L}^{(1)}$ and ${\cal O}^{(1)}$ LBK-Violating  Contributions at Tree Level}

\begin{figure}
\begin{center}
  \includegraphics[width=15cm]{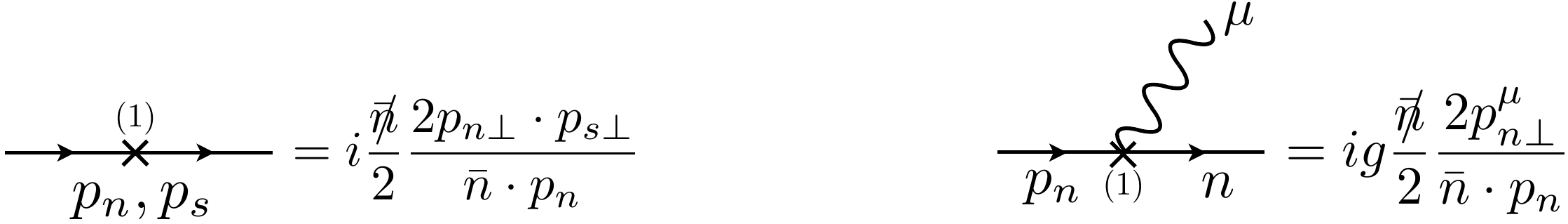}
\end{center}
\caption{
Subleading power Feynman rules for the coupling of a soft gluon to a collinear fermion.  The $\times$ symbol denotes the subleading Lagrangian insertion, $(1)$ denotes it is from ${\cal L}^{(1)}$, and $p_{s\perp}^\mu$ is the component of the off-shell collinear fermion's momentum from subsequent soft gluon emissions.
}
\label{fig:sub_feyn}
\end{figure}

First consider the coupling of soft gluons to collinear fermions through the SCET Lagrangian ${\cal L}^{(1)}$, \Eq{eq:sublagcollq}.  To do this explicitly requires the SCET Feynman rules from the subleading collinear fermion Lagrangian ${\cal L}^{(1)}$ in \Eq{eq:sublagcollq}, which are shown in \Fig{fig:sub_feyn}. 
Note that there are two new relevant pieces at subleading power: a propagator insertion with a leading power soft gluon coupling, and a subleading coupling of the soft gluon to the collinear fermion.  Thus to calculate the coupling of an $n_i$-collinear fermion to a soft gluon at subleading power there are two simple diagrams to evaluate:
\begin{align}\label{eq:subampcalc}
 \raisebox{-0.23\height}{\includegraphics[width=2.5cm]{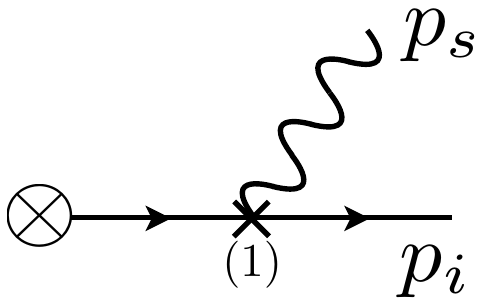}}&+
\raisebox{-0.23\height}{\includegraphics[width=2.5cm]{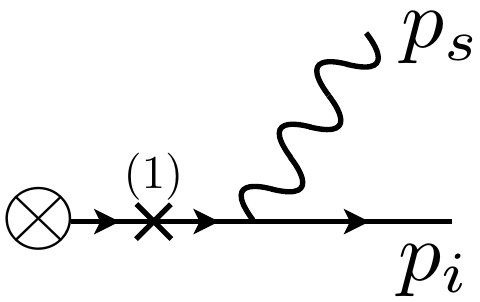}}
\nonumber \\*
&\quad=\bar u(p_i)\, \frac{i\frac{\slashed{n}_i}{2}p_i^-}{p_i^-(n_i\cdot p_s)} \Big(ig\frac{\slashed{\bar{n}}_i}{2}\frac{2p_{i\perp} \cdot \epsilon_s}{p_i^-} 
 \Big)
  \nonumber \\
&\quad\quad 
+\bar u(p_i)\, \frac{i\frac{\slashed{n}_i}{2}p_i^-}{p_i^-(n_i\cdot p_s)} 
\Big(i\frac{\slashed{\bar{n}}_i}{2}\frac{2p_{i\perp}\cdot p_{s\perp}}{p_i^-}
 \Big)
\frac{i\frac{\slashed{n}_i}{2} p_i^-}{p_i^- (n_i\cdot p_s)}
\Big(i g n_i\cdot \epsilon_s \frac{\slashed{\bar{n}}_i}{2} \Big)\nonumber \\
&\quad=\bar u(p_i)\cdot(-g)\frac{\epsilon_{s \mu} p_{s\nu}}{p_i^- (n_i\cdot p_s)}\left(
2p_{i\perp}^\mu\frac{n^\nu_i}{n_i\cdot p_s}-2p_{i\perp}^\nu \frac{n_i^\mu}{n_i\cdot p_s}
\right) \ .
\end{align}
As in \Sec{subsec:simplecalcs}, the symbol $\otimes$ denotes the tree-level amplitude ${\cal A}_n$ stripped of the external spinor $u(p_i)$.  To get the final line of \Eq{eq:subampcalc}, we used the collinear projection identity: 
\begin{align}
\bar{u}(p_i)\frac{\slashed{\bar{n}}_i\slashed{n}_i}{4}=\bar{u}(p_i) \, .
\end{align}
The soft momentum $p_s\sim \lambda^2$ and the $\perp$ component of the collinear momentum scales like $p_{i\perp}\sim \lambda$, so the factor in \Eq{eq:subampcalc} scales like $\lambda^{-1}$ as expected.  Also, \Eq{eq:subampcalc} is gauge invariant by itself due to the antisymmetry of the factor on the right of the final line.  

Since we are working at tree-level and constraining the external particles to be well-separated in angle, the collinear sector is composed of a single fermion.  Therefore, we can choose a coordinate system for our collinear momenta where $p_{i\perp}^\mu = 0$ for every $i$, namely by choosing $n_i^\mu$ so that each 
\begin{align}\label{eq:largelabelmom}
  p_i^\mu = \bn_i\cdot p\: \frac{n_i^\mu}{2} \,.
\end{align}  
This choice with $p_{i\perp}^\mu=0$ clearly makes the tree level contribution from the ${\cal L}^{(1)}$ term in \Eq{eq:subampcalc} manifestly zero.
If a different choice for $n_i^\mu$ was made, then this zero result is obtained by an RPI transformation on the vector $n_i$.   Therefore, due to RPI about the collinear fermion's direction, we see that the contribution of the subleading soft factor at ${\cal O}(\lambda^{-1})$ from ${\cal L}^{(1)}$ is a coordinate artifact that can be set to zero.

This vanishing by RPI implies that the expression for the amplitude in \Eq{eq:subampcalc} is not independent of the leading order soft factor. Indeed, expanding out the factor appearing in $S^{(0)}(s)$ using a generic coordinate basis
gives 
\begin{align}\label{eq:gencoorgauge}
 \frac{p_i \cdot \epsilon_s}{p_i\cdot p_s} 
  &= \frac{\bn_i\cdot p_i\, n_i\cdot \epsilon_s + 2 p_i^\perp\cdot \epsilon_s^\perp+n_i\cdot p_i\, \bn_i\cdot \epsilon_s  }
  {\bn_i\cdot p_i\, n_i\cdot p_s + 2 p_i^\perp\cdot p_s^\perp+n_i\cdot p_i\, \bn_i\cdot p_s} 
   \nonumber\\
 &= \frac{n_i\cdot \epsilon_s}{n_i\cdot p_s}  
 + \frac{\epsilon_{s\mu}p_{s\nu}}{(\bn_i\cdot p_i)\,(n_i\cdot p_s)} 
  \bigg( 2 p_{i\perp}^\mu \frac{ n_i^\nu}{n_i\cdot p_s}  - 
   2 p_{i\perp}^\nu\frac{ n_i^\mu}{n_i\cdot p_s}  \bigg) 
 + \ldots \,.
\end{align} 
Thus we immediately see that the subleading amplitude appearing in \Eq{eq:subampcalc} is simply an artifact of the coordinate system, and can be absorbed into $S^{(0)}(s)$.  

For collinear scalars or gluons, it can also be easily verified from the subleading SCET Lagrangians ${\cal L}_{\phi_n}^{(1)}$ and ${\cal L}_{A_n}^{(1)}$ that the contributions to the subleading soft factor at ${\cal O}(\lambda^{-1})$ can be set to zero by RPI.  For ${\cal L}^{(1)}_{\phi_n}$ the coupling of a soft gluon to scalars again always involves $p_{i\perp}^\mu$ so the argument is the same. For collinear gluons with ${\cal L}^{(1)}_{A_n}$ these contributions vanish by either being proportional to $p_{i\perp}^\mu$,  to $p_i^\mu \epsilon_{i\mu}=0$, or they vanish by gauge invariance of the base amplitude which produced the original collinear gluon. 

Next, consider potential tree-level contributions from ${\cal O}_N^{(1)}$ in a time-ordered product with the leading SCET Lagrangian ${\cal L}_{n,{\rm soft}}^{(0)}$.  We either have a contribution proportional to $p_{n_i\perp}^\mu=0$ from ${\cal O}_N^{[1](1,\partial)}$ or a contribution that must generate two particles in the same collinear sector from ${\cal O}_N^{[1](1,X)}$ which has a vanishing matrix element for the states in \Eq{eq:smelt}.  Therefore, matrix elements of the operator ${\cal O}_N^{(1)}$ vanish and so all possible contributions at ${\cal O}(\lambda^{-1})$ are zero. We can summarize this by saying the soft factor
\begin{align}
  S^{[0](1)} & = 0 \,.
\end{align}
Here the $[0]$ indicates that it is tree-level and the $(1)$ indicates that this vanishing soft factor includes all terms suppressed by $\lambda^1$ relative to $S^{(0)}$.

This argument using RPI to set the $\perp$ components to zero requires that there is a single collinear particle in each collinear sector.  If there is more than one particle in a collinear sector, then RPI could be used to set the $\perp$ component of one of the particles' momentum to zero, but the other particles would have a non-zero $\perp$ component.  Thus, these arguments only hold at tree-level, when all particles can be forced to be widely separated in angle.  This point has important consequences for the structure of the soft theorems with collinear emissions where a single insertion of the ${\cal L}^{(1)}$ Lagrangians or ${\cal O}_N^{(1)}$ operators do contribute.  We will show this explicitly in \Sec{subsec:splitsub}.

One can also consider the effect of collinear fermion masses on this result, since masses are straightforward to treat in SCET~\cite{Rothstein:2003wh}, and the corresponding SCET Lagrangian ${\cal L}_{\xi_n,m}^{(0)}$ is known~\cite{Leibovich:2003jd}. The subleading power Lagrangian involving fermion masses is again generated by replacing one $D_{n\perp}\to W_n D_{s\perp}W_n^\dagger $.  However, this does not generate an ${\cal O}(\lambda)$ Lagrangian  since we get
\begin{align}
{\cal L}_{\xi_n,m}^{(1)} &=  m (\bar{\xi}_n W_n) \bigg(
i \slashed{D}_{s\perp} \frac{1}{i \bar{n}\cdot \partial_n}-\frac{1}{i \bar{n}\cdot \partial_n} i \slashed{D}_{s\perp} 
\bigg)\frac{\slashed{\bar{n}}}{2}(W_n^\dagger \xi_n) 
  = 0
   \ .
\end{align}
Here $i\bn\cdot\partial_n$ commutes with $i D_{s\perp}$ due to the multipole expansion in SCET.
Therefore, fermion mass effects do not change the result that there is no contribution to the subleading soft factor at ${\cal O}(\lambda^{-1})$. A non-zero sub-subleading Lagrangian at ${\cal O}(\lambda^2)$ involving the fermion mass does appear, ${\cal L}_{\xi_n,m}^{(2)}\ne 0$.

\subsubsection{${\cal L}^{(2)}$ Contributions to LBK}  \label{subsec:L2LBK}

\begin{figure}
\begin{center}
  \includegraphics[width=15cm]{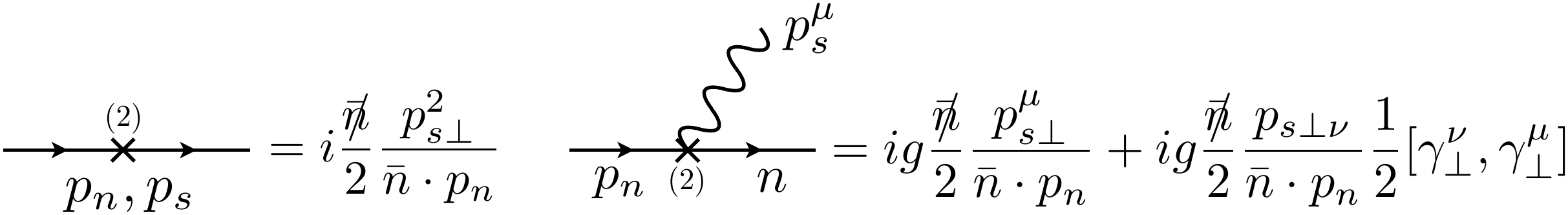}
\end{center}
\caption{
Sub-subleading power Feynman rules for the coupling of a soft gluon to a collinear fermion.  The $\times$ symbol denotes the sub-subleading Lagrangian insertion, $(2)$ denotes it is from ${\cal L}^{(2)}$, and $p_{s\perp}^\mu$ is the component of the off-shell collinear fermion's momentum from subsequent soft gluon emissions.  For simplicity, here the vertex rule assumes that the outgoing fermion is on-shell.
}
\label{fig:subsub_feyn}
\end{figure}

Next, we continue to sub-subleading order in SCET.  We will first consider the contribution at this order from subleading Lagrangians. One possibility is to have two insertions of the ${\cal O}(\lambda)$ Lagrangian, ${\cal L}^{(1)} {\cal L}^{(1)}$, but at tree-level this vanishes by the same logic used to show that one such insertion vanishes. That leaves the ${\cal O}(\lambda^2)$ 
SCET Lagrangian ${\cal L}^{(2)}$, \Eq{eq:subsublag}, where the term involving collinear fermions was
\begin{align}
{\cal L}_{\xi_n}^{(2)} &= \bar\chi_n \bigg(
 i \slashed{D}_{s\perp}\frac{1}{ i\bar{n}\cdot \partial_n}i \slashed{D}_{s\perp}
  - i \slashed{\cal D}_{n\perp} \frac{i\bar{n}\cdot D_s}{(i\bar{n}\cdot \partial_n)^2}  i \slashed{\cal D}_{n\perp}
 \bigg)\frac{\slashed{\bar{n}}}{2} \chi_n 
 \,.
\end{align}
At tree-level, with no extra (on-shell) collinear gluon emissions, we can set the second term in ${\cal L}_{\xi_n}^{(2)}$ to zero by RPI by taking $p_{i\perp}^\mu=0$.  The first term, however, cannot be set to zero and contributes to the subleading soft factor.  This contribution is expected: at this order, the collinear fermion is sensitive to the momentum carried by the soft gluon.  In \Fig{fig:subsub_feyn}, we show the Feynman rules from this term in the sub-subleading collinear fermion Lagrangian.  At sub-subleading power with $p_{i\perp}=0$, there are then two diagrams to compute for soft emission off of a collinear fermion:
\begin{align}\label{eq:subsub_diagsq}
& \raisebox{-0.23\height}{\includegraphics[width=2.5cm]{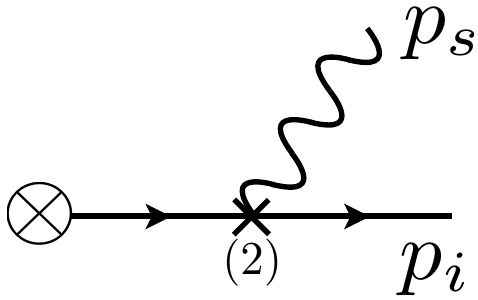}}+
\raisebox{-0.23\height}{\includegraphics[width=2.5cm]{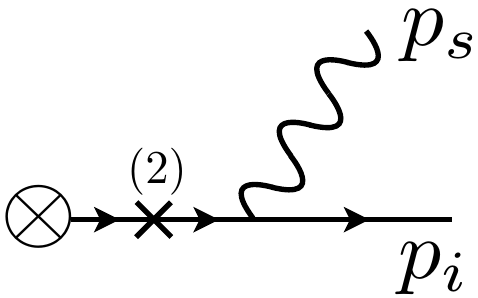}}
\nonumber \\
&
=\bar u(p_i)\cdot(-g)\frac{\epsilon_{s \mu} p_{s\nu}}{p_i^- (n_i\cdot p_s)}\left(
p_{s\perp}^\mu\frac{n^\nu_i}{n_i\cdot p_s}-p_{s\perp}^\nu \frac{n_i^\mu}{n_i\cdot p_s}
+\frac{1}{2}[\gamma_\perp^\nu,\gamma_\perp^\mu]
\right) \ .
\end{align}
We discuss the analogous results with a collinear scalar or gluon below in \Sec{sec:totalSCETLBK}.

\subsubsection{${\cal O}_N^{(2)}$ Operator Contributions to LBK}\label{sec:opcontribs}

The final contribution to the subleading soft factor are the operators ${\cal O}_N^{(2\delta,2r)}$, which are given in \Eqs{eq:subsub_op_delta}{eq:subsub_op_rpi}. Time ordered products of ${\cal O}_N^{(1)} {\cal L}^{(1)}$ again vanish using RPI.  Matrix elements of the operators ${\cal O}_N^{(2X/\partial)}$ vanish at tree-level for the final state we are considering, either by RPI or since they involve more than one collinear particle in at least one sector. 

To determine the contribution of ${\cal O}_N^{(2\delta,2r)}$ at tree-level to the Low-Burnett-Kroll theorem, we must take their matrix elements with $N$ hard partons and one soft parton.  
Recall that the matrix element in \Eq{eq:leadwilsoncoeff},
$\langle 0 | {\cal O}_N^{(0)}| p_1,\dotsc,p_N\rangle = {\cal A}_N^{[0]} [ 1 + {\cal O}(g^2) ]$, determines the matching coefficient as the polarization and color stripped amplitude at lowest order, ${\cal A}_N^{[0]}=C_N^{[0](0)} e_1\cdots e_N$.  The operator ${\cal O}_N^{(2,\delta)}$ is written in a factorized form in \Eq{eq:subsub_op_delta} as an operator involving $\bar n_i\cdot B_s^{(n_i)}$  multiplying and acting on $C_N^{(0)}$. Writing down the tree level Feynman rule for this operator we get
\begin{equation}\label{eq:subsubop_firline}
\langle 0 | {\cal O}_{N}^{(2,\delta)}| p_1,\dotsc,p_N,p_s\rangle = e_1\cdots e_N \sum_{i=1}^N gT^i\frac{2\epsilon_{s\mu} p_{s\nu} }{(\bar{n}_i\cdot p_i)(n_i\cdot p_s)}n_i^{[\mu}\bar{n}_i^{\nu]} \frac{\bar{n}_i\cdot p_i}{2} \frac{\partial}{\partial (\bar{n}_i\cdot p_i)} C_N^{[0](0)}\ .
\end{equation}
This is precisely the orbital angular momentum contribution to the Low-Burnett-Kroll theorem at tree-level, which was displayed above in \Eq{eq:lambda0_exp}.  This result is independent of the identity of the collinear fields $X_{n_i}$ in the operator, so it applies equally well for collinear fermions, scalars, and gluons.

Note that ${\cal O}_N^{(2,r)}$ in \Eq{eq:subsub_op_rpi} is also given by a sum of terms, one for each collinear sector.  ${\cal O}_N^{(2,r)}$ is only nonzero when $X_{n_i}$ is a $n_i$-collinear fermion field.  Using the operator in \Eq{eq:subsub_op_rpi} the matrix element is
\begin{align}\label{eq:subsubop_secline}
\left.\langle 0 | {\cal O}_{N}^{(2,r)}| p_1,\dotsc,p_N,p_s\rangle\right|_\text{$i$,fermion} 
 = (-g)\frac{\epsilon_s^\mu p_s^\nu}{p_i^- (n_i\cdot p_s)}\bar{u}(p_i) T_i \Big(
\gamma_{\perp\nu}n_{i\mu} \frac{\slashed{\bar{n}}_i}{2}-\gamma_{\perp\mu}n_{i\nu} \frac{\slashed{\bar{n}}_i}{2}
\Big)\tilde{\cal A}_N \,.
\end{align}
Here, we have explicitly inserted the $\gamma$ matrices between the external spinor $\bar{u}(p_i)$ and the amplitude  that is stripped of this polarization $\tilde{\cal A}_N$ to emphasize how the $\gamma$ matrices act.  The full amplitude with $N$ collinear particles is
\begin{equation}
{\cal A}_N = \bar{u}(p_i) \tilde {\cal A}_N \ .
\end{equation}

\subsubsection{Total tree level SCET Calculation for LBK} \label{sec:totalSCETLBK}

For collinear fermions putting all the pieces together from the ${\cal L}^{(2)}$ Lagrangian, \Eq{eq:subsub_diagsq} and the operators ${\cal O}_N^{(2,\delta)}$ in \Eq{eq:subsubop_firline}, and ${\cal O}_N^{(2,r)}$ in \Eq{eq:subsubop_secline}, we have
\begin{align}\label{eq:LBKinSCET}
S^{\text{(2)}}_{i\psi}{\cal A}_N&=g \frac{2\epsilon_{s\mu} p_{s\nu} }{(\bar{n}_i\cdot p_i)(n_i\cdot p_s)}\bar{u}(p_i) T_i \bigg\{n_i^{[\mu}\bar{n}_i^{\nu]} \frac{\bar{n}_i\cdot p_i}{2} \frac{\partial}{\partial (\bar{n}_i\cdot p_i)} \nonumber\\
&\qquad\qquad\qquad\qquad\quad+\gamma_{\perp}^{[\mu}n_i^{\nu]} \frac{\slashed{\bar{n}}_i}{4}
+p_{s\perp}^{[\mu} \frac{n_i^{\nu]}}{2(n_i\cdot p_s)}
+\frac{1}{4}[\gamma_\perp^\mu,\gamma_\perp^\nu]
\bigg\}\tilde{\cal A}_N \ ,
\end{align}
where, as earlier, we explicitly pull out the external collinear spinor from the amplitude to emphasize the action of the factor in braces.  As mentioned earlier, the derivative term corresponds to the orbital angular momentum contribution to the Low-Burnett-Kroll operator.  To show that \Eq{eq:LBKinSCET} is equivalent to the full LBK result in \Eq{eq:lambda0_exp}, we also need to show that the second line is the spin angular momentum contribution.

The spin angular momentum operator for fermions is
\begin{equation}
\Sigma_q^{\mu\nu} = \frac{1}{4}[\gamma^\mu,\gamma^\nu] \ .
\end{equation}
We can decompose the action of this operator on a collinear fermion spinor as:
\begin{align}
\bar{u}(p_i)\Sigma_q^{\mu\nu} &=\bar{u}(p_i) \frac{1}{4}[\gamma^\mu,\gamma^\nu] \nonumber \\
&=\bar{u}(p_i)\frac{1}{4}\left(
2\gamma_\perp^{[\mu}n_i^{\nu]}\frac{\slashed{\bar{n}}_i}{2}+n_i^{[\mu}\bar{n}_i^{\nu]}+\gamma_\perp^{[\mu}\gamma_\perp^{\nu]}
\right) \ .
\end{align}
To get the second line, we have used $n^2=\bar{n}^2=0$ and the projection identities:
\begin{equation}
\bar{u}(p_i)\slashed{n}_i = 0 \ , \qquad \bar{u}(p_i)\frac{\slashed{\bar{n}}_i}{2}\frac{\slashed{n}_i}{2} = \bar{u}(p_i) \ .
\end{equation}
This expression can be further simplified by dotting with the soft gluon's momentum and polarization vector.  Note that
\begin{equation}
\epsilon_s^\mu p_s^\nu \left(
n_i^{[\mu} \bar{n}_i^{\nu]} + \gamma_\perp^{[\mu}\gamma_\perp^{\nu]}
\right) = 2 (n_i\cdot \epsilon_s)(\bar{n}_i\cdot p_s)+2\slashed{\epsilon}_{s\perp}\slashed{p}_{s\perp} \, ,
\end{equation}
where we have used $2\epsilon_s\cdot p_s=(n_i\cdot \epsilon_s)(\bar{n}_i\cdot p_s)+(n_i\cdot p_s)(\bar{n}_i\cdot \epsilon_s)+2\epsilon_{s\perp}\cdot p_{s\perp} = 0$.
Then, using $p_s^2=0$, we finally produce
\begin{align}
\bar{u}(p_i)\epsilon_{s\mu}p_{s\nu}\Sigma_q^{\mu\nu} &= \bar{u}(p_i)\epsilon_{s\mu}p_{s\nu}\left(
\gamma_{\perp}^{[\mu}n_i^{\nu]} \frac{\slashed{\bar{n}}_i}{4}
+p_{s\perp}^{[\mu} \frac{n_i^{\nu]}}{2(n_i\cdot p_s)}
+\frac{1}{4}[\gamma_\perp^{\mu},\gamma_\perp^{\nu}]
\right) \ ,
\end{align}
which agree with the terms on the second line of \Eq{eq:LBKinSCET}.
Therefore, the result appearing in \Eq{eq:LBKinSCET} for a collinear fermion indeed corresponds to precisely the contribution of the LBK theorem, with the total angular momentum, orbital plus spin.

So far we have only presented an explicit calculation for collinear fermions, but this result holds for collinear scalars and gluons as well. The contribution from ${\cal O}_N^{(2,\delta)}$ which gave the orbital angular momentum was manifestly independent of the choice of fermions, scalars, or gluons.  For scalars we mentioned earlier that there was no contribution to ${\cal O}_N^{(2)}$ from expanding the scalar field to subleading power.  This is consistent with scalars having no spin. We also must consider a potential contribution from the ${\cal L}^{(2)}$ collinear scalar Lagrangian, which was
\begin{align}
{\cal L}_{\phi_n}^{(2)} &= 2 \, {\rm Tr} \Big[
\Phi_n^* \Big( 
\frac12  n\cdot {\cal D}_{ns}\, \bn\cdot D_s  
+\frac12  \bn\cdot D_s\, n\cdot {\cal D}_{ns}
+  D_{s\perp}^2  
\Big) \Phi_n \Big]  \,.
\end{align}
At tree-level, we assume that the soft gluon is on-shell and so $p_s^2=0$.  Also, we can use RPI to set the $\perp$ component of the scalar's momentum to zero: $p_{n\perp}=0$.  Because the external scalar is on-shell, this then implies that $n\cdot p_n = 0$.  These two constraints, which we can only apply at tree-level and for an on-shell soft gluon, then imply that inserting ${\cal L}_{\phi_n}^{(2)}$ gives zero at ${\cal O}(g^1)$.  Therefore, for a collinear scalar, neither the sub-subleading Lagrangian nor the operator ${\cal O}_N^{(2)}$ generate ``spin'' contributions to the subleading soft factor, exactly as predicted by the Low-Burnett-Kroll theorem.  

For collinear gluons the ${\cal L}^{(2)}_{A_n}$ Lagrangian insertion does produce the proper spin term, and we present the details of this calculation  in \App{app:gluonLBK}.  Together these results constitute a proof of the Low-Burnett-Kroll theorem in SCET for scalars, fermions, and gauge bosons.\footnote{ Although we presented the proof using modes with soft$_{\rm 1}$ scaling, the same result also applies for SCET$_{\rm II}$ where the soft modes have soft$_{\rm 2}$ scaling. To quickly see this, we simply match the final tree level SCET$_{\rm I}$ result onto SCET$_{\rm II}$ by lowering the $p_\perp$ and $n\cdot p$ momenta of collinear particles. } Key ingredients in our proof are gauge invariance and the use of RPI, which is the remnant of Lorentz invariance for this situation.

Importantly, the only restriction of the above effective theory analysis to tree-level was the accuracy to which we calculated matrix elements.  The subleading soft factor can be calculated to arbitrary perturbative order in SCET, but beyond tree-level the Low-Burnett-Kroll form of the subleading soft factor will generally not hold.  The proof of the Low-Burnett-Kroll theorem also delicately required that the terms at ${\cal O}(\lambda^{-1})$ in the expansion of the soft factor could be set to zero with RPI.  This is not true when the collinear sector contains more than one particle.  We will discuss these points further  in our analysis in \Sec{sec:amps_nonzero}.

The Low-Burnett-Kroll theorem in gauge theory, \Eq{eq:subsoft_intro}, is agnostic as to the exact hard interaction that sources the soft radiation. All it requires is the power counting given in \Eq{eq:lows_theorem_pc}. Thus by changing the power counting due to a different kinematic limit of gauge theory, the subleading soft factor for asymptotic soft radiation changes its form, while still probing the components of the angular momentum.  For example, rather than considering massless energetic final state particles, we could consider massive non-relativistic particles.  In non-relativistic QED or QCD, for ultrasoft radiation the dipole interactions starts at order $v$ (and is the leading interaction for neutral systems) and spin enters at order $v^2$ in the velocity power counting~\cite{Grinstein:1997gv,Manohar:1997qy}.  If instead we analyze a two-particle system in non-relativistic classical electromagnetism, then spin is suppressed and the magnetic field couples to the orbital angular momentum (see Sec.~71 of \Ref{landau}).  Both of these results follow from a consistent power counting of the dominant modes and multipole expanding \cite{Grinstein:1997gv,Beneke:2002ni} the appropriate operator or current in the system.

\subsection{Infinitesimal M\"obius Transforms, RPI, and Asymptotic Symmetries}
\label{sec:mobius}

The RPI of gauge theory SCET was vital for the proof of the Low-Burnett-Kroll theorem at tree-level for amplitudes involving well-separated energetic particles.  It is well-known that the action of the Lorentz group on the boundary of spacetime is locally isomorphic to the M\"obius group, PSL$(2,\mathbb{C})$ \cite{J.L.Synge:1960zz,Penrose:1959vz,Penrose:1960eq}. Here, we explore the interpretation of RPI further and explicitly connect the reparametrization transformations from \Eq{eq:RPI_Trans} to the generators of the M\"obius group.

To do this, we map any complex number $z=x+i\,y$ to the null vector:
\begin{align}\label{eq:nullcomplex}
n[z]=\frac{1}{1+x^2+y^2}\left(\begin{array}{c}1+x^2+y^2\\2x\\-2y\\1-(x^2+y^2)\end{array}\right)
\end{align}
In particular \Eq{eq:nullcomplex} allows us to see that $n[0]=n=(1,0,0,1)$, and that $n[\infty]=\bar{n}=(1,0,0,-1)$ where $z=\infty$ is the point at infinity on the Riemann sphere $\hat{\mathbf{C}}$ . Further, since we have taken the time component to be 1, we have conformally mapped all parallel light-cone vectors to a representative point on the sphere of spatial directions. The M\"obius group is isomorphic to PSL$(2,\mathbb{C})$, the group of complex-valued $2\times 2$ matrices with determinant 1, modulo $\pm 1$. The matrix $M$ defines the map $f_M:\hat{\mathbf{C}}\rightarrow \hat{\mathbf{C}}$ via the correspondence:
\begin{align}
M&=\left(\begin{array}{c c}a &b\\  c&d\end{array}\right) \leftrightarrow f_M(z)=\frac{a z+b}{c z+d}.
\end{align}
The elements of the M\"obius group are referred to as hyperbolic, parabolic, elliptic and inversion transformations, depending on their action on the Riemann sphere.

First we consider the transformations that leave null lines defined by $n$ and $\bar{n}$ fixed. These correspond to rotations about the $z$-axis in the transverse plane, and because of our normalization convention for the time component in \Eq{eq:nullcomplex} also the RPI-III transformations. These are given by the hyperbolic and elliptic M\"obius transformations respectively:
\begin{align}
& \text{RPI-III $\longleftrightarrow$ Hyperbolic: }M_H =\left(\begin{array}{c c} e^{\frac{\alpha}{2}}&0\\  0 & e^{-\frac{\alpha}{2}}\end{array}\right)\text{ and }z\rightarrow e^{\alpha}z,\\
& \text{$\perp$-Rotation $\longleftrightarrow$ Elliptic: }M_E =\left(\begin{array}{c c} e^{i\frac{\theta}{2}} & 0\\  0 & e^{-i\frac{\theta}{2}}\end{array}\right)\text{ and }z\rightarrow e^{i\theta}z.
\end{align}
The corresponding null vectors map to:
\begin{align}
\text{Hyperbolic: }n[f_{M_H}(z)]&=\frac{1}{1+e^{\alpha}(x^2+y^2)}\left(\begin{array}{c}1+e^{\alpha}(x^2+y^2)\\2e^{\alpha/2}x\\-2e^{\alpha/2}y\\1-e^{\alpha}(x^2+y^2)\end{array}\right),\\
 \text{Elliptic: }n[f_{M_E}(z)]&=\frac{1}{1+x^2+y^2}\left(\begin{array}{c}1+x^2+y^2\\2(x\text{cos}\,\theta-y\text{sin}\,\theta)\\-2(x\text{sin}\,\theta+y\text{cos}\,\theta)\\1-x^2-y^2\end{array}\right).
\end{align}
Both of these transformations clearly leave $n=n[0]$ and $\bar{n}=n[\infty]$ fixed. The Elliptic transformation is clearly the $\perp$-rotation. Recalling that RPI-III is a passive boost transformation for back-to-back vectors $n$ and $\bn$, one can demonstrate the equivalence to the Hyperbolic transformation by boosting a null vector along the $z$-direction with a velocity $v=\tanh\frac{\alpha}{2}$, and then conformally mapping the boosted vector to the corresponding point on the unit sphere: $(t,x,y,z)\rightarrow(1,\frac{x}{t},\frac{y}{t},\frac{z}{t})$. 

The RPI-I transformations correspond to the parabolic transformations, which are translations in the complex plane:
\begin{align}
 \text{RPI-I $\longleftrightarrow$ Parabolic: }M_P&=\left(\begin{array}{c c} 1 & a\\  0 &1\end{array}\right)\text{ and }z\rightarrow z+a.
\end{align}
Writing the shift as $a=u+i\,v$, the action on a null vector is:
\begin{align}
\text{Parabolic: }n[f_{M_P}(z)]&=\frac{1}{1+((x+u)^2+(y+v)^2)}\left(\begin{array}{c}1+((x+u)^2+(y+v)^2)\\2(x+u)\\-2(y+v)\\1-((x+u)^2+(y+v)^2)\end{array}\right) .
\end{align}
One can see how this corresponds to an RPI-I transformation on the vector $n$ when $|a|\ll 1$, so that to first order $n\rightarrow n+(0,2u,-2v,0)$.  Furthermore, $\bar{n}$ is left unchanged, since no finite translation in the complex plane moves the point at infinity. To connect to RPI-II transformations we will need inversions:
\begin{align}
\text{Inversions: }M&=\left(\begin{array}{c c} 0 &1\\  c & 0\end{array}\right)\text{ and }z\rightarrow \frac{1}{cz}.
\end{align}
These interchange $n$ and $\bar{n}$. To achieve a correspondence with RPI-II, we invert, perform a parabolic transformation, and invert back. This translates $\bar{n}$ and leaves $n$ fixed.

To connect with the full algebra of M\"obius transformations, we would also need the action of inversions alone. However, these act non-locally, and RPI transformations are restricted to a local region. In SCET the inversions simply interchange back-to-back collinear sectors, $n\leftrightarrow \bn$, which is an allowed relabeling in SCET due to the sum over all collinear sectors. 

RPI symmetry is local to each collinear momentum region (or jet region), and the full spacetime symmetry of the scattering process in SCET is the completion of the individual RPI transformations about each of the jets.  In light of recent emphasis of the symmetries at the boundary of spacetime in gauge theories \cite{Strominger:2013lka} and gravity \cite{He:2014laa,Kapec:2014opa}, this can be put into a more suggestive language.  The asymptotic spacetime boundary of the full theory is the effective theory, and RPI is a symmetry that acts on the effective theory. This correspondence is most apparent in position space. Each collinear field at coordinate $x$ has support in position space determined by the requirement $x\cdot p\sim 1$ for any collinear momentum $p$. From the power counting of \Tab{tab:scaling}, we see $n\cdot x\sim\frac{1}{Q}\ll  x_\perp\sim\frac{1}{Q\lambda}\ll \bar{n}\cdot x\sim\frac{1}{Q\lambda^2}$. Hence the field has support in a region far from the hard interaction, localized near the light-cone. This is perhaps unsurprising, since in a gauge theory, jets or collinear particles are the asymptotic states with large energy. These RPI transformations act on the $N$-jet operators that reorganize the full theory S-matrix. Though each operator connects only a finite number of collinear sectors, operators with arbitrarily many collinear sectors are allowed.\footnote{The number of possible distinct collinear sectors increases as $\lambda$ is decreased because of the requirement $n_i\cdot n_j\gg \lambda^2$ for distinct collinear directions. Perturbatively the formalism is valid for an arbitrarily small $\lambda$. For a hard energy scale $Q$ there is a practical upper limit in QCD set by $\lambda \sim \Lambda_{\rm QCD}/Q$, due to confinement.} Since each collinear sector has it own independent RPI symmetry, and the full S-matrix of SCET could contain an arbitrary number of ``jets'' in the final state as $\lambda\to 0$, so the full RPI symmetry of the S-matrix is effectively infinite dimensional. From the effective theory alone, we anticipate an infinite dimensional asymptotic symmetry of the S-matrix corresponding to the completion of RPI transformations about each jet in the final state.  In \App{app:gauge}, we emphasize that allowing arbitrarily small $\lambda$, the gauge symmetry of the effective theory is also effectively  infinite dimensional, because there are distinct transformations for each collinear sector.

\section{A Soft Theorem at One-Loop and with Collinear Emissions}
\label{sec:amps_nonzero}

The SCET proof of the Low-Burnett-Kroll theorem in gauge theory hinged on the fact that each collinear sector contained only a single particle.  In other words, if the soft momentum flows through propagators with hard momenta, we can Taylor expand:
\begin{align}
\frac{1}{(p_h+p_s)^2}=\frac{1}{p_h^2}-\frac{2 p_h\cdot p_s}{p_h^4}+\dotsc
  \,.
\end{align}
Here the soft momentum is completely removed from the denominator in the leading term. However, for any collection of $n$-collinear momenta $p_c$, the soft momentum still leaves the collinear propagator on-shell (see \Eq{eq:merging_soft_collinear_momenta}), and hence there is a component that does not cancel from the denominator of the leading term:
\begin{align}\label{eq:expanding_soft_collinear_prop}
\frac{1}{(p_c+p_s)^2}&=\frac{1}{\bar{n}\cdot p_c\,n\cdot p_s+p_c^2}+\dotsc
\end{align}
The dots here refer to subleading contractions with $p_s^\mu$, for example those with the soft transverse momenta, $p_s^\perp\cdot p_c^\perp$. Critically, the soft momentum still flows through the propagator. The denominator on the right-hand-side is homogenous in the power counting, and nothing further can be done to simplify it. This implies that the soft momenta can probe collinear poles of the amplitude. 

In this section, we study the effect of on-shell corrections to the Low-Burnett-Kroll soft theorem.  In particular, if the dynamics of the collinear sector become non-trivial, through either loops or collinear real emissions, then \Eq{eq:lows_theorem_pc} is violated, which was a requirement for the LBK theorem.  In this case new subleading soft contributions arise. The general structure and universality of these new terms will be derived with SCET.
We can generically express an amplitude through next-to-leading order (NLO) in the form
\begin{align}
{\cal A} = {\cal A}_{\text{LO}}+{\cal A}_{\text{NLO,loop}}+{\cal A}_{\text{NLO,emission}} \,.
\end{align}
The tree level amplitude, ${\cal A}_{\text{LO}}$, is defined to have all energetic particles well-separated in phase space.  In the previous section, we showed that the subleading soft behavior of this amplitude is given by the Low-Burnett-Kroll theorem.  The NLO loop amplitude, ${\cal A}_{\text{NLO,loop}}$, contains a single virtual loop.  The momentum in this loop ranges over all momentum regions and includes those regions where soft momentum cannot be expanded from propagators, as illustrated in \Eq{eq:expanding_soft_collinear_prop}.  We will show that these regions are responsible for modifications to the Low-Burnett-Kroll theorem at loop-level.  Finally, ${\cal A}_{\text{NLO,emission}}$ is the NLO amplitude for real emission in the singular regions of phase space.  This includes collinear emissions which spoil the scaling requirement of the Low-Burnett-Kroll theorem, \Eq{eq:lows_theorem_pc}.  We will show that collinear emissions manifestly violate the Low-Burnett-Kroll theorem at tree-level.  Note that ${\cal A}_{\rm NLO,loop}$ will generically require an infrared regulator, and this dependence will only cancel with the infrared regulator dependence in ${\cal A}_{\rm NLO,emission}$ once we square the amplitudes and integrate over phase space.

\subsection{Revisiting the single-minus amplitude}
\label{subsec:single_minus_amp_redux}

As noted in \Sec{sec:singminus}, the single-minus one-loop amplitude ${\mathcal A}^{[1]}(1^-,2^+,3^+,4^+,5^+)$ does not agree with the Low-Burnett-Kroll theorem's expectations. Nevertheless, the soft limit of particle 5 still exhibits an interesting structure.  We can write the expansion of the large $N_c$ primitive amplitude in the suggestive form
\begin{align}
{\mathcal A}^{[1]}(1^-,2^+,3^+,4^+,5_s^+)&= \bigg[\frac{\sab{4}{1}}{\sab{4}{5}\sab{5}{1}} \bigg]
  \bigg( \frac{i}{48\pi^2}\frac{\sab{1}{3}^3\sab{2}{4}\ssb{1}{2}}{\sab{2}{3}^2\sab{3}{4}^3} \bigg)
   \\
&+\bigg[\frac{\ssb{5}{2}}{\sab{5}{1}\ssb{1}{2}}\bigg] \bigg(\frac{i}{48\pi^2}\frac{\sab{1}{3}^3\sab{2}{4}\ssb{1}{2}}{\sab{2}{3}^2\sab{3}{4}^3} \bigg)
  \nonumber\\
& + \bigg[  \frac{-i}{48\pi^2}\frac{\sab{3}{5}\ssb{4}{5}}{\sab{3}{4}\sab{4}{5}^2} \bigg]
 \bigg( \frac{\sab{1}{4}^4}{\sab{1}{2}\sab{2}{3}\sab{3}{4}\sab{4}{1}} \bigg) 
 +{\cal O}(\lambda^1)
 \nonumber \,,
\end{align}
where we ignore the ${\cal O}(\lambda^1)$ terms that are higher order in the momentum of particle $5$.
From the bracketed terms one can explicitly see that the leading and subleading terms in the soft expansion take the factorized form:
\begin{align}\label{eq:subl_soft_factored_one_minus}
{\mathcal A}^{[1]}(1^-,2^+,3^+,4^+,5_s^+)
&= S^{[0](0)}(5^+){\mathcal A}^{[1]}(1^-,2^+,3^+,4^+)
  \\
&\quad+S^{[0](2)}(5^+){\mathcal A}^{[1]}(1^-,2^+,3^+,4^+)
\nonumber \\
&\quad+\text{Split}_{\bn_4=3}^{[1](2)}(P^+\rightarrow 4^+,5^+){\mathcal A}^{[0]}(1^-,2^+,3^+,P^-)
 +{\cal O}(\lambda^1)  \,.
 \nonumber
\end{align}
Here the superscripts $[0]$ or $[1]$ denote the loop order, while the superscripts $(0)$ and $(2)$ denote the order in the soft or $\lambda$ expansion (scaling like $\lambda^{-2}$ and $\lambda^{0}$, repsectively).  $S^{[0](0)}(5^+)$ is just the tree-level leading soft factor and ${\mathcal A}^{[1]}(1^-,2^+,3^+,4^+)$ is the finite one-loop amplitude, so
\begin{equation}
S^{[0](0)}(5^+){\mathcal A}^{[1]}(1^-,2^+,3^+,4^+)=\bigg[\frac{\sab{4}{1}}{\sab{4}{5}\sab{5}{1}} \bigg]
  \bigg( \frac{i}{48\pi^2}\frac{\sab{1}{3}^3\sab{2}{4}\ssb{1}{2}}{\sab{2}{3}^2\sab{3}{4}^3} \bigg) \, .
\end{equation}
Thus the leading term obeys the leading soft theorem as one might have anticipated.\footnote{As discussed in \Sec{sec:subsoftsingminus}, soft loops give vanishing contribution for this amplitude.}

At subleading order in the expansion, there are two terms that exist.  $S^{[0](2)}(5^+)$ is the subleading soft factor, and this contribution corresponds to the Low-Burnett-Kroll theorem:
\begin{align}
S^{[0](2)}(5^+){\mathcal A}^{[1]}(1^-,2^+,3^+,4^+)&=S^{\text{(sub)}}(5^+){\mathcal A}^{[1]}(1^-,2^+,3^+,4^+)\nn\\
&=\bigg[\frac{\ssb{5}{2}}{\sab{5}{1}\ssb{1}{2}}\bigg] \bigg(\frac{i}{48\pi^2}\frac{\sab{1}{3}^3\sab{2}{4}\ssb{1}{2}}{\sab{2}{3}^2\sab{3}{4}^3} \bigg) \, .
\end{align}
However, in \Eq{eq:subl_soft_factored_one_minus} there is a term that explicitly violates Low-Burnett-Kroll that comes from the one-loop splitting amplitude $\text{Split}^{[1](2)}_{\bn_4=3}(P^+\rightarrow 4^+,5^+)$,
\begin{align} \label{eq:Split1}
\text{Split}^{[1](2)}_{\bn_4=3}(P^+\rightarrow 4^+,5^+)&=\frac{-i}{48\pi^2}\frac{\sab{3}{5}}{\sab{3}{4}}\frac{\ssb{4}{5}}{\sab{4}{5}^2} \ ,
\end{align}
which multiplies the tree-level MHV amplitude ${\mathcal A}^{[0]}(1^-,2^+,3^+,P^-)$ in \Eq{eq:subl_soft_factored_one_minus}. Here $P$ denotes the intermediate particle in the collinear splitting.  In \Eq{eq:Split1} the presence of the momentum of particle 3 is simply to define the longitudinal momentum fractions of particles $4$ and $5$.  One can check that this one-loop splitting amplitude agrees with \Ref{Kosower:1999rx} for our choice of $\bar{n}_4\propto p_3$ in the soft limit, and once one rescales the momenta of particle $4$ to carry the total large component of the momenta involved in the splitting. Note that this splitting amplitude starts at ${\cal O}(\lambda^0)$ in the soft limit of particle $5$, and hence is the same order as $S^{[0](2)}$.

In \Sec{subsec:oneloop} below we will see that the factorized structure with a one-loop splitting amplitude appearing in \Eq{eq:subl_soft_factored_one_minus} can be derived as an immediate consequence of the factorized structure of subleading power one-loop soft amplitudes in SCET. 

This factorization structure of the subleading soft singularities nicely illustrates why the Low-Burnett-Kroll soft theorem is violated.  As the splitting amplitude $\text{Split}(P^+\rightarrow 4^+,5^+)$ is only non-zero beginning at one loop there are actually three particles involved in the splitting.  Two of these are real (4 and 5) and the third is the virtual particle $l$ circulating in the loop.  No matter how soft particle 5 is, or at what angle with respect to particle 4, there is always a region of loop integration where $p_4\cdot p_5\sim p_l\cdot p_4\sim p_l\cdot p_5$. This is the collinear scaling for the splitting amplitude where the particle in the loop $l$ and particle 4 form a collinear system.  This manifestly invalidates the requirement of the Low-Burnett-Kroll theorem in \Eq{eq:lows_theorem_pc}, since $p_4\cdot p_5 \sim p_4\cdot p_l$. Thus there are situations where the requirements of the Low-Burnett-Kroll theorem do not hold in quantum field theory, and these are even common momentum configurations for a gauge theory like QCD.   

We should emphasize that the violation occurs because the power counting assumption of the Low-Burnett-Kroll theorem is not satisfied, so this violation should be thought of as explicit rather than anomalous. Indeed, below  in \Sec{subsec:splitsub} we will demonstrate the same failure at tree-level with two particles in the collinear region of phase space.

\subsection{Power Counting for Loop and Emission Corrections to the Soft Theorems}
\label{subsec:loopsub}

To study corrections to the soft theorems from emissions or loops with SCET is conceptually not much more difficult than our analysis of widely separated energetic particles at tree level. For collinear emissions we consider Lagrangian insertions that cause a collinear splitting or higher order operators with two collinear fields in the same sector. For loops there are more technical details, but essentially matching the gauge theory onto the effective theory precisely maps the full theory loop integration into hard/ultraviolet corrections that modify the Wilson coefficient (hard amplitude) $C_N$, plus infrared loop corrections that are the loop integrals in the effective theory. The infrared part of the loops are precisely equal to the effective theory loops since $C_N$ is defined by the IR finite difference between the full theory and effective theory loop calculations.\footnote{A similar decomposition of loops is performed when the ``method of regions''  \cite{Beneke:1997zp} is used to calculate full theory loop integrals using dimensional regularization. Here the integrand is expanded according to a specific scaling for the loop momentum before integration, and then the results from all contributing scalings are added after integration. Schematically the hard loop momentum will correspond to the loop corrections in $C_N$, the collinear loop momentum will correspond with collinear loops in SCET, and the soft loop momentum will correspond with soft loops in SCET.  More precisely one must be careful to consider various choices for the momentum routing since it is the propagators that are determining whether a scaling contributes, rather than the loop momentum for a particular momentum routing. Also, the threshold expansion and SCET results may differ for loop amplitudes when the equations of motion in the effective theory have been used to simplify the structure of operators.  While often helpful in analyzing a process or obtaining the full theory result, this should not be confused with the effective theory. Finally, depending on the choice of IR regulators the effective theory fields may simultaneously encode several different distinct region results obtained from the threshold expansion (since the effective theory need not involve a single scale for all possible IR regulators).  For more information, \Ref{Bonocore:2014wua} applies the method of regions to the subleading soft limit of Drell-Yan production.}
The loop integrals in the effective theory will either involve an $n$-collinear loop momentum (if all particles in the loop are $n$-collinear), or a soft loop momentum if one or more of the propagators in the loop is for a soft particle.\footnote{The effective theory loops avoid potential double counting from overlap of the integration regions due to terms that are referred to as ``zero-bin subtractions''~\cite{Manohar:2006nz}.}
The effective theory reproduces the IR physics of the full theory in an on-shell and gauge-invariant manner.

From our analysis of the tree-level subleading soft factor in SCET, we see that to treat one external soft gluon up to ${\cal O}(\lambda^2)$ at loop-level or with collinear emissions we should consider the general ingredients that can show up at this order in the power counting:
\begin{enumerate}
\item insertions of the ${\cal L}^{(j)}$ Lagrangians with $j=0,1$, or $2$ that may create the soft gluon, pick out the subleading term in the multipole expansion of the soft momentum in a propagator, or create vertices in loop diagrams, 
\item a single hard scattering operator ${\cal O}_N^{(k)}$ with $k=0,1,2$.
\end{enumerate} 
To indicate the product of leading Lagrangians we will use $\Pi {\cal L}^{(0)}$ where
\begin{align}
 \Pi {\cal L}^{(0)} & 
  = \prod \int\!\! d^dx\, \bigg( \sum_i \Big[ {\cal L}_{\xi_{n_i}}^{(0)}(x) 
        + {\cal L}_{A_{n_i}}^{(0)}(x) +  {\cal L}_{\phi_{n_i}}^{(0)}(x) \Big]
        + {\cal L}_{\rm soft}^{(0)}(x) \bigg) 
  \,.
\end{align}
Here the number of terms $K$ in the product is not specified and we implicitly sum over $K$ (in a given diagram the value of $K$ is determined by the number of insertions we need, and may even be zero).  If we wish to denote that a soft gluon is extracted from a Lagrangian, then we will use the subset of terms ${\cal L}_{ns}^{(j)}\subset {\cal L}^{(j)}$, where
\begin{align}
 {\cal L}_{ns}^{(j)} 
  &= \sum_i \Big[ {\cal L}_{\xi_{n_i},{\rm soft}}^{(j)} 
        + {\cal L}_{A_{n_i},{\rm soft}}^{(j)} 
        +  {\cal L}_{\phi_{n_i},{\rm soft}}^{(j)} \Big] 
     + \delta_{j0}\: {\cal L}_{\rm soft,soft}^{(0)}
   \,.
\end{align}

Just accounting for the power counting we can then enumerate the classes of operator contributions for soft emission at various orders in $\lambda$ and any loop order: 
\begin{align} \label{eq:softpc}
& {\cal O}(\lambda^{-2}):
  & {\cal O}_N^{(0)} & \: \Pi {\cal L}^{(0)} 
  \\*[5pt]   
& {\cal O}(\lambda^{-1}):
  & {\cal O}_N^{(0)} &\: \Pi {\cal L}^{(0)} \: {\cal L}^{(1)} 
  \nn\\*   
&  & {\cal O}_N^{(1)} &\: \Pi {\cal L}^{(0)} 
  \nn\\*[5pt]  
& {\cal O}(\lambda^{0}):
  & {\cal O}_N^{(0)}& \: \Pi {\cal L}^{(0)} \: {\cal L}^{(2)} 
  \nn\\*  
&  & {\cal O}_N^{(0)}& \: \Pi {\cal L}^{(0)} \: {\cal L}^{(1)} \: {\cal L}^{(1)} 
  \nn\\*  
&  & {\cal O}_N^{(1)}&\: \Pi {\cal L}^{(0)} \: {\cal L}^{(1)} 
  \nn\\* 
&  & {\cal O}_N^{(2X/\partial)}&\: \Pi {\cal L}^{(0)}  
  \nn\\* 
&  & {\cal O}_N^{(2\delta,2r)}&\: \Pi {\cal L}^{(0)}  
  \nn 
\end{align}
Here ${\cal O}^{(1)}$ denotes contributions from both the operators ${\cal O}^{(1,X)}$ and ${\cal O}^{(1,\partial)}$, which were discussed in \Sec{sec:sub_ope}. For the operators ${\cal O}_N^{(2\delta,2r)}$ defined in \Eqs{eq:subsub_op_delta}{eq:subsub_op_rpi} the soft gluon may either come directly from the operator or from a ${\cal L}^{(0)}$  insertion, whereas for all other terms the soft gluon comes from a Lagrangian insertion (when we are considering the terms prior to the BPS field redefinition). The contributions from the category in each of the rows of \Eq{eq:softpc} are gauge invariant on their own. They generically include the contribution from several different types of Feynman diagrams. For example, at tree level we can extract a single soft gluon from different Lagrangians in the second row of \Eq{eq:softpc}, as
\begin{align}
 & {\cal O}_N^{(0)} \: \Pi {\cal L}^{(0)} \: {\cal L}^{(1)}  
  \to 
  {\cal O}_N^{(0)} \:  {\cal L}_{ns}^{(1)}  
 + {\cal O}_N^{(0)}  \: {\cal L}^{(1)}  
    \: {\cal L}_{ns}^{(0)}  \,.
\end{align}
These two terms exactly correspond with the two graphs calculated in \Eq{eq:subampcalc}, the first with the soft gluon from ${\cal L}^{(1)}$ and the second with ${\cal L}^{(1)}$ inserted on a propagator and the soft gluon extracted from an ${\cal L}^{(0)}$. Only the sum of these two contributions is gauge invariant. 

We will make use of the general classes in \Eq{eq:softpc} when deriving a soft theorem that is valid at one-loop in \Sec{subsec:oneloop}, and in discussing situations with two collinear emissions in \Sec{subsec:splitsub}.  For these purposes the key property of \Eq{eq:softpc} is that it gives a complete enumeration of contributions to the amplitudes at these orders in the $\lambda$ power expansion and any order in the coupling constant.

\subsection{Soft Theorem at One-Loop}
\label{subsec:oneloop}

In this section we will derive a one-loop level power suppressed soft theorem for the amplitude for a soft gauge boson emitted from well-separated energetic particles.  This result follows from the factorized structure of the amplitude which we will derive with SCET. From this analysis we will see how the factorized result for the single-minus amplitude ${\mathcal A}^{[1]}(1^-,2^+,3^+,4^+,5^+)$ in \Eq{eq:subl_soft_factored_one_minus} arises as a special case of the more general result.

The soft expansion of the one-loop amplitude can be decomposed as
\begin{align}
  {\cal A}_{N+1_s}^{[1]} & = {\cal A}_{N+1_s}^{[1](0)} 
    + {\cal A}_{N+1_s}^{[1](1)}  + {\cal A}_{N+1_s}^{[1](2)} + \ldots
   \,,
\end{align}
where the terms are all one-loop but leading, subleading, and sub-subleading in the $\lambda$ expansion respectively. The term ${\cal A}_{N+1_s}^{[1](2)}$ is the same order as the non-zero tree level LBK soft theorem. The ellipses denote terms that are even higher order in $\lambda$, which we neglect.  

\subsubsection{Leading power one-loop soft theorem}

For the leading term, using the notation of \Eq{eq:softpc} and splitting out the term emitting the soft gluon to distinguish whether it occurs inside or outside the loop, we have
\begin{align} \label{eq:ANs10}
  {\cal A}_{N+1_s}^{[1](0)} 
  &= \big\langle T\, {\cal O}_N^{[1](0)} {\cal L}_{ns}^{(0)}  \big\rangle  
   + \big\langle T\, \big[ {\cal O}_N^{[0](0)}  \Pi {\cal L}^{(0)} \big]^{[1]} {\cal L}_{ns}^{(0)}  \big\rangle 
   + \big\langle T\,  \big[  {\cal O}_N^{[0](0)}\Pi {\cal L}^{(0)} {\cal L}_{ns}^{(0)}   \big]^{[1]} \big\rangle  \,,
\end{align}
where the matrix elements all involve the same states and are in the interaction picture
\begin{align} \label{eq:smelt}
 \langle  \cdots \rangle 
   & = \langle 0 | \cdots | p_1,p_2,\ldots,p_N,p_s\rangle_{\rm int} \,.
\end{align}
Hard momentum loops enter through the first term in \Eq{eq:ANs10} which involves the one-loop operator ${\cal O}_N^{[1](0)}$. The ${\cal O}_N^{[1](0)}$ notation means that we evaluate the Wilson coefficient $C_N^{(0)}$ that appears inside ${\cal O}_N^{(0)}$  at one-loop order, namely we use $C_N^{[1](0)}$ in \Eq{eq:ON_soft_factorized}. The second term in \Eq{eq:ANs10} includes soft and collinear loops formed from vertices of ${\cal L}^{(0)}$, where the soft gluon emission Lagrangian ${\cal L}_{ns}^{(0)}$ acts on collinear particles that are external to the loop. For simplicity our notation $T \big[ {\cal O}_N^{[0](0)}  \Pi {\cal L}^{(0)} \big]^{[1]}$ includes contractions that involve fields in ${\cal O}_N^{[0](0)}$ inside the loop, as well as contractions that do not such as the collinear self-energy graphs. For the soft loops we have interactions between one or two collinear directions in a general gauge. For the purely collinear loops these contractions give the $1\to 1$ splitting function
\begin{align} \label{eq:splitkk}
 \text{Split}^{[1](0)}(k\to k) & = \big\langle 0 \big|   T \big[\delta(Q_k - i\bar n_k\cdot \partial_{n_k})  X_{n_k}^{\kappa_k} \big] \Pi {\cal L}_{n_k}^{(0)} \big| p_k \big\rangle^{[1]} \,.
\end{align}
For later convenience we also define
\begin{align} 
 \text{Split}^{[1](0)}  &\equiv \sum_{k=1}^N \: \text{Split}^{[1](0)}(k\to k) \,.
%    = \big\langle 0 \big| \hat {\cal I}^{(0)}_{N} \big| p_1,p_2,\ldots p_N \big\rangle^{[1]}\,.
\end{align}  
Together the first two terms of \Eq{eq:ANs10} give the sum of hard, soft, and collinear loops, and hence simply yield the leading power $N$-jet amplitude at one-loop order. The external soft emission from ${\cal L}_{ns}^{(0)}$ then just gives the leading power tree-level soft factor, so
\begin{align}
\big\langle T\, {\cal O}_N^{[1](0)} {\cal L}_{ns}^{(0)}  \big\rangle  
   + \big\langle T\, \big[{\cal O}_N^{[0](0)}  \Pi {\cal L}^{(0)} \big]^{[1]} {\cal L}_{ns}^{(0)}  \big\rangle  & = S^{[0](0)}(s)\, {\cal A}_N^{[1](0)} \,.
\end{align}
The third term in \Eq{eq:ANs10} includes soft loops where the soft gluon emission is coupled to the soft particles. Due to the all-orders factorized structure of \Eq{eq:ON_soft_factorized} in terms of soft Wilson lines, this type of one-loop correction only involves the soft sector, giving a one-loop correction to the soft factor $S^{[1](0)}$,
\begin{align}
  S^{[1](0)} &= \Big\langle 0 \Big| {\rm T} \Big\{ \prod_{i=1}^N Y_{n_i}^{\kappa_i}(0) \Big\} \Big| g_s \Big\rangle ^{[1]} \,.
\end{align}
The remaining collinear part of the matrix element is evaluated at tree level and just gives the $N$-point amplitude, so we have
\begin{align}
  \big\langle T\, \big[{\cal O}_N^{[0](0)}  \Pi {\cal L}^{(0)} {\cal L}_{ns}^{(0)}   \big]^{[1]} \big\rangle  &= S^{[1](0)}(s)\, {\cal A}_N^{[0](0)} \,.
\end{align}
All together, for the leading power one-loop soft emission amplitude we have
\begin{align}\label{eq:leadsoft_oneloop}
  {\cal A}_{N+1_s}^{[1](0)}  &= S^{[0](0)}(s)\, {\cal A}_N^{[1](0)} 
    +S^{[1](0)}(s)\, {\cal A}_N^{[0](0)} \,.
\end{align}
These two contributions are consistent with our discussion of the all-loop order factorized structure of the leading order amplitude in \Sec{subsec:simplecalcs}.

\subsubsection{Vanishing of the ${\cal O}(\lambda)$ one-loop soft theorem}
\label{sec:Olamsoftthm}

For the one-loop soft emission amplitude that is suppressed by ${\cal O}(\lambda)$, the decomposition of contributions is
\begin{align}\label{eq:ANs11}
  {\cal A}_{N+1_s}^{[1](1)} 
  &= \big\langle T\, {\cal O}_N^{[1](0)} \Pi {\cal L}^{(0)} {\cal L}^{(1)}  \big\rangle  
   + \big\langle T\, \big[{\cal O}_N^{[0](0)}  \Pi {\cal L}^{(0)} \big]^{[1]} (\Pi {\cal L}^{(0)} {\cal L}^{(1)})_{ns}  \big\rangle 
   + \big\langle T\,  \big[ {\cal O}_N^{[0](0)} \Pi {\cal L}^{(0)} {\cal L}^{(1)}   \big]^{[1]} \big\rangle  
  \nn \\*
  &+ \big\langle T\,  \big[ {\cal O}_N^{[0](0)} \Pi {\cal L}^{(0)} {\cal L}^{(1)}   \big]^{[1]} {\cal L}_{ns}^{(0)} \big\rangle  
   + \big\langle T\, {\cal O}_N^{[1](1)} {\cal L}_{ns}^{(0)}  \big\rangle  
   + \big\langle T\, \big[ {\cal O}_N^{[0](1)}  \Pi {\cal L}^{(0)}  \big]^{[1]} {\cal L}_{ns}^{(0)}  \big\rangle 
  \nn\\*
  &
   + \big\langle T\, \big[ {\cal O}_N^{[0](1)} \Pi {\cal L}^{(0)} 
       \big]^{[1]} \big\rangle  
   \,. 
\end{align}
Here the soft gluon is created by either a term from ${\cal L}^{(1)}$ or ${\cal L}^{(0)}$ and in the 1st, 2nd, 4th, 5th, and 6th terms the  soft emission is from outside of the loop. 
The 1st and 2nd terms involve an ${\cal L}^{(1)}$ outside of the loop, and vanish by the same RPI argument used for the tree-level amplitude at this order in $\lambda$ (taking $p_{n_i\perp}^\mu=0$ for external collinear particles). In the 5th term with ${\cal O}_N^{[1](1)}$, the loop correction is contained in the Wilson coefficient of the ${\cal O}_N^{(1)}$ operator. Here we either have a contribution proportional to $p_{n_i\perp}^\mu=0$ from ${\cal O}_N^{[1](1,\partial)}$ or a contribution that must generate two particles in the same collinear sector from ${\cal O}_N^{[1](1,X)}$ which has a vanishing matrix element for the states in \Eq{eq:smelt}.  This leaves
\begin{align}\label{eq:ANs11rpi}
  {\cal A}_{N+1_s}^{[1](1)} 
  &= \big\langle T\, \big[ {\cal O}_N^{[0](0)}  \Pi {\cal L}^{(0)} {\cal L}^{(1)}   \big]^{[1]} \big\rangle  
   + \big\langle T\, \big[  {\cal O}_N^{[0](0)} \Pi {\cal L}^{(0)} {\cal L}^{(1)} \big]^{[1]} {\cal L}_{ns}^{(0)}  \big\rangle 
  \nn \\
  & 
   + \big\langle T\, \big[ {\cal O}_N^{[0](1)}  \Pi {\cal L}^{(0)}  \big]^{[1]} {\cal L}_{ns}^{(0)}  \big\rangle 
  + \big\langle T\, \big[ {\cal O}_N^{[0](1)}  \Pi {\cal L}^{(0)}  \big]^{[1]}   \big\rangle 
  \, . 
\end{align}

\begin{figure}
	\begin{center}
		\includegraphics[width=3.5cm]{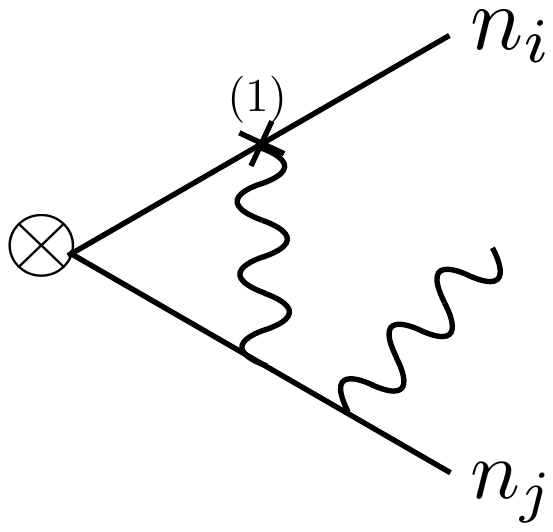}
		$\qquad\qquad\quad$
		\includegraphics[width=3.5cm]{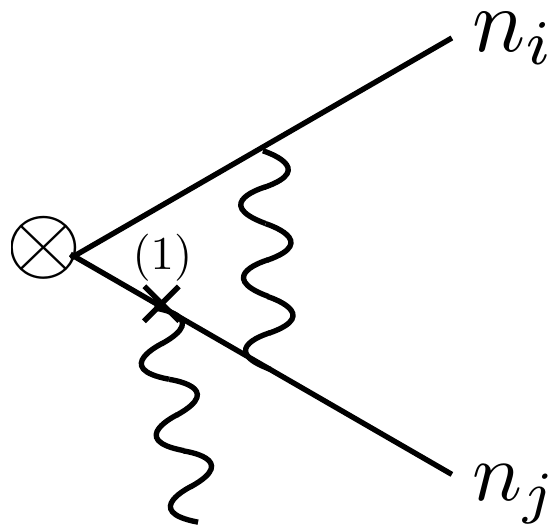}
				$\qquad\qquad\quad$
		\includegraphics[width=3.7cm]{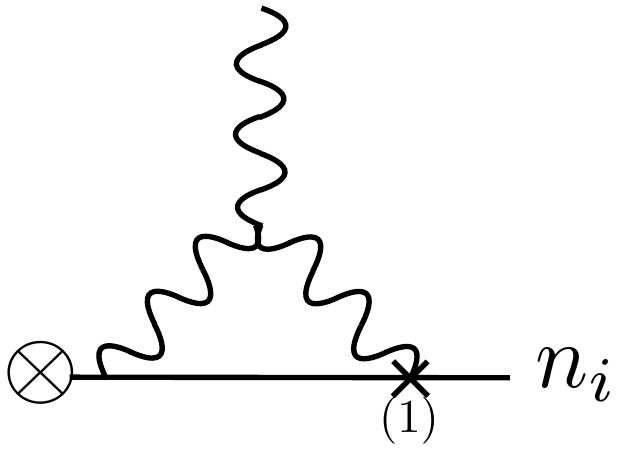}
	\end{center}
	\caption{
		Examples of one-loop diagrams with soft loops that enter for computing the correction to the subleading soft theorem.  $n_i$ and $n_j$ are two collinear directions with $n_i\cdot n_j \gg \lambda^2$ and $\stackrel{(1)}{\times}$ denotes the coupling of the soft gluon via the subleading SCET Lagrangian, ${\cal L}^{(1)}$. Using RPI to set $p_i^\perp=0$, all such %${\cal O}(\lambda)$ 
		soft loop graphs vanish.
	}
	\label{fig:loop_soft}
\end{figure}

All of the remaining contributions, given in \Eq{eq:ANs11rpi}, involve either a soft loop (with one or more soft gluon propagators) or a $n$-collinear loop (with only $n$-collinear propagators). Here the ${\cal O}(\lambda)$ vertex from ${\cal L}^{(1)}$ or ${\cal O}_N^{(1)}$ potentially participates in the loop.  We will show that these terms also all vanish at one-loop order.  First consider the soft loop with loop momentum $k_s$.  We show some representative diagrams in \Fig{fig:loop_soft}.  Here the emission of the soft gluon can either be from a collinear particle inside the loop, a collinear particle external to the loop, or from the soft gluon in the loop. The operators ${\cal L}^{(1)}$ and ${\cal O}_N^{(1,\partial)}$ both always introduce a collinear momentum $p_{n_i\perp}^\mu$ somewhere in the graph, and due to the multipole expansion ($p_{n_i\perp}\gg k_s$) this factor can be freely moved outside the soft loop. Because each collinear sector here only contains a single particle, we can use RPI to set $p_{i\perp}^\mu = 0$ for all hard external particles.  Therefore, all soft loop contributions to this order vanish, for essentially the same reason observed at tree-level.

\begin{figure}
	\begin{center}
		\includegraphics[width=4.5cm]{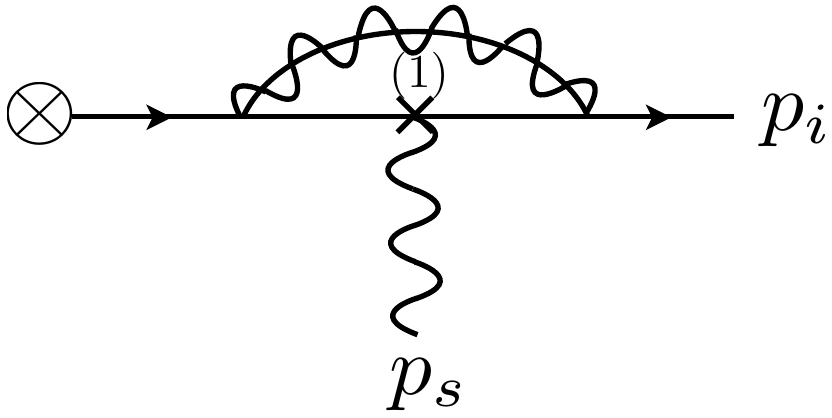}
				$\qquad$
\includegraphics[width=4.5cm]{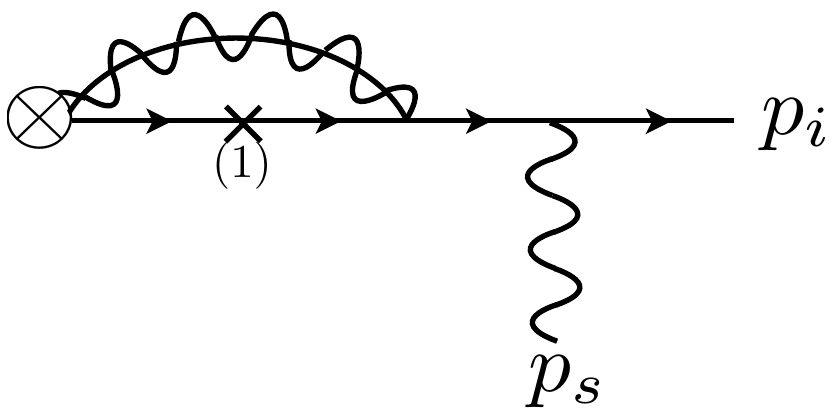}				
$\qquad$
\includegraphics[width=4.5cm]{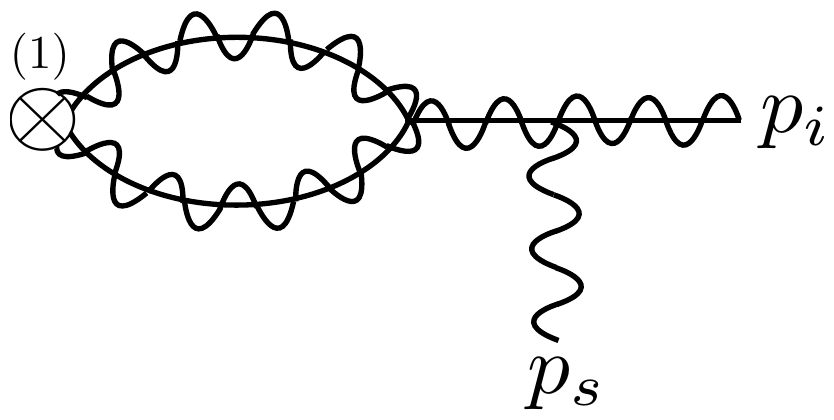}
	\end{center}
	\caption{
		Examples of a diagrams of collinear loops coupling to a soft gluon at subleading power.  $p_i$ is the momentum of the external collinear fermion and $\stackrel{(1)}{\times}$ denotes the subleading collinear operator ${\cal O}^{(1,X)}$ or the coupling of the soft gluon via the subleading SCET Lagrangian, ${\cal L}^{(1)}$. In the middle diagram, the collinear gluon couples to the $W_n$ collinear Wilson line present in $\otimes$. Other contributing diagrams are not shown, for example the soft gluon can couple to other collinear lines in the diagram, and also in a four point interaction with the collinear gluon. 
	}
	\label{fig:loop_soft_collinear}
\end{figure}

Collinear loops are more subtle.  In this case, RPI does not immediately guarantee that these corrections vanish.  Three example collinear loop diagrams are shown in \Fig{fig:loop_soft_collinear} for an external collinear fermion or gluon.  Even using RPI to set $p_{i\perp}^\mu = 0$ for the external collinear particle does not force these diagrams to be zero because the collinear gluon in the loop has non-zero $p_\perp$.  By explicitly computing a diagram, however, we can see the resolution.  For example, consider the diagram on the left of \Fig{fig:loop_soft_collinear}.  Setting $p_{i\perp}=0$, this evaluates to
\begin{align}\label{eq:subleading_coll_loop}
&\raisebox{-0.66\height}{\includegraphics[width=3.5cm]{./figures/one_loop_coll.pdf}}\nonumber \\*
&\quad= \bar{u}(p_i)\left[
-2i(d-2)
\right]\frac{\epsilon_{s \mu}}{n\cdot p_s}\int \frac{d^d\ell}{(2\pi)^d} \frac{\ell_\perp^2 \ell_\perp^\mu}{\ell^2 [\bar{n}\cdot (p_i-\ell)][(\bar{n}\cdot (p_i-\ell-p_s))(n\cdot (p_i-\ell))+\ell_\perp^2]}\nonumber \\
&\quad=0 \ ,
\end{align} 
where we have suppressed color and coupling factors, $\ell$ is the momentum of the virtual collinear gluon and $d$ is the space-time dimension.  Because the integrand is linear in $\ell_\perp^\mu$, the diagram vanishes.  Indeed, all one-loop diagrams with collinear loops from either ${\cal L}^{(1)}$ SCET Lagrangian insertions or matrix elements of ${\cal O}^{(1)}$ vanish either by RPI or because the integrand is linear in the $\perp$ component of loop momentum.  Therefore, we have proven that the one-loop ${\cal O}(\lambda)$-suppressed amplitude vanishes,
\begin{equation}
  {\cal A}_{N+1_s}^{[1](1)}  = 0 \,.
\end{equation}

\subsubsection{General One-Loop Soft Theorem at ${\cal O}(\lambda^2)$ }\label{sec:order_lambda_squ_one_loop}

Finally, we turn to the terms that will generate a subleading soft theorem valid at one-loop order.  For the one-loop soft emission amplitude that is suppressed by ${\cal O}(\lambda^2)$, the decomposition of contributions is
\begin{align}\label{eq:ANs12}
  {\cal A}_{N+1_s}^{[1](2)} 
% L^(2)
  &= \big\langle T\, {\cal O}_N^{[1](0)} \Pi {\cal L}^{(0)} {\cal L}^{(2)}  \big\rangle  
   + \big\langle T\, \big[{\cal O}_N^{[0](0)}  \Pi {\cal L}^{(0)} \big]^{[1]} \Pi {\cal L}^{(0)} {\cal L}^{(2)}  \big\rangle 
   + \big\langle T\,  \big[ {\cal O}_N^{[0](0)} \Pi {\cal L}^{(0)} {\cal L}^{(2)}   \big]^{[1]} \big\rangle  
  \nn \\
  &\hspace{-0.6cm}
   + \big\langle T\,  \big[ {\cal O}_N^{[0](0)} \Pi {\cal L}^{(0)} {\cal L}^{(2)}   \big]^{[1]} {\cal L}_{ns}^{(0)} \big\rangle  
% L^(1) L^(1)
   + \big\langle T\, {\cal O}_N^{[1](0)}  \Pi {\cal L}^{(0)}
    ({\cal L}^{(1)})^2  \big\rangle  
   + \big\langle T\, \big[ {\cal O}_N^{[0](0)} \Pi {\cal L}^{(0)} 
     ({\cal L}^{(1)})^2  \big]^{[1]} {\cal L}_{ns}^{(0)} \big\rangle  
  \nn\\
  &\hspace{-0.6cm}
   + \big\langle T\, \big[{\cal O}_N^{[0](0)}  \Pi {\cal L}^{(0)}  \big]^{[1]} \Pi {\cal L}^{(0)} ({\cal L}^{(1)})^2  \big\rangle 
   + \big\langle T\, \big[{\cal O}_N^{[0](0)}  \Pi {\cal L}^{(0)} {\cal L}^{(1)} \big]^{[1]} \Pi {\cal L}^{(0)} {\cal L}^{(1)}  \big\rangle 
  \nn\\
  &\hspace{-0.6cm}
   + \big\langle T\, \big[ {\cal O}_N^{[0](0)} \Pi {\cal L}^{(0)} 
     ({\cal L}^{(1)})^2   \big]^{[1]} \big\rangle  
% O^(1)
   + \big\langle T\, {\cal O}_N^{[1](1)} \Pi {\cal L}^{(0)}{\cal L}^{(1)}  \big\rangle  
   + \big\langle T\, \big[ {\cal O}_N^{[0](1)}  \Pi {\cal L}^{(0)}  \big]^{[1]} \Pi {\cal L}^{(0)} {\cal L}^{(1)}  \big\rangle 
  \nn\\
  &\hspace{-0.6cm}
   + \big\langle T\, \big[ {\cal O}_N^{[0](1)} \Pi {\cal L}^{(0)} 
     {\cal L}^{(1)}   \big]^{[1]}  {\cal L}_{ns}^{(0)} \big\rangle  
   + \big\langle T\, \big[ {\cal O}_N^{[0](1)} \Pi {\cal L}^{(0)} 
     {\cal L}^{(1)}   \big]^{[1]} \big\rangle  
% L^(2X,2partial)
   + \big\langle T\, {\cal O}_N^{[1](2X/\partial)} {\cal L}_{ns}^{(0)}  \big\rangle  
  \nn\\
  &\hspace{-0.6cm}
   + \big\langle T\, \big[ {\cal O}_N^{[0](2X/\partial)}  \Pi {\cal L}^{(0)}  \big]^{[1]} {\cal L}_{ns}^{(0)}  \big\rangle 
   + \big\langle T\, \big[ {\cal O}_N^{[0](2X/\partial)} \Pi {\cal L}^{(0)} 
     \big]^{[1]} \big\rangle  
% L^(2delta,2r)
   + \big\langle  {\cal O}_N^{[1](2\delta,2r)}  \Pi {\cal L}^{(0)}  \big\rangle   
 \nn\\
  &\hspace{-0.6cm}
   + \big\langle T\, \big[ {\cal O}_N^{[0](2\delta,2r)}  \Pi {\cal L}^{(0)}  \big]^{[1]} {\cal L}_{ns}^{(0)}  \big\rangle 
   + \big\langle T\, \big[ {\cal O}_N^{[0](2\delta,2r)} \Pi {\cal L}^{(0)} 
       \big]^{[1]} \big\rangle  
  \,.
\end{align}
Again terms involving a ${\cal L}^{(1)}$ outside of the loop vanish by RPI with the external $p_{n_i\perp}^\mu=0$, and terms with ${\cal O}_N^{[1](1)}$ or ${\cal O}_N^{[1](2X/\partial)} $ vanish by having an extra collinear particle that vanishes for the state in \Eq{eq:smelt}. Using these two properties leaves
\begin{align}\label{eq:ANs12rpi}
  {\cal A}_{N+1_s}^{[1](2)} 
% L^(2)
  &= \big\langle T\, {\cal O}_N^{[1](0)} \Pi {\cal L}^{(0)} {\cal L}^{(2)}  \big\rangle  
   + \big\langle T\, \big[{\cal O}_N^{[0](0)}  \Pi {\cal L}^{(0)} \big]^{[1]} \Pi {\cal L}^{(0)} {\cal L}^{(2)}  \big\rangle 
   + \big\langle T\,  \big[ {\cal O}_N^{[0](0)} \Pi {\cal L}^{(0)} {\cal L}^{(2)}   \big]^{[1]} \big\rangle  
  \nn \\
  &\hspace{-0.6cm}
   + \big\langle T\,  \big[ {\cal O}_N^{[0](0)} \Pi {\cal L}^{(0)} {\cal L}^{(2)}   \big]^{[1]} {\cal L}_{ns}^{(0)} \big\rangle  
% L^(1) L^(1) 
   + \big\langle T\, \big[ {\cal O}_N^{[0](0)} \Pi {\cal L}^{(0)} 
     ({\cal L}^{(1)})^2  \big]^{[1]} {\cal L}_{ns}^{(0)} \big\rangle  
  \nn\\
  &\hspace{-0.6cm}
   + \big\langle T\, \big[ {\cal O}_N^{[0](0)} \Pi {\cal L}^{(0)} 
     ({\cal L}^{(1)})^2   \big]^{[1]} \big\rangle  
% O^(1)
   + \big\langle T\, \big[ {\cal O}_N^{[0](1)} \Pi {\cal L}^{(0)} 
     {\cal L}^{(1)}   \big]^{[1]}  {\cal L}_{ns}^{(0)} \big\rangle  
   + \big\langle T\, \big[ {\cal O}_N^{[0](1)} \Pi {\cal L}^{(0)} 
     {\cal L}^{(1)}   \big]^{[1]} \big\rangle  
% L^(2X,2partial)
  \nn\\
  &\hspace{-0.6cm}
   + \big\langle T\, \big[ {\cal O}_N^{[0](2X/\partial)}  \Pi {\cal L}^{(0)}  \big]^{[1]} {\cal L}_{ns}^{(0)}  \big\rangle 
   + \big\langle T\, \big[ {\cal O}_N^{[0](2X/\partial)} \Pi {\cal L}^{(0)} 
     \big]^{[1]} \big\rangle  
% L^(2delta,2r)
   + \big\langle  {\cal O}_N^{[1](2\delta,2r)}  \Pi {\cal L}^{(0)}  \big\rangle   
 \nn\\
  &\hspace{-0.6cm}
   + \big\langle T\, \big[ {\cal O}_N^{[0](2\delta,2r)}  \Pi {\cal L}^{(0)}  \big]^{[1]} {\cal L}_{ns}^{(0)}  \big\rangle 
   + \big\langle T\, \big[ {\cal O}_N^{[0](2\delta,2r)} \Pi {\cal L}^{(0)} 
       \big]^{[1]} \big\rangle  
  \,.
  \end{align}
Note that with our notation that $\Pi {\cal L}^{(0)}$ could include zero, one, two or more insertions of ${\cal L}^{(0)}$.

A couple of the terms in \Eq{eq:ANs12rpi} can be recognized as involving the Low-Burnett-Kroll soft factor, $S^{\text{(sub)}}(s)=S^{[0](2)}(s)$ times a one-loop amplitude.  Ideally we would like to isolate a term involving the full one-loop amplitude times this LBK soft factor. To do this we would have to be able to ensure that the terms causing the ${\cal O}(\lambda^2)$ suppression in the SCET decomposition of the full theory result act as the LBK angular momentum operator acting on the result of hard, soft, or collinear loops. Since the hard loops are encoded in Wilson coefficients in the SCET operators, they automatically satisfy this criteria.  We can also easily group terms with subleading Lagrangian insertions into cases where the power suppression occurs outside of a soft or collinear loop integral. However, there is no obvious way to do this for the operators ${\cal O}_N^{[0](2\delta,2r)}$ in the presence of soft and collinear loops. In both cases the $D_s^\mu$ in the operators are generically internal to a soft loop involving this operator. Further, for the operator ${\cal O}_N^{[0](2,\delta)}$, the derivative on the Wilson coefficient also does not generically lead to a straightforward interpretation as the angular momentum derivative on the full amplitude, since the presence of a collinear loop can modify the $\bn\cdot p$ dependence. One cannot simply commute the $\partial/\partial(\bn\cdot p)$ derivative through so that it acts on both the hard amplitude and the result of the collinear loop. Thus, unlike at tree-level, it is only easy to separate out the LBK soft factor acting on the hard loop contribution, which is given by the terms
\begin{align}\label{eq:LBK_loop_term}
{\cal A}_{N+1_s}^{[1,\text{hard}](2)} 
 = \big\langle T\, {\cal O}_N^{[1](0)} \Pi {\cal L}^{(0)} {\cal L}^{(2)}  \big\rangle  
   + \big\langle  {\cal O}_N^{[1](2\delta,2r)}  \Pi {\cal L}^{(0)}  \big\rangle  =S^{[0](2)}(s){\cal A}^{[1,\text{hard}](0)}_N \,.
\end{align}
Here ${\cal A}^{[1,\text{hard}](0)}_N$ is the contribution to the one-loop amplitude from hard loops at leading power.  ${\cal A}^{[1,\text{hard}](0)}_N$ is infrared finite. It is in general not equal to the full one-loop amplitude ${\cal A}^{[1](0)}_N$, which often also has contributions from collinear and soft loops and has infrared divergences. The $\overline{\rm MS}$ result for  ${\cal A}^{[1,\text{hard}](0)}_N$ can be obtained using the standard matching trick of evaluating the full theory renormalized amplitude using $\epsilon$ to regulate both IR and UV divergences, and then simply dropping the $1/\epsilon_{\rm IR}$ poles (since the bare SCET graphs are scaleless with this regulator, the SCET UV counterterms cancel the full theory IR divergences when we subtract full and EFT results, and set $\epsilon_{\rm IR}=\epsilon_{\rm UV}$). 

For the soft expansion of the single-minus 5-point amplitude discussed below in \Sec{sec:subsoftsingminus} we will show that ${\cal A}^{[1,\text{hard}](0)}(1^-,2^+,3^+,4^+)={\cal A}^{[1](0)}(1^-,2^+,3^+,4^+)$, and so \Eq{eq:LBK_loop_term} will be the LBK soft factor acting on a full one loop amplitude.

The remaining ${\cal O}(\lambda^2)$  suppressed terms enumerated in \Eq{eq:ANs12rpi} must include either a soft or a collinear loop within SCET, and have hard coefficients evaluated at tree-level. Since there is not a significant benefit to separating cases where the soft attachments are inside or outside the loops we now make the BPS field redefintion, which absorbs the terms with an ${\cal L}_{ns}^{(0)}$ into Wilson lines appearing in other categories. This leaves only five types of terms
\begin{align}\label{eq:ANs12sc}
{\cal A}_{N+1_s}^{[1 \text{(soft,coll)}](2)} 
% L^(2)
&=  \big\langle T\,  \big[ {\cal O}_N^{[0](0)} \Pi {\cal L}^{(0)} {\cal L}^{(2)}   \big]^{[1]} \big\rangle  
% L^(1) L^(1) 
+ \big\langle T\, \big[ {\cal O}_N^{[0](0)} \Pi {\cal L}^{(0)} 
({\cal L}^{(1)})^2   \big]^{[1]} \big\rangle  
 \\
&\hspace{-0.8cm}
% O^(1)
+ \big\langle T\, \big[ {\cal O}_N^{[0](1)} \Pi {\cal L}^{(0)} 
{\cal L}^{(1)}   \big]^{[1]} \big\rangle  
% L^(2X,2partial)
+ \big\langle T\, \big[ {\cal O}_N^{[0](2X/\partial)} \Pi {\cal L}^{(0)} 
\big]^{[1]} \big\rangle  
%\nn\\
%&\hspace{-0.8cm}
% L^(2delta,2r)
+ \big\langle T\, \big[ {\cal O}_N^{[0](2\delta,2r)} \Pi {\cal L}^{(0)} 
\big]^{[1]} \big\rangle  
\,, \nn
\end{align}
where (soft,coll) denotes that the loops are soft or collinear.  Since we are not distinguishing soft attachments outside/inside the loop we have
here absorbed $ \big\langle T\, \big[{\cal O}_N^{[0](0)}  \Pi {\cal L}^{(0)} \big]^{[1]} \Pi {\cal L}^{(0)} {\cal L}^{(2)}  \big\rangle$  into the notation used in the first term of \Eq{eq:ANs12sc}. In the operators in \Eq{eq:ANs12sc} the soft and collinear fields appear in factorized blocks, connected only by global color and Lorentz indices. The subleading operators have non-Wilson line soft fields only through $g B_{s(n)}^{A\mu}$, see \Eqs{eq:subsub_op_delta}{eq:subsub_op_rpi}.
For the subleading Lagrangian insertions the soft and collinear fields were written in a factorized form above in \Eq{eq:LKB}, where the soft fields also appear in $g B_{s(n)}^{A\mu}$.   All terms in \Eq{eq:ANs12sc} therefore involve one of the following products of soft fields and soft Wilson lines:
\begin{align}\label{eq:soft_emit_simp1}
 \hat {\cal E}_{s(n)}^{\mu\,\vec \kappa}(x) &\equiv T\, \prod_i\, Y_{n_i}^{\kappa_i}(0)\, T^{\kappa A} g B_{s(n)}^{A\mu} (x)  \,, 
 \qquad\qquad  
  \hat {\cal E}_{s(n)}^{\mu\,\vec \kappa}(0)  \,,
  \\
\hat {\cal E}_{s(n)(n')}^{\mu\nu\,\vec \kappa}(x,y) 
   &\equiv  T\,  \prod_i\, Y_{n_i}^{\kappa_i}(0)\, T^{\kappa A} T^{\kappa' B}
    g B_{s(n)}^{A\mu} (x)\,   g B_{s(n')}^{B\mu} (y) \,,
   \qquad
    \hat {\cal E}_{s(n)(n)}^{\mu\nu\,\vec \kappa}(x,x)  
    \,, \nn\\
\hat E_{s[n]N_X}^{\,\vec \kappa} & \equiv T \prod_{i,n_i\ne n} Y^{\kappa_i}_{n_i}(0) \: \prod_{j=1}^{N_X} Y^{\kappa_j}_n(0) \,,
  \qquad
\hat {\cal E}_{s(n)[n']}^{\mu \,\vec \kappa}(x) \equiv T\, Y_{n'}^{\kappa_{n'}}(0) \prod_i\, Y_{n_i}^{\kappa_i}(0)\, g B_{s(n)}^{A\mu} (x) 
  \: , \nn \\
\hat E^{\,\vec \kappa}_{s[n_j][n_k]} & \equiv T\: Y^{\kappa_j'}_{n_j}(0) 
   Y^{\kappa_k'}_{n_k}(0)   \prod_{i} Y^{\kappa_i}_{n_i}(0) 
  \,,\nn
\end{align}
where $N_X=1, 2,$ or $3$ and $T$ denotes time-ordering.  The notation $\vec \kappa$ on the LHS of these structures encodes the possible color representations. We will denote $\hat E_s= \hat E_{s[n]1}$ since it is independent of $n$. Note that the leading soft factor is given by the matrix element of $\hat E_s$ at any loop order. For the other soft operators we will simply drop the hat when denoting the one soft gluon matrix elements, so
at $l$ loop level
\begin{align}
  S^{[l](0)} &= \big\langle g(s) \big| \hat E_s \big| 0\big\rangle^{[l]} \,,
  & E_{\cdots}^{[l]\cdots \vec\kappa}&
    =  \big\langle g(s) \big| 
     \hat E_{\cdots}^{\cdots \vec\kappa} \big| 0\big\rangle^{[l]} 
  \,,
  & {\cal E}_{\cdots}^{[l]\cdots \vec\kappa}(\cdots) &
    =  \big\langle g(s) \big| 
     \hat {\cal E}_{\cdots}^{\cdots \vec\kappa}(\cdots) \big| 0\big\rangle^{[l]}
  \,.
\end{align}    
These matrix elements must be considered with insertions of the soft Lagrangian, $\Pi {\cal L}_{\rm soft}^{(0)}$, which we suppress for simplicity. Most often the soft loops involve attachments between Wilson lines in multiple collinear directions, and hence are not related to one-particle splitting functions.    If we consider the classes in \Eq{eq:ANs12sc} then the terms that are generated by the various contributions include
\begin{align} \label{eq:Ematch}
{\cal L}^{(2)}: &  \qquad\qquad  
 \hat {\cal E}_{s(n)}^{\mu\,\vec \kappa}(x) \,, \
 \hat {\cal E}_{s(n)(n)}^{\mu\nu\,\vec \kappa}(x,x) \,,\
 \hat E_{s} \,,
  \\
\big( {\cal L}^{(1)}\big)^2: &  \qquad\qquad 
 \hat {\cal E}_{s(n)}^{\mu\,\vec \kappa}(x) \,, \
 \hat {\cal E}_{s(n)(n')}^{\mu\nu\,\vec \kappa}(x,y) \,, \
 \hat E_{s} \,, \
  \nn \\
{\cal O}_N^{(1,X)} {\cal L}^{(1)}: & \qquad \qquad
%  \hat {\cal E}_{s(n)}^{\mu\,\vec \kappa}(x) \,,\
  \hat {\cal E}_{s(n)[n']}^{\mu\,\vec \kappa}(x) \,,\
  \hat E_{s[n]2} \,,\
%  \hat E_{s} \,, \
  \nn\\
{\cal O}_N^{(2,X\partial)}: & \qquad\qquad
  \hat E_{s[n]2} \,,\
  \nn\\
{\cal O}_N^{(2,X^2)}: & \qquad\qquad
  \hat E_{s[n]3} \,,\
  \nn\\
{\cal O}_N^{(2,XX)}: & \qquad\qquad
  \hat E_{s[n_j][n_k]} \,,\
  \nn\\
{\cal O}_N^{(2\delta,2r)}: & \qquad\qquad
  \hat {\cal E}_{s(n)}^{\mu\,\vec \kappa}(0) 
\,. \
  \nn
\end{align}
Note that $\hat E_s$ is the same operator that appeared at leading power. Here the situations with an $\hat E_s$ occur when the subleading Lagrangians contribute only $i\partial_s^\mu$ rather than a $g B_{s(n)}^{A\mu}$. The situations with one of $\hat E_{s[n]2}$, $\hat E_{s[n]3}$, or $\hat E_{s[n_j][n_k]}$ occur when we have multiple collinear building blocks $X_n$ in the same collinear direction. The fact that only these fairly simple soft operators appear in the one-loop subleading soft theorem will imply a particular form for the universality of contributions to the amplitude at this order.

The cases in \Eq{eq:ANs12sc} with a soft or collinear loop are individually gauge invariant (with appropriate IR regulators like dimensional regularization) and hence can be considered separately.  Let's first consider the contribution from soft loops.  Because soft emissions cannot affect the $\perp$ component of collinear momentum, all soft loop contributions involving terms with an insertion of ${\cal L}^{(1)}$ can be set to zero by RPI.  Also, insertions of ${\cal O}_N^{[0](2X/\partial)}$ from the operators in \Eq{eq:O2nonrpi} have an extra external collinear particle that cannot match onto the state with soft gluons or are proportional to $p_{i\perp}=0$. The only terms contributing with a soft loop are therefore
\begin{align}\label{eq:ANs12rpi_soft}
  {\cal A}_{N+1_s}^{[1,\text{soft}](2)} 
  &= \big\langle T\,  \big[ {\cal O}_N^{[0](0)} \Pi {\cal L}^{(0)} {\cal L}^{(2)}   \big]^{[1,\text{soft}]} \big\rangle  
   + \big\langle T\, \big[ {\cal O}_N^{[0](2\delta,2r)} \Pi {\cal L}^{(0)} 
       \big]^{[1,\text{soft}]} \big\rangle  
  \,.
\end{align}
From \Eq{eq:Ematch} this implies that the only soft operators that can contribute to ${\cal A}_{N+1_s}^{[1,\text{soft}](2)}$ are:
\begin{align}
 \hat {\cal E}_{s(n)}^{\mu\,\vec \kappa}(0) \,, \qquad
 \hat {\cal E}_{s(n)}^{\mu\,\vec \kappa}(x) \,, \qquad
 \hat {\cal E}_{s(n)(n)}^{\mu\nu\,\vec \kappa}(x,x) \,,\qquad
  \hat E_{s} \,.
\end{align}

Using the generic form of the subleading SCET Lagrangians after the BPS field redefinition from \Eq{eq:LKB}, we can then express each contribution in \Eq{eq:ANs12rpi_soft} in terms of the universal factors composed of soft fields from \Eq{eq:Ematch} and hard operators that depend on the specific lower point amplitude.  We have
\begin{align}\label{eq:loop_structure}
T\,  {\cal O}_N^{(0)} \Pi {\cal L}^{(0)} {\cal L}^{(2)} 
&=C_N^{(0)}\,\bigg\{
\sum_k \int\!\! d^dx\, \Big[T\, \hat{\cal O}_N^{(0)}(0)\, \hat K_{n_k}^{(2)}(x)\, \Pi {\cal L}^{(0)}_\text{coll}\Big] \hat E_{s} 
 \\*
&\quad+ \sum_k \int\!\! d^dx\, \Big[
T\, \hat{\cal O}_N^{(0)}(0)\, \hat K_{n_k\mu}^{(2) \kappa}(x)\, \Pi {\cal L}^{(0)}_\text{coll}
\Big]  \hat {\cal E}_{s(n_k)}^{\mu\,\vec \kappa}(x) 
\nn \\*
&\quad+ \sum_k \int\!\! d^dx\, \Big[
T\, \hat{\cal O}_N^{(0)}(0)\, \hat K_{n_k\mu\nu}^{(2) \kappa\kappa'}(x)\, \Pi {\cal L}^{(0)}_\text{coll}
\Big]  \hat {\cal E}_{s(n_k)(n_k)}^{\mu\nu\,\vec \kappa}(x,x)
\bigg\}
  \,,\nn\\*
T\,  {\cal O}_N^{(2,\delta)} \Pi {\cal L}^{(0)} 
&= - \sum_k 
 \frac{\partial C_N^{(0)} }{\partial \bar{n}_k\cdot Q_k} 
 \, \left[ T\, \hat{\cal O}_N^{(0)}(0) 
  \: \Pi {\cal L}^{(0)}_\text{coll}\: \bar{n}_{k\, \mu}
\right] \hat {\cal E}_{s(n_k)}^{\mu\,\vec \kappa}(0)\,,
\nn\\
T\, {\cal O}_N^{(2,r)}(0) 
  \Pi {\cal L}^{(0)} 
&=C_N^{(0)}\,
 \sum_k \left[ T\, \hat{\cal O}_{N\mu}^{(0,r_k)}(0)
 \: \Pi {\cal L}^{(0)}_\text{coll} 
\right] \hat {\cal E}_{s(n_k)}^{\mu\,\vec \kappa}(0) \,,\nn
\end{align}
where ${\cal L}_\text{coll}^{(0)}$ are the terms in the leading power Lagrangian that only involve collinear fields. 

By defining collinear time ordered products of the operators in \Eq{eq:loop_structure}, we can then write a fully factorized expression for these terms in the subleading soft expansion.  We define
\begin{align} \label{eq:Idefs}
& {\cal L}^{(2)}:
& \hat {\cal I}_N^{(2L)} 
&\equiv \sum_k  \int \!\! d^dx\: T\, \hat {\cal O}_N^{(0)}(0)\, 
  \hat K_{n_k}^{(2)}(x)\, \Pi {\cal L}^{(0)}_\text{coll} 
 \,, \\
& & \hat  {\cal I}^{(2L)k}_{N\, \mu}(x) 
&\equiv 
T\, \hat{\cal O}_N^{(0)}(0)\, \hat K_{n_k\, \mu}^{(2) \kappa}(x)\, \Pi {\cal L}^{(0)}_\text{coll}
 \,,
\nn\\[5pt]
& & \hat  {\cal I}^{(2L)k}_{N\, \mu\nu}(x) 
&\equiv  
T\, \hat{\cal O}_N^{(0)}(0)\, \hat K_{n_k\, \mu\nu}^{(2) \kappa\kappa'}(x)\, \Pi {\cal L}^{(0)}_\text{coll}
 \,,
\nn\\[5pt]
&{\cal O}_N^{(2,\delta)}: 
& \hat  {\cal I}^{(0)}_{N} \bn_{k\mu} &\equiv 
  T\, \hat{\cal O}_N^{(0)}(0) \,
   \Pi {\cal L}^{(0)}_\text{coll}\: \bar{n}_{k\, \mu}
  \,,\nn\\[5pt]
&{\cal O}_N^{(2,r)}:  
& \hat  {\cal I}^{(0r)k}_{N\, \mu} &\equiv 
T\, \hat{\cal O}_{N\mu}^{(0,r_k)}(0) \,  \Pi {\cal L}^{(0)}_\text{coll}
 \,,\nn
\end{align}
for the collinear operators appearing in these terms. For the matrix elements of all these collinear operators with $N$ well separated collinear particles we simply use the same notation, but drop the hats, 
\begin{align}
 {\cal I}_{N\cdots}^{[l] \cdots} 
   & = \big\langle 0 \big| \hat {\cal I}_{N\cdots}^{\cdots} 
      \big| p_1,p_2,\ldots,p_N\big\rangle^{[l]}
  \,,
  & {\cal I}_{N\cdots}^{[l]\cdots}(x)
    & = \big\langle 0 \big| \hat {\cal I}_{N\cdots}^{\cdots}(x) 
      \big| p_1,p_2,\ldots,p_N\big\rangle^{[l]}
  \,. 
\end{align}
With these definitions we find the following all orders factorization theorems
\begin{align}
  T\,  {\cal O}_N^{(0)} \Pi {\cal L}^{(0)} {\cal L}^{(2)} &= C_N^{(0)} \hat{\cal I}_{N}^{(2L)}  \hat E_s 
   \\
  &\qquad
   + C_N^{(0)} \sum_k \int \!\! d^dx\, 
    \Big\{ \hat{\cal I}^{(2L)k}_{N\, \mu}(x) \hat{\cal E}_{s(n_k)}^{\mu\,\vec \kappa}(x)
   + 
    \hat{\cal I}^{(2L)k}_{N\, \mu\nu}(x) \hat{\cal E}_{s(n_k)(n_k)}^{\mu\nu\,\vec \kappa}(x,x)
  \Big\}
  , \nn \\
  T\,  {\cal O}_N^{(2,\delta)} \Pi {\cal L}^{(0)} 
  &= - \sum_k 
     \frac{\partial C_N^{(0)} }{\partial \bar{n}_k\cdot Q_k} \,
    {\cal I}^{(0)}_{N}\, 
    \hat{\cal E}_{s(n_k)}^{\mu\,\vec \kappa}(0)\: \bn_{k\mu}
  \,, \nn \\[5pt]
  T\,  {\cal O}_N^{(2,r)} \Pi {\cal L}^{(0)}
  &= C_N^{(0)}  \sum_k 
    \hat{\cal I}^{(0r)k}_{N\, \mu} \,
    \hat{\cal E}_{s(n_k)}^{\mu\,\vec \kappa}(0)
 \, .\nn
\end{align}
For one soft loop, putting the above results into \Eq{eq:ANs12rpi_soft} gives
\begin{align} \label{eq:Factsubsoft}
  {\cal A}_{N+1_s}^{[1,\text{soft}](2)} 
  &=C_N^{[0](0)} {\cal I}_{N}^{[0](2L)}  \, 
   S^{[1](0)}(s)
  + 
 C_N^{[0](0)} \sum_k\int \!\! d^dx\,
  \Big\{ 
 {\cal I}^{[0](2L)k}_{N\, \mu}(x)  {\cal E}_{s(n_k)}^{[1]\mu\,\vec \kappa}(x)
\nn\\
& +
{\cal I}^{[0](2L)k}_{N\, \mu\nu}(x)
{\cal E}_{s(n_k)(n_k)}^{[1]\mu\nu\,\vec \kappa}(x,x)
 \Big\}
 - \sum_k 
     \frac{\partial C_N^{[0](0)} }{\partial \bar{n}_k\cdot Q_k} \,
 \text{Split}^{[0](0)}\,  \bn_{k\mu}\, 
   {\cal E}_{s(n_k)}^{[1]\mu\,\vec \kappa}(0) 
    \nn \\*
 &\
 + C_N^{[0](0)}  \sum_k   
    {\cal I}^{[0](0r)k}_{N\, \mu}(0)
     {\cal E}_{s(n_k)}^{[1]\mu\,\vec \kappa}(0) 
\, .
  \end{align}
Here we have used the fact that the leading power $1\to 1$ splitting amplitude $\text{Split}^{(0)}     = \langle\hat {\cal I}^{(0)}_{N}\rangle$. The matrix element ${\cal I}^{[0](0r)k}_{N\, \mu}$, which contributes only when its  $X_{n_k}^{\kappa_k}$ is a collinear fermion, also corresponds to a leading power splitting amplitude, but with a different spin contraction involving $t_k^\mu$ in \Eq{eq:tk}. In \Eq{eq:Factsubsoft} the terms with a superscript $[1]$ are matrix elements of purely soft fields and denote the soft loop, while the remaining terms are tree level hard coefficients or collinear matrix elements.  Thus, we have factorized the universal soft contribution from hard and collinear physics in all the terms which include a soft loop. 

For collinear loops, contributions from the terms enumerated in \Eq{eq:ANs12sc} are all generically non-zero.  These contributions can be divided into two possible classes, those with ${\cal O}_N^{(0)}$ and those with operators that are higher order in the power counting.  First, consider the contributions that contain the $N$-jet operator ${\cal O}_N^{(0)}$ and Lagrangian insertions. The matching coefficient of this operator is evaluated at tree-level and the collinear loops involve the various possibilities from the Lagrangian insertions.  In these contributions, the collinear loop and the emission of the soft gluon can either be from the same collinear direction or from different collinear directions.  If from the same collinear direction, because these collinear loop contributions involve only a single external collinear direction and the hard coefficient $C_N$ of the operator is a tree-level amplitude, these contributions can always be re-expressed in terms of the soft limit of a one-loop splitting amplitude times a lower-point tree-level amplitude.  If they are from different collinear directions, then the emission of the soft gluon is exclusively from an external collinear leg at tree-level, and is also given by a splitting amplitude.  In these contributions, we can set the $\perp$ component of momentum of the leg off of which the soft gluon is emitted to zero. Using these observations, we then have
\begin{align}\label{eq:ANs12rpi_coll}
%  {\cal A}_{N+1_s}^{[1,\text{coll}](2)} \supset 
% L^(2)
&   \big\langle T\,  \big[ {\cal O}_N^{[0](0)} \Pi {\cal L}^{(0)} {\cal L}^{(2)}   \big]^{[1,\text{coll}]} \big\rangle  
 + \big\langle T\, \big[ {\cal O}_N^{[0](0)} \Pi {\cal L}^{(0)} 
     ({\cal L}^{(1)})^2   \big]^{[1,\text{coll}]} \big\rangle  
     \nn  \\*[5pt]
& \quad
  =  C_N^{[0](0)}  \sum_{k} \bigg(  \int\!\! d^dx \, d^dy\, \bigg\{
   \Big\langle T\,
  \hat{K}_{n_k}^{(1)}(x)\, \hat{K}_{n_k}^{(1)}(y)\,\hat{\cal O}_N^{(0)}(0)\, 
 \Pi {\cal L}^{(0)}_{\text{coll}, n= n_k} \Big\rangle^{[1]} \, S^{[0](0)}(k)
 \nn \\[5pt]
 &\qquad\quad 
+2\Big\langle T\,
  \hat{K}_{n_k\mu}^{(1) \kappa}(x)\, \hat{K}_{n_k}^{(1)}(y)\,  \hat{\cal O}_N^{(0)}(0)\, 
 \Pi {\cal L}^{(0)}_{\text{coll}, n= n_k} \Big\rangle^{[1]}  {\cal E}_{s(n_k)}^{[0]\mu\,\vec \kappa}(x)
 \bigg\}  
 \nn \\[5pt]
 & \qquad\quad
  +  \int\!\! d^dx\,  \bigg\{ 
\Big\langle T\,   \hat{K}_{n_k}^{(2)}(x)\, \hat{\cal O}_N^{(0)}(0)\, 
 \Pi {\cal L}^{(0)}_{\text{coll}, n= n_k} \Big\rangle^{[1]} \, S^{[0](0)}(k)
 \nn \\[5pt]
& \qquad\quad
+  \Big\langle T\,  \hat{K}_{n_k\mu}^{(2) \kappa}(x)\, 
  \hat{\cal O}_N^{(0)}(0)\, 
\Pi {\cal L}^{(0)}_{\text{coll}, n= n_k} \Big\rangle^{[1]} \,
  {\cal E}_{s(n_k)}^{[0]\mu\,\vec \kappa}(x)
 \bigg\}
   \bigg)
 \nn \,\\[5pt]
& \quad \ \
  + C_N^{[0](0)}\sum_k \int\!\! d^dx\, \bigg\{ 
\Big\langle T\,   \hat{K}_{n_k}^{(2)}(x) \,  \hat{\cal O}_N^{(0)}(0)\, 
 \Pi {\cal L}^{(0)}_{\text{coll}, n\ne n_k} \Big\rangle^{[1]}\,   S^{[0](0)}(k)
 \nn \\
& \qquad\quad 
+ \Big\langle T\,  \hat{K}_{n_k\mu}^{(2) \kappa}(x)\,  
  \hat{\cal O}_N^{(0)}(0)\, 
\Pi {\cal L}^{(0)}_{\text{coll}, n\ne n_k} \Big\rangle^{[1]}\,  
 {\cal E}_{s(n_k)}^{[0]\mu\,\vec \kappa}(x)\bigg\}
  \,.
\end{align}
On the right of the equality, the terms in the first four lines run over all possible legs $k$ from which the soft gluon can be emitted, fully accounting for color, and contains a collinear loop correction from the sector emitting the soft gluon. In the fifth and sixth lines the loop correction is in a collinear sector not participating in the soft emission. These loop contributions are simply the leading order collinear virtual corrections. All of these terms can be recognized as soft limits of splitting amplitude contributions, giving
\begin{align} \label{eq:ANs12rpi_coll2}
&   \big\langle T\,  \big[ {\cal O}_N^{[0](0)} \Pi {\cal L}^{(0)} {\cal L}^{(2)}   \big]^{[1,\text{coll}]} \big\rangle  
 + \big\langle T\, \big[ {\cal O}_N^{[0](0)} \Pi {\cal L}^{(0)} 
     ({\cal L}^{(1)})^2   \big]^{[1,\text{coll}]} \big\rangle  
       \\*[5pt]
  & \quad
=\sum_k \Splitbar^{[1](2)}_{\bar n_k}(P_k\to k,s)  {\cal A}^{[0](0)}(1,\dotsc,P_k,\dotsc,N) \nn\\
& \ \ 
\quad + \sum_{k,\, l\neq k} \Splitbar^{[0](2)}_{\bar n_k}(P_k\to k,s)
  \text{Split}^{[1](0)}(l\to l) {\cal A}^{[0](0)}(1,\dotsc,l,\dotsc,P_k,\dotsc,N)
\,. \nn
\end{align}
Here the first four lines of \Eq{eq:ANs12rpi_coll} give the $\Splitbar^{[1](2)}_{\bar n_k}(P_k\to k,s)$ term, while the fifth and sixth lines give the $\Splitbar^{[0](2)}_{\bar n_k}(P_k\to k,s)$ term. The subscript $\bn_k$ indicates the reference vector that is used to define momentum fractions in the splitting amplitude. The bar here indicates that these splitting amplitudes are both collinear and soft gauge invariant, in contrast to the QCD splitting amplitudes which are only collinear gauge invariant. The $\Splitbar$ amplitudes can be calculated from gauge invariant antennae \cite{Kosower:1997zr,Kosower:2003bh,GehrmannDeRidder:2005cm}, only keeping the terms that are at ${\cal O}(\lambda^2)$ and with an appropriate decomposition and choice for the vector $\bar n_k$.  These antennae functions correctly describe soft wide angle emissions from a dipole.
The result for $\Splitbar^{[0](2)}_{\bar n_k}(P_k\to k,s)$ was calculated for fermions in \Sec{subsec:L2LBK} and for gluons in \App{app:gluonLBK}. Neither contribution changes the hard matching coefficient, which is the leading power tree-level amplitude $C_N^{[0](0)}={\cal A}^{[0](0)}$.  The last line of \Eq{eq:ANs12rpi_coll2} includes virtual corrections at one loop order from $ \text{Split}^{[1](0)}(k\to k)$ defined in \Eq{eq:splitkk}. For the $1\to 2$ splitting amplitudes $\Splitbar^{[j](2)}_{\bar n_k}(P_k\to k,s)$ the intermediate particle $P_k$ is located in the amplitude at $k$ and the sum over spins and color is implicit. These contributions are only non-zero if the tree-level amplitudes are non-zero. For a color-ordered amplitude where the soft gluon is emitted between particles $N$ and 1, the $1\to 2$ splitting amplitude contribution at 1-loop becomes
\begin{align}
\Splitbar^{[1](2)}_{\bar n_N}(P\to N,s){\cal A}^{[0]}(1,\dotsc,N-1,P)+ \Splitbar^{[1](2)}_{\bar n_1}(P\to s,1){\cal A}^{[0]}(P,2,\dotsc,N) \,.
\end{align}

In addition to the splitting amplitude contributions, there are also contributions that contain collinear loops involving the higher power $N$-jet operators in \Eq{eq:ANs12sc}, given by:
\begin{align}\label{eq:ANs12_coll}
%  {\cal A}_{N+1_s}^{[1,\text{coll}](2)} \supset
% L^(2)
  &
% O^(1)
  \big\langle T\, \big[ {\cal O}_N^{[0](1)} \Pi {\cal L}^{(0)} 
     {\cal L}^{(1)}   \big]^{[1,\text{coll}]} \big\rangle  
 + \big\langle T\, \big[ {\cal O}_N^{[0](2\delta,2r)} \Pi {\cal L}^{(0)} 
       \big]^{[1,\text{coll}]} \big\rangle
 \\ 
 &\ \qquad 
   + \big\langle T\, \big[ {\cal O}_N^{[0](2X/\partial)} \Pi {\cal L}^{(0)} 
     \big]^{[1,\text{coll}]} \big\rangle  
  \,.\nn
\end{align}
For the ${\cal O}_N^{[0](2\delta,2r)}$ term we can use the all orders factorized result given above in \Eq{eq:loop_structure}. The remaining  terms are reduced by using RPI to set the total $\perp$ component of momenta in each collinear sector to zero, which leaves only the operators listed in \Eq{eq:O2nonrpi}. For these terms we define the factorized expressions that appear for the collinear loop contribution to the loop-level subleading soft theorem as 
\begin{align}\label{eq:appf_def2}
& {\cal O}_N^{(1)} {\cal L}^{(1)}: 
& \hat{\cal J}^{(2X_k L)}_N &\equiv 
  C_N^{(1X_k)} \otimes \int \!\! d^dx\,  
T\, \hat{\cal O}_N^{(1X_{n_k})}(0) \sum_{n'} \hat{K}^{(1)}_{n'}(x)
 \,, \\
&& 
\hat{\cal J}^{(2\,X_k L_{k'})\, \mu}_N(x) &\equiv 
 C_N^{(1X_k)}\otimes
 T\, \hat{\cal O}_N^{(1 X_{n_k})}(0)\,\hat{K}_{{n_{k'}}}^{(1)\kappa\, \mu}(x)
 \,,
\nn\\
& {\cal O}_N^{(2,X\partial)}: &
\hat{\cal J}^{(2X_k\partial)}_N &\equiv
C_N^{(2X_k\partial)} \otimes \hat{\cal O}_N^{(2 X_{n_k}\partial)}(0)
\,,\nn
\\
& {\cal O}_N^{(2,X^2)}: &
\hat{\cal J}^{(2X^2_k)}_N &\equiv 
C_N^{(2X_k^2)}\otimes \hat{\cal O}_N^{(2,X_{n_k}^2)}(0)\,,
\nn\\
& {\cal O}_N^{(2,XX)}: &
\hat{\cal J}^{(2\,X_k X_{k'})}_N &\equiv 
C_N^{(2X_k X_{k'})}\otimes
  \hat{\cal O}_N^{(2,X_{n_k} X_{n_{k'}})\,\vec \kappa} (0)
 \,.\nn
\end{align}
As mentioned at the end of \Sec{subsec:simplecalcs}, the hatted notation means that the operators only involve collinear fields.  $\hat K_n^{(1)}$ and  $\hat K_{n\mu}^{(1) \kappa}$ are from ${\cal L}^{(1)}$, and were defined by \Eq{eq:LKB}. These terms in general involve convolution integrals over the large momentum fractions carried by the multiple collinear objects in a given sector. Therefore, in the various $\hat {\cal J}_N$s we leave both the Wilson coefficients and collinear operators, so that both of the potential terms participating as integrands in these convolution integrals are left together.   When taking matrix elements, we again use the notation:
\begin{align}
{\cal J}^{[l]...}_N =\langle 0|\hat{\cal J}^{...}_N |p_1,p_2,...,p_N\rangle^{[l]} \,.
\end{align}
The result for one collinear loop with power suppressed operators is then
\begin{align} \label{eq:weareoutofnames}
%  {\cal A}_{N+1_s}^{[1,\text{coll}](2)} 
% L^(2)
 & \Big\langle
 T\, \big[ {\cal O}_N^{[0](2\delta,2r)} \Pi {\cal L}^{(0)} 
       \big]^{[1,\text{coll}]}
  +   T\, \big[ {\cal O}_N^{[0](1)} \Pi {\cal L}^{(0)} 
     {\cal L}^{(1)}   \big]^{[1,\text{coll}]} 
  +  T\, \big[ {\cal O}_N^{[0](2X/\partial)} \Pi {\cal L}^{(0)} 
     \big]^{[1,\text{coll}]} \Big\rangle  
  \nn \\
  & 
  \quad=
 \sum_k\bigg\{ 
 -\frac{\partial C_N^{[0](0)} }{\partial \bar{n}_k\cdot Q_k} \,
 \text{Split}^{[1](0)}\, {\cal E}_{s(n_k)}^{[0]\mu\,\vec \kappa}(0)\:\bn_{k\mu}
 +C_N^{[0](0)}  {\cal I}^{[1](0r)k}_{N\, \mu} \,
 {\cal E}_{s(n_k)}^{[0]\mu\,\vec \kappa}(0)
 \bigg\}
\nonumber\\
&
 \qquad+ \sum_k\bigg\{
 \Big( {\cal J}^{[1](2X_k L)}_N\, 
 +  {\cal J}^{[1](2X_k\partial)}_N \Big)  E_{s[n_k]2}^{[0]\,\vec \kappa} 
 +  {\cal J}^{[1](2X^2_k)}_N  E_{s[n_k]3}^{[0]\,\vec \kappa}
\bigg\}
  \nn \\
  &
 \qquad+ \sum_{k,k'}
% \bigg\{ 
% {\cal J}^{[1](2\,X_k X_k')}_N  E^{[0]\,\vec \kappa}_{s[n_k][n_{k'}]} 
% + 
 \int \!\! d^dx\,
 {\cal J}^{[1](2\,X_k L_{k'})\, \mu}_N(x)\,  
   {\cal E}_{s(n_{k'})[n_k]\mu}^{[0]\vec \kappa}(x)  
% \bigg\}
  \,.
\end{align}
We will call the functions ${\cal J}^{\cdots}_N$ fusion terms since their parent operators ${\cal O}_N^{(1,X)}$, ${\cal O}_N^{(2,X\partial)}$, or ${\cal O}_N^{(2,X^2)}$ create or annihilate two or three $n$-collinear partons at tree-level. These collinear partons must then fuse back together in a collinear loop in order to produce a single $n$-collinear particle for the state in our subleading soft theorem. Since the operator ${\cal O}^{[1](2\,X_k X_k')}_N$ has a pair of collinear fields in each of two distinct collinear directions, it can not have both pairs fuse to a single collinear field at one-loop order, and hence $\hat {\cal J}^{[1](2\,X_k X_k')}_N$ does not contribute in \Eq{eq:weareoutofnames}.  (From the point of view of a two-loop subleading soft theorem, this operator would be a new contribution that did not already appear at one-loop.) The full contribution from collinear loops, ${\cal A}_{N+1_s}^{[1,\text{coll}](2)}$, is given by the sum of \Eqs{eq:ANs12rpi_coll2}{eq:weareoutofnames}.

\begin{figure}[t!]
	\begin{center}
		\includegraphics[width=14cm]{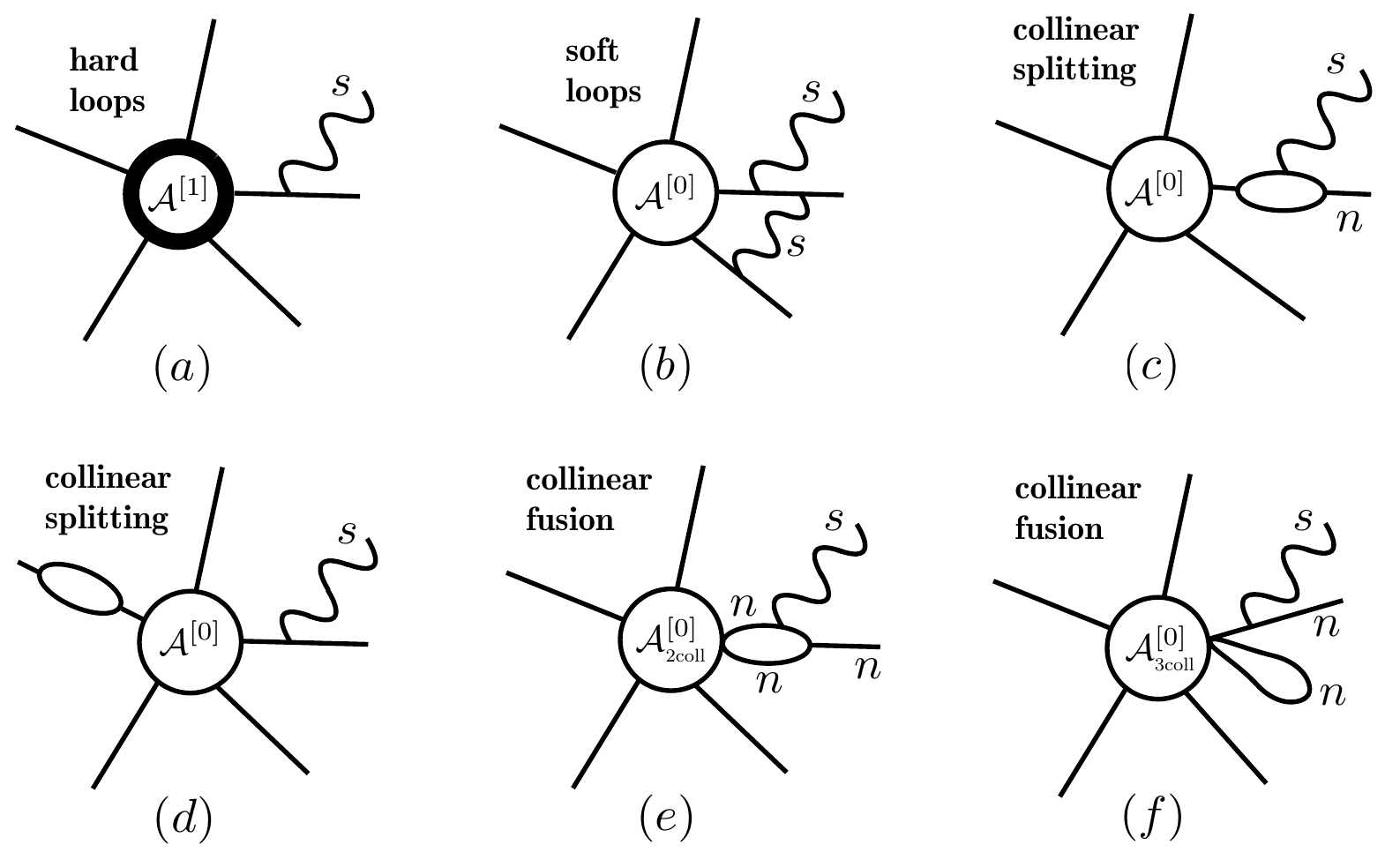}
	\end{center}
	\caption{
Illustrations of the various contributions to the one-loop subleading soft theorem at ${\cal O}(\lambda^2)$.  Fig.~$(a)$ are the hard loops, $(b)$ are soft loops, and $(c)$, $(d)$, $(e)$ and $(f)$ are collinear loops.  $(c)$ and $(d)$ are the collinear loops arising from splitting amplitudes, while $(e)$ and $(f)$ represent collinear loops from the fusion terms in the one-loop subleading soft theorem.  In each figure, the matching coefficient is written, with ${\cal A}^{[0]}$ and ${\cal A}^{[1]}$ the tree-level or one-loop amplitude.  For the fusion terms in $(e)$ and $(f)$, we use the short-hand ${\cal A}_\text{2coll}^{[0]}\equiv C^{[0](1X)}$ and ${\cal A}_\text{3coll}^{[0]}\equiv C^{[0](2X^2)}$.
}
	\label{fig:pic_1loop}
\end{figure}

Putting all the contributions together, our fully factorized, one-loop subleading soft theorem for an arbitrary amplitude with $N$ well separated particles plus one soft particle is
\begin{empheq}[box=\fbox]{align}\label{eq:fulloneloopsubsoft}
  {\cal A}_{N+1_s}^{[1](2)} 
 &= 
S^{\text{[0](2)}}(s){\cal A}^{[1,\text{hard}](0)}_N\\
&
+{\mathcal A}_{N}^{[0](0)} {\cal I}_{N}^{[0](2L)}  \, 
   S^{[1](0)}(s)
   \nn\\
%
% Two lines with remaining Soft terms:
%
&  + 
 {\mathcal A}_{N}^{[0](0)} \sum_{k=1}^N\int \!\! d^dx\,
  \Big\{ 
 {\cal I}^{[0](2L)k}_{N\, \mu}(x)\,  {\cal E}_{s(n_k)}^{[1]\mu\,\vec \kappa}(x)
 +
{\cal I}^{[0](2L)k}_{N\, \mu\nu}(x)\,
{\cal E}_{s(n_k)(n_k)}^{[1]\mu\nu\,\vec \kappa}(x,x)
 \Big\}
   \nn\\
& % \hspace{-0.8cm}
 +\sum_{k=1}^N 
     \bigg\{-\frac{\partial {\mathcal A}_{N}^{[0](0)} }{\partial \bar{n}_k\cdot Q_k} \,
   \text{Split}^{[0](0)} \, 
   {\cal E}_{s(n_k)}^{[1]\mu\,\vec \kappa}(0) \,
    \bn_{k\mu}
 + {\mathcal A}_{N}^{[0](0)}   
   {\cal I}^{[0](0r)k}_{N\, \mu}(0)\,
    {\cal E}_{s(n_k)}^{[1]\mu\,\vec \kappa}(0) 
    \bigg\} \nn\\[5pt]
%+{\cal A}_{N+1_s}^{[1,\text{soft}](2)}
%
% Three lines with Splitting terms:
%
& % \hspace{-0.8cm} 
+\sum_{k=1}^N \Splitbar^{[1](2)}_{\bar n_k}(P_k\to k,s)  {\cal A}_N^{[0](0)}(1,\dotsc,P_k,\dotsc,N)
  \nn\\[5pt]
&
  +\sum_{\substack{k=1\\ l\neq k}}^N\Splitbar^{[0](2)}_{\bar n_k}(P_k\to k,s)\,\text{Split}^{[1](0)}(l\to l)\, {\cal A}_N^{[0](0)}(1,\dotsc,l,\dotsc,P_k,\dotsc,N)
  \nn\\[5pt]
&%\hspace{-0.8cm}
+\sum_{k=1}^N\bigg\{ 
     -\frac{\partial {\mathcal A}_{N}^{[0](0)} }{\partial \bar{n}_k\cdot Q_k} \,
    \text{Split}^{[1](0)}\, {\cal E}_{s(n_k)}^{[0]\mu\,\vec \kappa}(0)\, \bn_{k\mu}
+{\mathcal A}_{N}^{[0](0)}  {\cal I}^{[1](0r)k}_{N\, \mu} \,
{\cal E}_{s(n_k)}^{[0]\mu\,\vec \kappa}(0)
\bigg\}
\nonumber\\[5pt]
&%\hspace{-0.8cm}
%
% second to last line:
%
+ \sum_{k=1}^N\bigg\{
\Big( {\cal J}^{[1](2X_k L)}_N  
 +   {\cal J}^{[1](2X_k\partial)}_N \Big) E_{s[n_k]2}^{\,[0]\vec \kappa} 
 +   {\cal J}^{[1](2X^2_k)}_N  E_{s[n_k]3}^{\,[0]\vec \kappa} 
\bigg\}
  \nn \\[5pt]
%
% last line:
%
  & % \hspace{-0.8cm} 
  + \sum_{k,k'=1}^N
%\bigg\{ 
%  {\cal J}^{[1](2\,X_k X_k')}_N  E^{[0]\,\vec \kappa}_{s[n_k][n_{k'}]}
%  +
  \int \!\! d^dx\,
{\cal J}^{[1](2\,X_k L_{k'})\, \mu}_N(x) \, 
{\cal E}_{s(n_{k'})[n_k]\mu}^{[0]\vec \kappa}(x)  
% \bigg\}
\nn
\end{empheq}
Here, we have explicitly indicated that the leading order tree-level matching is the tree level amplitude:
\begin{equation}
C_N^{[0](0)}={\mathcal A}_{N}^{[0](0)}\,.
\end{equation}
The different contributions to \Eq{eq:fulloneloopsubsoft} were derived above in Eqs.~(\ref{eq:LBK_loop_term}), (\ref{eq:soft_emit_simp1}), (\ref{eq:Idefs}), (\ref{eq:ANs12rpi_coll}) and (\ref{eq:appf_def2}).  The first line contains the hard loop contributions, the next three lines are soft loops, and the final five lines are collinear loops. It is worth noting that all terms, except for some of the possible fusion terms in the last two lines, are connected via RPI to the leading order amplitude.
In \Eq{eq:fulloneloopsubsoft} we have suppressed color-index contractions and the sum over helicities in the $\Splitbar^{[l](2)}$ dependent terms.  The various terms contributing to the soft theorem in \Eq{eq:fulloneloopsubsoft} are schematically illustrated in \Fig{fig:pic_1loop}.  Hard loop and soft loop contributions are illustrated in Figs.~\ref{fig:pic_1loop}$(a)$ and $(b)$, respectively. The terms with one-loop collinear splitting functions, involving a Split$^{[1](k)}$  or ${\cal I}^{[1](0r)k}_{N\, \mu}$ factor, are illustrated in Figs.~\ref{fig:pic_1loop}$(c,d)$, and the collinear fusion terms involving a ${\cal J}^{[1]}$ factor, are illustrated in Figs.~\ref{fig:pic_1loop}$(e,f)$.   \Eq{eq:fulloneloopsubsoft} is a central result of this paper.

In general the factorized one-loop soft and collinear matrix elements in \Eq{eq:fulloneloopsubsoft} will have UV and IR divergences, and we implicitly have been assuming a regulator like dimensional regularization with $d=4-2\epsilon$ that does not spoil any symmetries of our gauge theory. The UV divergences from soft and collinear loops are exactly canceled by the same EFT counterterms that were used to obtain UV finite results for the hard matching coefficients/amplitudes, plus coupling renormalization. Accounting for this, \Eq{eq:fulloneloopsubsoft} yields a UV finite result. This amplitude will still contain IR divergences, which appear as $1/\epsilon^2$ and $1/\epsilon$ poles at this loop order. These IR divergences will only cancel when we consider the phase space integrated amplitude squared for a physical cross section, which also contains additional real emission diagrams. The real emission diagrams are not part of \Eq{eq:fulloneloopsubsoft}, but can be factorized and treated in a similar manner, as discussed below in \Sec{subsec:splitsub}.

Generically, all of the terms in \Eq{eq:fulloneloopsubsoft} will be non-zero, but some contributions may vanish for special cases with particular helicity or color choices. First consider soft dynamics.  At leading power the soft gluon couplings preserve collinear helicity at any loop order (which is explicit in Feynman-'t Hooft gauge). The power suppressed ${\cal O}(\lambda^2)$ soft couplings also preserve helicity at tree level, as was explicitly seen in our discussion of LBK.  Therefore there are no helicity flips in the first two terms of \Eq{eq:fulloneloopsubsoft}. Due to the connection between chirality and helicity there are also no helicity flips for collinear fermions in the presence of soft loops, which are contributions in the 3rd and 4th lines of \Eq{eq:fulloneloopsubsoft}. Determining whether there are spin flips to the collinear gluon terms in the 3rd and 4th lines of \Eq{eq:fulloneloopsubsoft} requires an investigation beyond those done here (due to the vector indices $\mu$ and $\nu$ in those terms).  Also, the hard coefficient in the $N$-jet operator $C_N$ is evaluated at tree-level in all soft loop contributions in \Eq{eq:fulloneloopsubsoft}.   Therefore, if the helicity configuration of the external collinear particles is such that the tree-level amplitude is zero, then all contributions from soft loops vanish. For the pure gluon all-plus and single-minus helicity amplitudes, because the hard coefficient $C_N^{[0](0)}$ for those amplitudes vanishes, there is no contribution from soft loops to the loop-level subleading soft theorem.  For MHV and beyond-MHV helicity configurations, soft loops contribute to the loop-level subleading soft theorem. 

For collinear particles, the total angular momentum for each collinear direction is preserved by the collinear dynamics at leading power. Again by the connection between chirality and helicity, the helicity is preserved for collinear fermions in the presence of collinear loops at an arbitrary order in the power expansion. For collinear gluons the tree level Lagrangian insertion preserves helicity (see \App{app:gluonLBK}), so the helicity of the original hard configuration is preserved in the 6th line of \Eq{eq:fulloneloopsubsoft}. For the other collinear loop contributions the helicity of the initial and outgoing collinear particle can be flipped, which includes the terms in the 5th, and and 7th-9th lines of  \Eq{eq:fulloneloopsubsoft}. For collinear loops, examining the loop level gluon splitting amplitudes given in \Refs{Bern:1998sc,Kosower:1999rx}, one finds that all splitting amplitudes are non-zero at subleading orders in the soft expansion.  At tree-level these splitting amplitudes in the soft limit do not flip the collinear particles helicity, so  $\Splitbar^{[0]}(P_k^{\pm}\to k^{\pm},s)=0$.

A recent proposal in the literature only considered the first term, $S^{\text{[0](2)}}(s){\cal A}^{[1,\text{hard}]}_N$, in a loop-level, subleading soft emission analysis \cite{Cachazo:2014dia}. In \Ref{DelDuca:1990gz}, a loop level subleading soft photon theorem was derived for the Sudakov form factor, explicitly taking into account collinear loop effects. They used the Grammer-Yennie \cite{Grammer:1973db} decomposition of the photon propagator and its relation to the Ward identities, after the factorization of collinear effects from hard loops. This result captures the LBK related terms in \Eq{eq:fulloneloopsubsoft}, as well as subleading time-ordered products due to soft particles interacting with collinear loops, like those generated by the subleading splitting amplitude terms in \Eqs{eq:ANs12rpi_coll}{eq:ANs12rpi_coll2}.  While a full comparision to this result is beyond the scope of this paper, it is worth noting that no higher order collinear operators like those of ${\mathcal O}_N^{(2X/\partial)}$ which generate the fusion terms ${\cal J}^{\cdots}_N$ were taken into account. This can be seen from the fact that all hard matching coefficients in \Ref{DelDuca:1990gz} are related to the leading order Sudakov form factor, or its derivatives, while generically there will be no such relation for these subleading collinear operators. For the dijet operator giving the quark Sudakov form factor, the ${\cal O}^{(1,X)}$ operators that appear at ${\cal O}(\lambda)$ are known to be disconnected from the leading power operator~\cite{Marcantonini:2008qn}.

\subsubsection{One-loop soft theorem for single-minus amplitude}
\label{sec:subsoftsingminus}

In \App{app:ampexs}, we apply the loop-level soft theorem of \Eq{eq:fulloneloopsubsoft} to several amplitude examples.  As it is quite simple, it is instructive to apply the one-loop subleading soft theorem in \Eq{eq:fulloneloopsubsoft} to the expansion of the color-ordred single-minus amplitude.  As we showed in  \Eq{eq:subl_soft_factored_one_minus}, the soft expansion of the single-minus large $N_c$ primitive amplitude for gauge bosons takes the form
\begin{align} \label{eq:subsoftoneminus}
{\mathcal A}^{[1]}(1^-,2^+,3^+,4^+,5_s^+)
&= S^{[0](0)}(5^+){\mathcal A}^{[1]}(1^-,2^+,3^+,4^+)
  \\*
&\quad
+S^{[0](2)}(5^+){\mathcal A}^{[1]}(1^-,2^+,3^+,4^+)
\nonumber \\
&\quad+\Splitbar^{[1](2)}_{\bar n_4=3}(P^+\rightarrow 4^+,5^+){\mathcal A}^{[0]}(1^-,2^+,3^+,P^-)
 +{\cal O}(\lambda^1)  \,.
 \nonumber
\end{align}
Order-by-order in $\lambda$, we can see how the effective theory exactly reproduces this soft expansion.

Starting with the leading soft factor, from \Eq{eq:leadsoft_oneloop}, we would expect the expansion
\begin{align}
  {\cal A}^{[1]}(1^-,2^+,3^+,4^+,5_s^+)  &= S^{[0](0)}(5^+)\, {\cal A}^{[1]}(1^-,2^+,3^+,4^+)
    +S^{[1](0)}(5^+)\, {\cal A}^{[0]}(1^-,2^+,3^+,4^+) 
    \nn\\
    &\qquad+{\cal O}(\lambda^{-1})\,,
\end{align}
where $S^{[0](0)}(5^+)$ and $S^{[1](0)}(5^+)$ are the tree-level and one-loop leading soft factors, respectively.  The one-loop soft factor multiplies a tree-level single-minus amplitude ${\cal A}^{[0]}(1^-,2^+,3^+,4^+)$, which is zero, and so does not contribute to the soft expansion. We also argued in \Sec{sec:Olamsoftthm} that all ${\cal O}(\lambda^{-1})$ contributions at one-loop vanish, and so do not appear in the soft theorem. Hence the first corrections are ${\cal O}(\lambda^0)$. Therefore, we find
\begin{align}
  {\cal A}^{[1]}(1^-,2^+,3^+,4^+,5_s^+)  &= S^{[0](0)}(5^+)\, {\cal A}^{[1]}(1^-,2^+,3^+,4^+)
 +{\cal O}(\lambda^{0})\,,
\end{align}
in agreement with the explicit expansion of the amplitude.

Now consider the ${\cal O}(\lambda^0)$ terms. From the general one-loop soft theorem result in \Eq{eq:fulloneloopsubsoft} there are several contributions to consider, involving hard, soft, and collinear loops.
Since ${\cal A}^{[1]}(1^-,2^+,3^+,4^+)$ is infrared finite, the matching for the hard loop amplitude on the first line of \Eq{eq:fulloneloopsubsoft} is given by
\begin{align} 
  {\cal A}^{[1,\text{hard}]}(1^-,2^+,3^+,4^+)={\cal A}^{[1]}(1^-,2^+,3^+,4^+)
\,.
\end{align}
At tree level this amplitude vanishes, ${\cal A}_4^{[0](0)}(1^-,2^+,3^+,4^+)=0$.  Therefore the soft loop terms on the 2nd to 4th lines of \Eq{eq:fulloneloopsubsoft}, which all contain a single minus ${\cal A}_4^{[0](0)}$ amplitude, vanish. This agrees with our discussion of helicity conservation below  \Eq{eq:fulloneloopsubsoft}.  The single minus ${\cal A}_4^{[0](0)}$ also appears in the terms in the 7th line, and hence they also vanish.  Due to color ordering the collinear splitting terms in the 5th and 6th line of \Eq{eq:ANs12rpi_coll2} must be adjacent to the soft emission.  The splitting functions therefore multiply one of four amplitudes, of which three vanish $0={\cal A}^{[0]}(1^-,2^+,3^+,P^+)={\cal A}^{[0]}(P^-,2^+,3^+,4^+)={\cal A}^{[0]}(P^+,2^+,3^+,4^+)$, and only ${\cal A}^{[0]}(1^-,2^+,3^+,P^-)\ne 0$. For this nonzero amplitude the splitting amplitude $\Splitbar^{[0](2)}_{\bar n_4}(P^+\to 4^+,5^+_s)=0$, while  $\Splitbar^{[1](2)}_{\bar n_4}(P^+\to 4^+,5^+_s)\ne 0$, hence there is only one non-zero splitting term. With these simplifications, the soft theorem becomes
\begin{align} \label{eq:A1minus}
  {\cal A}^{[1]}(1^-,2^+,3^+,4^+,5_s^+)  
   &= S^{[0](0)}(5^+_s)\, {\cal A}^{[1]}(1^-,2^+,3^+,4^+) 
% O^(1)
    +S^{\text{(sub)}}(5_s^+){\cal A}^{[1]}(1^-,2^+,3^+,4^+)
    \nn \\
    &
    + \Splitbar^{[1](2)}_{\bar n_4}(P^+\to 4^+,5_s^+){\cal A}^{[0]}(1^-,2^+,3^+,P^-) 
  \nn\\
    &
    + F_{N}(1^-,2^+,3^+; 4^+,5_s^+)   
   \,.
\end{align}
Here the fusion term is 
\begin{align} 
 F_{N} &= \sum_Y C_{N=4}^{[0](Y)}\otimes (\cdots)
   +{\cal O}(\lambda)
   \,,
\end{align}
where $Y$ ranges over $(Y)=(1,X)$, $(2,X\partial)$, $(2,X^2)$, %$(2,XX)$, 
and the $(\cdots)$ represents the appropriate matrix elements of collinear and soft operators from the full one-loop soft theorem, \Eq{eq:fulloneloopsubsoft}. \Eq{eq:A1minus} reproduces the expansion in \Eq{eq:subsoftoneminus} except for the presence of the $C_4^{[0](Y)}$ fusion terms. The fusion terms are generated by the operators ${\cal O}_4^{[0](1,X)}$ and ${\cal O}_4^{[0](2,X\partial)}$ which produce 5 particles, where 2 are in a single collinear sector and fuse through a loop graph, and by the operator ${\cal O}_4^{[0](2,X^2)}$ which produces 6 particles, where 3 are in a single collinear sector and two of these fuse without producing another particle. This gives 4 energetic particles, and the 5th particle is soft and generated by the soft part of each operator.  

For this single minus amplitude, the need for these fusion terms only becomes apparent if we choose a generic reference vector to define the momentum fraction of soft gluon 5 instead of $\bar{n}_4=3$ in \Eq{eq:subl_soft_factored_one_minus}. In that case:
\begin{align}
F_{N}(1^-,2^+,3^+; 4^+,5_s^+)&\equiv \fuse(1^-,2^+,3^+; P^+\rightarrow 4^+,5_s^+){\mathcal A}^{[0]}(1^-,2^+,3^+,P^-)\\
&
\hspace{-2cm}
=\left[\Splitbar^{[1](2)}_{\bar n_4=3}(P^+\to 4^+,5_s^+)-\Splitbar^{[1](2)}_{\bar n_4}(P^+\to 4^+,5_s^+)\right]{\cal A}^{[0]}(1^-,2^+,3^+,P^-) \,. \nn
\end{align}
This difference of splitting amplitudes can be expressed as a fusion term. This follows from the fact that the splitting amplitude with $\bar{n}_4=3$ can be calculated from the collinear operator $B_{n_4\perp}^\mu$ with the specific choice of reference vector in the collinear Wilson line. This operator with a specific choice can be expanded in terms of the generic $\bar{n}_4$, as was done in \Ref{Marcantonini:2008qn}. The subtraction kills the leading term, and the subleading terms all have the required structures for the operators ${\cal O}_4^{[0](1,X)}$ and ${\cal O}_4^{[0](2,X\partial)}$ contributing to the fusion terms. The fact that making the specific choice $\bar{n}_4=3$ eliminates these fusion terms highlights that the subleading soft theorem has RPI connections beyond those we have exploited in the analysis of LBK. In \App{app:ampexs}, we will present examples where such a simple choice of reference vector cannot eliminate the fusion terms. However, all fusion terms encountered can be expressed as differences of QCD splitting amplitudes with different external legs' momenta acting as the reference vectors.

It is also instructive to compare the finite single-minus amplitude to the case of the one-loop finite all-plus helicity amplitudes ${\cal A}(1^+,\dotsc,k^+)$, which are known to satisfy the Low-Burnett-Kroll theorem \cite{He:2014bga,Bern:2014oka}.  For essentially the same arguments as the single-minus amplitude, the soft expansion of the all-plus amplitude does not contain soft loops nor terms from the operators ${\cal O}_N^{[0](1)}$, ${\cal O}_N^{[0](2\delta,2r)}$, or ${\cal O}_N^{[0](2X/\partial)}$.  The expansion of the all-plus helicity amplitude in the soft limit can then be expressed as
\begin{align}
{\cal A}^{[1]}(1^+,\dotsc,k_s^+)&= S^{[0](0)}(k^+){\mathcal A}^{[1]}(1^+,\dotsc,(k-1)^+) 
 \\*
&\quad +S^{\text{[0](2)}}(k^+){\mathcal A}^{[1]}(1^+,\dotsc,(k-1)^+) \nonumber \\*
&\quad + \Splitbar^{[1](2)}_{\bar n_{k-1}}(P^+\to (k-1)^+,k_s^+) {\mathcal A}^{[0]}(1^+,\dotsc,(k-2)^+,P^-)\nonumber \\*
&\quad + \Splitbar^{[1](2)}_{\bar n_{k-1}}(P^-\to (k-1)^+,k_s^+) {\mathcal A}^{[0]}(1^+,\dotsc,(k-2)^+,P^+)\nonumber \\*
&\quad + \Splitbar^{[1](2)}_{\bar n_1}(P^+\to k_s^+,1^+) {\mathcal A}^{[0]}(P^-,2^+,\dotsc,(k-1)^+) \nonumber \\*
&\quad + \Splitbar^{[1](2)}_{\bar n_1}(P^-\to k_s^+,1^+) {\mathcal A}^{[0]}(P^+,2^+,\dotsc,(k-1)^+)  \nonumber \\*
&\quad + {\cal O}(\lambda^1)  
   \nonumber \\*
 &= S^{[0](0)}(k^+){\mathcal A}^{[1]}(1^+,\dotsc,(k-1)^+) 
   \nonumber \\*
   &\quad +S^{\text{(\text{sub})}}(k^+){\mathcal A}^{[1]}(1^+,\dotsc,(k-1)^+)
   + {\cal O}(\lambda^1)  \,,
    \nonumber
\end{align}
where $ \Splitbar^{[1](2)}$ denotes the one-loop splitting amplitudes.  In the first equality, the splitting amplitudes $\Splitbar^{[1](2)}$ each multiply a tree-level amplitude; however, these amplitudes are either all-plus or single-minus amplitudes, which are zero at tree-level.  Therefore, the contributions which violate the Low-Burnett-Kroll theorem vanish for this amplitude.  This is nicely consistent with the interpretation in \Ref{Bianchi:2014gla} that the all-plus amplitudes satisfy the Low-Burnett-Kroll theorem because of the conformal symmetry of the self-dual field configuration of all-plus helicity gluons.

It is important to stress that the fact that we could argue that the fusion terms were either zero or related to difference of splitting amplitudes, relied heavily on the special kinematics of the leading color single-minus, all-plus, and tree-level MHV amplitudes.  In particular, for considering the subleading collinear limits of NMHV or higher helicity configurations, we do not generically expect that the fusion terms can be expressed as a difference of splitting amplitudes with different choices for reference vectors.  In particular, $C_N^{[0](1X)}$ and other coefficients appearing in the fusion terms will not generically be a tree-level amplitude.  An example with such a matching coefficient $C_N^{[0](1X)}$, would be a subleading color amplitude obtained from fusing non-adjacent collinear particles.  Therefore, for an arbitrary amplitude, RPI alone does not suffice to determine the operators like ${\cal O}_N^{(1,X)}$ appearing from the subleading collinear expansion.

\subsection{Non-Trivial Subleading Soft Limit of Splitting Amplitudes at ${\cal O}(\lambda)$}
\label{subsec:splitsub}

At tree level for well separated collinear particles in gauge theory, the leading soft factor enters at ${\cal O}(\lambda^{-2})$, potential ${\cal O}(\lambda^{-1})$ contributions vanish, and the Low-Burnett-Kroll theorem applies at ${\cal O}(\lambda^{0})$.  This result is special to this order in 
in perturbation theory, as seen from the one-loop soft theorem for well separated particles discussed in the previous section. At one loop the leading terms are ${\cal O}(\lambda^{-2})$, the ${\cal O}(\lambda^{-1})$ contributions still vanish, and the nonzero terms predicted by our more sophisticated subleading one-loop soft theorem in \Eq{eq:fulloneloopsubsoft} are the same order as LBK, namely ${\cal O}(\lambda^0)$.

The pure LBK result is not just violated by loops, since it also depends on the kinematics of the final state particles which must be taken to be parametrically separated in angle.  Because of collinear singularities in a gauge theory, an energetic particle will preferentially  emit other parametrically close collinear particles, yielding a situation that violates this kinematic constraint at tree level.  At tree level for well-separated particles, the terms at ${\cal O}(\lambda^{-1})$ in the expansion of the subleading soft factor vanished  because RPI allowed us to set the $\perp$ component of the momentum of the one particle in each collinear sector in the scattering process to zero.  If a collinear sector has more than one particle, then RPI can be used to set the total $\perp$ momentum in a collinear sector to zero, but the individual $\perp$ momenta of the particles in the collinear sector will generically be non-zero.  This implies that there can be contributions to the subleading soft factor suppressed just by ${\cal O}(\lambda)$ that manifestly violate the Low-Burnett-Kroll theorem, even at tree-level.  

In this section, we will check this explicitly for a collinear sector with two particles in it.  The two collinear particles $1$ and $2$ will be parametrically close, obeying $p_1\cdot p_2\sim p_1\cdot p_s\sim p_2\cdot p_s$ for a soft momentum $p_s$. The soft expansion of the tree-level amplitude with a single collinear splitting can then be decomposed as
\begin{align} \label{eq:ANcoll}
{\cal A}_{{N+1}_s}^{[0,\text{coll}]} = {\cal A}_{{N+1}_s}^{[0,\text{coll}](0)}+ {\cal A}_{{N+1}_s}^{[0,\text{coll}](1)}+ {\cal A}_{{N+1}_s}^{[0,\text{coll}](2)}+\dotsc\,.
\end{align}
Here ${\cal A}_{{N+1}_s}^{[0,\text{coll}]}$ involves 2 particles that are both $n$-collinear, $N-2$ additional particles that are well separated from each other and from those two particles, and $1$ soft particle. Each term on the RHS of \Eq{eq:ANcoll} is at tree-level, and of increasing order in the $\lambda$ power expansion.   Using RPI, we will set the total $\perp$ momentum of the two $n$-collinear sector to zero, as well the $\perp$ momenta of each of the individual well separated particles.

For the language we use in this section we will take the one collinear emission amplitude as the base result, and hence use a counting where the leading power ${\cal A}_{{N+1}_s}^{[0,\text{coll}](0)}$ is said to be ${\cal O}(\lambda^{-2})$, which refers to the counting for the additional $1/n_i\cdot p_s$ eikonal propagator. (This does not count $\lambda^{-1}$ factors generated by the splitting itself.) At leading power both the collinear splitting and soft emission are described by an ${\cal L}^{(0)}$ Lagrangian insertion. Therefore the collinear splitting does not change the color structure of the amplitude from the point of view of the soft gluon, and hence it is still described by $\hat E_s$.  The final result is therefore given by the leading-power soft factor for the $N-1$ distinct collinear sectors, multiplied by the tree-level amplitude with the collinear splitting:
\begin{align}
{\cal A}_{{N+1}_s}^{[0,\text{coll}](0)} 
 = \big\langle T\, {\cal O}_{N-1}^{(0)} \: \Pi {\cal L}^{(0)} 
\big\rangle_{\rm coll+s}^{[0]} 
 = S_{N-1}^{[0](0)}(s)\,{\cal A}_{N}^{[0,\text{coll}](0)} \, .
\end{align}
Here the subscript on the matrix element $\langle \cdots \rangle_{\rm coll+s}$ indicates that the particles are taken in the kinematic situation described below \Eq{eq:ANcoll}, and hence differ from the matrix elements used in \Sec{subsec:oneloop}.  If particles $1$ and $2$ are collinear to each other in the $n$ direction, with $3$ to $N$ well separated, then we have
\begin{align}
   S_{N-1}^{[0](0)}(s) =  T_{1+2} \frac{n\cdot\epsilon_s}{n\cdot p_s} + 
     \sum_{i=3}^N \, T_i \, \frac{n_i\cdot\epsilon_s}{n_i\cdot p_s } \,,
\end{align}
where $T_{1+2}$ is the color matrix for the composition of particles $1$ and $2$ (equivalent to the color matrix for their parent particle).  For simplicity we will continue to use the convention that particles $1$ and $2$ are collinear for the analysis at subleading order below.

At one higher order in the power expansion, ${\cal O}(\lambda^{-1})$, we have to consider two contributions:
\begin{align}\label{eq:sub_split_terms}
{\cal A}_{{N+1}_s}^{[0,\text{coll}](1)}
 &=
  \big\langle T\, {\cal O}_{N-1}^{(1)} \: \Pi {\cal L}^{(0)} 
  \big\rangle_{\rm coll+s}^{[0]}
 +
  \big\langle T\, {\cal O}_{N-1}^{(0)} \: \Pi {\cal L}^{(0)} \: {\cal L}^{(1)} \big\rangle_{\rm coll+s}^{[0]} 
  \,.
\end{align}
The first term has a subleading operator insertion ${\cal O}_{N-1}^{(1)}$ which is either ${\cal O}_{N-1}^{(1,\partial)}$ or ${\cal O}_{N-1}^{(1,X)}$. For ${\cal O}_{N-1}^{(1,\partial)}$ we can make its matrix element vanish by using RPI to set the total perp momentum in the $n$-collinear direction to zero, $p_{n1}^\perp + p_{n2}^\perp = 0$. The operator ${\cal O}_{N-1}^{(1,X)}$ does not vanish by RPI and describes the collinear splitting at subleading power through the matrix element
\begin{align} \label{eq:O1Xemission}
 \big\langle T\, {\cal O}_{N-1}^{(1,X)} \: \Pi {\cal L}^{(0)} 
  \big\rangle_{\rm coll+s}^{[0]}
  &= C_{N}^{[0](1X)} \ 
  \Big\langle \hat {\cal O}_{N-1}^{(1,X)} \Big\rangle_{\rm coll}^{[0]} 
  \ E_{s[n]2}^{[0](N-1)\vec \kappa} \,.
\end{align}
Here the purely collinear matrix element $\big\langle \hat {\cal O}_{N-1}^{(1,X)}\big\rangle_{\rm coll}^{[0]}$ involves $N$ final state particles, two of which are collinear, for $N-1$ collinear directions. It produces no inverse powers of $\lambda$ because it does not have a parent collinear propagator that produces particles $1$ and $2$ in the $n$-collinear direction. Therefore this matrix element is suppressed by one power of $\lambda$ relative to the amplitude that appeared at leading power, ${\cal A}_{N}^{[0,\text{coll}](0)}$. The soft emission matrix element in \Eq{eq:O1Xemission} involves the same operator defined in \Eq{eq:soft_emit_simp1} except that there are only $N-1$ distinct collinear directions, so we have the tree level matrix element
\begin{align} \label{eq:Esn2forNminus1}
  E_{s[n]2}^{[0](N-1)\vec \kappa}   
   &= \Big\langle g_s \Big| T\,  Y^{\kappa_1}_n(0)\, Y^{\kappa_2}_n(0) 
   \prod_{i=3}^N Y^{\kappa_i}_{n_i}(0)
  \Big| 0 \Big\rangle^{[0]}   
  \\
 & =  (T_1+ T_2) \frac{n\cdot\epsilon_s}{n\cdot p_s} + 
     \sum_{i=3}^N \, T_i \, \frac{n_i\cdot\epsilon_s}{n_i\cdot p_s } 
  \,. \nn
\end{align}  
Here if the color indices of $Y^{\kappa_1}_n(0)$ and $Y^{\kappa_2}_n(0)$ are contracted then $T_1+T_2 = T_{1+2}$.

The second term in \Eq{eq:sub_split_terms} includes a subleading Lagrangian insertion, and is given by the factorized matrix element
\begin{align} \label{eq:L1emission}
  \big\langle T\, {\cal O}_{N-1}^{(0)} \: \Pi {\cal L}^{(0)} \: {\cal L}_n^{(1)} \big\rangle_{\rm coll+s}^{[0]}  
  &=  C_N^{[0](0)} \int \!\! d^dx \: 
  \Big\langle T\, \hat {\cal O}_{N-1}^{(0)}(0) \hat K_{n\mu}^{(1)\kappa}(x) 
    \Big\rangle^{[0]}_{\rm coll}\
   {\cal E}_{s(n)}^{[0](N-1)\mu\,\vec \kappa}(x) \,.
\end{align}
The operator $\hat K_{n\mu}^{(1)\kappa}$ contains the relevant terms without soft fields from ${\cal L}_n^{(1)}$ and was defined in \Eq{eq:LKB}. As indicated, this collinear matrix element is only nonzero if ${\cal L}^{(1)}$ acts on one of the two $n$-collinear particles. If it acts on any of the remaining individual well separated $N-2$ particles then it is zero by RPI, since $p_{i>2}^\perp =0$.   The soft matrix element in \Eq{eq:L1emission} involves the same operator $\hat {\cal E}_{s(n)}^{[0](N-1)\mu\,\vec \kappa}(x)$ defined in \Eq{eq:soft_emit_simp1}, but here with only $(N-1)$ distinct collinear directions, so
\begin{align}
  {\cal E}_{s(n)}^{(N-1)\mu\,\vec \kappa}(x) 
  &= 
  \Big\langle g_s \Big| 
  T\, g B_{s(n)}^{A\mu}(x)\, T^{\kappa A}  Y^{\kappa_1+\kappa_2}_n(0)\
   \prod_{i=3}^N\, Y_{n_i}^{\kappa_i}(0)
   \Big| 0 \Big\rangle
  \,.
\end{align}
Here $Y^{\kappa_1+\kappa_2}_n$ is in the color representation of the parent of particles $1$ and $2$.  At tree level, since $gB_{s(n)}^{A\mu}$ starts with one gluon field, we can drop the other Wilson lines to obtain
\begin{align}
  {\cal E}_{s(n)}^{[0](N-1)\mu\,\vec \kappa}(x) 
  &= \big\langle g_s \big| T^{\kappa A}
    g B_{s(n)}^{A\mu}(x)
   \big| 0 \big\rangle^{[0]}   
  \,.
\end{align}
The factorization properties of the real collinear emission contributions from this Lagrangian insertion are the same as those discussed above for the analogous terms in the collinear loop contribution.  The collinear and soft matrix elements are therefore given by the soft limit of a $1\to 3$ splitting amplitude,
\begin{align} \label{eq:2collsubfactor}
 & \big\langle 0 \big| 
T\, {\cal O}_{N-1}^{(0)} \: \Pi {\cal L}^{(0)} \: {\cal L}_n^{(1)} 
\big| 1^{a_1}_{/\!\!/}, 2^{a_2}_{/\!\!/}, 3^{a_3}, \ldots, N^{a_N}, s^{a_s} 
\big\rangle^{[0]}   
%\nn\\
%&
= \Splitbar^{[0](1)}(P\to 1,2,s)
{\cal A}^{[0]}(P,3,\cdots,N)
\, ,
\end{align}
where particles $1$ and $2$ are in the same collinear sector and we have made the color indices explicit. If the soft emission is not from either $1$ or $2$ then the contribution is zero by RPI, for the reasons explained earlier.

The full result for the subleading soft emission from the collinear splitting amplitude is given by the sum of \Eqs{eq:O1Xemission}{eq:L1emission}, 
\begin{align} \label{eq:split_full_fact}
{\cal A}_{{N+1}_s}^{[0,\text{coll}](1)}(1_{/\!\!/}, 2_{/\!\!/}, 3,\ldots)
 &=  C_{N}^{[0](1X)} \ 
  \Big\langle \hat {\cal O}_{N-1}^{(1,X)} \Big\rangle_{\rm coll}^{[0]} 
  \ E_{s[n]2}^{[0](N-1)\vec \kappa}  \nn\\
 &\quad + \Splitbar^{[0](1)}(P\to 1,2,s)
{\cal A}^{[0]}(P,3,\cdots,N) \,,
\end{align}
and both of these terms give nonzero contributions.  Unlike the situation with well separated particles at tree level or at one-loop these amplitudes for subleading soft with collinear emission are only suppressed by a single power of $\lambda$.

As an explicit example of the above general results we will compute the subleading soft gluon coupling to a jet sector with net fermion flavor for the more non-trivial contribution
$\big\langle T\, {\cal O}_{N-1}^{(0)} \: \Pi {\cal L}^{(0)} \: {\cal L}_n^{(1)} \big\rangle_{\rm coll+s}^{[0]}$.  If the parent $n$-collinear particle has momentum $p$, our convention that $p_\perp =0$ forbids the subleading soft emission or propagator insertion from appearing on the parent propagator. Furthermore, we will make the BPS field redefinition everywhere, so that graphs with a soft gluon generated by ${\cal L}^{(0)}$ are absorbed into diagrams where the soft gluon is emitted from the ${\cal L}^{(1)}$ directly. (This removes the need to separately consider the propagator insertion diagrams.) Here, we consider only one of the two possible color orderings $T^A T^B$ where the emission of the soft gluon comes either from a $n$-collinear gluon through ${\cal L}_{A_n}^{(1)}$ or the soft and collinear gluons are emitted simultaneously by ${\cal L}_{\xi_n}^{(1)}$. The expression for the amplitude with the other color ordering $T^B T^A$ is given in \App{app:123}.  For the color ordering considered here there are only three non-zero diagrams which give
\begin{align}\label{eq:2collsubcalc}
&
\raisebox{-0.305\height}{\includegraphics[width=3.5cm]{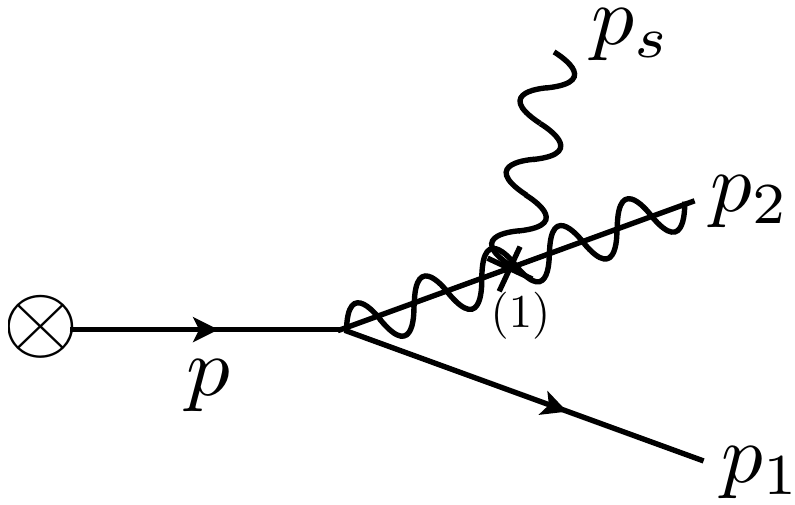}}
\ + \
\raisebox{-0.305\height}{\includegraphics[width=2.35cm]{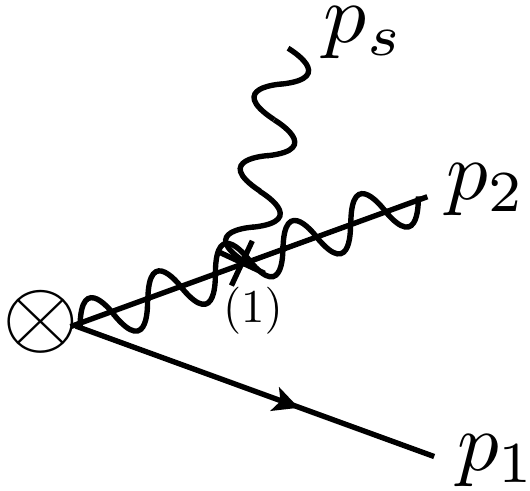}}
\ + \  
\raisebox{-0.35\height}{\includegraphics[width=3.5cm]{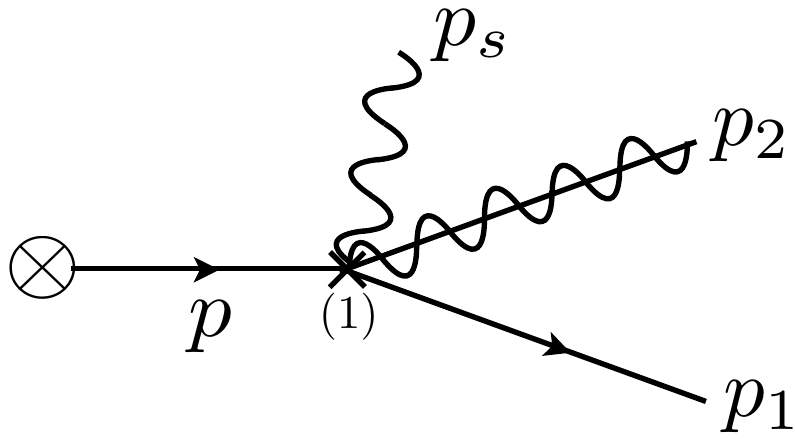}}
 \\*
&
\qquad\qquad
 = g^2\,  \bar u(p_1)  T^A T^B\, \bigg[\bigg(
\frac{n\cdot\epsilon_2}{n\cdot p}
-\frac{\bar n\cdot \epsilon_2}{\bar n\cdot p_2}
+\frac{\slashed{p}_{1\perp} \slashed{\epsilon}_{2\perp}}
   {n\cdot p\:\bar n\cdot p_1}
\bigg)2p_{2\perp}^\rho 
+2\epsilon_{2\perp}^\rho \frac{n\cdot p_s}{n\cdot p}
\nn\\
&\qquad\qquad\quad
-\bigg(\slashed \epsilon_{2\perp} \frac{\bar n \cdot p_2}{\bar n \cdot p}+\slashed p_{1\perp}\frac{\bar n\cdot \epsilon_2}{\bar n \cdot p} \bigg)\frac{n\cdot p_s}{n \cdot p}\gamma_\perp^\rho
\biggl]
\frac{\epsilon_s^{\mu}\, p_s^\nu}{(\bar n\cdot p_2)(n\cdot p_s)}\left(
g^\perp_{\mu\rho}\frac{n_\nu}{n\cdot p_s} - g^\perp_{\nu\rho}\frac{n_\mu}{n\cdot p_s}
\right)\nn \,.
\end{align}
Here $p=p_1+p_2+p_s$ is the total momentum,  $p_1$ ($p_2$) is the outgoing momentum of the collinear fermion (gluon), and $p_s$ is the outgoing momentum of the soft gluon.  Due to our choice of $p_\perp=0$ we have $p_{1\perp} = -p_{2\perp}$. We have explicitly included the coupling and color factors, where the collinear gluon has color index $A$ and the soft gluon has color index $B$.
The first graph in \Eq{eq:2collsubcalc} involves a collinear splitting from a ${\cal L}^{(0)}_n$ and a vertex insertion from ${\cal L}^{(1)}_{A_n}$, and is proportional to $1/n\cdot p$. The second diagram involves the same subleading Lagrangian but produces the collinear gauge particle $p_2$ from a collinear Wilson line $W_n$ that sits inside the $X_n$ in ${\cal O}_{N-1}^{(0)}$.  The third diagram consists of the emission of the soft and collinear gauge bosons from the same vertex on the fermion, using a term in ${\cal L}^{(1)}_{\xi_n}$.
The result in \Eq{eq:2collsubcalc} is also manifestly gauge invariant for both the soft gluon and the collinear gluon.  The gauge invariance of the soft gluon follows because of the anti-symmetry of final factor on the last line of \Eq{eq:2collsubcalc}.  The gauge invariance of the collinear gluon follows using the on-shell conditions for $p_1$ and $p_2$. Replacing $\epsilon_2\to p_2$, and setting $p_{2\perp} = -p_{1\perp}$, the factor in square brackets from \Eq{eq:2collsubcalc} becomes 
\begin{align}
&\bigg(
1-2\frac{n\cdot p_2}{n\cdot p}+2\frac{p^2_{1\perp} }{(n\cdot p)(\bar n\cdot p_1)}
+\frac{n\cdot(p_2+p_s)}{n\cdot p}
-\frac{p^2_{1\perp} }{(n\cdot p)(\bar n \cdot p_1)}
-2\frac{n\cdot p_s}{n\cdot p}
\bigg)p_{2\perp}^\rho 
\nn \\*
&\qquad\qquad= \frac{ p_{2\perp}^\beta(
  n\cdot p -n\cdot p_1-n\cdot p_2- n\cdot p_s )}{n\cdot p} 
=0 \ .
\end{align}

Since we have already used RPI to set $p_\perp=0$, \Eq{eq:2collsubcalc} cannot be set to zero using RPI.  Further, we explicitly see that the propagator carrying the total momentum $p$, contains $\bn\cdot p=\bn\cdot p_1+\bn\cdot p_2\sim \lambda^0$ and $n\cdot p=n\cdot p_1+n\cdot p_2+n\cdot p_s\sim \lambda^2$. It has the soft $n\cdot p_s$ momentum flowing through it, but is homogeneous in power counting. It is therefore an on-shell propagator where the soft momentum can not be expanded away. It can be explicitly checked that \Eq{eq:2collsubcalc} reproduces the soft limit of QCD splitting amplitudes (often defined via limits of full scattering amplitudes \cite{Mangano:1987xk,Bern:1995ix,Bern:1998sc,DelDuca:1999ha,Birthwright:2005ak,Birthwright:2005vi,Bern:2004cz}) by appropriately expanding the light-cone gauge Berends-Giele current \cite{Berends:1987me,Kosower:1999rx,Kosower:2002su}.  Our SCET result has additional terms required by soft gauge invariance that are proportional to $n\cdot p_s$, which do not appear in the expansion of the splitting amplitudes since the splitting amplitudes in full QCD are only collinear gauge invariant.  If one calculated the analogous splitting from gauge invariant antennae \cite{Kosower:1997zr,Kosower:2003bh,GehrmannDeRidder:2005cm} which correctly describe soft wide angle emissions from a dipole, it would agree exactly with the SCET result, with an appropriate decomposition and choice for the vector $\bar n$.

Although we presented above the calculation for the soft attachment with one color ordering, the soft attachment for the other color ordering (which involve more diagrams with fermion attachments) are also nonzero. The more complicated result for this second color ordering is presented in \App{app:123}. It is clear that the two do not cancel in general, since the gluon attachments do not exist for an abelian gauge theory, and in the non-abelian case they have a different color structure. 
These non-zero terms at ${\cal O}(\lambda^{-1})$ manifestly violate the Low-Burnett-Kroll theorem at tree-level.
Contributions at ${\cal O}(\lambda^0)$, which is ${\cal O}(\lambda^2)$ supressed relative to the leading soft factor for a collinear emission, can also be computed. This would require considering analogous matrix elements involving the operators enumerated in the ${\cal O}(\lambda^0)$ list in \Eq{eq:softpc}, much as we did for the one-loop subleading soft theorem.  Generically, terms at this order will be non-zero and would include further modifications to the Low-Burnett-Kroll theorem in the presence of collinear splittings.

When considering the contribution to the cross section, we must square the amplitude and multiply the contributions at different powers with one another.  Importantly, the contribution to the cross section from \Eq{eq:split_full_fact} does not interfere with the leading power splitting amplitude with a soft emission, so there is no amplitude squared contribution at ${\cal O}(\lambda^{-3})$. When summed over spins, this product of the leading and subleading amplitudes vanishes because it is proportional to $n\cdot n =0$, $n\cdot p_{2\perp}=0$, $n\cdot \epsilon_{2\perp}=0$ or  $n\cdot p_{s\perp}=0$. 
Therefore \Eq{eq:split_full_fact} first contributes at ${\cal O}(\lambda^{-2})$, corresponding to the square of this splitting amplitude.  The squared amplitude's contribution to the total cross section is non-zero when summed over the spins of the soft gluon because $p_{1\perp}$ and $p_{2\perp}$ are non-zero for this configuration.  At this order there are also nonzero interference terms between the ${\cal O}(\lambda^{-2})$ leading soft emission with collinear splitting amplitude, and the ${\cal O}(\lambda^0)$ tree level subleading soft splitting amplitude.

The fact that amplitude terms at ${\cal O}(\lambda^{-1})$ in the soft gluon expansion are generically non-zero for a collinear sector with more than one particle implies that typically the corresponding collinear loop corrections will diverge.  By the KLN theorem \cite{Kinoshita:1962ur,Lee:1964is}, real emission divergences are exactly canceled by IR divergences in loop integrals for sufficiently inclusive observables, called infrared and collinear safe.\footnote{The S-matrix elements are an important ingredient for the calculation of a physical observable. That they are in general IR divergent is not really problematic since these cancel by the KLN theorem or are replaced by physical scales when calculating observables in the field theory. Observables can also be related to expectation values of operators (for instance see \Refs{Sveshnikov:1995vi,Cherzor:1997ak}) to which the KLN theorem directly applies. In SCET, observables are expressed as expectation values of measurement weighted squares of operators, such as the $N$-jet operators.} Since the two collinear particles in \Eq{eq:2collsubcalc}, $p_1$ and $p_2$, are parametrically close in phase space they will be grouped together for the computation of the infrared safe observable.  Integrating over the phase space of the collinear splitting in \Eq{eq:2collsubcalc}, we see that the collinear splitting correction to the subleading soft theorem diverges, because of the singular collinear limit which allows both $n\cdot p\to 0$ and $n\cdot p_s\to 0$.  

These results are nicely consistent with \Sec{subsec:loopsub} where loop-level splitting functions contributed to the soft expansion at one-loop order.  When summed coherently and squared, the subleading soft terms for the collinear emission, given by \Eq{eq:2collsubfactor} and the analog from interference terms between the ${\cal O}(\lambda^{-2})$ and ${\cal O}(\lambda^0)$ tree level subleading soft splitting amplitudes, plus the the collinear loop correction from \Eq{eq:fulloneloopsubsoft}, will yield an infrared finite result for an infrared safe observable.  

\section{Conclusions}
\label{sec:conc}

In this paper, we have applied the methods of effective field theory to the problem of understanding the subleading soft limit of gauge theory amplitudes.  Soft-collinear effective field theory provides, among other things, insights into the coupling of soft gluons at subleading power, the structure of soft theorems at one-loop, the nature of asymptotic symmetries, and a formalism to interpret, cancel and/or remove divergences from loop integrals.  Once the IR structure of the theory is organized with the appropriate power counting, the asymptotic symmetries of RPI (a descendant of Lorentz Symmetry) and soft-collinear gauge invariance are true symmetries of the infrared theory. They are therefore directly represented as symmetries in SCET, and lead to explicit invariances and relations for the Lagrangians and $N$-jet operators that reorganize the full-theory S-matrix. 

In this paper we discussed several interesting results for soft theorems.  We used SCET to prove the Low-Burnett-Kroll theorem for soft gluon emission at subleading power from well-separated energetic particles, providing complementary understanding to other proofs already in the literature. In particular, the crucial role of the RPI symmetry is highlighted by the SCET proof. We also derived a subleading soft theorem that is valid for $N$ point amplitudes at one-loop order, given in \Eq{eq:fulloneloopsubsoft}. The terms in this one-loop result include those from hard loops multiplied by an LBK soft factor, terms from soft loops which are given by matrix elements of operators involving Wilson lines and the soft field object $g B_{s(n)}^{A\mu}$, terms from collinear loops which can be associated to one-loop $1\to 2$ and $1\to 1$ splitting amplitudes, and additional terms from collinear loops which involve fusion of collinear particles in $2\to 1$ and $3\to 1$ transitions.
Finally, we formulated a subleading soft theorem for tree level amplitudes where two particles are collinear (hence not well-separated), \Eq{eq:2collsubfactor}, which involves both a direct production term and the soft limit of $1\to 3$ splitting amplitudes. This is the real emission counterpart to the one-loop soft theorem. It starts at one lower order in the power expansion at the amplitude level, but contributes at the same order as the one-loop soft theorem when the amplitude is squared. Our general results for tree-level and one-loop level subleading soft amplitudes in gauge theory are consistent with the explicit example of the subleading factorization theorem for the $b\to s\gamma$ decay rate in QCD derived in Ref.~\cite{Lee:2004ja}, which involves subleading hard, soft, and collinear matrix elements.

Effective theories are especially powerful for making statements that hold to all-orders in the coupling at a particular power. To adequately confront the observables of a weakly-coupled gauge theory, soft and collinear regions of phase space must be under control. The correlated dynamics of these sectors often dominate the values of physical observables that are directly testable at modern colliders. Indeed, renormalization in the effective theory allows for a summation of logarithms of ratios of hierarchical scales in a field theory to all orders in the coupling. Thus the renormalization of the $N$-jet operators give a precise definition of the renormalization for scattering amplitudes describing the IR region physics. 

Beyond the results derived here, there are many other potential applications of effective theories to an understanding of the all-orders, perturbative S-matrix.  Progress can be made in many different directions working to higher powers of $\lambda$, studying the SCET of gravity, and further exploring loop amplitudes.  As we discussed in \Sec{subsec:loopsub}, effective theory techniques allow for a systematic study of the hard, soft and collinear regions of loop integrals to arbitrary order in the coupling.  SCET for gravity \cite{Beneke:2012xa} has so far only been formulated to leading power in the weak-field expansion. The study of this formalism at higher powers could provide techniques and insights for understanding the soft theorems in gravity.  We touched on the connection between RPI and the infinite-dimensional asymptotic symmetries of the S-matrix. It would be especially pleasing to see the covariantization of the RPI generators naturally forming the extended BMS group of asymptotic gravity.  Clearly there are many interesting directions for the study of subleading power amplitudes, and effective field theory analyses will be an important and powerful tool to move forward.

\begin{acknowledgments}
We thank Zvi Bern for useful feedback, Ian Moult and Dan Kolodrubetz for helpful comments on the manuscript, and Matthew Schwartz, Andy Strominger, Vyacheslav Lysov, Ira Rothstein,  Anastasia Volovich and Yu-tin Huang for discussions.  This work is supported by the U.S. Department of Energy (DOE) and the Office of Nuclear Physics and High Energy Physics under DE-SC0011090 and DE-SC00012567.  I.S. was also supported in part by the Simons Foundation through the Investigator grant 327942.  D.N. is also supported by an MIT Pappalardo Fellowship.
\end{acknowledgments}

\appendix

\section{Review of Derivation of the Fermionic SCET Lagrangian}
\label{app:colq}

In this appendix, we provide a schematic derivation of the collinear fermion SCET Lagrangian.  The first step to deriving the SCET Lagrangian is to construct the fields that create and annihilate the collinear and soft modes.  For concreteness and simplicity, consider the fermion part of the Lagrangian
\begin{equation}
{\cal L}_q = \bar{\psi} \, i\, \slashed{D}\, \psi \ ,
\end{equation}
where here $D_\mu$ is the gauge covariant derivative.  We would like to expand this Lagrangian to leading power in $\scetscale$ to describe collinear fermions in the $n$ direction.  To do this, it is useful to define the projection operators
\begin{equation}
\frac{\slashed{n}\slashed{\bar{n}}}{4}+\frac{ \slashed{\bar{n}}\slashed{n}}{4} = \mathbbm{1} \ ,
\end{equation}
which follows from $\{\gamma^\mu,\gamma^\nu\} = 2g^{\mu\nu}$.  Then, we can decompose $\psi$ as
\begin{equation}
\psi = \varphi_{\bar{n}} + \xi_n  \ ,
\end{equation}
where
\begin{equation}
\frac{\slashed{n}\slashed{\bar{n}}}{4}\psi = \xi_n \ ,\qquad  \frac{ \slashed{\bar{n}}\slashed{n}}{4} \psi = \varphi_{\bar{n}} \ .
\end{equation}
Note that this implies that $\slashed{n} \xi_n = 0$ and $\slashed{\bar{n}} \varphi_{\bar{n}} = 0$.  We can also expand the covariant derivative in the $n,\bar{n}$ basis as
\begin{equation}
\slashed{D} = \frac{\slashed{n}}{2}\bar{n}\cdot D +\frac{ \slashed{\bar{n}}}{2} n\cdot D + \slashed{D}_\perp \ .
\end{equation}
Exploiting the properties of the projection onto the spinors $\varphi_{\bar{n}}$ and $\xi_n$ we then can write the fermion Lagrangian as
\begin{equation}
{\cal L}_q = 
\bar{\xi}_n \frac{ \slashed{\bar{n}}}{2} in\cdot D \xi_n+\bar{\varphi}_{\bar{n}} i\slashed{D}_\perp\xi_n+\bar{\xi}_n i\slashed{D}_\perp \varphi_{\bar{n}} + \bar{\varphi}_{\bar{n}} \frac{\slashed{n}}{2}i\bar{n}\cdot D \varphi_{\bar{n}} \ .
\end{equation}

So far this is just a rewriting of the original Lagrangian.  However, consider the equation of motion for the Dirac spinor $\psi$:
\begin{equation}
\slashed{p} \psi =\left( \frac{p^-}{2}\slashed{n} +  \frac{p^+ }{2}\slashed{\bar{n}} + \slashed{p}_\perp\right)\psi = 0 \ .
\end{equation}
For a collinear fermion in the $n$ direction, the largest component of momentum from \Tab{tab:scaling} is $p^- \sim Q$, and so the leading-power equations of motion are just
\begin{equation}
 \frac{p^- }{2}\slashed{n} \psi = 0 \ .
\end{equation}
Note that this is just the equation of motion for the spinor $\xi_n$.  Thus, we associate $\xi_n$ with the leading power spinor and $\varphi_{\bar{n}}$ is the spinor whose components are subleading.  Therefore, we can integrate out $\varphi_{\bar{n}}$, which is exact quantum mechanically because it appears quadratically in the action.  Doing this, we then find,
\begin{equation}\label{eq:scetlag_again}
{\cal L}_{\xi} = \bar{\xi}_n \left(
i n\cdot D + i\slashed{D}_\perp \frac{1}{i\bar{n}\cdot D}i\slashed{D}_\perp
\right)\frac{\slashed{\bar{n}}}{2} \xi_n \ .
\end{equation}
The results of decomposing the collinear and soft fields in a manner consistent with gauge symmetry, and of carrying out the multipole expansion between collinear momentum $i\partial_n^\mu$ in $i D_n$, and soft momentum $i\partial^\mu$ in $i D_s$, can be expressed by writing
\begin{align}  \label{eq:scetDs}
  in\cdot D &= in\cdot\partial + g n\cdot A_n + g n\cdot A_s \,,
 & i D_\perp^\mu & =  i D_{n\perp}^\mu +  W_n i D_{s\perp}^\mu W_n^\dagger\,,
 \nonumber\\
 i \bn\cdot D &= i \bn\cdot D_n + W_n i\bn\cdot D_s W_n^\dagger \,.
\end{align}
All terms in $in\cdot D$ are ${\cal O}(\lambda^2)$, whereas in $iD_\perp^\mu$ and $i\bn\cdot D$ the $n$-collinear terms are leading order and terms with soft covariant derivatives are suppressed by one or two powers of $\lambda$ respectively. Plugging in \Eq{eq:scetDs} into \Eq{eq:scetlag_again} gives the leading and subleading power collinear fermion Lagrangians in gauge theory presented in \Eqs{eq:scetlag}{eq:sublagcollq}.

\section{LSZ Reduction and the Low-Burnett-Kroll Theorem}\label{app:LSZ}

In this appendix we write the Low-Burnett-Kroll theorem in terms of time-ordered Green's functions via the LSZ reduction formula. This clearly illustrates the action of the addition of a soft particle on each external amputated leg, before the final contraction with the charge and polarizations. By working with the LSZ formula, we avoid derivatives of polarizations. To begin we need the parent S-matrix element in terms of LSZ reduction:
\begin{align}\label{eq:LSZ}
{\mathcal A}(p_1^{f_1},p_2^{f_2},...,p_{N}^{f_N})&=\,_{\text{in}}\langle 0 |p_1^{f_1},p_2^{f_2},...,p_{N}^{f_N}\rangle_{\text{out}}\nonumber\\
&=\prod_{i=1}^{N}\lim_{p_i^2\rightarrow m_i^2}\int d^dx_i e^{i\,p_i\cdot x_i} g_i\epsilon_{i}\cdot G^{-1}_i(x_i)\cdot G^{C}(x_1,...,x_N)\, ,\\
G^{C}(x_1,...,x_N)&=\frac{\,_{\text{in}}\langle 0|T\phi^{f_1}(x_1)...\phi^{f_N}(x_N)|0\rangle_{\text{out}}}{\,_{\text{in}}\langle 0|0\rangle_{\text{out}}}.
\end{align}
The $f_i$ denotes flavor, charge, and spin indicies, and $\epsilon_{i}$ is the spinor or polarization vector of the asymptotic particle, and $g_i$ is its charge. $G_i(x_i)$ is the time-ordered two-point function for the $i$-th particles. We take the connected time-ordered Green's function $G^C$ for fields $\phi^{f_i}$ to be only at tree level. Finally, the ``$\cdot$'' denotes spin/charge index contractions between the amputated Green's function and the external particle states. Then the statement of Low-Burnett-Kroll Theorem is:
\begin{align}\label{eq:LSZ_LBK}
&\,_{\text{in}}\langle 0 |p_1^{f_1},p_2^{f_2},...,p_{N}^{f_N},p_s^{f_s}\rangle_{\text{out}}\nonumber \\*
&\qquad=\sum_{j=1}^N\prod_{i=1}^{N}\lim_{p_i^2\rightarrow m_i^2}\int d^dx_i e^{i\,p_i\cdot x_i} g_i\epsilon_{i}\cdot S_{ij}(p_s^{f_s})\cdot G^{-1}_i(x_i)\cdot G^{C}(x_1,...,x_N) \ ,
\end{align}
where $|p_s^{f_s}\rangle$ is a single soft gluon state. The soft factor matrix $S_{ij}(p_s^{f_s})$ is defined as:
\begin{align}\label{eq:Soft_Matrix_on_LSZ_LBK}
S_{ij}(p_s^{f_s})=\begin{cases}
& \idop_c \otimes \idop_s \text{ , if }i\neq j \\
& S^{(0)a}(p_s^{f_s};p_i){\mathbf T}_i^a\otimes\idop_s +S^{(2)a\mu\nu}(p_s^{f_s};p_i){\mathbf T}_i^{a}\otimes\Big\{\idop_s \,ip_{i[\mu}x_{i\nu]}+ \Sigma_{\mu\nu}\Big\}\text{ , if }i = j 
\end{cases}
\end{align}
The $\idop_{c,s}$ are identity matricies on the charge/spin indicies. The $\Sigma_{\mu\nu}$ are the spin-generators for the given flavor of the external leg, and carry spin indicies for that leg. The soft kinematic factors are given by:
\begin{align}  \label{eq:Soft_LSZ_onegluon}
S^{(0)a}(p_s^{f_s};p_i)&=\,_{\text{in}}\langle 0|\Bigg(\int_{0}^{\infty}d\lambda \,p_i\cdot A^{a}(\lambda\,p_i)\Bigg)|p_s^{f_s}\rangle_{\text{out}} \ , \nn \\
S^{(2)a\mu\nu}(p_s^{f_s};p_i)&=\,_{\text{in}}\langle 0|\Bigg(\int_{0}^{\infty}d\lambda \,F^{a\mu\nu}(\lambda\,p_i)\Bigg)|p_s^{f_s}\rangle_{\text{out}} \ .
\end{align}

%%%---------------------------------------------------------------------
\section{RPI Expansion of the $N$-jet operator ${\cal O}_N$}\label{app:rpi}
%%%---------------------------------------------------------------------

In the effective theory expansion of the full theory, one in general considers all possible gauge invariant operators at each order in the power expansion. Further, these operators are constrained by the reparametrization invariance symmetry of the effective theory. RPI mixes operators of differing power counting orders, so that terms higher order in the expansion are connected to lower order terms. These constraints can be implemented in one of two ways:
\begin{description}
\item[A:] Construct all gauge invariant operators that are explicitly reparameterization invariant. These operators must be expanded in $\lambda$, which generates a series of operators that are homogeneous in $\lambda$ and gauge invariant. This series of operators will be connected to each other by RPI as long as they are linearly independent from the operators obtained by expanding other RPI-invariants. 
\item[B:] Construct all gauge invariant operators that are homogeneous in the power counting. These operators may transform under a reparameterization transformation, and linear combinations of them are then grouped into RPI-invariants.
\end{description} 
Approach B has been used to constrain weak decay operators in Refs.~\cite{Chay:2002vy,Manohar:2002fd,Pirjol:2002km,Becher:2004kk,Lee:2004ja,Arnesen:2005nk,Chay:2005ck}, while the simpler, but more technically involved, Approach A was used to constrain operators in Ref.~\cite{Marcantonini:2008qn}. 
These methods are often complimentary. In this appendix, we will write the leading order $N$-jet operator in a RPI form, and show that the ${\cal O}(\lambda^2)$ subleading $N$-jet operators that describe soft gluon emission follow from the expansion of the leading order RPI operator. Then we will consider other possible homogenous operators that could appear at this order, and show they have no RPI completion, thus ruling them out as possible contributions to all orders in the coupling constant at this order in $\lambda$.

There is a unique RPI operator whose expansion starts with the $N$-jet operator ${\cal O}_N^{(0)}$. Before the BPS field redefinition it is:
\begin{align}
& {\cal O}_N^{\text{RPI}}
 =\int \prod_{i=1}^N (d\bar{n}_i\cdot Q_idn_i\cdot Q_i d^2\vec{Q}_{i\perp })
  \, C^{\{\kappa_k\}}(\{Q_k\})
\\*
&\qquad\quad \times \prod_{i=1}^N \Big[\delta(\bar{n}_i\cdot Q_i-i\bar{n}_i\cdot\partial_{n_i}^{\text{RPI}})\delta^{(2)}\Big(\vec{Q}_{i\perp }-i\vec{\partial}_{n_i\perp}^{\,\text{RPI}}\Big)\delta(n_i\cdot Q_i-in_i\cdot D_{s})X_{n_i}^{\kappa_i\,\text{RPI}}(0)\Big]
 \,,\nn
\end{align}
where
\begin{align}
\bar{n}_i\cdot{\partial}_{n_i}^{\text{RPI}}&=\bar{n}_i\cdot\partial_{n_i}+\bar{n}_i\cdot D_{s},\nn\\
{\partial}_{n_i\perp}^{\mu\,\text{RPI}}&=\partial^{\mu}_{n_i\perp}+D_{s\perp}^{\mu}.
\end{align}
The operators $X_{n_i}^{\kappa_i\,\text{RPI}}$ are taken as the full RPI invariant field operators appearing in the first line of \Eq{eq:rpiops}. In the $\bar{n}_i$ and transverse momentum $\delta$-functions, the derivatives contains a soft component according to \Eqs{eq:RPI_to_soft_covariant_der_I}{eq:RPI_to_soft_covariant_der_II} that must be expanded out for homogenuous power counting. The $n_i$ $\delta$-function will have its soft gauge field removed from $n_i\cdot D_s$ once we perform the BPS field redefinition to factorize soft and collinear operators. Focusing on the terms with an expansion, we have:
\begin{align}
& \delta\big(\bar{n}_i\cdot Q_i-i\bar{n}_i\cdot\partial_{n_i}^{\text{RPI}}\big)
  \delta^{(2)}\Big(\vec{Q}_{i\perp}-i\vec{\partial}_{n_i\perp}^{\,\text{RPI}}
   \Big) \\*
&\quad 
 = \Big(1+i\bar{n}_i\cdot D_{s}\frac{\partial}{\partial\bar{n}_i\cdot Q_i}+i D_{s\perp_i}^\mu\frac{\partial}{\partial Q_{i\perp}^{\mu}}\Big)\delta(\bar{n}_i\cdot Q_i-i\bar{n}_i\cdot\partial_{n_i})\delta^{(2)}\Big(\vec{Q}_{i\perp}-i\vec{\partial}_{n_i\perp}\Big)+...
\,, \nn
\end{align}
where we have dropped terms that are even higher order in $\lambda$ than those displayed. Integrating by parts, we can move the derivatives onto the hard interaction Wilson coefficient.  Expanding also the $X_{n_i}^{\kappa_i\,\text{RPI}}= X_{n_i}^{\kappa_i} + \ldots$ to ${\cal O}(\lambda^2)$ where we consider only terms involving soft field components as in \Eq{eq:subsubfields}, we find:
\begin{align}
{\cal O}_N^\text{RPI}&={\mathcal O}^{(0)}_N+{\mathcal O}^{(2,r)}_N+{\mathcal O}^{(2,\delta)}_N+...
\end{align}
After performing the BPS field redefinition, and using RPI the operator ${\mathcal O}^{(2,r)}_N$ has precisely the form given in \Eq{eq:subsub_op_rpi}. Since the logic is the same we will carry out these steps only for ${\mathcal O}^{(2,\delta)}_N$. Immediately after the expansion the sub-subleading operator ${\mathcal O}^{(2,\delta)}_N$ is:
\begin{align}
{\mathcal O}^{(2,\delta)}_N 
 &=-\sum_{j=1}^N\int 
 \prod_{i=1}^N (d\bar{n}_i\cdot Q_i dn_i\cdot Q_i d^2\vec{Q}_{i\perp })
 \,\Big(\bar{n}_j^{\mu}\frac{\partial}{\partial\bar{n}_j\cdot Q_j}+\frac{\partial}{\partial Q_{j\perp}^{\mu}}\Big) 
 C^{\{\kappa_k\}}(\{Q_k\})
 \\*
& \times\Big[\delta(\bar{n}_j\cdot Q_j-i\bar{n}_j\cdot \partial_{n_j})\delta^{(2)}\Big(\vec{Q}_{j\perp }-i\vec{\partial}_{n_j\perp}\Big)\,iD_{s}^\mu\delta(n_j\cdot Q_j-in_j\cdot D_{s})X_{n_j}^{\kappa_j}(0)\Big]
 \nn \\
& \times \prod_{i\neq j}\Big[\delta(\bar{n}_i\cdot Q_i-i\bar{n}_i\cdot \partial_{n_i})\delta^{(2)}\Big(\vec{Q}_{i\perp }-i\vec{\partial}_{n_i\perp}\Big)\delta(n_i\cdot Q_i-in_i\cdot D_{s})\,X_{n_i}^{\kappa_i}(0)\Big]
 \,. \nn
\end{align}
Performing the BPS field redefinition to give Wilson lines $Y_{n}^\kappa$ and using RPI, the soft derivative $D_s^\mu$ from the expansion only acts on the soft Wilson line in the $n_j$ direction, $Y_{n_j}^{\kappa_j}$. The $n_i\cdot D_s$ soft derivatives in the $\delta$-function act only on $Y_{n_i}^{\kappa_i}$, which can be commuted through all $\delta$-functions to leave simply a partial derivative $n_i\cdot \partial$. This renders both the $Q^{\mu}_{n_i\perp}$ and $n_i\cdot Q_i$ integrals trivial, since for any state that this operator can produce we can perform an RPI transformation to set these total momenta for the corresponding $X_{n_i}^{\kappa_i}$ to zero.  

We can make use of the RPI$_\text{II}$ transformations to change $\bar n_j^\mu$ and eliminate the ${\partial}/{\partial Q_{j\perp}^{\mu}}$ derivative on the hard matching coefficient $C(\{ Q_i\})$. Making the BPS field redefinition, using \Eq{eq:YDY}, then dropping the term proportional to $i\bn_j\cdot i\partial_s X_{n_j}^{\kappa_j}(0)$ by use of RPI,  this then gives
\begin{align}
{\mathcal O}^{(2,\delta)}_N &=
 -\sum_{j=1}^N\int \prod_{i=1}^N (d\bar{n}_i\cdot Q_i) \,\frac{\partial}{\partial\bar{n}_j\cdot Q_j} C^{\{\kappa_k\}}(\{Q_k\})\bigg|_{Q_k=\frac{\bar{n}_k\cdot Q_k}{2}n_k}\:
\nn \\
& \qquad\qquad\times
 \prod_{i=1}^N \Big[\delta(\bar{n}_i\cdot Q_i-i\bar{n}_i\cdot \partial_{n_i})\,X_{n_i}^{\kappa_i}(0)\Big]\:
  T\Big\{\Big(\prod_{i\neq j}Y_{n_i}^{\kappa_i}\Big)\,i\bar{n}_j\cdot D_{s} Y_{n_j}^{\kappa_j} \Big\}
 \nn\\
&=
 -\sum_{j=1}^N\int \prod_{i=1}^N (d\bar{n}_i\cdot Q_i) \,\frac{\partial}{\partial\bar{n}_j\cdot Q_j} C^{\{\kappa_k\}}(\{Q_k\})\bigg|_{Q_k=\frac{\bar{n}_k\cdot Q_k}{2}n_k}\:
\nn \\
& \qquad\qquad\times
 \prod_{i=1}^N \Big[\delta(\bar{n}_i\cdot Q_i-i\bar{n}_i\cdot \partial_{n_i})\,X_{n_i}^{\kappa_i}(0)\Big]\:
  T\Big\{\Big(\prod_{i}Y_{n_i}^{\kappa_i}\Big) \bn_j \cdot g B_s^{(n_j)A} 
  T^{\kappa_j A} \Big\}
\,.
\end{align}
This form agrees exactly with \Eq{eq:subsub_op_delta}. 

Alternatively, if we do not transform the transverse derivative away, we can make use of the fact that $n_j\cdot D_s Y_j=0$ to write the sub-subleading operator as:
\begin{align}\label{eq:RPI_descend_for_LBK_Compare}
{\mathcal O}^{(2,\delta)}_N
   &=-\sum_{j=1}^N\int \prod_{i=1}^N (d\bar{n}_i\cdot Q_i) \,\frac{\partial}{\partial Q_j^{\mu}} C^{\{\kappa_k\}}(\{Q_k\})\bigg|_{Q_k=\frac{\bar{n}_k\cdot Q_k}{2}n_k}\:
  \prod_{i=1}^N  \Big[\delta(\bar{n}_i\cdot Q_i-i\bar{n}_i\cdot \partial_{n_i})\,X_{n_i}^{\kappa_i}(0)\Big]
  \nn \\*
  & \qquad\qquad\times
T\Big\{\Big(\prod_{i\neq j}Y_{n_i}^{\kappa_i}\Big)\,i D_{s}^{\mu} Y_{n_j}^{\kappa_j} \Big\}\,.
\end{align}
Note that the derivative is taken before we set $Q_k=\frac{\bar{n}_k\cdot Q_k}{2}n_k$. Using this form of the operator, we can connect to the LSZ reduced description of the scattering amplitude using the identity:
\begin{align}
 iD_{s}^{\mu} Y_{n_j} 
    &=\text{P}\,g\int_0^\infty d\lambda\, n_{j\nu} F^{\mu\nu}_s(\lambda\, n_j)Y_{n_j} \,.
\end{align}
For states containing a single soft gluon, we see that at lowest order in perturbation theory:
\begin{align}
\big\langle 0\big|T\Big\{\Big(\prod_{i\neq j}Y_{n_i}\Big)\,i D_{s}^{\mu} Y_{n_j}\Big\}\big|p_s\big\rangle
  &=\big\langle 0\big| g\int_0^\infty d\lambda\, n_{j\nu} F^{\mu\nu}_s(\lambda n_j) \big|p_s\big\rangle\,.
\end{align}
Thus comparing with \Eqs{eq:Soft_Matrix_on_LSZ_LBK}{eq:Soft_LSZ_onegluon} we see that the tree-level matrix element of ${\mathcal O}^{(2,\delta)}_N$ reproduces the orbital angular momentum contribution (irrespective of the precise identity of the various $X_n^\kappa$ fields).

Lastly, we can ask whether any other ${\cal O}(\lambda^2)$ suppressed operators involving soft fields exist, other than ${\cal O}_N^{(2,r)}$ and ${\cal O}_N^{(2,\delta)}$. The only possible such operator at this order involves a single $D_s^\mu$.  For example, we can consider operators of the form:
\begin{align}\label{eq:arb_soft_operator}
{\mathcal O}^{(2,?)}_N=&\sum_{j=1}^N\int 
 \prod_{i=1}^N (d\bar{n}_i\cdot Q_i) \ C_{j\mu}^{\{\kappa_k\}}(\{Q_k\})\bigg|_{Q_k=\frac{\bar{n}_k\cdot Q_k}{2}n_k}\:
 \prod_{i=1}^N \Big[\delta(\bar{n}_i\cdot Q_i-i\bar{n}_i\cdot \partial_{n_i})\,X_{n_i}^{\kappa_i}(0)\Big]
\nonumber\\*
&\qquad\times
T\Big\{\Big(\prod_{i\neq j}Y_{n_i}^{\kappa_i}\Big)\,i D_{s}^{\mu} Y_{n_j}^{\kappa_j} \Big\}\,.
\end{align}
Here we have written an arbitrary matching coefficient $C_{j\mu}$ that contracts with the explicit soft derivative, allowing for more general kinematic dependence of this index. We must also consider all operators where the index $\mu$ can be contracted within the collinear field structures, 
\begin{align} \label{eq:arb_soft_operator2}
{\mathcal O}^{(2,??)}_N
=&\sum_{j,l=1}^N\int \prod_{i=1}^N (d\bar{n}_i\cdot Q_i) \,
  C_{jl}^{\{\kappa_k\}} (\{Q_k\})\bigg|_{Q_k=\frac{\bar{n}_k\cdot Q_k}{2}n_k}\prod_{i\neq l}\Big[\delta(\bar{n}_i\cdot Q_i-i\bar{n}_i\cdot \partial_{n_i})\,X_{n_i}^{\kappa_i}(0)\Big]
 \nonumber\\
&\qquad \times
  \bigg[\delta(\bar{n}_l\cdot Q_l-i\bar{n}_l\cdot \partial_{n_l})\,\frac{t_l^{\mu}}{\bn_l\cdot Q_l} \, 
   X_{n_l}^{\kappa_l}(0)\bigg]
  \:  T\Big\{\Big(\prod_{i\neq j}Y_i\Big)\,i D_{s\mu} Y_{n_j}\Big\}\,.
\end{align}
The factor of $1/\bn_l\cdot Q_l$ matches the mass dimension. The vector $t_l^{\mu}$ cannot have $D_{n_l\perp}$ or $n_l\cdot D_{n_l}$ collinear derivatives, since this would be beyond the power counting order, while it must contain an $\bar n_l^\nu$ by RPI-III invariance. We also allow the possibility that $t_l^\mu$ facilitates the contraction of a vector index in an $X_{n_l}^{\kappa_l}$ with the $D_s^\mu$. 
In order for new operators of the forms in \Eqs{eq:arb_soft_operator}{eq:arb_soft_operator2} to appear, they must occur in the expansion of an RPI operator whose leading term starts at subleading order in the power expansion. However in SCET there is no RPI operator that at lowest order that can be expanded to start with a single $D_s^\mu$.\footnote{This differs from the situation in HQET~\cite{Luke:1992cs}, where the RPI combination $(v+i D/m)^2-1$ can be expanded to start with a single soft derivative. } This implies that any operators ${\cal O}_N^{(2,?)}$ or ${\cal O}_N^{(2,??)}$   can only be given by ${\cal O}_N^{(2,\delta)}$ and ${\cal O}_N^{(2,r)}$ themselves. Thus  in \Eq{eq:arb_soft_operator} we have $C_{j\mu}^{\{\kappa_k\}}(\{Q_k\})= -\partial/\partial Q_j^\mu C^{\{\kappa_k\}} (\{Q_k\})$ and in \Eq{eq:arb_soft_operator2} we have $t_l^\mu=\gamma_{\perp l}^{\mu}\frac{\bar{n}_l\!\!\!\!\!\slash}{2}$ for a quark field $X_{n_l}^{\kappa_l}$, and $t_l^\mu=0$ otherwise.  Thus RPI is sufficient to demonstrate that all operators with a single $D_s^\mu$ are connected to the leading N-jet operator.

\section{LBK for Gluons using SCET}\label{app:gluonLBK}

In this appendix, we show that at tree-level, the Low-Burnett-Kroll operator for gluons is reproduced in SCET.  In \Sec{sec:opcontribs}, we showed that matrix elements of the sub-subleading operator ${\cal O}_N^{(2,\delta)}$ produce the orbital angular momentum for any external particle: scalars, fermions, or gluons.  To show that the Low-Burnett-Kroll operator is reproduced at tree-level in SCET for gluons, we must show that the matrix element of  insertions of the sub-subleading Lagrangian ${\cal L}_{A_n}^{(2)}$ reproduce the spin angular momentum of a gluon. Specifically, we will calculate
\begin{align}
\big\langle 0\big| 
T\, {B}_{n\perp}^{\nu}(0){\cal L}_{A_{n}}^{(2)}
\big|\epsilon_n,p_n;\epsilon_s,p_s\big\rangle\,.
\end{align}
We have a sum over this type of matrix element for each external collinear particle and this is the only T-product term that needs to be considered.

To do so, we first rewrite the sub-subleading Lagrangian from \Eq{eq:subsublag} in a more convenient form.  Using $n\cdot {\cal D}_{ns}=n\cdot D_s+n\cdot {\cal D}_n-n\cdot \partial$ and including the covariant gauge fixing terms, we have 
\begin{align} \label{eq:LAn2again}
{\cal L}_{A_n}^{(2)}&=  \frac{1}{g^2}\text{Tr}\Big(
\big[
i {\cal D}_{ns}^\mu, i D_s^{\nu} 
\big]\big[
i {\cal D}_{ns\mu}, i D_{s\nu}
\big]
\Big)
+\frac{1}{g^2}\text{Tr}
\Big( \big[
i D_{s\perp}^{\mu},  iD_{s\perp}^{\nu} 
\big]\big[
i {\cal D}_{n\mu}^{\perp},  i {\cal D}^\perp_{n\nu}
\big] \Big)
\nonumber \\*
& 
\qquad+\frac{1}{g^2}\text{Tr}
\Big( \big[
i {\cal D}_{ns}^{\mu}, in\cdot {\cal D}_n-in\cdot \partial
\big]\big[
i {\cal D}_{ns\mu},  i \bar{n}\cdot D_{s}
\big] \Big)
+\frac{1}{g^2}\text{Tr}
\Big( \big[
 i D_{s\perp}^{\mu},  i{\cal D}_{n\perp}^{\nu} 
\big]\big[
i {\cal D}_{n\mu}^{\perp},  i D^\perp_{s\nu}
\big] \Big) \nonumber \\*
&\qquad+\frac{1}{\alpha}\text{Tr}
\Big(
\big[
 i D_{s\perp}^{\mu}, A_{n\perp \mu}
\big]
\big[
 i D_{s\perp}^{\nu}, A_{n\perp \nu}
\big]
\Big)+
\frac{1}{\alpha}\text{Tr}
\Big(
\big[
 i \bar{n}\cdot D_{s},n\cdot A_n
\big]
\big[
 i \partial_{ns}^{\mu},A_{n\mu}
\big]
\Big) \,.
\end{align} 
Insertions of the first term
$$
\frac{1}{g^2}\text{Tr}\Big(
\big[
i {\cal D}_{ns}^\mu, i D_s^{\nu} 
\big]\big[
i {\cal D}_{ns\mu}, i D_{s\nu}
\big]
\Big)
$$
 always leads to terms proportional to $p_s^2$ or $p_s\cdot \epsilon_s$, both of which vanish for the on-shell soft gluon. The third term in \Eq{eq:LAn2again},
$$
\frac{1}{g^2}\text{Tr}
\Big( \big[
i {\cal D}_{ns}^{\mu}, in\cdot {\cal D}_n-in\cdot \partial
\big]\big[
i {\cal D}_{ns\mu},  i \bar{n}\cdot D_{s}
\big] \Big)
$$ 
also vanishes, since the combination $in\cdot {\cal D}_n-in\cdot \partial$ always gives an $n$ contracted into either the external collinear gluon's polarization vector which vanishes (since $p_n \cdot \epsilon_n = \frac12 \bn\cdot p_n n\cdot \epsilon = 0$ for our RPI choice)  or with the field operator ${B}_{n\perp}^{\nu}$ from the parent amplitude which vanishes in Feynman gauge ($\alpha=1$).  In Feynman gauge insertions of the sixth (gauge-fixing) term in \Eq{eq:LAn2again}, 
$$
\frac{1}{\alpha}\text{Tr}
\Big(
\big[
 i \bar{n}\cdot D_{s},n\cdot A_n
\big]
\big[
 i \partial_{ns}^{\mu},A_{n\mu}
\big]
\Big)
$$
vanishs for the same reason. 

Therefore, the terms in ${\cal L}_{A_n}^{(2)}$ that produce a non-zero matrix element are
\begin{align}
{\cal L}_{A_n}^{(2)}&\supset \frac{1}{g^2}\text{Tr}
\Big( \big[
i D_{s\perp}^{\mu},  iD_{s\perp}^{\nu} 
\big]\big[
i {\cal D}_{n\mu}^{\perp},  i {\cal D}^\perp_{n\nu}
\big] \Big)
+\frac{1}{g^2}\text{Tr}
\Big( \big[
 i D_{s\perp}^{\mu},  i{\cal D}_{n\perp}^{\nu} 
\big]\big[
i {\cal D}_{n\mu}^{\perp},  i D^\perp_{s\nu}
\big] \Big) \nonumber \\
&\qquad +\frac{1}{\alpha}\text{Tr}
\Big(
\big[
 i D_{s\perp}^{\mu}, A_{n\perp \mu}
\big]
\big[
 i D_{s\perp}^{\nu}, A_{n\perp \nu}
\big]
\Big) \,.
\end{align} 
Computing the matrix element with these terms we find
\begin{align} \label{eq:SCETLBKglue}
\big\langle 0\big|
T{B}_{n\perp}^{\nu}(0){\cal L}_{A_{n}}^{(2)}
\big|\epsilon_n,p_n;\epsilon_s,p_s\big\rangle &=\epsilon_{n\mu}\frac{2\epsilon_{s\rho}p_{s\sigma}}{p_n^-(n\cdot p_s)}\Big(g_\perp^{ \mu\rho}g_\perp^{\sigma \nu}-g_\perp^{ \mu\sigma}g_\perp^{\rho \nu}\Big) \,.
\end{align} 
Here, $g_\perp^{\mu\nu}$ represents the $\perp$-components of the flat space metric.

To see how LBK matches with this SCET calculation, we need the spin angular momentum operator for gluons:
\begin{equation}
\Sigma^{\mu\nu\rho\sigma} = g^{\mu\rho}g^{\sigma\nu}-g^{\mu\sigma}g^{\rho\nu} \, .
\end{equation}
Then, the spin contribution to the Low-Burnett-Kroll subleading soft factor for emission of a soft gluon $s$ off of collinear gluon $n$ is
\begin{equation}\label{eq:spingluelbk}
S^{\text{(sub)}}_\text{spin} = \epsilon_{n\mu}\frac{\epsilon_{s\rho}p_{s\sigma}}{p_n\cdot p_s}\left(
 g^{\mu\rho}g^{\sigma\nu}-g^{\mu\sigma}g^{\rho\nu}
\right)\tilde{\cal A}_\nu \, .
\end{equation}
Here, we have stripped the external polarization from the lower point amplitude; that is
\begin{equation}
{\cal A} = \epsilon_n\cdot \tilde {\cal A}\, .
\end{equation}
Setting $p_{n\perp} = 0$ and using $p_n^2=0$ so that $p_n^\mu = \frac12 n^\mu \bn\cdot p_n$, we see that $p_n\cdot \epsilon_n=0$ implies $n\cdot \epsilon_n=0$, and $p_n\cdot \tilde {\cal A}=0$ implies $n\cdot \tilde {\cal A}=0$. Decomposing $g^{\mu\nu} = g_\perp^{\mu\nu} + \frac12 n^\mu \bn^\nu+\frac12 \bn^\mu n^\nu$ and using these results, \Eq{eq:spingluelbk} becomes
\begin{equation} \label{eq:LBKspinglue}
S^{\text{(sub)}}_\text{spin} = \epsilon_{n\mu}\frac{2\epsilon_{s\rho}p_{s\sigma}}{p_n^-(n\cdot p_s)}\left(g_\perp^{ \mu\rho}g_\perp^{\sigma \nu}-g_\perp^{ \mu\sigma}g_\perp^{\rho \nu}
\right)\tilde{\cal A}_\nu \, .
\end{equation}
Thus we see that the Lagrangian insertion of SCET which gives \Eq{eq:SCETLBKglue}, directly reproduces the spin contribution to the Low-Burnett-Kroll theorem at tree-level, given by \Eq{eq:LBKspinglue}.

Note that \Eq{eq:LBKspinglue} does not generate a spin flip for the collinear gluon since the helicity states are eigenstates of this spin operator. The products of helicity terms appearing here are the same as those appearing in the helicity conserving $\epsilon_{n\mu} g_\perp^{\mu\nu} \tilde {\cal A}_\nu$.

%%%%%%%%%%%%%%%%%%%%%%%%%%%%%%%%%%%%%%%%%%%%%%%%%%
\section{SCET Gauge Symmetries as Asymptotic Gauge Symmetries }
\label{app:gauge}
%%%%%%%%%%%%%%%%%%%%%%%%%%%%%%%%%%%%%%%%%%%%%%%%%%

As an effective theory, we expect the space of symmetries of SCET to be larger than that in the full gauge theory.  We have already seen the manifestation of this in the RPI transformations.  Similarly each collinear direction has its own independent set of gauge transformations~\cite{Bauer:2001ct,Bauer:2001yt}, and we review these results in this appendix. From the full theory perspective, these appear to be supported asymptotically far away from the hard scattering, and localized close to the given collinear direction. For any number of collinear jet directions, the Lagrangian breaks up into a sum of terms for each collinear direction.  Then, for example, because each collinear direction is a gauge theory of SU$(N_c)$, the SCET Lagrangian has an SU$(N_c)^N$ gauge symmetry, for a system with $N$ jets (plus an additional gauge symmetry for the soft fields).  As mentioned earlier, because we have assigned a definite power counting to the collinear and soft gauge fields, the gauge transformations of the gauge fields must respect this scaling.  Also, the standard collinear gauge transformations of the collinear gluon field mixes collinear and soft gluons.

In the full gauge theory, the gauge group element $U(x)$ is
\begin{equation}
U(x) = \exp \left[
i\alpha^A(x) \cdot T^A
\right] \ ,
\end{equation}
where $T^A$ are the generators of the gauge group.  The transformation rule of the gauge field $A^\mu(x)$ is
\begin{align}\label{eq:gaugetrans}
-igA^\mu(x)&\to U(x)\left(
-igA^\mu(x) + \partial^\mu
\right)U^\dagger(x)
  =U(x)D^\mu(x)U^\dagger(x) \ ,
\end{align}
where $iD^\mu(x)=i\partial^\mu+gA^\mu(x)$ is the covariant derivative.  In the effective theory, because we have assigned a strict power counting to the components of the collinear gauge field, we restrict the set of possible gauge transformations $U(x)$ to those that respect the power counting.  For the transformation of the collinear gauge field $A_n$, we require the collinear gauge transformation $U_n(x)$ with the scaling
\begin{equation}\label{eq:collscalegauge}
\partial^\mu U_n(x)\sim Q(\scetscale^2,1,\scetscale)U_n(x) \ ,
\end{equation}
in the $+$, $-$ and $\perp$ components, respectively.  Note that this gauge transformation is defined in the $n$-collinear direction.  So, if there are multiple jet directions, there are collinear gauge transformations defined about each of them.

Restricting to gauge group elements with the scaling in \Eq{eq:collscalegauge}, it is now clear how to define the gauge transformation of the collinear gauge field $A_n$~\cite{Bauer:2001ct}.  In \Eq{eq:gaugetrans} we replace the full covariant derivative with the $n$-collinear covariant derivative:  
\begin{equation}\label{eq:collgaugetrans}
-ig A_n^\mu(x)\to U_n(x)\left(
\frac{n^\mu}{2}\bar{n}\cdot D_n + \frac{\bar{n}^\mu}{2}n\cdot D+D_{n\perp}^\mu
\right)U_n^\dagger (x) \ .
\end{equation}
Recall that the component $n\cdot D$ contains both the collinear and soft gauge fields. It is these soft fields that are removed with the BPS field redefinition, \Eq{eq:bps}. The gauge transformation for the collinear fermion field $\xi_n(x)$ follows:
\begin{equation}
\xi_n(x)\to U_n(x) \xi_n(x) \ .
\end{equation}

In the SCET Lagrangian there are also soft gauge transformations~\cite{Bauer:2001ct}, where $i\partial^\mu U_s(x) \sim \lambda^2 U_s(x)$.  Under these soft gauge transformations all collinear fields, including $A_n^\mu$, transform like matter fields, while the soft gauge fields themselves transform as in \Eq{eq:gaugetrans}.

\section{Applying the One-Loop Soft Theorem}\label{app:ampexs}

In this appendix, we apply the one-loop soft theorem of \Eq{eq:fulloneloopsubsoft} to several other amplitude examples.  Each term in \Eq{eq:fulloneloopsubsoft} is defined as appropriate matrix elements of operators and Lagrangian insertions, and as such, can be calculated directly in SCET.  As with the single-minus helicity amplitude example from \Sec{sec:subsoftsingminus}, we will demonstrate consistency of the one-loop soft theorem with the explicit expansion of amplitudes in the soft limit. Rather than carrying out new computations, such as subleading SCET amplitudes, we will instead seek to exploit as much as possible objects already computed in the literature, such as one-loop splitting amplitudes, whose role in the factorization theorem is known.  This guides somewhat our choice of examples.  However, without complete calculations in the effective theory, there will exist factors in the subleading soft limits whose precise form is only determined by explicitly expanding the amplitude.  Nevertheless, they are highly constrained and we will show that they satisfy all properties as predicted by the factorization structure of the effective theory.

%%%%%%%%%%%%%%%%%%%%%%%%%%%%%%%%%%%%%%%%%%%%%%%

\subsection{${\cal A}^{[1]}(1^+,2^-,3^+,4^+,5_s^+)$ Pure Gluon Amplitude}\label{app:5ptother}

%%%%%%%%%%%%%%%%%%%%%%%%%%%%%%%%%%%%%%%%%%%%%%%

In \Sec{sec:subsoftsingminus}, we showed that the one-loop soft theorem was consistent with the limit of the one-loop amplitude ${\cal A}^{[1]}(1^-,2^+,3^+,4^+,5^+)$ as the energy of gluon 5 goes soft.  Here, we will show that the one-loop soft theorem also correctly describes the soft limit when the minus helicity is not adjacent to the soft particle.  The relevant amplitude can be found by simply permuting the labels of the amplitude ${\cal A}^{[1]}(1^-,2^+,3^+,4^+,5^+)$:
\begin{equation}\label{eq:other1m}
{\cal A}^{[1]}(1^+,2^-,3^+,4^+,5^+)=\frac{i}{48\pi^2}\frac{1}{\sab{4}{5}^2} \left(
-\frac{\sab{2}{4}^3\ssb{4}{3}\sab{5}{3}}{\sab{2}{1}\sab{1}{5}\sab{4}{3}^2}+\frac{\sab{2}{5}^3\ssb{5}{1}\sab{4}{1}}{\sab{2}{3}\sab{3}{4}\sab{5}{1}^2}-\frac{\ssb{3}{1}^3}{\ssb{2}{3}\ssb{1}{2}}
\right) \,.
\end{equation}

At leading power, ${\cal O}(\lambda^{-2})$, the one-loop soft theorem predicts that this amplitude factorizes as
\begin{align}
{\cal A}^{[1]}(1^+,2^-,3^+,4^+,5^+_s) &= S^{[0](0)}(5_s^+){\mathcal A}^{[1]}(1^+,2^-,3^+,4^+)
+S^{[1](0)}(5_s^+){\mathcal A}^{[0]}(1^+,2^-,3^+,4^+) \nonumber \\
&\qquad+
{\cal O}(\lambda^{-1})
\,.
\end{align}
The single-minus amplitude ${\mathcal A}^{[0]}(1^+,2^-,3^+,4^+)$ is zero at tree-level, and so to this order, the amplitude factorizes as
\begin{align}
{\cal A}^{[1]}(1^+,2^-,3^+,4^+,5^+_s) &= S^{[0](0)}(5^+){\mathcal A}^{[1]}(1^+,2^-,3^+,4^+)
+
{\cal O}(\lambda^{-1})
\,.
\end{align}
Explicitly expanding \Eq{eq:other1m} to ${\cal O}(\lambda^{-2})$, we find
\begin{equation}
{\cal A}^{[1]}(1^+,2^-,3^+,4^+,5^+_s) =\frac{\sab{4}{1}}{\sab{4}{5}\sab{5}{1}}\cdot \frac{i}{48\pi^2}\frac{\sab{2}{4}^3\sab{3}{1}\ssb{2}{3}}{\sab{3}{4}^2\sab{4}{1}^3}+{\cal O}(\lambda^{-1}) \,.
\end{equation}
Since the single-minus four-point amplitude is
\begin{equation}\label{eq:4ptsingmin}
{\cal A}^{[1]}(1^+,2^-,3^+,4^+) = \frac{i}{48\pi^2}\frac{\sab{2}{4}^3\sab{3}{1}\ssb{2}{3}}{\sab{3}{4}^2\sab{4}{1}^3} \,,
\end{equation}
this agrees with the one-loop soft theorem at leading order in $\lambda$, as expected.

The one-loop soft theorem predicts that there are no contributions at ${\cal O}(\lambda^{-1})$. Going to ${\cal O}(\lambda^0)$, there are again potentially several contributions.  However, many can be shown to vanish.  First, unlike the expansion of ${\cal A}^{[1]}(1^-,2^+,3^+,4^+,5^+)$, there is actually no contribution to the expansion of ${\cal A}^{[1]}(1^+,2^-,3^+,4^+,5^+)$ from the Low-Burnett-Kroll operator.  This follows from the fact that the four-point amplitude ${\cal A}^{[1]}(1^+,2^-,3^+,4^+)$ as written in \Eq{eq:4ptsingmin} is independent of the anti-holomorphic spinors of particles 4 and 1.\footnote{If the amplitude instead was written as
\begin{equation}\label{eq:other4pt}
{\cal A}^{[1]}(1^+,2^-,3^+,4^+) = \frac{i}{48\pi^2}\frac{\sab{2}{4}^3\ssb{4}{2}}{\sab{3}{4}^2\sab{4}{1}^2} \,,
\end{equation}
then there would be a contribution from the Low-Burnett-Kroll operator, in addition to the contributions from the splitting amplitude terms.  This contribution arises from re-expressing the leading power soft theorem with respect to this four-point amplitude.  That is, using five-point momentum conservation,
\begin{equation}
\frac{\sab{4}{1}}{\sab{4}{5}\sab{5}{1}}\cdot \frac{i}{48\pi^2}\frac{\sab{2}{4}^3\sab{3}{1}\ssb{2}{3}}{\sab{3}{4}^2\sab{4}{1}^3}=\frac{\sab{4}{1}}{\sab{4}{5}\sab{5}{1}}\cdot \frac{i}{48\pi^2}\frac{\sab{2}{4}^3\ssb{4}{2}}{\sab{3}{4}^2\sab{4}{1}^2}+\frac{i}{48\pi^2}\frac{\ssb{5}{2}\sab{2}{4}^3}{\sab{4}{5}\sab{3}{4}^2\sab{4}{1}^2} \,,
\end{equation}
where the second term on the right is exactly that found by acting the Low-Burnett-Kroll operator on the amplitude in \Eq{eq:other4pt}.
}  By the same arguments as with ${\cal A}^{[1]}(1^-,2^+,3^+,4^+,5^+)$, there are no soft loop contributions through ${\cal O}(\lambda^0)$ of the amplitude ${\cal A}^{[1]}(1^+,2^-,3^+,4^+,5^+)$.  Therefore, the only contributions to the soft expansion at ${\cal O}(\lambda^{0})$ are from collinear splitting and fusion terms.  
Due to the helicity configuration, the prediction of the one-loop soft theorem through ${\cal O}(\lambda^0)$ takes the form
\begin{align}\label{eq:oneloopsingminus2}
{\cal A}^{[1]}(1^+,2^-,3^+,4^+,5^+_s) &= S^{[0](0)}(5_s^+){\mathcal A}^{[1]}(1^+,2^-,3^+,4^+) \\
& \ \ + \Splitbar^{[1](2)}_{\bar n_4}(P^+\rightarrow 4^+,5_s^+){\mathcal A}^{[0]}(1^+,2^-,3^+,P^-)\nonumber \\
&\ \ + \Splitbar^{[1](2)}_{\bar n _1}(P^+\rightarrow 5_s^+,1^+){\mathcal A}^{[0]}(P^-,2^-,3^+,4^+) \nonumber \\
&\ \ + \fuse(1^+,2^-,3^+;P^+\rightarrow 4^+,5_s^+){\mathcal A}^{[0]}(1^+,2^-,3^+,P^-)\nonumber \\
&\ \ + \fuse(2^-,3^+,4^+;P^+\rightarrow 5_s^+,1^+){\mathcal A}^{[0]}(P^-,2^-,3^+,4^+) \nonumber \\
&\ \ +
{\cal O}(\lambda^{1})
\,.\nonumber
\end{align}
All other collinear splitting and fusion terms vanish.

As discussed in the text near \Eq{eq:Split1}, to the power to which we work, the collinear splitting amplitude $\Splitbar^{[1](2)}_{\bar n_4}(P^+\rightarrow 4^+,5_s^+)$ can be written in the form
\begin{align}\label{eq:splitbar4}
\Splitbar^{[1](2)}_{\bar n_4}(P^+\rightarrow 4^+,5_s^+) &= \frac{-i}{48 \pi^2}\frac{z^{1/2}\ssb{4}{5}}{\sab{4}{5}^2} \\
&=\frac{-i}{48 \pi^2}\frac{\sab{\bar n_4}{5}}{\sab{\bar n_4}{4}}\frac{\ssb{4}{5}}{\sab{4}{5}^2} \,, \nonumber
\end{align}
where the momentum fraction $z$ is defined so that the product of $\Splitbar^{[1](2)}_{\bar n_4}(P^+\rightarrow 4^+,5_s^+)$ and the amplitude has the correct little group scaling for all particles.  The vector $\bar n_4$ is arbitrary and in the exact collinear limit of gluons 4 and 5 $\Splitbar^{[1](2)}_{\bar n_4}(P^+\rightarrow 4^+,5_s^+)$ is independent of $\bar n_4$.  Similarly, the collinear splitting amplitude $\Splitbar^{[1](2)}_{\bar n_1}(P^+\rightarrow 5_s^+,1^+)$ can be written in the form
\begin{equation}\label{eq:splitbar1}
\Splitbar^{[1](2)}_{\bar n_1}(P^+\rightarrow 5_s^+,1^+) = \frac{-i}{48 \pi^2}\frac{\sab{\bar n_1}{5}}{\sab{\bar n_1}{4}}\frac{\ssb{1}{5}}{\sab{1}{5}^2} \,.
\end{equation}
The fusion terms in \Eq{eq:oneloopsingminus2}, $ \fuse(1^+,2^-,3^+;P^+\rightarrow 4^+,5_s^+)$ and $ \fuse(2^-,3^+,4^+;P^+\rightarrow 5_s^+,1^+)$ are linear combinations of differences of the appropriate splitting amplitudes with different choices for the $\bar n$ vectors.  That is, in the exact collinear limit, the fusion terms vanish, but away from the collinear limit, these terms contribute to this power.

This result can be verified explicitly by expanding the amplitude in \Eq{eq:other1m}.  To organize the ${\cal O}(\lambda^0)$ terms in the expansion, we use momentum conservation and other spinor identities to rewrite the amplitude in such a way that the terms that appear have poles in gluon 5 with only a single other particle,  motivated by \Eq{eq:oneloopsingminus2}. This is achieved with two terms, one with $1/\sab{4}{5}^2$ and one with  $1/\sab{1}{5}^2$. Through ${\cal O}(\lambda^0)$, expanding ${\cal A}^{[1]}(1^+,2^-,3^+,4^+,5^+)$ therefore yields
\begin{align}\label{eq:1min2exp}
{\cal A}^{[1]}(1^+,2^-,3^+,4^+,5^+_s) &=\frac{\sab{4}{1}}{\sab{4}{5}\sab{5}{1}}\cdot \frac{i}{48\pi^2}\frac{\sab{2}{4}^3\sab{3}{1}\ssb{2}{3}}{\sab{3}{4}^2\sab{4}{1}^3} \\
&
\hspace{-2cm}
 + \frac{-i}{48\pi^2} \frac{\sab{\bar n_4}{5}}{\sab{\bar n_4}{4}}\frac{\ssb{4}{5}}{\sab{4}{5}^2}
 \cdot\frac{\sab{2}{4}^4}{\sab{1}{2}\sab{2}{3}\sab{3}{4}\sab{4}{1}}\nonumber \\
&
\hspace{-2cm}
 +\frac{-i}{48\pi^2} \frac{\sab{\bar n_1}{5}}{\sab{\bar n_1}{1}}\frac{\ssb{5}{1}}{\sab{5}{1}^2}
\cdot\frac{\sab{1}{2}^4}{\sab{1}{2}\sab{2}{3}\sab{3}{4}\sab{4}{1}}\nonumber \\
&
\hspace{-2cm}
 + \frac{-i}{48\pi^2}\frac{\ssb{4}{5}}{\sab{4}{5}^2}
 \left[
 \left(
 \frac{\sab{3}{5}}{\sab{3}{4}}-\frac{\sab{\bar n_4}{5}}{\sab{\bar n_4}{4}}
 \right)
 +2\left(
 \frac{\sab{2}{5}}{\sab{2}{4}}-\frac{\sab{1}{5}}{\sab{1}{4}}
 \right)
 \right]
 \cdot\frac{\sab{2}{4}^4}{\sab{1}{2}\sab{2}{3}\sab{3}{4}\sab{4}{1}}\nonumber \\
 &
\hspace{-2cm}
 +\frac{-i}{48\pi^2}\frac{\ssb{5}{1}}{\sab{5}{1}^2}\left[
\left(
 \frac{\sab{2}{5}}{\sab{2}{1}}- \frac{\sab{\bar n_1}{5}}{\sab{\bar n_1}{1}}
 \right)
 +2\left(
 \frac{\sab{2}{5}}{\sab{2}{1}}-\frac{\sab{4}{5}}{\sab{4}{1}}
 \right)
\right]\cdot\frac{\sab{1}{2}^4}{\sab{1}{2}\sab{2}{3}\sab{3}{4}\sab{4}{1}}\nonumber \\
&
\hspace{-2cm}
+{\cal O}(\lambda^1)
\,. \nonumber
\end{align}
In this form, we can immediately identify terms with the one-loop soft theorem, \Eq{eq:oneloopsingminus2}.  The second and third lines of \Eq{eq:1min2exp} correspond to the terms with splitting amplitudes defined in \Eqs{eq:splitbar4}{eq:splitbar1}.  The fourth and fifth lines of \Eq{eq:1min2exp} are the fusion terms, and manifest the form as a difference of splitting amplitudes.  In particular, 
\begin{equation}
\fuse(1^+,2^-,3^+;P^+\rightarrow 4^+,5_s^+)= \frac{-i}{48\pi^2}\frac{\ssb{4}{5}}{\sab{4}{5}^2}
 \left[
 \left(
 \frac{\sab{3}{5}}{\sab{3}{4}}-\frac{\sab{\bar n_4}{5}}{\sab{\bar n_4}{4}}
 \right)
 +2\left(
 \frac{\sab{2}{5}}{\sab{2}{4}}-\frac{\sab{1}{5}}{\sab{1}{4}}
 \right)
 \right]\,,
\end{equation}
which vanishes in the limit where gluons 4 and 5 are exactly collinear.  Similarly, 
\begin{equation}
\fuse(2^-,3^+,4^+;P^+\rightarrow 5_s^+,1^+)=\frac{-i}{48\pi^2}\frac{\ssb{5}{1}}{\sab{5}{1}^2}\left[
\left(
 \frac{\sab{2}{5}}{\sab{2}{1}}- \frac{\sab{\bar n_1}{5}}{\sab{\bar n_1}{1}}
 \right)
 +2\left(
 \frac{\sab{2}{5}}{\sab{2}{1}}-\frac{\sab{4}{5}}{\sab{4}{1}}
 \right)
\right]\,,
\end{equation}
vanishes in the limit where gluons 5 and 1 are exactly collinear.  Therefore, the one-loop soft theorem of \Eq{eq:fulloneloopsubsoft} exactly reproduces the explicit expansion of the amplitude ${\cal A}^{[1]}(1^+,2^-,3^+,4^+,5^+)$  in \Eq{eq:1min2exp}.

%%%%%%%%%%%%%%%%%%%%%%%%%%%%%%%%%%%%%%%%%%%%%%%%%%%

\subsection{${\cal A}^{[1]}(1^-,2^+,3^+,4^+,5^+,6_s^+)$ Pure Gluon Amplitude}\label{app:6pt}

%%%%%%%%%%%%%%%%%%%%%%%%%%%%%%%%%%%%%%%%%%%%%%%%%%%

Five-point kinematics is fairly constraining, and so it is useful to consider higher-point amplitudes to further exhibit the one-loop subleading soft theorem in action.  Here, we will consider the limit of the  six-point single-minus helicity amplitude ${\cal A}^{[1]}(1^-,2^+,3^+,4^+,5^+,6^+)$ as gluon 6 goes soft.  This amplitude is \cite{Bern:2005hs}
\begin{align}
{\cal A}^{[1]}(1^-,2^+,3^+,4^+,5^+,6^+) &= \\
&
\hspace{-4cm}
\frac{i}{48\pi^2}\left[
\frac{\langle 1|2+3|6]^3}{\sab{1}{2}\sab{2}{3}\sab{4}{5}^2(1+2+3)^2\langle 3|1+2|6]}+\frac{\langle 1|3+4|2]^3}{\sab{3}{4}^2\sab{5}{6}\sab{6}{1}(2+3+4)^2\langle 5|3+4|2]} \right.\nonumber \\
&
\hspace{-3cm}
+\frac{\ssb{2}{6}^3}{\ssb{1}{2}\ssb{6}{1}(3+4+5)^2}\left(
\frac{\ssb{2}{3}\ssb{3}{4}}{\sab{4}{5}\langle 5|3+4|2]}-\frac{\ssb{4}{5}\ssb{5}{6}}{\sab{3}{4}\langle3|1+2|6]}+\frac{\ssb{3}{5}}{\sab{3}{4}\sab{4}{5}}
\right)\nonumber \\
&
\hspace{-3cm}
-\frac{\sab{1}{3}^3\ssb{2}{3}\sab{2}{4}}{\sab{2}{3}^2\sab{3}{4}^2\sab{4}{5}\sab{5}{6}\sab{6}{1}}+\frac{\sab{1}{5}^3\sab{4}{6}\ssb{5}{6}}{\sab{1}{2}\sab{2}{3}\sab{3}{4}\sab{4}{5}^2\sab{5}{6}^2}\nonumber \\
&
\hspace{-3cm}
\left.
-\frac{\sab{1}{4}^3\sab{3}{5}\langle 1|2+3|4]}{\sab{1}{2}\sab{2}{3}\sab{3}{4}^2\sab{4}{5}^2\sab{5}{6}\sab{6}{1}}
\right]\nonumber \,.
\end{align}

At leading power in $\lambda$, the one-loop soft theorem predicts that this amplitude factorizes as
\begin{align}
{\cal A}^{[1]}(1^-,2^+,3^+,4^+,5^+,6^+_s) &= S^{[0](0)}(6^+){\mathcal A}^{[1]}(1^-,2^+,3^+,4^+,5^+)
\\
&
\qquad+S^{[1](0)}(6^+){\mathcal A}^{[0]}(1^-,2^+,3^+,4^+,5^+) \nonumber \\
&\qquad+
{\cal O}(\lambda^{-1})
\,.\nonumber
\end{align}
The single-minus amplitude ${\mathcal A}^{[0]}(1^-,2^+,3^+,4^+,5^+)$ is zero at tree-level, and so to this order, the amplitude factorizes as
\begin{align} \label{eq:factF2lo}
{\cal A}^{[1]}(1^-,2^+,3^+,4^+,5^+,6^+_s) &= S^{[0](0)}(6^+){\mathcal A}^{[1]}(1^-,2^+,3^+,4^+,5^+)
+
{\cal O}(\lambda^{-1})
\,.
\end{align}
Explicitly expanding \Eq{eq:other1m} to ${\cal O}(\lambda^{-2})$, we find
\begin{align}
{\cal A}^{[1]}(1^-,2^+,3^+,4^+,5^+,6^+_s) &=\frac{\sab{5}{1}}{\sab{5}{6}\sab{6}{1}}\cdot \frac{i}{48\pi^2}\frac{1}{\sab{3}{4}^2}
  \left(
  -\frac{\sab{1}{3}^3\ssb{3}{2}\sab{4}{2}}{\sab{1}{5}\sab{5}{4}\sab{3}{2}^2}
  +\frac{\sab{1}{4}^3\ssb{4}{5}\sab{3}{5}}{\sab{1}{2}\sab{2}{3}\sab{4}{5}^2}
  -\frac{\ssb{2}{5}^3}{\ssb{1}{2}\ssb{5}{1}}
   \right) \nonumber \\
   &\qquad+{\cal O}(\lambda^{-1}) \,,
\end{align}
where the structure of $S^{[0](0)}(6^+)$ times the single-minus five-point amplitude predicted in \Eq{eq:factF2lo} is apparent.

The one-loop soft theorem predicts that there are no contributions at ${\cal O}(\lambda^{-1})$. As with previous amplitude examples, beginning at ${\cal O}(\lambda^0)$, there are potentially several contributions.  The contribution from a hard loop corresponding to the Low-Burnett-Kroll operator will be non-zero because the five-point amplitude ${\mathcal A}^{[1]}(1^-,2^+,3^+,4^+,5^+)$ has non-trivial dependence on the anti-holomorphic spinors of gluons 5 and 1.  Contributions from soft loops will again be zero by the special helicity configuration of single-minus amplitudes.  Similarly, due to the helicity configuration of the amplitude, there will be only one contribution from a one-loop collinear splitting amplitude, when gluon 5 splits collinearly producing gluon 6.  There will also be corresponding fusion terms describing the subleading collinear limits of gluons 5 and 6.

Thus, the one-loop soft theorem predicts that through ${\cal O}(\lambda^0)$, the 6-point amplitude has the following expansion:
\begin{align}
{\cal A}^{[1]}(1^-,2^+,3^+,4^+,5^+,6_s^+) &= S^{[0](0)}(6_s^+){\mathcal A}^{[1]}(1^-,2^+,3^+,4^+,5^+) \\
&
\ 
+S^{[0](2)}(6_s^+){\mathcal A}^{[1]}(1^-,2^+,3^+,4^+,5^+)\nn\\
&
\ 
+\Splitbar^{[1](2)}_{\bar n_5}(P^+\rightarrow 5^+,6_s^+){\mathcal A}^{[0]}(1^-,2^+,3^+,4^+,P^-)
\nonumber \\
&
\ 
+\fuse(1^-,2^+,3^+,4^+;P^+\rightarrow 5^+,6_s^+){\mathcal A}^{[0]}(1^-,2^+,3^+,4^+,P^-)
\nonumber \\
&
\ 
+{\cal O}(\lambda^1) \,.\nonumber
\end{align}

This decomposition can be explicitly verified by expanding the amplitude through ${\cal O}(\lambda^0)$.  We have
\begin{align}\label{eq:6ptexp}
{\cal A}^{[1]}(1^-,2^+,3^+,4^+,5^+,6_s^+) &=\frac{\sab{5}{1}}{\sab{5}{6}\sab{6}{1}}\cdot \frac{i}{48\pi^2}\frac{1}{\sab{3}{4}^2}
  \left(
  -\frac{\sab{1}{3}^3\ssb{3}{2}\sab{4}{2}}{\sab{1}{5}\sab{5}{4}\sab{3}{2}^2}
  +\frac{\sab{1}{4}^3\ssb{4}{5}\sab{3}{5}}{\sab{1}{2}\sab{2}{3}\sab{4}{5}^2}
  -\frac{\ssb{2}{5}^3}{\ssb{1}{2}\ssb{5}{1}}
   \right) \nonumber \\
   &
   \hspace{-4cm}
   +\frac{i}{48\pi^2}\frac{1}{\sab{3}{4}^2}\left[
   \frac{\ssb{5}{2}^3\langle 6|1+5|6]}{\sab{5}{6}\sab{6}{1}\ssb{5}{1}^2\ssb{1}{2}}+3\frac{\ssb{5}{2}^2\ssb{6}{2}}{\sab{5}{6}\ssb{1}{2}\ssb{5}{1}}-\frac{\ssb{5}{2}^3\ssb{6}{2}}{\sab{6}{1}\ssb{5}{1}\ssb{1}{2}^3}-\frac{\sab{1}{4}^3\sab{3}{5}\ssb{6}{4}}{\sab{1}{2}\sab{2}{3}\sab{4}{5}^2\sab{5}{6}}
   \right]\nonumber \\
 &
 \hspace{-4cm}
+\frac{-i}{48\pi^2}\frac{\sab{\bar n_5}{6}}{\sab{\bar n_5}{5}}\frac{\ssb{5}{6}}{\sab{5}{6}^2}
\cdot\frac{-\sab{5}{1}^4}{\sab{1}{2}\sab{2}{3}\sab{3}{4}\sab{4}{5}\sab{5}{1}}
 \nonumber \\
  &
 \hspace{-4cm}
+\frac{-i}{48\pi^2}\frac{\ssb{5}{6}}{\sab{5}{6}^2}
\left(
\frac{\sab{4}{6}}{\sab{4}{5}}-\frac{\sab{\bar n_5}{6}}{\sab{\bar n_5}{5}}
\right)
\cdot\frac{-\sab{5}{1}^4}{\sab{1}{2}\sab{2}{3}\sab{3}{4}\sab{4}{5}\sab{5}{1}}
 \nonumber \\
 &
 \hspace{-4cm}
 +{\cal O}(\lambda^1) \,.
\end{align}
It it straightforward to verify that the second line of \Eq{eq:6ptexp} is precisely the action of the Low-Burnett-Kroll operator on the five-point amplitude:
\begin{align}
S^{[0](2)}(6_s^+){\mathcal A}^{[1]}(1^-,2^+,3^+,4^+,5^+)&=\\
&
\hspace{-4cm}
\frac{i}{48\pi^2}\frac{1}{\sab{3}{4}^2}\left[
   \frac{\ssb{5}{2}^3\langle 6|1+5|6]}{\sab{5}{6}\sab{6}{1}\ssb{5}{1}^2\ssb{1}{2}}+3\frac{\ssb{5}{2}^2\ssb{6}{2}}{\sab{5}{6}\ssb{1}{2}\ssb{5}{1}}-\frac{\ssb{5}{2}^3\ssb{6}{2}}{\sab{6}{1}\ssb{5}{1}\ssb{1}{2}^3}-\frac{\sab{1}{4}^3\sab{3}{5}\ssb{6}{4}}{\sab{1}{2}\sab{2}{3}\sab{4}{5}^2\sab{5}{6}}
   \right] \,. \nonumber
\end{align}
The third line of \Eq{eq:6ptexp} is precisely the splitting amplitude contribution:
\begin{align}
\Splitbar^{[1](2)}_{\bar n_5}(P^+\rightarrow 5^+,6_s^+){\mathcal A}^{[0]}(1^-,2^+,3^+,4^+,P^-)&=\\
&
\hspace{-1cm}
\frac{-i}{48\pi^2}\frac{\sab{\bar n_5}{6}}{\sab{\bar n_5}{5}}\frac{\ssb{5}{6}}{\sab{5}{6}^2}
\cdot\frac{-\sab{5}{1}^4}{\sab{1}{2}\sab{2}{3}\sab{3}{4}\sab{4}{5}\sab{5}{1}} \,,\nonumber
\end{align}
while the fourth line is the fusion term contribution:
\begin{equation}
\fuse(1^-,2^+,3^+,4^+;P^+\rightarrow 5^+,6_s^+)=\frac{-i}{48\pi^2}\frac{\ssb{5}{6}}{\sab{5}{6}^2}
\left(
\frac{\sab{4}{6}}{\sab{4}{5}}-\frac{\sab{\bar n_5}{6}}{\sab{\bar n_5}{5}}
\right)
\cdot\frac{-\sab{5}{1}^4}{\sab{1}{2}\sab{2}{3}\sab{3}{4}\sab{4}{5}\sab{5}{1}} \,,
\end{equation}
which vanishes in the exact collinear limit of gluons 5 and 6.  Therefore, the one-loop subleading soft theorem of \Eq{eq:fulloneloopsubsoft} exactly reproduces the explicit expansion of the amplitude ${\cal A}^{[1]}(1^-,2^+,3^+,4^+,5^+,6^+)$ as gluon 6 goes soft.

%%%%%%%%%%%%%%%%%%%%%%%%%%%%%%%%%%%%%%%%%%%%%%%%%%%

\subsection{${\cal A}^{[1]}_q(1^-,2^-,3^+,4_s^+,5^+)$ Quark Loop Amplitude}\label{app:5ptirdiv}

%%%%%%%%%%%%%%%%%%%%%%%%%%%%%%%%%%%%%%%%%%%%%%%%%%%

All previous examples to which we applied the one-loop subleading soft theorem of \Eq{eq:fulloneloopsubsoft} were finite loop amplitudes.  Since the one-loop soft theorem applies to any one-loop amplitude, it is also useful to explicitly check that it holds for an amplitude with infrared divergences.  In this appendix, we will apply the one-loop subleading soft theorem to the five-gluon, one-loop amplitude ${\cal A}^{[1]}_q(1^-,2^-,3^+,4^+,5^+)$ with only quarks running in the loop.  This amplitude is only infrared divergent due to collinear physics, and so in this analysis, we will be able to use known results in the literature to test our one-loop soft theorem.\footnote{We could test this with an amplitude  that is also soft divergent, such as ${\cal A}^{[1]}(1^-,2^-,3^+,4^+,5^+)$ with gluons in the loop, but this would require a new calculation of subleading power one-loop soft currents.  Such a result would have important consequences for Drell-Yan processes, Higgs physics, and others and deserves a dedicated study.}  The  dimensionally-regulated and $\overline{\text{MS}}$-renormalized leading-color amplitude ${\cal A}^{[1]}_q(1^-,2^-,3^+,4^+,5^+)$ is \cite{Bern:1993mq}:
\begin{align}
{\cal A}^{[1]}_q(1^-,2^-,3^+,4^+,5^+)&=\frac{i}{16\pi^2}\frac{\sab{1}{2}^4}{\sab{1}{2}\sab{2}{3}\sab{3}{4}\sab{4}{5}\sab{5}{1}}\left[
\frac{5}{3\epsilon}+\frac{1}{3}\left(
\log\frac{\mu^2}{-s_{23}}+\log\frac{\mu^2}{-s_{15}}
\right)+\frac{10}{9}
\right]\nonumber\\
&
\hspace{-3cm}
+\frac{i}{16\pi^2}\left[
\frac{1}{3}\frac{\sab{1}{2}^2(\sab{2}{3}\ssb{3}{4}\sab{4}{1}+\sab{2}{4}\ssb{4}{5}\sab{5}{1})}{\sab{2}{3}\sab{3}{4}\sab{4}{5}\sab{5}{1}}\frac{\log\frac{s_{23}}{s_{15}}}{s_{15}-s_{23}}\right.\nonumber\\
&
\hspace{-2.5cm}
\left.
+\frac{1}{3}\frac{\ssb{3}{4}\sab{4}{1}\sab{2}{4}\ssb{4}{5}(\sab{2}{3}\ssb{3}{4}\sab{4}{1}+\sab{2}{4}\ssb{4}{5}\sab{5}{1})}{\sab{3}{4}\sab{4}{5}}\frac{\log\frac{s_{23}}{s_{15}}-\frac{s_{23}}{2s_{15}}+\frac{s_{15}}{2s_{23}}}{(s_{15}-s_{23})^3}\right.\nonumber\\
&
\hspace{-2.5cm}
\left.
+\frac{1}{3}\frac{\sab{3}{5}\ssb{3}{5}^3}{\ssb{1}{2}\ssb{2}{3}\sab{3}{4}\sab{4}{5}\ssb{5}{1}}-\frac{1}{3}\frac{\sab{1}{2}\ssb{3}{5}^2}{\ssb{2}{3}\sab{3}{4}\sab{4}{5}\ssb{5}{1}}-\frac{1}{6}\frac{\sab{1}{2}\ssb{3}{4}\sab{4}{1}\sab{2}{4}\ssb{4}{5}}{\sab{2}{3}\ssb{3}{2}\sab{3}{4}\sab{4}{5}\sab{5}{1}\ssb{1}{5}}
\right] \,. 
\end{align}
Here, $\epsilon$ is the dimensional regularization parameter defined by $d=4-2\epsilon$ and $\mu^2$ is the dimensional regularization scale. Terms at ${\cal O}(\epsilon)$ or higher have been dropped.  

We now expand this amplitude in the limit that the momentum of gluon $4$ becomes small.  The one-loop soft theorem predicts that, up to ${\cal O}(\lambda^{0})$, this amplitude expands as
\begin{align}
{\cal A}^{[1]}_q(1^-,2^-,3^+,4_s^+,5^+)&=  S^{[0](0)}(4_s^+){\mathcal A}^{[1]}_q(1^-,2^-,3^+,5^+)+S^{[1](0)}(4_s^+){\mathcal A}^{[0]}(1^-,2^-,3^+,5^+) \nonumber \\
&\qquad+
{\cal O}(\lambda^{0})
\,.
\end{align}
For fermions in the loop, the one-loop soft amplitude $S^{[1](0)}(4^+)$ is zero \cite{Bern:1998sc}.  The one-loop four-point amplitude is \cite{Ellis:1985er,Bern:1991aq,Kunszt:1993sd}
\begin{align}\label{eq:4qmhvloop}
{\mathcal A}^{[1]}_q(1^-,2^-,3^+,5^+) &=\frac{i}{16\pi^2}\frac{\sab{1}{2}^4}{\sab{1}{2}\sab{2}{3}\sab{3}{5}\sab{5}{1}}\left[
\frac{5}{3\epsilon}+\frac{2}{3}\log\frac{\mu^2}{-s_{23}}+\frac{10}{9}
\right] \,. 
\end{align}
Therefore, the expansion of ${\cal A}^{[1]}_q(1^-,2^-,3^+,4^+,5^+)$ to leading power in the energy of gluon 4 is
\begin{align}
{\cal A}^{[1]}_q(1^-,2^-,3^+,4_s^+,5^+)&=  S^{[0](0)}(4_s^+){\mathcal A}^{[1]}_q(1^-,2^-,3^+,5^+) \\
&=\frac{\sab{3}{5}}{\sab{3}{4}\sab{4}{5}}\cdot\frac{i}{16\pi^2}\frac{\sab{1}{2}^4}{\sab{1}{2}\sab{2}{3}\sab{3}{5}\sab{5}{1}}\left[
\frac{5}{3\epsilon}+\frac{2}{3}\log\frac{\mu^2}{-s_{23}}+\frac{10}{9}
\right] \,. \nonumber
\end{align}
It is easy to verify that this agrees with the leading term in the expansion of the five-point amplitude ${\cal A}^{[1]}_q(1^-,2^-,3^+,4^+,5^+)$.

We can continue to higher power in $\lambda$.  The soft theorem predicts several possible contributions.  There are non-zero contributions at ${\cal O}(\lambda^0)$ from the Low-Burnett-Kroll operator, collinear splitting amplitudes and collinear fusion terms, while contributions from soft loops vanish because soft fermions only first contribute beyond ${\cal O}(\lambda^0)$.  Thus the soft theorem predicts that the amplitude expands as
\begin{align}
{\cal A}^{[1]}_q(1^-,2^-,3^+,4_s^+,5^+)
  &=S^{[0](0)}(4_s^+){\mathcal A}^{[1]}_q(1^-,2^-,3^+,5^+)
    \\
&
\ 
  + S^{[0](2)}(4_s^+){\mathcal A}^{[1,\text{hard}]}_q(1^-,2^-,3^+,5^+)\nonumber \\
&
\ 
+\Splitbar^{[1](2)}_{\bar n_3}(P^-\rightarrow 3^+,4_s^+){\mathcal A}^{[0]}(1^-,2^-,P^+,5^+)\nonumber\\
&
\ 
+\Splitbar^{[1](2)}_{\bar n_5}(P^-\rightarrow 4_s^+,5^+){\mathcal A}^{[0]}(1^-,2^-,3^+,P^+)
\nonumber \\
&
\ 
+\fuse(5^+,1^-,2^-;P^-\rightarrow 3^+,4_s^+){\mathcal A}^{[0]}(1^-,2^-,P^+,5^+)\nonumber\\
&
\ 
+\fuse(1^-,2^-,3^+;P^-\rightarrow 4_s^+,5^+){\mathcal A}^{[0]}(1^-,2^-,3^+,P^+)
\nonumber \\
& 
\ 
 +{\cal O}(\lambda^1)\nn \,.
\end{align}
Unlike other amplitudes we have studied which were infrared finite, the contribution from the Low-Burnett-Kroll operator acts only on the hard, infrared finite part of the four-point amplitude ${\mathcal A}^{[1,\text{hard}]}_q(1^-,2^-,3^+,5^+)$.  The hard part of the amplitude can be read off from \Eq{eq:4qmhvloop}:
\begin{equation}
{\mathcal A}^{[1,\text{hard}]}_q(1^-,2^-,3^+,5^+) = \frac{i}{16\pi^2}\frac{\sab{1}{2}^4}{\sab{1}{2}\sab{2}{3}\sab{3}{5}\sab{5}{1}}\left[\frac{2}{3}\log\frac{\mu^2}{-s_{23}}+\frac{10}{9}
\right] \,.
\end{equation}

Using this expression for the hard function, the contribution to the subleading soft expansion from the Low-Burnett-Kroll operator is
\begin{equation}
S^{[0](2)}(4_s^+){\mathcal A}^{[1,\text{hard}]}_q(1^-,2^-,3^+,5^+)=-\frac{i}{16\pi^2}\frac{2}{3}\frac{1}{\sab{3}{4}}\frac{\ssb{2}{4}}{\ssb{2}{3}}\frac{\sab{1}{2}^4}{\sab{1}{2}\sab{2}{3}\sab{3}{5}\sab{5}{1}} \,.
\end{equation}
For the collinear splitting amplitude contribution, we need the corresponding one-loop amplitudes.  The splitting amplitude with a quark in the loop to this power is \cite{Bern:1998sc,Kosower:1999rx}
\begin{equation}\label{eq:splitbar5pt3}
\Splitbar^{[1](2)}_{\bar n_3}(P^-\rightarrow 3^+,4_s^+) = \frac{-i}{48\pi^2}\frac{\ssb{\bar n_3}{4}}{\ssb{\bar n_3}{3}}\frac{1}{\sab{3}{4}}\,,
\end{equation}
where the vector $\bar n_3$ is arbitrary and defines the momentum fraction of gluon 4.  The form of the momentum fraction $\ssb{\bar n_3}{4}/\ssb{\bar n_3}{3}$ is such that $\Splitbar^{[1](2)}_{\bar n_3}(P^-\rightarrow 3^+,4_s^+)$ has the correct little group properties.  Similarly, the other splitting amplitude is 
\begin{equation}\label{eq:splitbar5pt5}
\Splitbar^{[1](2)}_{\bar n_5}(P^-\rightarrow 4_s^+,5^+) = \frac{-i}{48\pi^2}\frac{\ssb{\bar n_5}{4}}{\ssb{\bar n_5}{5}}\frac{1}{\sab{4}{5}}\,.
\end{equation}
The fusion terms $\fuse(5^+,1^-,2^-;P^-\rightarrow 3^+,4_s^+)$ and $\fuse(1^-,2^-,3^+;P^-\rightarrow 4_s^+,5^+)$ vanish in the exactly collinear limit and are defined as differences between splitting amplitudes with different definitions of the momentum fraction.

This can be verified by explicitly expanding the amplitude in the limit that particle four becomes soft.  We find
\begin{align}\label{eq:5mhvexp1loop}
{\cal A}^{[1]}_q(1^-,2^-,3^+,4_s^+,5^+)&=\frac{\sab{3}{5}}{\sab{3}{4}\sab{4}{5}}\cdot\frac{i}{16\pi^2}\frac{\sab{1}{2}^4}{\sab{1}{2}\sab{2}{3}\sab{3}{5}\sab{5}{1}}\left[
\frac{5}{3\epsilon}+\frac{2}{3}\log\frac{\mu^2}{-s_{23}}+\frac{10}{9}
\right] \\
&
\hspace{-2cm}
-\frac{i}{16\pi^2}\frac{2}{3}\frac{1}{\sab{3}{4}}\frac{\ssb{2}{4}}{\ssb{2}{3}}\frac{\sab{1}{2}^4}{\sab{1}{2}\sab{2}{3}\sab{3}{5}\sab{5}{1}}\nonumber \\
&
\hspace{-2cm}
+\frac{-i}{48\pi^2}\frac{\ssb{\bar n_3}{4}}{\ssb{\bar n_3}{3}}\frac{1}{\sab{3}{4}}
\cdot \frac{\sab{1}{2}^4}{\sab{1}{2}\sab{2}{3}\sab{3}{5}\sab{5}{1}} \nonumber \\
&
\hspace{-2cm}
+\frac{-i}{48\pi^2}\frac{\ssb{\bar n_5}{4}}{\ssb{\bar n_5}{5}}\frac{1}{\sab{4}{5}}
\cdot\frac{\sab{1}{2}^4}{\sab{1}{2}\sab{2}{3}\sab{3}{5}\sab{5}{1}}\nonumber \\
&
\hspace{-2cm}
+\frac{-i}{48\pi^2}\frac{1}{\sab{3}{4}}\left[
\left(
\frac{\ssb{5}{4}}{\ssb{5}{3}}-\frac{\ssb{\bar n_3}{4}}{\ssb{\bar n_3}{3}}
\right)+2\left(
\frac{\ssb{5}{4}}{\ssb{5}{3}}-\frac{\ssb{2}{4}}{\ssb{2}{3}}
\right)
\right]\cdot \frac{\sab{1}{2}^4}{\sab{1}{2}\sab{2}{3}\sab{3}{5}\sab{5}{1}} \nonumber \\
&
\hspace{-2cm}
+\frac{-i}{48\pi^2}\frac{1}{\sab{4}{5}}\left[
\left(
\frac{\ssb{3}{4}}{\ssb{3}{5}}-\frac{\ssb{\bar n_5}{4}}{\ssb{\bar n_5}{5}}
\right)+2\left(
\frac{\ssb{3}{4}}{\ssb{3}{5}}-\frac{\ssb{1}{4}}{\ssb{1}{5}}
\right)
\right]\cdot\frac{\sab{1}{2}^4}{\sab{1}{2}\sab{2}{3}\sab{3}{5}\sab{5}{1}}\nonumber \\
& 
\hspace{-2cm}
 +{\cal O}(\lambda^1) \,.\nonumber
\end{align}
To match with the prediction of the subleading soft theorem, note that the second line of \Eq{eq:5mhvexp1loop} is the Low-Burnett-Kroll contribution, and the third and fourth lines are the splitting amplitude contributions, as defined in \Eqs{eq:splitbar5pt3}{eq:splitbar5pt5}.  The fifth and sixth lines of \Eq{eq:5mhvexp1loop} are the collinear fusion terms with 
\begin{equation}
\fuse(5^+,1^-,2^-;P^-\rightarrow 3^+,4_s^+) =\frac{-i}{48\pi^2}\frac{1}{\sab{3}{4}}\left[
\left(
\frac{\ssb{5}{4}}{\ssb{5}{3}}-\frac{\ssb{\bar n_3}{4}}{\ssb{\bar n_3}{3}}
\right)+2\left(
\frac{\ssb{5}{4}}{\ssb{5}{3}}-\frac{\ssb{2}{4}}{\ssb{2}{3}}
\right)
\right]\,,
\end{equation}
which vanishes in the exactly collinear limit of gluons 3 and 4.  Similarly, 
\begin{equation}
\fuse(1^-,2^-,3^+;P^-\rightarrow 4_s^+,5^+) = \frac{-i}{48\pi^2}\frac{1}{\sab{4}{5}}\left[
\left(
\frac{\ssb{3}{4}}{\ssb{3}{5}}-\frac{\ssb{\bar n_5}{4}}{\ssb{\bar n_5}{5}}
\right)+2\left(
\frac{\ssb{3}{4}}{\ssb{3}{5}}-\frac{\ssb{1}{4}}{\ssb{1}{5}}
\right)
\right] \,,
\end{equation}
which vanishes in the exactly collinear limit of gluons 4 and 5.  Therefore, the one-loop subleading soft theorem of \Eq{eq:fulloneloopsubsoft} exactly reproduces the explicit expansion of the amplitude ${\cal A}^{[1]}_q(1^-,2^-,3^+,4^+,5^+)$ as gluon 4 goes soft.

It is important to note that the expansion in \Eq{eq:5mhvexp1loop} requires a non-trivial re-association of terms.  The second, third and fifth lines of \Eq{eq:5mhvexp1loop} sum to
\begin{align}
&-\frac{i}{16\pi^2}\frac{2}{3}\frac{1}{\sab{3}{4}}\frac{\ssb{2}{4}}{\ssb{2}{3}}\frac{\sab{1}{2}^4}{\sab{1}{2}\sab{2}{3}\sab{3}{5}\sab{5}{1}} 
+\frac{-i}{48\pi^2}\frac{\ssb{\bar n_3}{4}}{\ssb{\bar n_3}{3}}\frac{1}{\sab{3}{4}}
\cdot \frac{\sab{1}{2}^4}{\sab{1}{2}\sab{2}{3}\sab{3}{5}\sab{5}{1}} \\
&
\hspace{1cm}
+\frac{-i}{48\pi^2}\frac{1}{\sab{3}{4}}\left[
\left(
\frac{\ssb{5}{4}}{\ssb{5}{3}}-\frac{\ssb{\bar n_3}{4}}{\ssb{\bar n_3}{3}}
\right)+2\left(
\frac{\ssb{5}{4}}{\ssb{5}{3}}-\frac{\ssb{2}{4}}{\ssb{2}{3}}
\right)
\right]\cdot \frac{\sab{1}{2}^4}{\sab{1}{2}\sab{2}{3}\sab{3}{5}\sab{5}{1}} \nn\\
&
=\frac{-i}{16\pi^2}\frac{1}{\sab{3}{4}}
\frac{\ssb{5}{4}}{\ssb{5}{3}}
\cdot \frac{\sab{1}{2}^4}{\sab{1}{2}\sab{2}{3}\sab{3}{5}\sab{5}{1}}\,. \nonumber
\end{align}
This demonstrates that the contribution from the Low-Burnett-Kroll operator to the expansion of the amplitude ${\cal A}^{[1]}_q(1^-,2^-,3^+,4^+,5^+)$ is vital for the correct factorization properties in the soft limit.  The terms in \Eq{eq:5mhvexp1loop} arrange themselves in precisely the correct way to ensure that the properties of the collinear fusion terms are maintained.

%%%%%%%%%%%%%%%%%%%%%%%%%%%%%%%%%%%%%%%%%%%%%%%%%%%

\section{Soft Limit of the $1\to 3$ Splitting Amplitude in SCET}\label{app:123}

%%%%%%%%%%%%%%%%%%%%%%%%%%%%%%%%%%%%%%%%%%%%%%%%%%%

\begin{figure}
	\begin{center}
		\includegraphics[width=15.5cm]{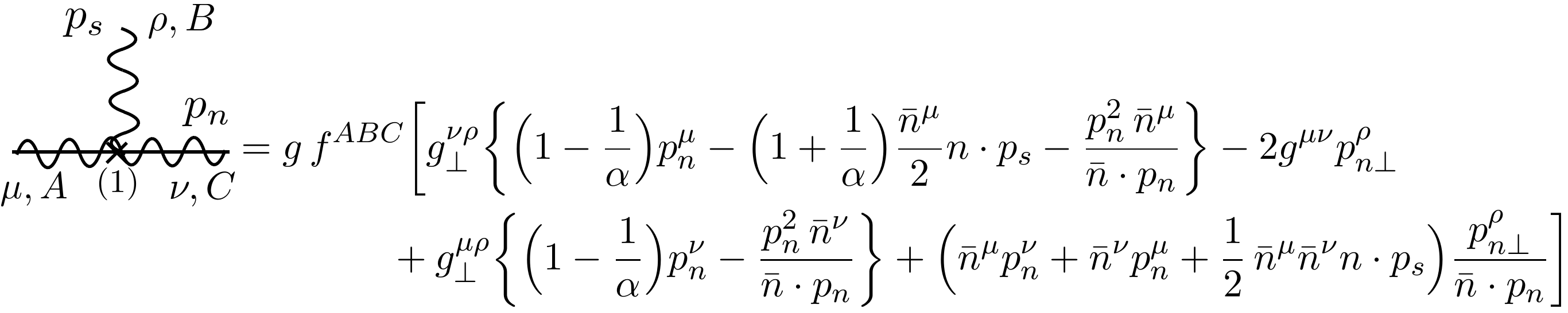}
	\end{center}
	\caption{Feynman rules for soft gluon emission from a collinear gluon at ${\cal O}(\lambda^1)$ in a general covariant collinear gauge.  Here, $\stackrel{(1)}{\times}$ denotes the subleading vertex from the Lagrangian ${\cal L}_{A_n}^{(1)}$ and the momenta $p_n$ and $p_s$ are outgoing. $A$, $B$ and $C$ denote the color indices of the gluons, and $f^{ABC}$ are the structure constants. Finally $\alpha$ is the covariant gauge parameter, and in Feynman-`t Hooft gauge $\alpha = 1$.
	}
	\label{fig:sub_glue}
\end{figure}

\begin{figure}
	\begin{center}
		\includegraphics[width=12.5cm]{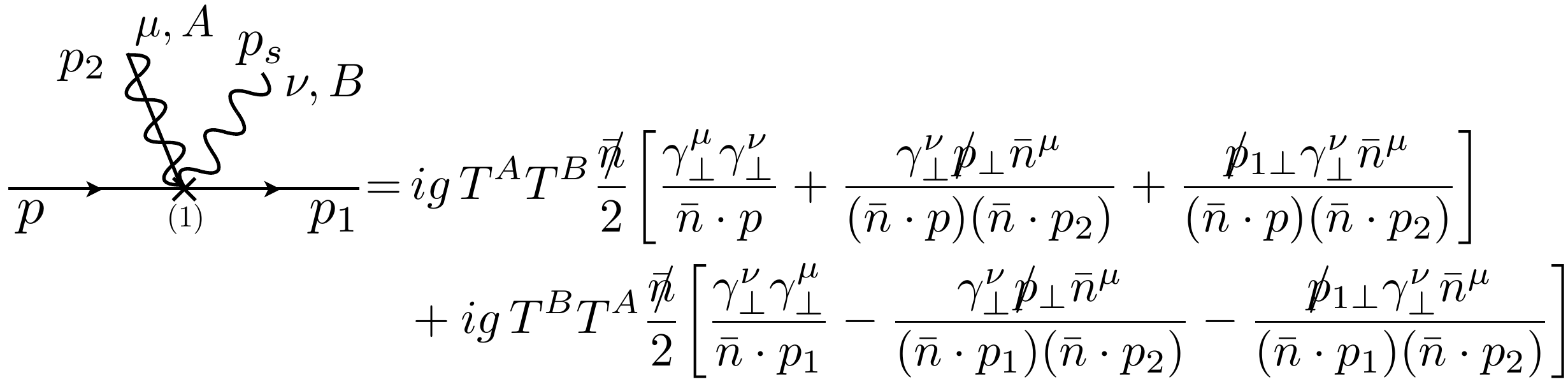}
	\end{center}
	\caption{
	Feynman rules for collinear and soft gluon emission from a collinear fermion line at ${\cal O}(\lambda^1)$.  Here, $\stackrel{(1)}{\times}$ denotes the subleading vertex from ${\cal L}_{\xi_n}^{(1)}$, the momenta $p_1$, $p_2$ and $p_s$ are all outgoing, and $p$ is incoming.  $A$ and $B$ are the color indices of the collinear gluon and soft gluon, respectively.
}
	\label{fig:sub_4pt}
\end{figure}

In this appendix, we present the expression for the other color ordering, $T^BT^A$, of the soft limit of the $1\to 3$ splitting amplitude studied in \Sec{subsec:splitsub}.  The color index of the collinear gluon is $A$ and the color index of the soft gluon is $B$. The result is computed using SCET, and the two most complicated Feynman rules required for this purpose are the subleading 3-gluon vertex displayed in \Fig{fig:sub_glue} and the four point fermion-gluon vertex shown in \Fig{fig:sub_4pt}. These Feynman rules are shown prior to making the BPS field redefinition. After making the BPS field redefinition the propagator insertion diagrams are removed, and the Feynman rules in both \Figs{fig:sub_glue}{fig:sub_4pt} should be multiplied by $\big( g^\perp_{\rho\alpha}  - \frac{n_\rho\, p^\perp_{s\alpha} }{n\cdot p_s} \big)$ where $\alpha$ is now the external index for the soft gluon.

For this $T^BT^A$ color ordering, the soft gluon can be emitted from either the collinear gluon or the collinear quark, and we again set the total $p_\perp$ of the collinear sector to zero.  There are then five diagrams to consider. As in \Sec{subsec:splitsub}, we are computing after using the BPS field redefinition, so soft gauge invariance is manifest in each diagram.  The result is:
\begin{align}
&
\raisebox{-0.45\height}{\includegraphics[height=1.5cm]{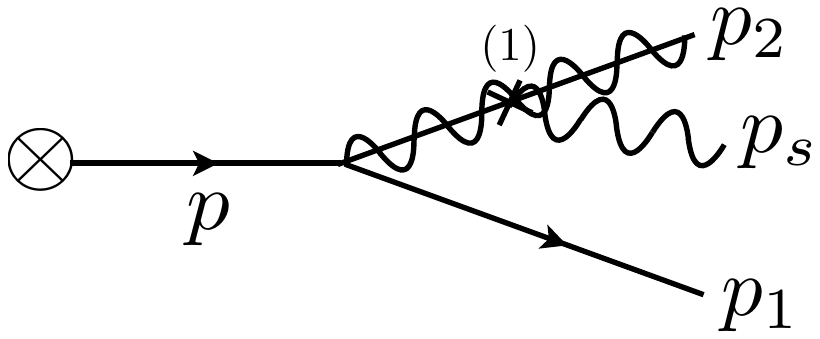}}
\ + \
\raisebox{-0.45\height}{\includegraphics[height=1.5cm]{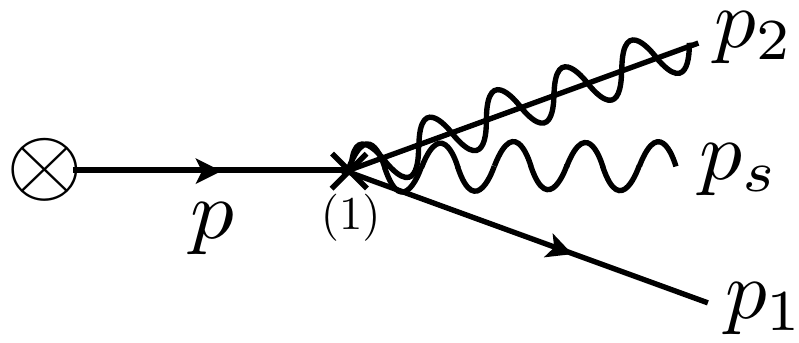}}
\ + \
\raisebox{-0.45\height}{\includegraphics[height=1.5cm]{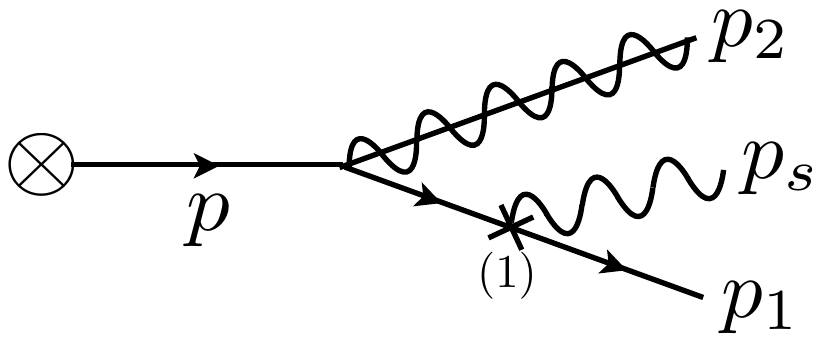}}
\\*
&
\qquad + \
\raisebox{-0.45\height}{\includegraphics[height=1.5cm]{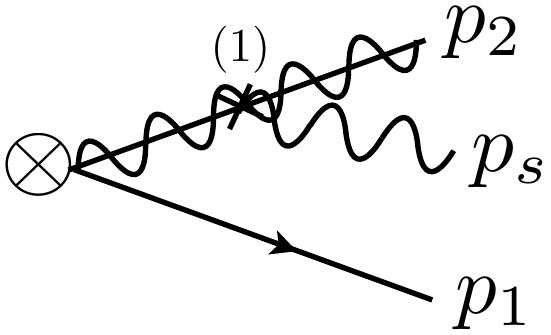}}
\ + \
\raisebox{-0.45\height}{\includegraphics[height=1.5cm]{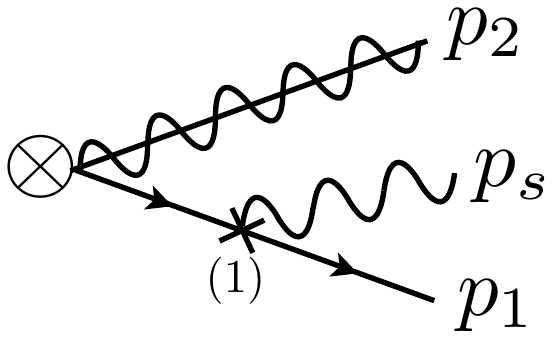}}
\nn \\*
& = 
g^2\, \bar{u}(p_1)\,  T^B T^A \bigg[ \bigg(
 \frac{n\cdot \epsilon_2}{n\cdot p}
 -\frac{\bn\cdot \epsilon_2}{\bn\cdot p_2} 
  +\frac{\slashed{p}_{1\perp}\slashed{\epsilon}_{2\perp}}
    {n\cdot p\: \bar{n}\cdot p_1}
  \bigg) \bigg( 2p_{1\perp}^{\rho} - 2 p_{2\perp}^\rho \frac{\bn\cdot p_1}{\bn\cdot p_2}\bigg)
\nn\\*
&\quad
+ \bigg( \slashed{p}_{1\perp}\!\gamma_{\perp}^{\rho} \frac{\bn\cdot \epsilon_2}{\bn\cdot p_2}
 - \gamma_{\perp}^{\rho} \slashed{\epsilon}_{2\perp} 
 - 2 \epsilon_{2\perp}^\rho \frac{\bn\cdot p_1}{\bn\cdot p_2} \bigg) \frac{n\cdot p_s}{ n\cdot p}
\bigg]
 \frac{\epsilon_s^{\mu}\, p_s^\nu}{(\bar n\cdot p_1)(n\cdot p_s)}\left(
g^\perp_{\mu\rho}\frac{n_\nu}{n\cdot p_s} - g^\perp_{\nu\rho}\frac{n_\mu}{n\cdot p_s}
\right)
\, .\nn
\end{align}
In this expression, we have explicitly included coupling and color factors, where the color index of the collinear gluon is $A$ and the color index of the soft gluon is $B$.
Collinear gauge invariance can be verified explicitly by taking $\epsilon_2 \to p_2$, using momentum conservation and the fact that all external particles are on-shell.  This result agrees with the soft expansion of the $1\to 3$ splitting amplitude, that is, it agrees with the full $1\to 3$ splitting amplitude up to terms suppressed by higher powers of $\lambda$ and terms in our result that enforce soft gauge invariance.

\bibliography{SCETforAmps}

\end{document}